\documentclass[12pt,a4,twoside]{book}
\pdfoutput=1

\usepackage{framed}
\usepackage{amsmath,amsfonts,amssymb} 
\usepackage{bm} 
\usepackage{theorem}

\usepackage{sfmath}

\usepackage{makeidx} 
\makeindex 

\usepackage{abbrevs}

\usepackage{nomencl} 
\renewcommand{\nomgroup}[1]{%
\ifthenelse{\equal{#1}{S}}{\vspace*{\baselineskip}\item[\textbf{Symbols}]\vspace*{\baselineskip}}{%
\ifthenelse{\equal{#1}{A}}{\vspace*{\baselineskip}\item[\textbf{Abbrivations}]\vspace*{\baselineskip}}{}}} 

\usepackage{ifpdf}
\ifpdf
\usepackage[pdftex]{graphicx}
\else
\usepackage{graphicx}
\fi

\usepackage[hyperindex]{hyperref} 

\usepackage{multicol}        		

\usepackage{sectsty}

\usepackage{units}		
\usepackage{mathrsfs}	

\usepackage{fancyhdr}
\usepackage{titlesec}
\usepackage{titletoc}

\usepackage{appendix}
\usepackage{helvet}

\sectionfont{\sffamily \mdseries}
\subsectionfont{\sffamily \mdseries}
\subsubsectionfont{\sffamily }

\chapternumberfont{ \sffamily \mdseries}

\makeatletter 
\def\@seccntformat#1{\protect\makebox[0pt][r]{\sf  \csname
 the#1\endcsname $\,|$ }} 
\makeatother

\titleformat{\chapter}[display] 
{\bfseries \Large} 
{\filleft\MakeUppercase{\chaptertitlename}  \fontsize{60pt}{4ex} \selectfont \thechapter} 
{4ex} 
{\titlerule 
\vspace{2ex}%
\filright} 
[\vspace{2ex}%
\titlerule
]

\renewcommand{\chaptermark}[1]{\markboth{Chapter \thechapter.\ {#1}}{}}
\renewcommand{\sectionmark}[1]{\markright{\thesection\ \boldmath {#1}\unboldmath}}
\newcommand{\rf}[1]{(\ref{#1})}
\newcommand{\beq}{\begin{equation}}
\newcommand{\eeq}{\end{equation}}
\newcommand{\bea}{\begin{eqnarray}}
\newcommand{\eea}{\end{eqnarray}}

\newcommand{\e}{\mbox{e}}
\renewcommand{\d}{\mbox{d}}
\newcommand{\g}{\gamma}

\renewcommand{\l}{\lambda}
\newcommand{\La}{\Lambda}
\renewcommand{\b}{\beta}
\renewcommand{\a}{\alpha}

\renewcommand{\th}{\theta}
\newcommand{\ep}{\varepsilon}

\newcommand{\del}{\delta}
\newcommand{\Del}{\Delta}

\renewcommand{\k}{\kappa}
\newcommand{\vph}{\varphi}

\newcommand{\oh}{\frac{1}{2}}

\newcommand{\dg}{\dagger}

\newcommand{\tr}{\mathrm{tr}\,}
\newcommand{\ra}{\rangle}
\newcommand{\la}{\langle}
\newcommand{\prt}{\partial}
\newcommand{\mi}{\!-\!}
\newcommand{\equ}{\!=\!}
\newcommand{\pl}{\!+\!}

\newcommand{\cD}{{\cal D}}

\newcommand{\cT}{{\cal T}}

\newcommand{\cN}{{\cal N}}
\newcommand{\cL}{{\cal L}}
\newcommand{\cO}{{\cal O}}

\newcommand{\cZ}{{\cal Z}}

\newcommand{\tG}{{\tilde{G}}}
\newcommand{\tg}{{\tilde{g}}}

\newcommand{\tL}{{\tilde{\La}}}
\newcommand{\tLe}{{\tilde{L}}}
\newcommand{\tX}{{\tilde{X}}}
\newcommand{\tY}{{\tilde{Y}}}

\newcommand{\tW}{{\tilde{W}}}
\newcommand{\trho}{{\tilde{\rho}}}

\newcommand{\tH}{{\tilde{H}}}
\newcommand{\tPsi}{{\tilde{\Psi}}}
\newcommand{\tT}{{\tilde{T}}}
\newcommand{\tsL}{{\sqrt{\tilde{\La}}}}

\newcommand{\hH}{{\hat{H}}}

\newcommand{\bX}{{\bar{X}}}

\newcommand{\no}{\nonumber}
\newcommand{\nn}{\no\\}
\newcommand{\non}{\nonumber \\}

\newcommand{\SL}{\sqrt{\La}}

\newcommand{\sL}{\sqrt{\Lambda}}

\newcommand{\hWg}{{{\hat{W}}_{\La,g_s}}}

\newcommand{\vac}{|0\ra}
\newcommand{\cav}{\la 0 |}

\newcommand{\Vol}{\mathrm{Vol}}

\newcommand{\M}{\mathcal{M}}
\newcommand{\Mink}{\mathbb{M}}
\newcommand{\C}{\mathcal{C}}
\newcommand{\Bcal}{\mathcal{B}}
\newcommand{\Scal}{\Sigma}

\newcommand{\card}{\mathbf{card}}
\newcommand{\past}{\mathbf{past}}
\newcommand{\fut}{\mathbf{future}}
\newcommand{\maxi}{\mathbf{max}}
\newcommand{\mini}{\mathbf{min}}

\newcommand{\gu}{\underline{g}}
\newcommand{\elem}{\!\in\!}

\newcommand{\expec}[1]{\left<#1\right>}

\newcommand{\ket}[1]{ \mid\! #1 \rangle}
\newcommand{\bra}[1]{ \langle #1 \!\mid}
\newcommand{\braket}[2]{ \langle #1 \mid #2 \rangle}
\newcommand{\scprod}[2]{ \langle #1 , #2 \rangle}

\newcommand{\Tri}{ \mathcal{T}}
\newcommand{\TriC}{ \mathcal{T}_c}

\newcommand{\ointz}{\oint \frac{dz}{2\pi i \, z}\;}

\newcommand{\dL}{\frac{\partial}{\partial L}}
\newcommand{\ddL}{\frac{\partial^2}{\partial L^2}}

\newcommand{\hcW}{{\hat{W}}}
\newcommand{\Zb}{{\bar{Z}}}
\newcommand{\Zbt}{{\bar{Z}(T,Z)}}
\newcommand{\bXt}{{\bar{X}(T,X)}}
\newcommand{\dT}{\frac{\partial}{\partial T}}

\newcommand{\dphi}{\frac{\partial}{\partial \varphi}}
\newcommand{\ddphi}{\frac{\partial^2}{\partial\varphi^2}}

\newcommand{\D}{ \mathcal{D}}
\newcommand{\Diff}{\mathrm{Diff}}
\newcommand{\Geom}{\mathrm{Geom}}
\newcommand{\atanh}{\mathrm{arctanh}}     

\usepackage{makeidx}
\makeindex

\begin{document}
%

\thispagestyle{empty}

\begin{center}
  {\sc{University of London}}\\[1.0cm]
  \centerline{\includegraphics[height=3.5cm]{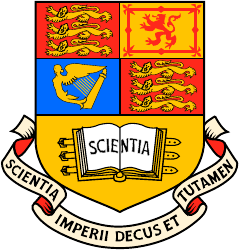}}

  \vspace{1.0cm}
  Imperial College of
  Science, Technology and Medicine\\ The
  Blackett Laboratory\\ Theoretical Physics Group\\[1cm]
  
  {\Huge A CAUSAL PERSPECTIVE ON}\\[0.5cm]
  {\Huge RANDOM GEOMETRY}\\[1.5cm]

  {\Large{Stefan Zohren}}\\[1cm]

  Thesis submitted in partial fulfilment of the \\
  requirements for the degree of\\
  Doctor of Philosophy\\
  of the University of London\\
  and the Diploma of Membership of Imperial College.\\[1cm]
  \vfill
October 2008
\end{center}

\newpage \thispagestyle{empty}
\noindent
\begin{tabular}{l l}
Supervisor: &  Dr.\ Helen Fay Dowker\\
&\\
Examiners: & Prof.\ Dr.\ Kellog Stelle (Imperial College)\\
& Prof.\ Dr.\ John Wheater (University of Oxford)\\
\end{tabular}

\newpage \thispagestyle{empty}

${}$

\vspace{4cm}
\noindent
 {\Huge A CAUSAL PERSPECTIVE ON}\\[0.5cm]
  {\Huge RANDOM GEOMETRY}\\[1.5cm]

\noindent
 {\Large{Stefan Zohren}}\\

\vspace{9cm}

\noindent
 \includegraphics[height=1.8cm]{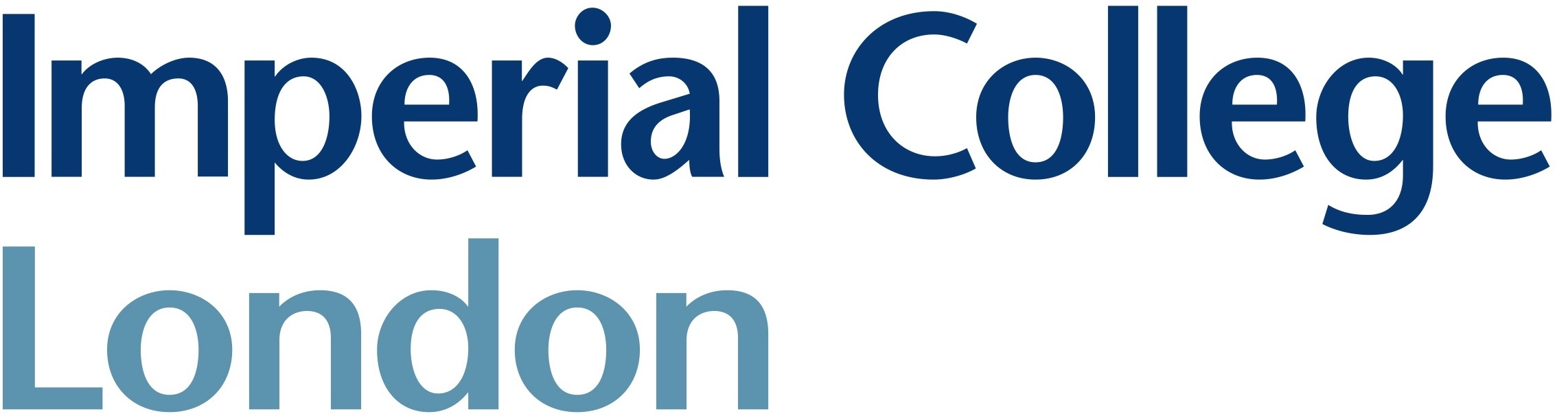}

\newpage \thispagestyle{empty}

\noindent
Stefan Zohren\\
Blackett Laboratory\\
Imperial College London\\
Prince Consort Road\\
SW7 2AZ London\\
United Kingdom

\vspace{11cm}

\noindent
\copyright \, Stefan Zohren, London, 2008\\\\
All rights reserved. No part of this publication may be reproduced in any form without prior written permission of the author.

\vspace{1cm}

\noindent
Printed in The Netherlands\\
PrintPartners Ipskamp B.V.\\\\
ISBN 978-90-9023894-4




\newpage \thispagestyle{empty}
\frontmatter
\addcontentsline{toc}{chapter}{Abstract}
\renewcommand{\chaptermark}[1]{\markboth{ ABSTRACT}{}}
\chapter*{Abstract}

One of the biggest challenges of present day theoretical physics is to find a consistent theory of quantum gravity. There are however obvious problems in constructing a quantum theory of general relativity. It has already been shown in the seventies by 't Hooft and Veltman that perturbative quantum gravity is non-renormalizable in four dimensions. However, this does not mean that it is impossible to formulate quantum gravity by use of field theoretic techniques, on the contrary, it suggests that special, so-called non-perturbative methods have to be applied. In this thesis we investigate the importance of causality in such non-perturbative approaches to quantum gravity. 

Firstly, causal sets are introduced as a simple kinematical model for causal geometry. It is shown how causal sets could account for the microscopic origin of the Bekenstein entropy bound. Holography and finite entropy emerge naturally from the interplay between causality and discreteness. 

Going beyond causal set kinematics is problematic however. It is a hard problem to find the right amplitude to attach to each causal set that one needs to define the non-perturbative quantum dynamics of gravity. One approach which is ideally suited to define the non-perturbative gravitational path integral is dynamical triangulation. Without causality this method leads to unappealing features of the quantum geometry though. It is shown how causality is instrumental in regulating this pathological behavior. In two dimensions this approach of causal dynamical triangulations has been analytically solved by transfer matrix methods. 

In this thesis considerable progress has been made in the development of more powerful techniques for this approach. The formulation through matrix models and a string field theory allow us to study interesting generalizations. Particularly, it has become possible to define the topological expansion. A surprising twist of the new matrix model is that it partially disentangles the large-N and continuum limit. This makes our causal model much closer in spirit to the original idea by 't Hooft than the conventional matrix models of non-critical string theory. 

The power of the here introduced techniques greatly extends the scope of analytical calculations for causal quantum gravity. This ranges from simple matter coupling to a new causal perspective on non-critical string theory.

\newpage 

\addcontentsline{toc}{chapter}{Acknowledgments}
\renewcommand{\chaptermark}[1]{\markboth{ ACKNOWLEDGEMENTS}{}}
\chapter*{Acknowledgments}

First of all I would like to thank my supervisor Fay Dowker for guiding me through this research, as well as for giving me the freedom to follow different themes during my research. Fay has been supporting me and my career in quantum gravity already since my time at Queen Mary in 2003.

Further, I would like to thank Jan Amb{\o}rn and Renate Loll at Utrecht University for supervising me during my regular visits to Utrecht. With respect to this I would like to thank the Theoretical Physics Institute at Utrecht University for the kind hospitality during my many visits there. 

Special thanks also to my main collaborator and good friend Willem Westra without whom my work would have been only half as enjoyable as it was now. Thanks also to my other collaborators: Richard Gill, Romuald Janik, David Rideout and Yoshiyuki Watabiki, as well as my friends and colleagues at Imperial College and Utrecht University. I am also grateful to Rafael Sorkin for stimulating discussions.

With respect to my visit in Japan I would like to thank everyone at the Particle Physics Group at Ochanomizu University as well as at the Theoretical High Energy Physics Group at Tokyo Institute of Technology for the nice time we spent together. I am particularly grateful to Akio Sugamoto and Yoshiyuki Watabiki. Many thanks also to Ishiro Oda for inviting me to the University of the Ryukyus in Okinawa.

I would also like to thank the Niels Bohr Institute at Copenhagen University, the Thiele Center For Applied Mathematics in Natural Science at \.Arhus University, the Perimeter Institute in Waterloo, the Department of Physics at Reykjavik University,  
the Service de Physique Th\'eorique at CEA Saclay, the 
Institute of Physics at Jagellonian University and the Mathematical Institute at Leiden University for hospitality during short visits.\\

Financial support through ENRAGE (European Network on
Random Geometry), a Marie Curie Research Training Network in the
European Community's Sixth Framework Programme, network contract
MRTN-CT-2004-005616 as well as through the JSPS (Japan Society for the Promotion of Science) is kindly acknowledged. Additional financial support to visit GRG18 and Perimeter Institute was provided through the Imperial College Trust and Perimeter Institute.\\

To my family and Maru. 
\newpage 

\renewcommand{\chaptermark}[1]{\markboth{ {#1}}{}}
\addcontentsline{toc}{chapter}{Contents}
\tableofcontents 
\newpage 
\addcontentsline{toc}{chapter}{List of Figures}
\listoffigures 
\newpage 

\mainmatter
\renewcommand{\chaptermark}[1]{\markboth{Chapter \thechapter.\ {#1}}{}}
\renewcommand{\sectionmark}[1]{\markright{\thesection\ \boldmath {#1}\unboldmath}}
\lhead[\fancyplain{}{\thepage}]         	{\fancyplain{}{\rightmark}}
\chead[\fancyplain{}{}]                 	{\fancyplain{}{}}
\rhead[\fancyplain{}{\leftmark}]       	{\fancyplain{}{\thepage}}

\chapter{Introduction \label{Chap:Introduction}}


Finding a consistent theory of quantum gravity is a notoriously hard problem. Such a theory should give a fundamental quantum description of space-time geometry with general relativity as a classical limit. However, more than ninety years have passed since Einstein's discovery of general relativity, and still very little is known about the ultimate structure of space-time at very small scales.

Quantum field theory one the other hand has proven to be a marvelously successful way to describe three of the four fundamental forces of nature: the electro-magnetic, the weak and the strong force. For the gravitational force however one does not have a well-defined predictive quantum field theoretic description yet, but one does have a very successful classical field theoretic description in the form of Einstein's theory of general relativity. Since the other forces can be described by quantum field theories, it seems natural that there also exists a quantum theory for the gravitational field.

Another reason which supports the existence of such a theory of quantum gravity is the fact that gravity couples to all forms of energy. Hence, one expects the energy fluctuations at small distances due to Heisenberg's uncertainty relations to induce also quantum fluctuations in the gravitational field. This leads to the prediction that space-time geometry has a highly non-trivial microstructure at extremely small scales close to the Planck length, $l_p\!=\!\sqrt{\hbar G_Nc^{-3}}\!\approx\!1.616\!\times\! 10^{-35}m$.

There are however obvious problems in constructing a quantum theory of general relativity (see \cite{Carlip:2001wq} for an overview). It has already been shown in the seventies by 't Hooft and Veltman that perturbative quantum gravity is non-renormalizable in four dimensions \cite{tHooft:1974bx}. There are several ways to confront this problem:

One of the most prominent approaches to quantum gravity is string theory (see standard text books \cite{Green:1987sp,Green:1987mn,Polchinski:1998rq,Polchinski:1998rr}). In this unifying theory the worldlines of propagating particles are replaced by worldsheets of propagating strings. The issue of non-renormalizability is thought to be resolved from the start in string theory, since the point-like interactions which cause the divergencies are replaced by extended interactions which cannot be localized. Whereas string theory has recently had some successes such as the calculation of strongly coupled gauge theories using the AdS/CFT correspondence (see for instance \cite{Alday:2008cg}), as a theory of quantum gravity it still relies on supersymmetry and  extra dimensions. 

Another approach is loop quantum gravity \cite{Nicolai:2005mc,Nicolai:2006id} in which one uses a Dirac (constraint) quantization \cite{Dirac1964} of Ashtekar's connection variables \cite{Ashtekar:1986yd}. Recent advances in this approach have been made by use of covariant spin-foam models \cite{Nicolai:2006id,Perez:2003vx}. These approaches suggest that the ultraviolet divergences can be resolved by the existence of a minimal length scale, commonly expressed in terms of the Planck length $l_p$. Finiteness of gravitational entropy can be seen as evidence for such fundamental discreteness of space-time.

A more radical approach is causal set theory \cite{Bombelli:1987aa,valdivia} which postulates fundamental discreteness from the outset. In this approach discrete sets of space-time events are viewed as fundamental objects underlying continuum space-times. Even though some interesting phenomenological models using causal sets have been proposed, there has been little progress in formulating a quantum path integral for causal sets. 

There are other approaches which define quantum gravity non-perturbatively. Renormalizability of the non-perturbative theory could be achieved by a nontrivial fixed point scenario as described by Weinberg \cite{Weinberg:1980gg}. One possible example of such a scenario is the exact renormalization group flow method for Euclidean quantum gravity in the continuum \cite{Lauscher:2005xz}. 

Another famous attempt to non-perturbatively define quantum gravity is dynamical triangulations (DT), a covariant path integral formulation, in which quantum gravity is obtained as a continuum limit of a superposition of simplicial space-time geometries (see for instance \cite{Ambjorn:1997di}). Two-dimensional DT has also been used as a worldsheet regularization of the bosonic string. A particular strength of this approach is that it expresses the theory as a statistical sum over entirely geometric objects. Since the path integral assigns a probability to each contributing geometry, we also call it a model of random geometry. To perform the statistical sum one usually starts off in the Euclidean sector including many ``acausal'' geometries. This leads to unappealing features of the quantum geometries in dimensions higher than two. It can be shown within the approach of causal dynamical triangulations (CDT) that causality is essential to regulate this pathological behaviour. In CDT one starts off with a sum over Lorentzian geometries and then analytically continues to the Euclidean sector to perform the summation \cite{Ambjorn:1998xu}. Numerical simulations indicate that CDT in contrast to DT leads to a well-behaved continuum limit in four dimensions \cite{Ambjorn:2006jf,Ambjorn:2008sw}. \\

In this thesis we focus on the implementation of causality in models of random geometry. 

In Part I of this thesis we concentrate on the approach of causal sets. In Chap.\ \ref{Chap:causets} causal sets are introduced as a simple and instructive model of causal random geometry. After defining causal sets we show how they can correspond to continuum space-times in the continuum approximation. Although we are not able to define a path integral for causal sets, we describe a simple phenomenological model where causal sets are randomly embedded in a fixed space-time. From this model we show in Chap.\ \ref{Chap:entropy} how causal sets could account for the microscopic origin of the Bekenstein or Susskind entropy bound. 

It seems that to define random geometry in terms of a non-perturbative path integral over geometries more structure is needed. Such structure is provided by causal dynamical triangulations (CDT) on which we focus in Part II, III and IV of this thesis.

Part II gives an introduction to CDT as a model of causal quantum gravity with emphasis on analytical results in two dimensions. In Chap.\ \ref{Chap:QG} we give a brief introduction into path integrals in general with emphasis on the gravitational path integral. In Chap.\ \ref{Chap:dt} we introduce dynamical triangulations (DT) as a non-perturbative definition of the path integral for two-dimensional Euclidean quantum gravity. At the end of this chapter we comment on the problems one encounters in higher-dimensional DT. In the next chapter, Chap.\ \ref{Chap:cdt}, CDT are introduced as a model of two-dimensional causal quantum gravity. A comparison to the Euclidean model (DT) is given as well as a brief outlook to results in higher dimensions. In Chap.\ \ref{Chap:Relating} several relations between DT and CDT in two dimensions are discussed. In particular, we describe how CDT and DT can be related by respectively introducing or ``integrating out'' baby universes, i.e.\ spatial topology changes. In Chap.\ \ref{Chap:emergence} we discuss CDT for the case of non-compact geometries. We see how a semi-classical background emerges from quantum fluctuations yielding a simple model of two-dimensional quantum cosmology. 

In Part III we introduce a generalized CDT model. This is formulated as a third quantization of two-dimensional CDT, i.e.\ a model in which spatial universes or strings can be annihilated and created from the vacuum. In Chap.\ \ref{Chap:cap} we show how by introducing a coupling constant to the process of off-splitting baby-universes one can non-perturbatively solve for the disc function with an arbitrary number of baby-universes. In Chap.\ \ref{Chap:sft} this result is embedded in the framework of a string field theory (SFT) model of CDT. As an application of the resulting Dyson-Schwinger equations (DSE) of the SFT we calculate amplitudes of higher genus. 

The last Part IV deals with the development of matrix models for CDT. After a brief introduction to matrix models in the context of DT in Chap.\ \ref{Chap:matrix}, we formulate a matrix model description for the contributing triangulations in CDT in Chap.\ \ref{Chap:loop}. In the continuum limit one recovers the results of the generalized CDT model introduced in Chap.\ \ref{Chap:cap} giving a combinatorial interpretation of this model. In Chap.\ \ref{Chap:contmatrix} we show that also the continuum CDT model is described by a matrix model. This is particularly interesting, since contrary to DT the size of the matrices $N$ does not have to scale with the cut-off in the so-called double scaling limit. Further, the expansion in $1/N^2$ reorganizes the power expansions in terms of topology, where terms with powers $N^{-2h+2}$ can be identified with the \emph{continuum} surfaces of genus $h$. This makes the new matrix model much closer to the original proposal by 't Hooft for QCD than the conventional matrix models of non-critical string theory. 

In Chap.\ \ref{Chap:Summary} we conclude this thesis by summarizing the results. Further, we comment on possible applications of the newly discovered matrix model to simple matter coupling in CDT. Interestingly, it might be possible to give a simple derivation of the Onsager exponents from CDT.

Finally App. \ref{Chap:app} and \ref{Chap:app2} provide supplementary information to the chapters on causal sets and causal dynamical triangulations respectively.
\newpage

Most of the novel results of this thesis have been published in research articles. In particular, the following chapters are based on the following articles:

\begin{itemize}

\item Chapter \ref{Chap:entropy} on D.~Rideout and S.~Zohren, {\em Class. Quantum Grav.} {\bf 23} (2006) \href{http://www.iop.org/EJ/abstract/0264-9381/23/22/008}{6195--6213};

\item Chapter \ref{Chap:emergence} on J.~Ambj{\o}rn, R.~Janik, W.~Westra, and S.~Zohren, {\em Phys. Lett. B} {\bf 641} (2006)
  \href{http://dx.doi.org/10.1016/j.physletb.2006.08.021}{94--98};
  
\item Chapter \ref{Chap:cap} on J.~Ambj{\o}rn, R.~Loll, W.~Westra, and S.~Zohren, {\em JHEP} {\bf 0712} (2007)  \href{http://dx.doi.org/10.1088/1126-6708/2007/12/017}{017};  

\item Chapter \ref{Chap:sft} on J.~Ambj{\o}rn, R.~Loll, W.~Westra, Y.~Watabiki and S.~Zohren, {\em JHEP} {\bf 0805} (2008)  \href{http://dx.doi.org/10.1088/1126-6708/2008/05/032}{032};

\item Chapter \ref{Chap:loop} on J.~Ambj{\o}rn, R.~Loll, W.~Westra, Y.~Watabiki and S.~Zohren,  {\em Phys. Lett. B} {\bf 670} (2008)
  \href{http://dx.doi.org/10.1016/j.physletb.2008.11.003}{224-230}; \\ 
 {\em idem}, to appear in {\em Acta Phys. Polon. B} {\bf 39} (2008) 3355-3364.
 
\item Chapter \ref{Chap:contmatrix} on J.~Ambj{\o}rn, R.~Loll, W.~Westra, Y.~Watabiki and S.~Zohren, {\em Phys. Lett. B} {\bf 665} (2008)
  \href{http://dx.doi.org/10.1016/j.physletb.2008.06.026}{252--256}.
  
\end{itemize}

The remaining chapters are mainly introductory parts which review existing literature. These are included to make the thesis self-contained, but do not represent new work done by the author of this thesis.  


\part{Motivating causality: Causal sets}

\chapter{Causal sets: Discrete causal geometry\label{Chap:causets}}
	In this chapter we introduce causal sets as a simple model of causal random geometry. After giving a mathematical definition of causal sets we show how those can correspond to continuum space-times in the continuum approximation. Even though we are not able to define a path integral for causal sets we describe a simple phenomenological model where causal sets are randomly embedded in a fixed space-time \cite{Bombelli:1987aa}. In the next chapter we will use this model to give a microscopic understanding of the Susskind or Bekenstein entropy bound.

\section{Introducing causal sets}

Causal order suggests itself as a fundamental principle for quantum gravity because of the enormous amount of topological
and geometrical information which it contains \cite{valdivia}. 

There are several reasons to support the assumption of causal structure as
 primary in quantum gravity. 
It has been shown that, given merely the causal
relations of events in a space-time manifold,
one needs only the volume measure to
recover the full geometry in the continuum \cite{Hawking:1976fe,malment}.

Being faced with the non-renormalizability of perturbative quantum gravity in four dimensions 't Hooft already proposed in 1978 several radical ideas to possibly define a theory of quantum gravity \cite{Hooft:1978id}, of which one was to view quantum gravity as a theory of causal sets of space-time events (see also \cite{Myrheim:1978ce}). This idea was later picked up again by Bombelli et al.\ \cite{Bombelli:1987aa} and further developed by many others to what is sometimes referred to as causal set theory (see for instance \cite{Sorkin:1990bh,Sorkin:1990bj} for some reviews).

\index{Causal sets}
Mathematically speaking, a \textit{causal set} is defined to be a locally finite partially
ordered set $\C=(C,\prec)$, 
namely a set $C$ together with a relation $\prec$, 
called ``precedes'', which satisfy the following axioms:
\paragraph{Transitivity:} If $x \prec y$ and $y \prec z$ then $x \prec z$, 
$\forall x,y,z \in C$;
\paragraph{Irreflexivity:} $x \not\prec x$;
\paragraph{Local Finiteness:} For any pair of fixed elements $x$ and $z$ of $C$, the set of elements lying between $x$ and $z$ is finite, $\card\{y | x \prec y \prec z \}<\infty$, where $\card X$ means the cardinality of the set $X$. \\

\begin{figure}[t]
\begin{center}
\includegraphics[width=3.5in]{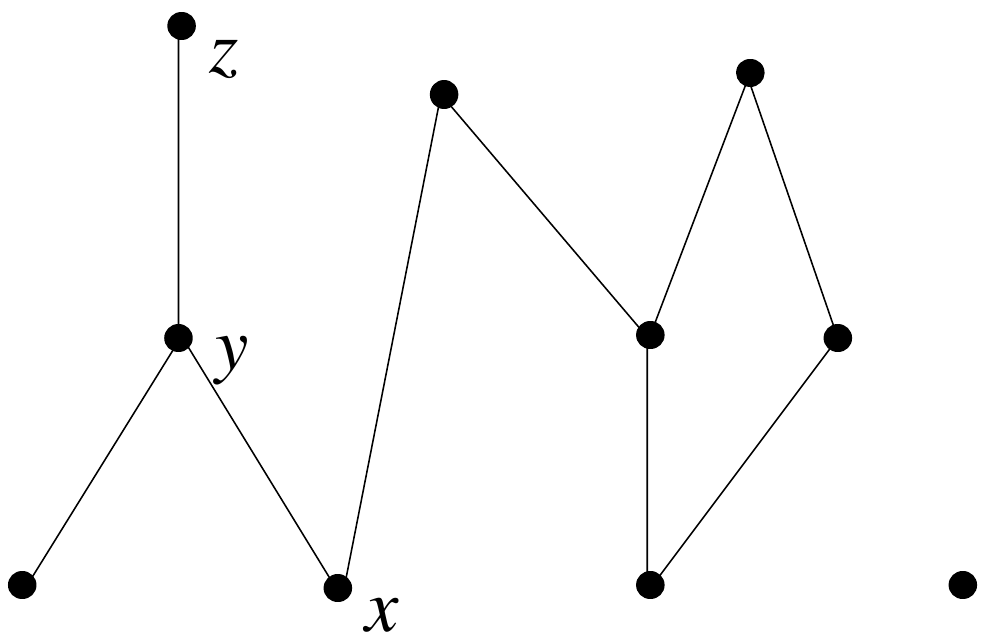}
\caption{A so-called Hasse diagram of a partially order set. In this example $x \prec y$ and $y \prec z$ and hence $x \prec z$ through transitivity.\index{Hasse diagram}}\label{fig:hasse}
\end{center}
\end{figure}

Fig.\ \ref{fig:hasse} shows a graphical representation of a causal set as an example.
Of these axioms, the first two say that $\C=(C,\prec)$ is a partially ordered
set or poset. The last expression, local finiteness, is related to the ``radical idea'' of this approach, namely that space-time is fundamentally discrete. The motivation for introducing fundamental discreteness is that it provides us with a way to account for the missing volume information. However, from a phenomenological point of view there are also other reasons, for example the finiteness of gravitational entropy which suggests that space-time might be fundamentally discrete.

In the following we summarize some basic definitions for intrinsic causal set quantities
of which most are ``discrete'' versions of analogous concepts used to
describe the causal structure of continuum space-times
(cf.\ App.\ \ref{app:causal}).

The \emph{past} of an element $x\in{C}$ is the subset
$\past(x)=\{y\in{C}\,|\,y\prec{x}\}$.\index{Causal sets!past} This corresponds to the causal past $J^-(x)$ in the continuum approximation. The past of a subset of $C$ is the union of the pasts of its elements.\index{Causal sets!future}  The \emph{future} of an element $x\in{C}$ is the subset $\fut(x)=\{y\in{C}\,|\,x\prec y\}$ which respectively corresponds to the causal future $J^+(x)$ in the continuum approximation.


An important concept for the formulation of the entropy bound from
causal set theory is that of a \emph{maximal element} in a causal set
$\C=(C,\prec)$.\index{Causal sets!maximal element} This is an element which has no successors, i.e. an element
$x$ for which $\nexists \, y \in C$ such that $x \prec y$. The set of all
maximal elements in $\C$ is denoted by $\maxi(\C)=\{x\in{C}\,|\nexists \, y
\in C \,\text{s.t.}\, x \prec y\}$.  In analogy, a \emph{minimal element} is
one which has no ancestors, i.e. an element $x$ for which $\nexists y \in C$
such that $y \prec x$ and the set of all minimal elements in $\C$ is denoted
by $\mini(\C)=\{x\in{C}\,|\nexists \, y \in C \,\text{s.t.}\, y \prec x\}$. \index{Causal sets!minimal element}

\section{Towards a continuum approximation}

Having given the precise definition of a causal set one might ask the
questions:  How could one actually be able to formulate a theory of quantum gravity using
causal sets, and how do causal sets relate to the known classical notion of
smooth Lorentzian manifolds? In
the following we give a definition of what it means for a causal set $\C$ to
be approximated by a Lorentzian space-time $(\M,g)$.

\index{Conformal embedding}
Consider a strongly causal space-time $(\M,g)$. The map $\phi : C \to \M$ from a causal set
$\C=(C,\prec)$ into a space-time $(\M,g)$ is called a \emph{conformal embedding} 
if $x \prec y \iff \phi(x) \in J^-(\phi(y)),
\;\;\forall x, y \in C$. Consider the Alexandrov neighborhood $J^+(p)\cap J^-(q)$, for every $p,q \in \M$, which forms a basis for the manifold topology of $(\M,g)$ if $(\M,g)$ is strongly causal, which we assume throughout.\index{Faithful embedding} \emph{sprinkling} The
map $\phi$ is called a \emph{faithful embedding} or \emph{sprinkling} if it has the following property: The number of elements $n$ mapped into an
Alexandrov neighborhood is equal to its space-time volume $V$ times the space-time density, up to
Poisson fluctuations. Thus, the probability of finding $n$ elements in
this region is given by the Poisson distribution \index{Poisson distribution}
\begin{equation}\label{eq:poisson}
P(n) = \frac{(\rho V)^n e^{-\rho V}}{n!} \,,
\end{equation}
where $\rho=l_f^{-d}$ is the density set by the fundamental length
scale $l_f$ in $d$ dimensions. In other words, this means that in a space-time region of volume $V$ one finds on average $N\!=\!V\rho$ elements of the causal set embedded into this region, and the fluctuations are typically of the order $\sqrt{N}$. Further, the properties of the Poisson distribution lead to local Lorentz invariance in the continuum space-time \cite{Dowker:2003hb,Bombelli:2006nm}.

Using the above definitions, we say that a space-time $(\M,g)$ \emph{approximates} a causal set $\C$ if
there exists a faithful embedding of $\C$ into $(\M,g)$.
This notion gives a correspondence between causal sets and continuum
space-times. In terms of this correspondence continuum causal structure arises from the microscopic order relations of the causal set elements and the continuum volume measure of a region arises from counting the number of elements comprising this region. Nevertheless, one still has to prove the uniqueness of this
correspondence (up to small fluctuations). In general the precise formulation
of such a uniqueness proof is a difficult mathematical problem. However, it has been proven to hold for the limiting case
$\rho\mapsto\infty$ \cite{Bombelli:1989mu} and certain progress has been made
in the generalization to large but finite $\rho$ \cite{Noldus:2003si}.

\section{Discussion and outlook}

The ultimate aim of causal set theory is to formulate a theory of quantum gravity using
a path integral approach, where the single histories are causal sets. Continuum physics would then be supposed to emerge in a suitable continuum approximation.
However, since there is basically no information on spatial distances of space-time events, there is no obvious way to find a discretized version of the gravitational action in this context. And even if such an action would be found it is not clear how the sum over different causal sets in the path integral could be performed. 

\index{Sequential growth models}
A different path in the search for causal set dynamics has been taken by analyzing certain sequential growth models of the causal sets \cite{Rideout:1999ub,Martin:2000js,Rideout:2000fh,Varadarajan:2005gg}. These growth models aimed to determine an effective \emph{classical} dynamics of causal sets. One interesting result is that the ``early universe'' phase of such growth models closely resembles de Sitter space, at least in the way that
volumes of causal intervals scale with their length \cite{Davidprivate}.

From the above discussion we can say that it is still very ambitious to view causal sets as a model for quantum gravity, but nevertheless, it still gives a very simple and instructive model of causal random geometry. In particular, we can take a fixed Lorentzian space time and view it as a continuum approximation of a causal set. Using the Poisson distribution, as mentioned in the previous section, we are then able to randomly draw causal sets which can be faithfully embedded in this space-time. From a phenomenological point of view we can now study certain properties of such an ensemble, as will be done in the next chapter with view on entropy bounds. Even though some of these phenomenological models give interesting predictions \cite{Ahmed:2002mj,Dowker:2003hb,Dou:2003af,Rideout,Johnston:2008za}, one should still treat them with a bit of care for the following reason: Already in the case of a quantum mechanical particle we know that even though the classical path is smooth and nice, single histories of the path integral look nothing like this and are not even described by differentiable functions. Hence, it is not a priori clear, that classical space-times should approximate a \emph{single} causal sets history. Keeping this in mind we use causal sets as a phenomenological model of fundamentally discrete space-time.

\chapter{Entropy bounds from causal sets \label{Chap:entropy}}
	In this chapter we propose a measure for the maximal entropy of spherically symmetric spacelike space-time regions in terms of the ensemble of causal sets approximating this continuum space-time \cite{Rideout,Rideout:2006ww}. A bound for the entropy contained in this region is obtained from a counting of potential ``degrees of
freedom'' associated to the Cauchy horizon of its future domain of
dependence. For different spherically symmetric spacelike regions in Minkowski
space-time of arbitrary dimension, we show that this proposal leads, in the
continuum approximation, to Susskind's or Bekenstein's well-known spherical entropy bound up to a numerical factor.

\section{Entropy bounds in gravity}

\index{Bekenstein-Hawking formula}\index{Black hole entropy}
In the history of general relativity there has been a long discussion regarding the thermodynamics of gravitational systems. One of the most famous examples is the \textit{Bekenstein-Hawking formula} for black hole entropy \cite{Bekenstein:1972tm,Bekenstein:1973ur,Bekenstein:1974ax,Hawking:1974rv}, stating that the entropy of a black hole is given by a quarter of the area of the event horizon in Planckian units\footnote{As a convention we will throughout the discussion regarding entropy bounds work in Planck units with
\begin{equation}
c=k_B=G_N=\hbar=1,\nn
\end{equation}
where $c,k_B,G_N,\hbar$ denote the speed of light, Boltzmann's constant, Newton's constant and Planck's constant respectively. 
In these units the Planck length is given by 
\begin{equation}
l_p=\left(\frac{\hbar G_N}{c^3}\right)^{\frac{d-2}{2}}=1,\quad\text{for}\quad d\geqslant 3.\nn
\end{equation}}
\begin{equation}\label{eq:BHbound}
S_{BH}=\frac{A}{4}.
\end{equation}
This bound is truly universal, meaning that it is independent of the characteristics of the matter system and can be derived without any knowledge of the actual microstates of the quantum statistical system.

\index{Susskind's spherical entropy bound}
Along this line there have been several generalizations of this entropy
bound. One is \textit{Susskind's spherical entropy bound}
\cite{Susskind:1994vu}, stating that the upper bound
for the entropy of the matter content of an arbitrary spherically symmetric
spacelike region (of finite volume) is,
\begin{equation}
S_{\mathrm{matter}}\leqslant\frac{A}{4},
\end{equation}
where $A$ is the area of the boundary of the region.\footnote{There was an earlier proposed bound by Bekenstein \cite{Bekenstein:1980jp} stating that the entropy of any weakly gravitating matter system obeys $S_{\mathrm{matter}}\leqslant 2\pi E R$, where $E$ is the energy of the matter system and $R$ the circumferential radius of the smallest sphere that contains it. If one further assumes that this bound is valid for strongly gravitating matter systems, then gravitational stability in four dimensions implies that $2E\leqslant R$ and hence $S_{\mathrm{matter}}\leqslant 2\pi E R \leqslant A/4$. One sees that in four dimensions the Bekenstein bound is stronger then the Susskind bound, however, in $d>4$ gravitational stability and the Bekenstein bound only imply that $S_{\mathrm{matter}}\leqslant(d-2) A/8$ \cite{Bousso:2000md}. Hence, the geometrical Susskind bound is arguably more fundamental. \index{Bekenstein's entropy bound}}
Even though this spacelike entropy bound cannot be generalized to arbitrary
non-spherically symmetric spacelike regions, there exists a generalization in
terms of light-sheets, namely the \textit{Bousso or covariant entropy bound}
\cite{Bousso:1999xy,Bousso:2002ju}. \index{Covariant entropy bound}\index{Bousso's entropy bound} More precisely, let $A(\Bcal)$ be the area
of any $(d-2)$-dimensional surface $\Bcal$, then the $(d-1)$ dimensional hypersurface
$L$ is called the light-sheet of $\Bcal$ if $L$ is generated by light rays which
begin at $\Bcal$, extend orthogonally away from $\Bcal$ and have everywhere
non-negative expansion. The entropy flux through the light-sheet is then
bounded by
\begin{equation}
S(L)\leqslant\frac{A(\Bcal)}{4}.
\end{equation}
This entropy bound is also widely regarded as evidence for the \textit{holographic principle} \cite{Hooft:1993gx,Susskind:1994vu,Bousso:2002ju}, stating that the maximum number of degrees of freedom carried by $L$ is given by $A(\Bcal)/4$. 

These entropy bounds 
suggest that an underlying theory
of quantum gravity should predict the bounds from a counting of
microstates (see for example \cite{Carlip:2008wv}). This verification of the thermodynamic laws is an important
consistency check for any approach to quantum gravity. Further, the finiteness
of the entropy might already give some indications about the actual
microstructure of space-time. There is a semi-classical argument that the
description of a quantum theory of gravity by a local quantum field theory in
the continuum, in the absence of a high frequency cut-off, leads to infinitely many degrees of freedom in a finite region,
and therefore to a divergence in the entropy of this region \cite{Dou:2003af,Sorkin:1997gi}.  The entropy
bounds therefore suggest that space-time might posses a fundamental discreteness
at scales of order of 
the Planck scale. Continuum physics would then have to
emerge from this fundamental theory when making a continuum approximation at
large scales. This suggests that to obtain a theory of quantum gravity one does not have to
quantize the metric fields of the continuum geometries, but should rather find a quantum theory of the discrete structure underlying those continuum geometries
\cite{Isham:1993ji}.

In the following we show how using causal sets as a phenomenological model of causal random geometry one can
derive a notion of maximum entropy from a counting of potential horizon ``degrees of
freedom''
of the fundamental theory reminiscent of former ideas in the context of black
hole entropy \cite{Sorkin:2005qx,Dou:2003af}. Using this measure we
formulate an entropy bound for spherically symmetric spacelike
regions within the causal set approach. We then show that in the continuum approximation, for different
spherically symmetric spacelike regions in Minkowski space-time of arbitrary
dimension, this leads to Susskind's
spherical entropy bound up to a numerical factor.

\section{An entropy bound from causal set theory}\label{sec:conjecture}

In the previous chapter we introduced causal sets as a phenomenological model of causal random geometry. As explained earlier space-time volume arises from a counting
of fundamental space-time elements. In the following we show how
entropy bounds could arise from a counting of potential horizon ``degrees of
freedom'' at the fundamental level, giving a microscopic origin for Susskind's
spherical entropy bound.

As already mentioned in the previous section, the spherical entropy bound states that the entropy of the matter content of a spherically symmetric spacelike region $\Scal$ (of finite volume) is bounded by a quarter of the area of the boundary of $\Scal$ in Planck units
\begin{equation}\label{eq:spericalbound}
S\leqslant \frac{A}{4},
\end{equation}
in full units $S\leqslant A k_B c^3/(4 G_N \hbar)$, where $A=\Vol (\Bcal(\Scal))$ is the area of the boundary of this region. 

Surprisingly, the maximum entropy in \eqref{eq:spericalbound} can be determined
without any knowledge of the microscopic properties of the thermodynamic
system. A theory of quantum gravity however should be able
to deduce \eqref{eq:spericalbound} purely from a counting of the fundamental
degrees of freedom at the microscopic level.
(One may wonder in what manner a
counting of degrees of freedom measures the entropy of a system.  In a
discrete context ``degrees of freedom'' are generally finite, and a state
counting can be expected to yield something proportional to the exponential of
the number of degrees of freedom.  Thus measuring the entropy as the logarithm of
the number of states can be seen to be equivalent to counting the number of
degrees of freedom of the system.)
In the following we want to give a notion of those fundamental degrees of
freedom in the context of causal set theory leading to the formulation of an
entropy bound within this approach.

\begin{figure}
\begin{center}
\includegraphics[width=3in]{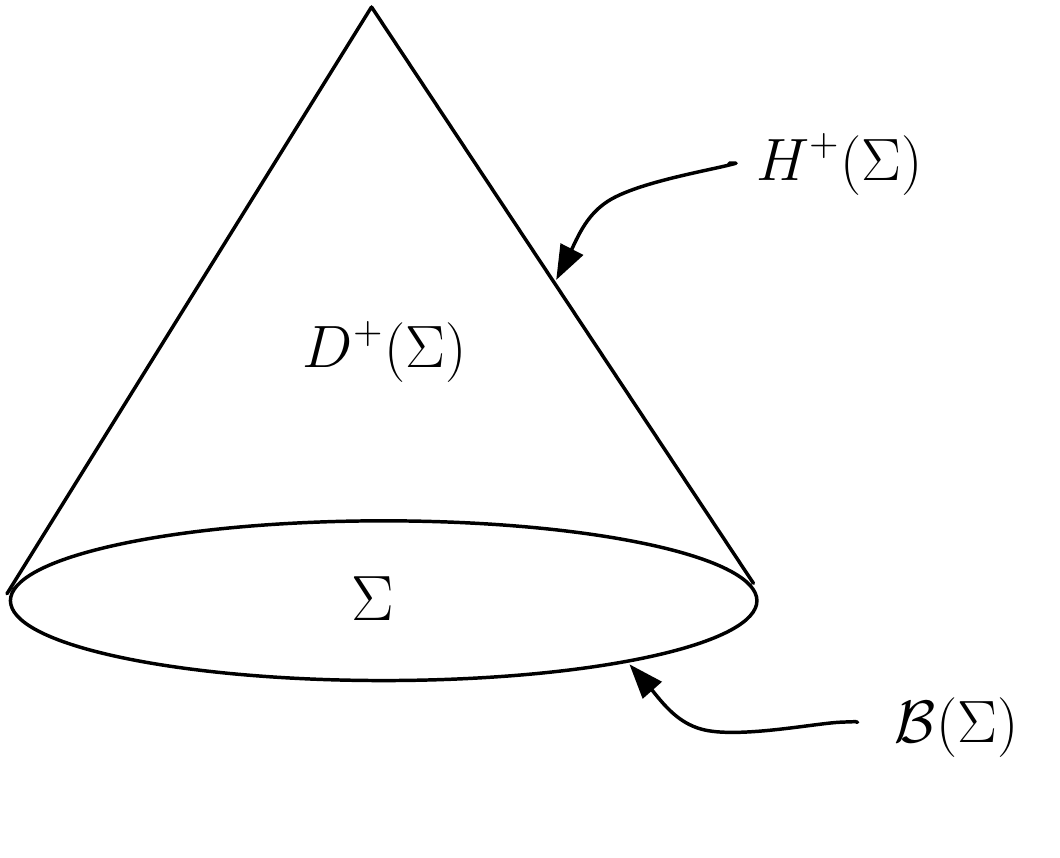}
\caption{Illustration of a spacelike hypersurface $\Scal$, its boundary $\Bcal(\Scal)$, its future domain of dependence $D^+(\Scal)$ and its future Cauchy horizon $H^+(\Scal)$.}\label{fig:domain}
\end{center}
\end{figure}

Consider a spherically symmetric spacelike region of space-time, potentially
containing some matter system.  We assume that the ``back reaction'' of the matter
content upon the space-time geometry can be neglected, so that different states
of the matter system lead to the same spherically symmetric space-time geometry.
The entropy of that system
must eventually ``flow out'' of the region by passing over the
boundary of its future domain of dependence.  But because space-time is
fundamentally discrete, the amount of such entropy flux is bounded above
by the number of discrete elements comprising this boundary.  This is a
fundamental limit imposed by discreteness on the amount of information flux
which can emerge from the region, and therefore on the amount of entropy which
it can contain. 
Thus we have argued that an entropy bound arises from causal sets.

\index{Causal sets!entropy bound}
\vspace{2mm}
\noindent\textbf{Proposal}\hspace{2mm}
\textit{Consider a spherically symmetric spacelike hypersurface $\Scal$ of
finite volume in a strongly causal space-time $\M$ of dimension $d\geqslant
3$. Denote the future domain of dependence of this hypersurface by
$D^+(\Scal)$ (c.f.\ App.\ \ref{app:causal} and Fig.\ \ref{fig:domain}). Let
$\C=(C,\prec)$ be a causal set which can be faithfully embedded into
$D^+(\Scal)$. Then the maximum entropy contained in $\Scal$ is given by the number of
maximal elements of $\C$,
\begin{equation}
S_{\max}=\card\{\maxi(\C)\}.
\label{eq:proposal}
\end{equation}
}
\vspace{2mm}
\noindent\textbf{Claim}\hspace{2mm}
\textit{This proposal leads to Susskind's entropy bound in the continuum approximation,
\begin{equation}
S_{\max}= \frac{A}{4},
\end{equation}
where $A=\Vol (\Bcal(\Scal))$ is the area of the boundary of this region $\Scal$, if the fundamental discreteness scale is fixed at a dimension dependent value to be calculated.
}

\section{Evidence for the claim}\label{sec:evidence}

In this section we provide analytical and numerical evidence for the claim that the entropy bound \eqref{eq:proposal} leads to Susskind's bound in the continuum.  In all discussed
examples we consider $(d-1)$-dimensional spherically symmetric spacelike
hypersurfaces $\Scal$ in $d$-dimensional Minkowski space $\Mink^d$, and
calculate the number of maximal elements in its domain of dependence
$D^+(\Scal)$. Since we assume that $D^+(\Scal)$ arises as a continuum
approximation of a causal set $\C$, we know from the phenomenological model described above 
that the elements of $\C$ are faithfully embedded into $D^+(\Scal)$
according to the Poisson distribution
\eqref{eq:poisson}. Hence the expected number of maximal elements in
$D^+(\Scal)$ is given by
\begin{equation}\label{eq:defentint}
\expec{n}=\rho \int_{D^+(\Scal)} dx^d \exp\left\{-\rho \,\Vol\left(J^+(x)\bigcap D^+(\Scal)\right)\right\},
\end{equation}
where again $\rho$ is the fundamental density of space-time. On the right-hand-side of \eqref{eq:defentint} one integrates over all points $x\in
D^+(\Scal)$, where every point is first weighted by the probability of finding
an element of $\C$ embedded at this point and further weighted by the
probability of not finding any other element of $\C$ embedded in
$J^+(x)\bigcap D^+(\Scal)$.
Note that because of the fundamentally random nature of the discrete-continuum
correspondence in causal sets we calculate the expected number of maximal
elements in a sprinkling, even though the proposal (\ref{eq:proposal}) is phrased in terms of a
fixed causal set.

\subsection{The ball in Minkowski space-time}

In this section we want to calculate the expected number of maximal elements
$\expec{n}$ in $D^+(\Scal)$, where $\Scal$ is chosen to be a
$(d-1)$-dimensional ball $S_{d-1}(R)$ with radius $R$ in Minkowski space-time
$\Mink^d$. Due to the spherical symmetry of the problem it is useful to
introduce spherical coordinates $x=(t,r,\theta_1,...,\theta_{d-2})$, where we
choose the origin to be the futuremost event of $D^+(\Scal)$. The volume element
$\Vol\left(J^+(x)\bigcap D^+(\Scal)\right)$ is equal to the volume of the
Alexandrov neighborhood of $x$ and $0$, $\Vol_d(\tau)\!\equiv\!\Vol(J^+(x)\bigcap
J^-(0))$, where we denote proper time by $\tau=\sqrt{t^2-r^2}$. Using the
result for $\Vol_d(\tau)$ as calculated in App.\ \ref{app:volume} and
integrating out the spherical symmetry in \eqref{eq:defentint} one obtains a
general expression for the expected number of maximal elements in
$D^+(S_{d-1}(R))$,
\begin{equation}\label{eq:minkd}
\expec{n}=\rho\, \frac{(d-1)\pi^{\frac{d-1}{2}}}{\Gamma(\frac{d+1}{2})} \int_0^R dt \int_0^t dr\, r^{d-2}e^{-\rho D_d(t^2-r^2)^\frac{d}{2}},
\end{equation}
where the dimension dependent constant $D_d$ is defined in App.\ \ref{app:volume}.

In the following we evaluate \eqref{eq:minkd} for various dimensions by analytical and numerical methods.

\subsubsection{2+1 dimensions}\label{sec:ball2+1}

In $d=3$ dimensions one can explicitly evaluate \eqref{eq:minkd}. It is useful
to express the result in terms of the expected total number of space-time elements
faithfully embedded into $D^+(S_{2}(R))$, $N=\rho V$, where $V=
\frac{\pi}{3}R^3$ is the volume of the domain of dependence
$D^+(S_{2}(R))$. 
The expected number of maximal elements is
\begin{equation}\label{eq:nmax2+1}
\expec{n}=8\left(e^{-\frac{N}{4}}-1\right)-2 N E_{\frac{1}{3}}\left(\frac{N}{4}\right)+4\Gamma\left(\frac{2}{3}\right)\sqrt[3]{2 N},
\end{equation}
where $E_n(x)$ is the exponential integral defined by
\begin{equation}
E_n(x)=\int_1^\infty t^{-n} e^{-xt}dt.
\end{equation}
\index{Exponential integral}

We also measured the expected number of maximal elements numerically by
``sprinkling'' a causal set into a 2+1-dimensional $2 \times 2 \times 1$ square box, which
contains $D^+(S_{2}(1))$.  By sprinkling we mean simply selecting $N_c$
elements at random with uniform distribution within the box, and computing the
causal relation between each pair from the Minkowski metric.  For each of 100 trials $i=1\ldots 100$, we deduce
the set of elements which fall within $D^+(S_{2}(1))$, compute its cardinality
$N_i$, and count the number of such elements $n_i$ which are maximal within that
region.  From these we compute the sample mean and its error, and 
repeat this
computation for a range of values of $N_c$.  These computations were greatly
facilitated by utilizing causal set and Monte-Carlo toolkits
within the Cactus computational framework \cite{cactus}.
\index{Cactus}

\begin{figure}[t]
\begin{center}
\includegraphics[width=5in]{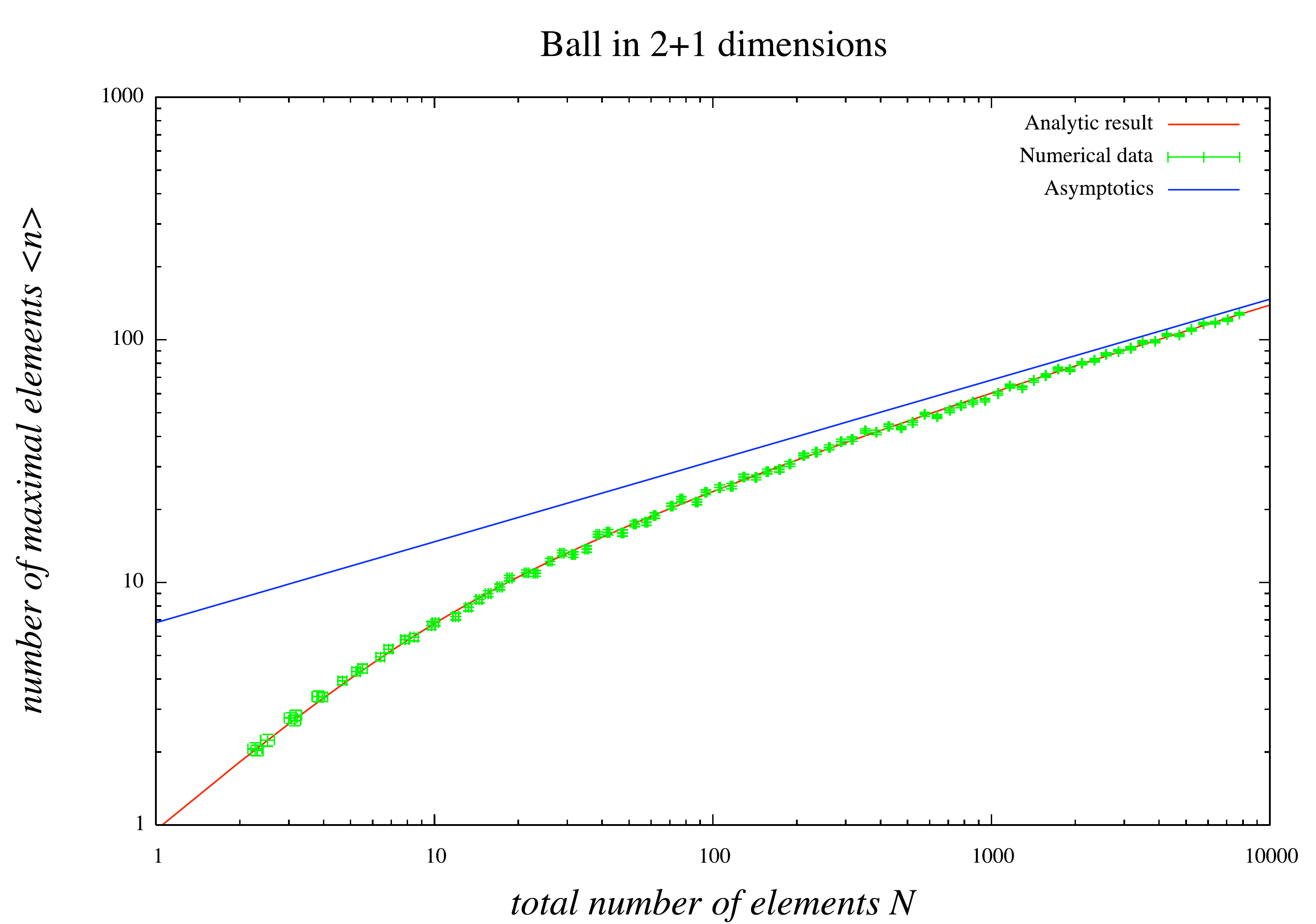}
\caption{Shown is the expected number of maximal elements $\expec{n}$ as a
  function of the total 
  number of elements in the domain of dependence of the
  unit disk in 2+1 dimensional Minkowski space-time, on a logarithmic
  scale. Besides the analytical result and its asymptotics, data points
  with error bars from Monte-Carlo simulations are also shown.}\label{fig:disc2+1}
\end{center}
\end{figure}

The plot of the expected number of maximal elements $\expec{n}$ as a function
of $N$ for the unit disk is shown in Fig.\ \ref{fig:disc2+1},
on a logarithmic scale. 
The
agreement of analytical and numerical results justifies the 
numerical methods. 

For a large number of elements $N\!\rightarrow\!\infty$ we can use the asymptotic expansion of the exponential integrals \index{Exponential integral!asymptotic expansion}
\begin{equation}\label{eq:Enasymp}
E_n(x)\propto \frac{e^{-x}}{x}\left(1-\frac{n}{x}+...\right)\quad\text{for}\quad |x|\rightarrow\infty
\end{equation}
yielding
\begin{equation}
\expec{n}=4\,\Gamma\left(\frac{2}{3}\right)\sqrt[3]{2 N}+\ldots\quad\text{for}\quad N\rightarrow \infty,
\end{equation}
where $+\ldots$ are lower order terms in $N$.
The asymptotics are also displayed in Fig.\ \ref{fig:disc2+1} together with the full expression for the expected number of maximal elements.
In terms of the density $\rho$ and the radius $R$ of $S_2(R)$ this result reads
\begin{equation}
\expec{n}=\frac{16\sqrt[3]{2\pi}}{3^{5/6}\Gamma\left(\frac{1}{3}\right)}\, \rho^{\frac{1}{3}} \frac{2\pi R}{4},
\end{equation}
up to lower order corrections.
It is important
to see that $\expec{n}\!\propto\!A/4$, where $A=2\pi R$ is the length of the
boundary of $S_{2}(R)$. This is highly non-trivial, as one can see by looking
at a
snapshot of a numerical simulation (Fig.\ \ref{fig:disc2+1}). There one observes that
the maximal elements do not align along the one-dimensional boundary
$\Bcal(S_2(R))$.  Instead they are distributed along a hyperbola close to the
two-dimensional Cauchy horizon $H^+(S_2(R))$, with a density of maximal
elements which \emph{decreases} with distance from the center. Hence the fact that the
expected number of maximal elements is indeed proportional to the length $A$
of the one-dimensional boundary $\Bcal(S_2(R))$ for large $A$ already gives very
non-trivial evidence for the proposed entropy bound.

\begin{figure}[t]
\begin{center}
\includegraphics[width=4.5in]{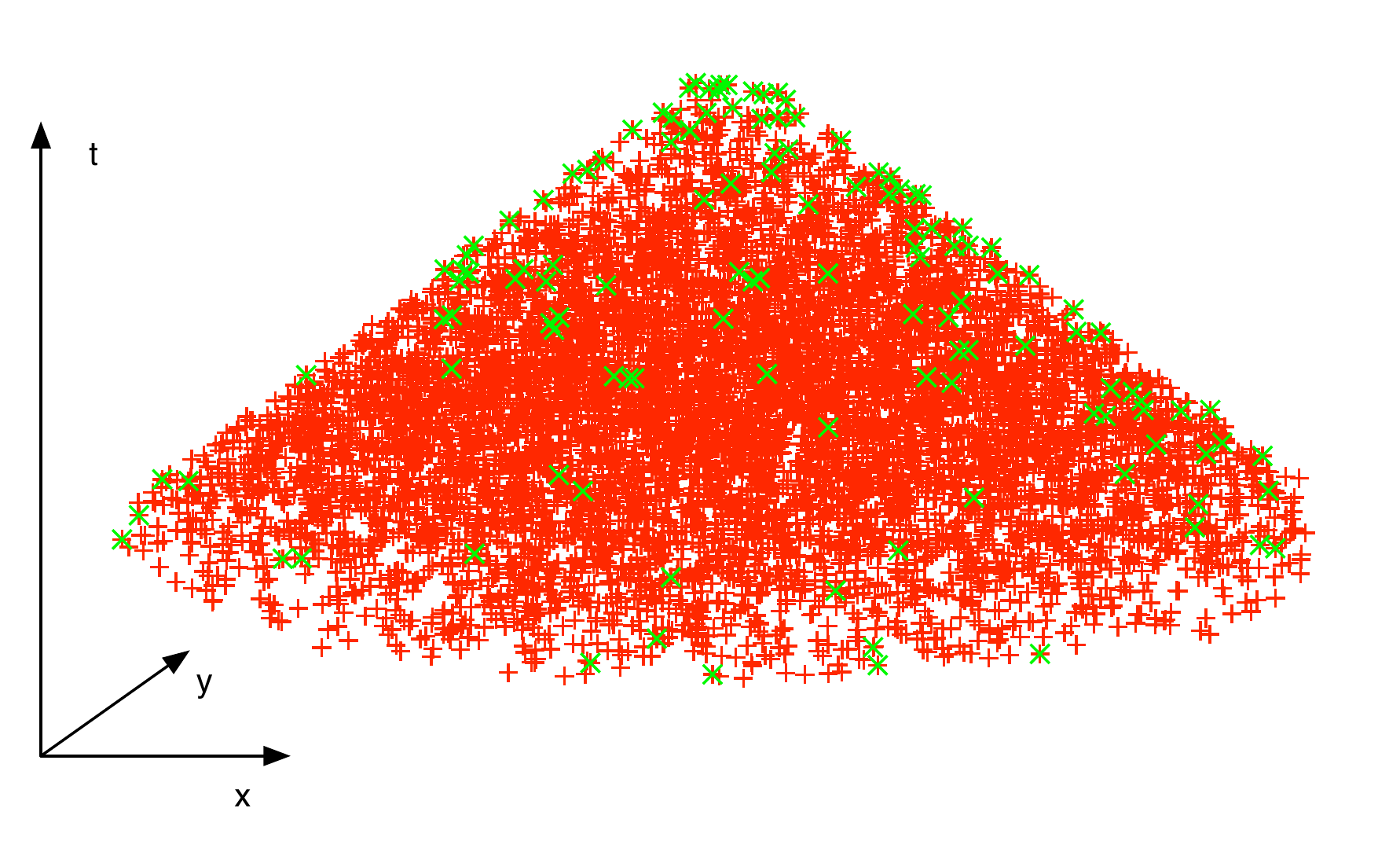}
\caption{A snapshot from a simulation showing $N=7806$ space-time elements
forming the domain of dependence of the two-dimensional ball in 2+1
dimensional Minkowski space-time (illustrated in red) and $n=126$
maximal elements therein (illustrated in green).}\label{}
\end{center}
\end{figure}

To have the precise confirmation of the $\mathcal{O}(1)$ constant in
$\expec{n}\!\propto\!A/4$ one has to choose the fundamental length scale
$l_f=\rho^{-1/3}$ to be $l_f=16\sqrt[3]{2\pi}/(3^{5/6}\Gamma(1/3))\approx4.41$. This gives support to the belief 
that the fundamental discreteness scale
of space-time is of the order of the Planck length. 

Hence, using this value for the fundamental discreteness scale we have confirmed our claim, namely in the continuum approximation we have
\begin{eqnarray}\label{eq:res2+1}
S_{\max}=\expec{n}=\frac{2\pi R}{4}
\end{eqnarray}
or $S_{\max}=k_B 2\pi R/(4\sqrt{\hbar G_N/c^3})$ in full units, when we set
$l_f=16\sqrt[3]{2\pi}/(3^{5/6}\Gamma(1/3))$.  
One now sees
that the assumption of large but finite $N$ used to obtain the asymptotic
behavior is equivalent to saying that $A$ is much larger than the Planck
length. However, one can see that even for relatively small values of the
length $A$, such as $10^3$ in Planck units, the approximation of $\expec{n}$
by its asymptotic expansion is already very accurate. For even smaller values
of the length scale the Planck corrections become significant. However, as
one can see in Fig.\ \ref{fig:disc2+1}, the corrections always decrease the
expected number of maximal elements, such that $\expec{n}$ never exceeds the
bound \eqref{eq:res2+1}.

\subsubsection{3+1 dimensions}

Clearly from a physical point of view the evaluation of \eqref{eq:minkd} in 3+1 dimensions is the most interesting case. For $d=4$ one can write \eqref{eq:minkd} as follows,
\begin{equation}\label{eq:3+1firststep}
\expec{n}=2\pi\rho\int_0^R dt\int_0^{t^2}dz \sqrt{t^2-z} e^{-\rho D_4 z^2}
\end{equation}
One can perform the first integration in \eqref{eq:3+1firststep} by using the following integral relation \cite{gradshteyn}
\begin{eqnarray}
&&\int_0^u x^{\nu-1}(u-x)^{\mu-1} e^{\beta x^n} dx = B(\mu,\nu) u^{\mu+\nu-1}\times\nonumber\\
&\times& {}_nF_n\left(\frac{\nu}{n},\frac{\nu+1}{n},...,\frac{\nu+n-1}{n};\frac{\mu+\nu}{n},\frac{\mu+\nu+1}{n},...,\frac{\mu+\nu+n-1}{n};\beta u^n\right), \label{eq:integrationrel1}
\end{eqnarray}
for $\Re(\mu)>0$, $\Re(\nu)>0$ and $n=2,3,...$, where $B(\mu,\nu)$ denotes Euler's beta function and ${}_pF_q(a_1,...,a_p;b_1,...,b_q;z)$ is the generalized hypergeometric function defined through \index{Euler's beta function}\index{Generalized hypergeometric function}
\begin{equation}
{}_pF_q(a_1,...,a_p;b_1,...,b_q;z)=\sum_{k=0}^\infty\frac{z^k}{k!}\frac{\prod_{i=1}^p (a_i)_k}{\prod_{j=1}^q (b_j)_k},
\end{equation}
and $(a)_n=\Gamma(a+n)/\Gamma(a)$ are the usual Pochhammer polynomials.\index{Pochhammer polynomials} The second integration, namely the one of the generalized hypergeometric function, can be obtained by use of the following relation \cite{gradshteyn}
\begin{equation} \label{eq:integrationrel2}
\int z^{\alpha-1} \,{}_pF_q(a_1,...,a_p;b_1,...,b_q;z) dz= \frac{z^\alpha}{\alpha}\, 
{}_{p+1}F_{q+1}(a_1,...,a_p,\alpha;b_1,...,b_q,\alpha+1;z).
\end{equation}
Expressed in terms of the number of causal set elements sprinkled into $D^+(S_{3}(R))$, $N=\rho V$, where the volume of $D^+(S_{3}(R))$ is given by $V= \frac{\pi}{3}R^4$, the final result reads
\begin{equation}\label{eq:nofN3+1}
\expec{n}=N\, {}_3F_3\left(\frac{1}{2},1,1;\frac{5}{4},\frac{7}{4},2;-\frac{1}{8}N \right).
\end{equation}

\begin{figure}[t]
\begin{center}
\includegraphics[width=5in]{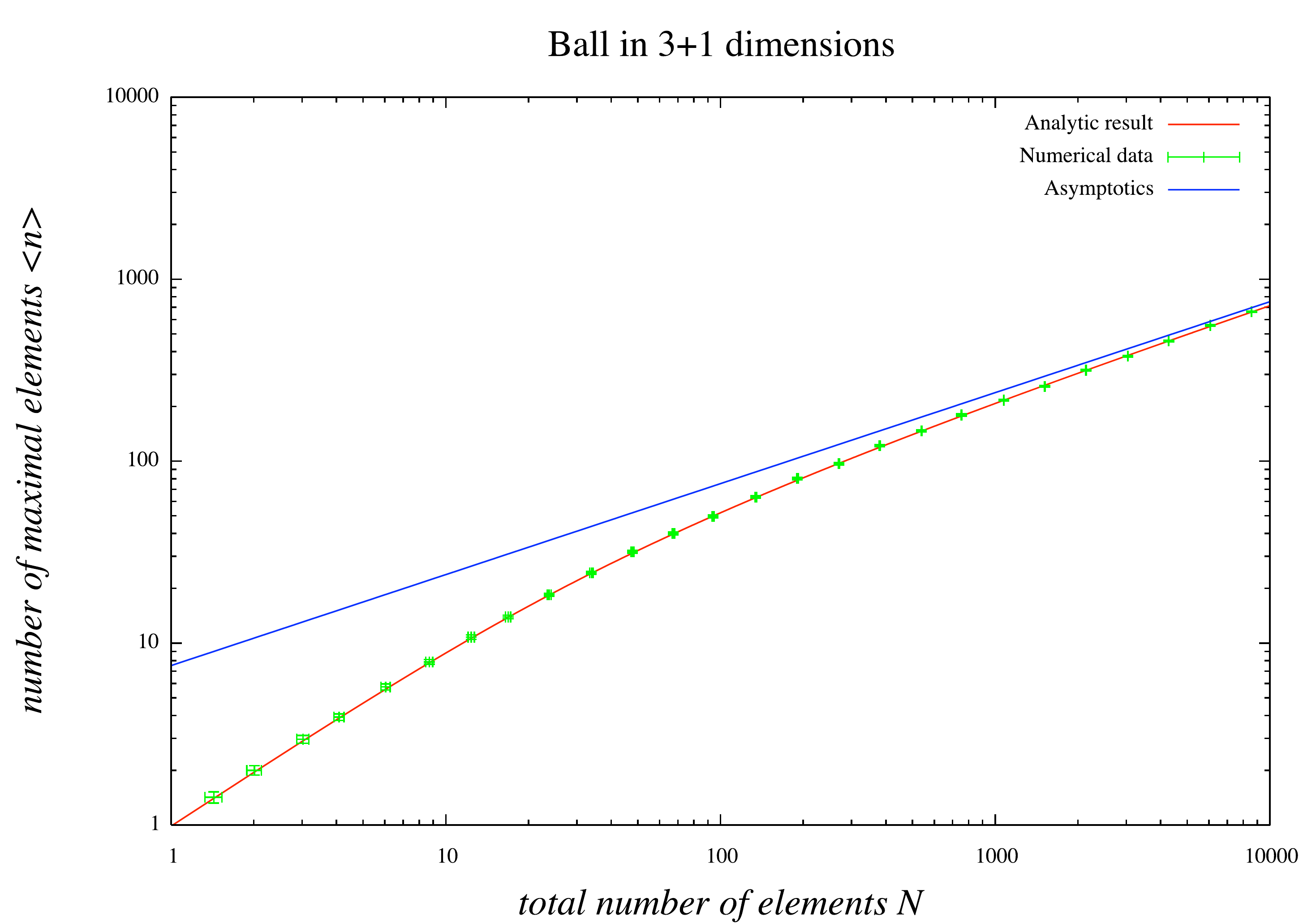}
\caption{Shown is the expected number of maximal elements $\expec{n}$ as a
  function of the total number of elements in the domain of dependence of the
  3-dimensional unit ball in 3+1 dimensional Minkowski space-time on a
  logarithmic scale. The plot shows the analytical result, its asymptotics,
  and numerical results from Monte-Carlo simulation.}\label{fig:disc3+1}
\end{center}
\end{figure}

For a large number of elements $N$ we can use the asymptotic expansion of the generalized hypergeometric functions (cf.\ App.\ \ref{app:asymp}), yielding the asymptotic expression for the expected number of maximal elements,
\begin{equation}
\expec{n}=3\sqrt{2\pi} N^{\frac{1}{2}}+...\quad\text{for }N\rightarrow \infty.
\end{equation}
In terms of the fundamental density of space-time $\rho$ and the radius $R$ of $S_3(R)$ this result translates into
\begin{equation}
\expec{n}=\sqrt{6}\rho^{\frac{1}{2}}\frac{4\pi R^2}{4}.
\end{equation}
As in the previous case the number of maximal elements follows the right scaling according to the entropy bound, namely $\expec{n}\!\propto\!A/4$, where the area of the boundary of 
$S_{3}(R)$ is given by $A=4\pi R^2$. The factor of proportionality is a $\mathcal{O}(1)$ constant as in the previous case, supporting the assumption that the fundamental length scale $l_f$ is proportional to the Planck length $l_p$. More precisely, to have an exact agreement with the Susskind bound the fundamental length scale in four dimensions will be given by $l_f\!=\!\sqrt[4]{6}$ in Planck units.
Using this value for the fundamental length scale the asymptotic expansion for the expected number of maximal elements reads
\begin{eqnarray}
S_{\max}=\expec{n}=\frac{4\pi R^2}{4}
\end{eqnarray}
or $S_{\max}=4\pi R^2 k_B c^3/(4 \hbar G)$ in full units. At first sight one
might feel uncomfortable in absorbing the order one constant into the
fundamental length scale to exactly confirm the Susskind bound. 
However, as discussed in the previous case, the scaling
$\expec{n}\!\propto\!A/4$ is nontrivial and already serves as a
confirmation of the bound. Taking the phenomenological law $S_{max}=A/4$
(spherical entropy bound) as ``data'' gives the
fundamental length scale in four dimensions to be
\begin{equation}
l_f=\sqrt[4]{6} \approx 1.57\quad\text{(in four dimensions)}.
\end{equation}
In comparison to the previous case of 2+1 dimensions one observes that the
fundamental discreteness scale $l_f$ depends on the dimension. In
Sec.\ \ref{sec:ballhigherd} we will derive an expression for the factor for
arbitrary (even) dimensions, showing that this factor tends exactly to one as
$d\!\rightarrow\!\infty$. It is important to check that $l_f$ is universal for
all cases with the same space-time dimension. In Sec. \ref{sec:hyp} we will
provide evidence that this is indeed the case.

At this point it is interesting to note that a similar method of fixing the
fundamental discreteness scale to obtain the right factor of proportionality
in the entropy bound is followed in loop quantum gravity in the context
of black hole entropy (cf.\ \cite{Meissner:2004ju}). There one fixes the
Immirzi parameter which can be regarded as a measure for the discreteness
scale (through its relation to the lowest eigenvalue of the area operator) to
obtain the right factor of a quarter in the black hole entropy. However, since
there are several ambiguities in the relation between the Immirzi parameter
and the fundamental discreteness scale in the sense one uses it in causal set
theory, it is hard to compare the numerical values in any sense.

\subsubsection{4+1 dimensions}

In $d=5$ dimensions one can also evaluate the integral in \eqref{eq:minkd} in a
similar way to the calculation in 2+1 dimensions. As in the previous cases the
result is expressed in terms of the number of causal set elements
 sprinkled into $D^+(S_{4}(R))$, $N=\rho V$, with the volume of $D^+(S_{4}(R))$ given by $V= \frac{\pi^2}{10}R^5$. The final result reads
\begin{eqnarray}
\expec{n}&=&\frac{4}{3}\left(32\left(1-e^{-\frac{N}{16}}\right)+3 N E_{\frac{1}{5}}\left(\frac{N}{16}\right)-N E_{\frac{3}{5}}\left(\frac{N}{16}\right)+\right.\nonumber\\
&&\left. +\, 5 (2N)^{\frac{3}{5}}\Gamma\left(\frac{7}{5}\right)-30 (2N)^{\frac{1}{5}}\Gamma\left(\frac{9}{5}\right)\right).\label{eq:max4+1}
\end{eqnarray}

\begin{figure}[t]
\begin{center}
\includegraphics[width=5in]{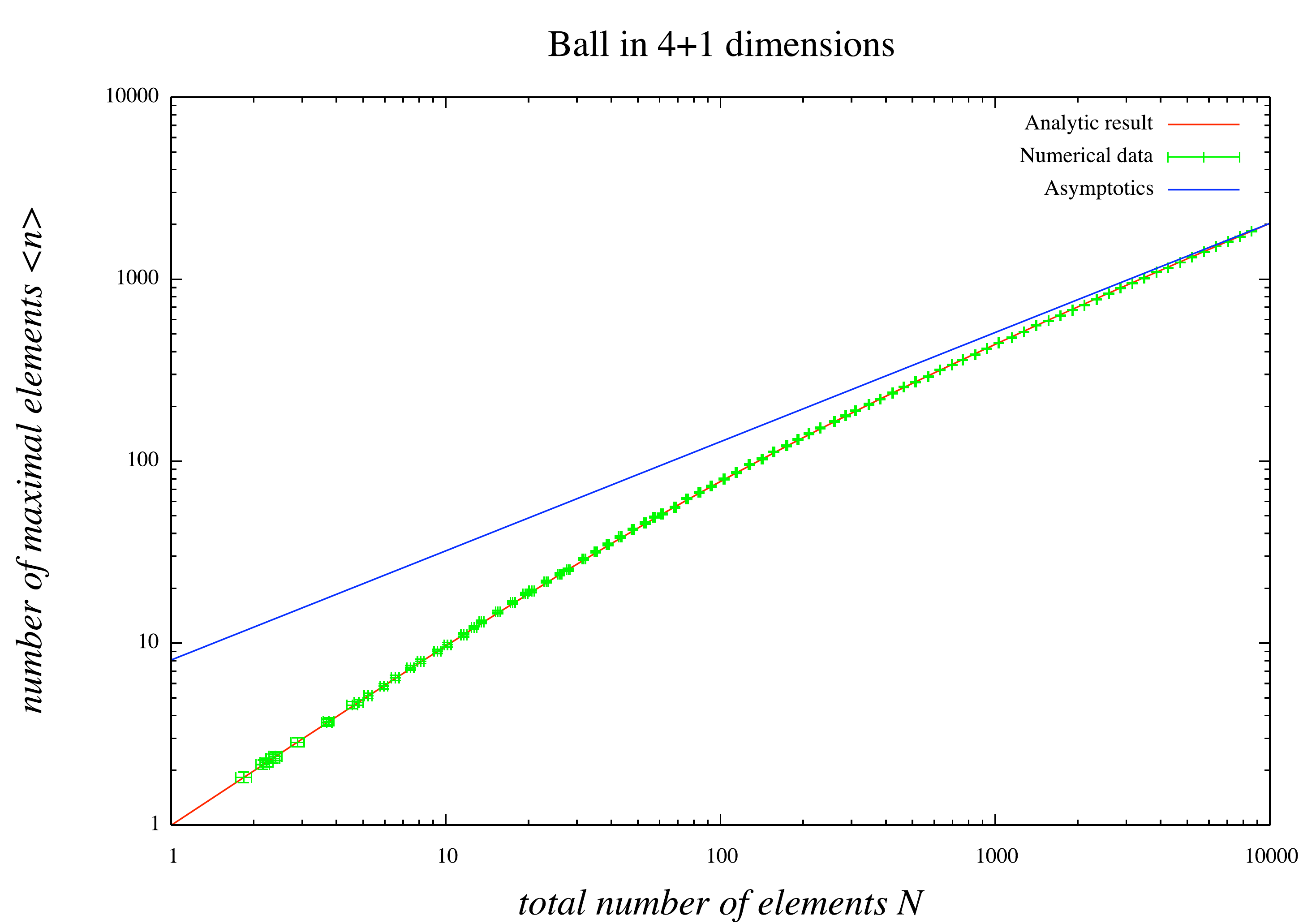}
\caption{Shown is the plot of the number of maximal elements $\expec{n}$ as a function of the total number of elements in the domain of dependence of the 4-dimensional ball in 4+1 dimensional Minkowski space-time on a logarithmic scale, its asymptotics, and numerical results.}\label{fig:disc4+1}
\end{center}
\end{figure}

The plot of $\expec{n}$ as a function of $N$ is shown in Fig.\ \ref{fig:disc4+1}
on a logarithmic scale as well as the numerical results obtained from
Monte-Carlo simulation and the asymptotic behavior. As in the previous cases
there is agreement between analytical and numerical results.

For  large $N$ one can expand \eqref{eq:max4+1} using the asymptotic expansion for the exponential integral \eqref{eq:Enasymp}, yielding
\begin{equation}
\expec{n}=\frac{20}{3}\Gamma\left(\frac{7}{5}\right)(2 N)^\frac{3}{5}+...\quad\text{for}\quad N\rightarrow\infty.
\end{equation}
Re-expressed in terms of the density $\rho$ and the radius $R$ of $S_4(R)$ this result reads
\begin{equation}\label{eq:max4+1exp}
\expec{n}=\frac{32\pi^{1/5}}{5^{8/5}\Gamma\left(\frac{8}{5}\right)}\sqrt{\frac{2}{5+\sqrt{5}}}\, \rho^{\frac{3}{5}} \frac{2\pi^2 R^3}{4}.
\end{equation}
As for the lower dimensional cases this result gives the correct behavior
$\expec{n}\!\propto\!A/4$, where $A=2\pi^2 R^3$ is the volume of the boundary of
$S_{4}(R)$, providing further evidence for the claim. Further,
we can use the coefficient in \eqref{eq:max4+1exp} to fix the fundamental
length scale in 4+1 dimensions $l_f=\rho^{-1/5}$ to be $l_f =2^{11/6}\pi^{1/15}/(5^{8/15}(5+\sqrt{5})^{1/6}\Gamma(8/5)^{1/3})\approx 1.22$ in Planck units. 
Using this fundamental length scale the number of maximal elements \eqref{eq:max4+1exp} reads
\begin{eqnarray}
S_{\max}=\expec{n}= \frac{2\pi^2 R^3}{4},
\end{eqnarray}
or in full units $S_{\max}=2\pi^2 R^3 k_B/(4 (\hbar G_N/c^3)^{3/2})$, which confirms our claim.

\subsubsection{Generalizations to higher dimensions}\label{sec:ballhigherd}

In the previous sections we have seen that causal set theory can provide a fundamental explanation for Susskind's entropy bound
for the spacelike hypersurface $S_{d-1}(R)$ in 2+1, 3+1 and 4+1-dimensional Minkowski space-time. In this section we want to generalize these
results to arbitrary dimensions. For the case of odd space-time dimension, it turns out
that one can easily calculate the expected number of maximal elements, but one cannot write the results in a closed
form. However, for even dimensions one can find a closed expression.

As in the previous cases we will express the result of the expected number of maximal elements in terms of $N\!=\!\rho V$, where $V$ is the volume of the domain of dependence of $S_{d-1}(R)$, given by
\begin{equation}
V\equiv\Vol\left(D^+(S_{d-1}(R))\right)= \frac{\pi^{\frac{d-1}{2}}}{d\,\Gamma(\frac{d+1}{2})} R^d.
\end{equation}
Using the integration relations \eqref{eq:integrationrel1} and \eqref{eq:integrationrel2} one can integrate \eqref{eq:minkd} for even dimensions, yielding
\begin{equation}\label{eq:maxgenerald}
\expec{n}=N\,{}_{\frac{d}{2}+1}F_{\frac{d}{2}+1}\left( \frac{2}{d},\frac{4}{d},...,\frac{d}{d},1; 
\frac{d+1}{d},\frac{d+3}{d},...,\frac{2d-1}{d},2;-2^{1-d}N
\right).
\end{equation}
Note that this result is also valid for the case of $d=2$ dimensions, however
it is not related to any entropy of the system, and is thus
excluded from the proposal.

\begin{figure}[t]
\begin{center}
\includegraphics[width=5in]{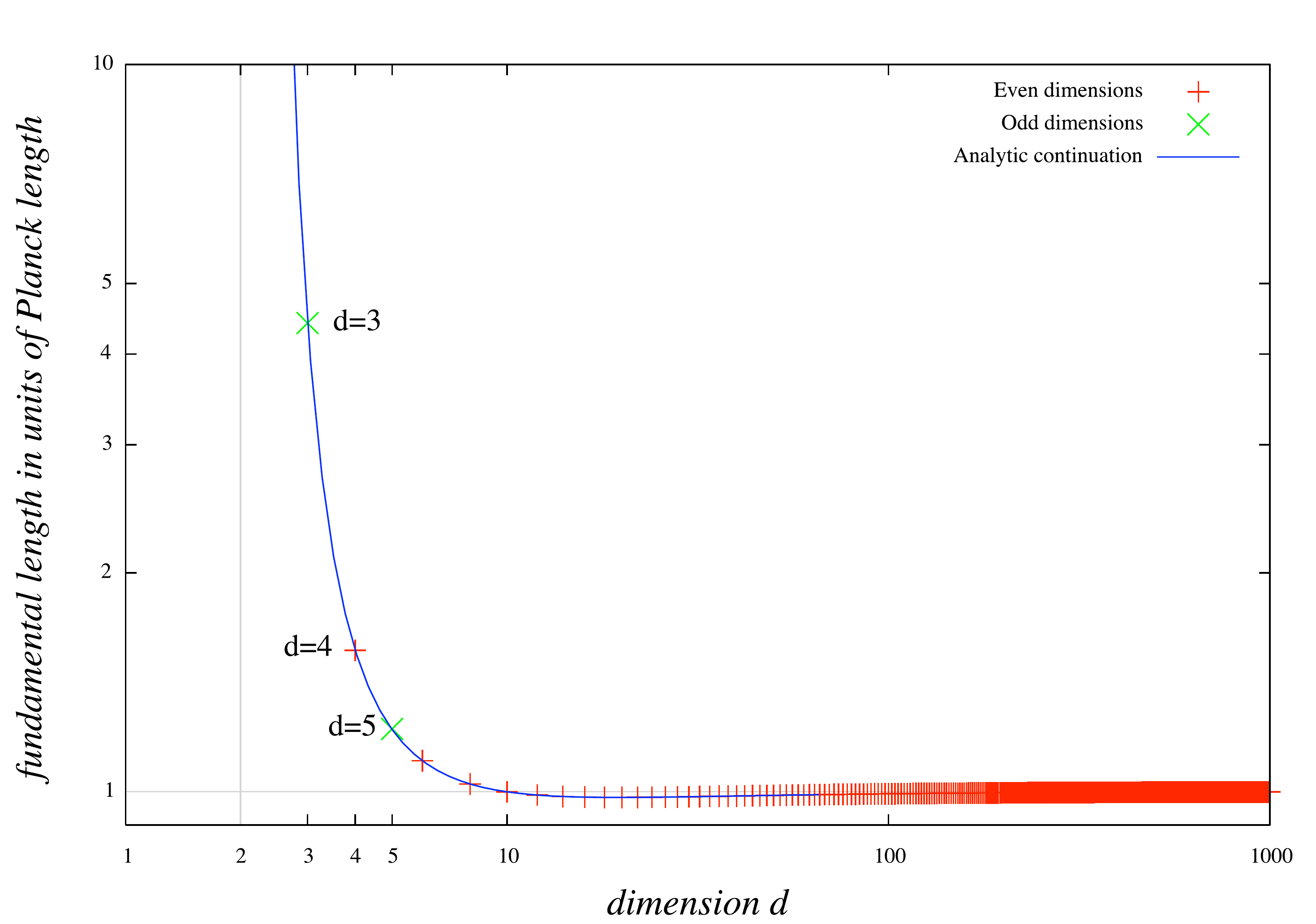}
\caption{Shown is the plot of the fundamental length scale $l_f$ in Planck units ($l_p\!=\!(\hbar G_N/c^3)^{(d-2)/2}$) as a function of the dimensions $d$ for even and odd dimensions and the analytic continuation.}\label{fig:dimension}
\end{center}
\end{figure}

For large $N$ one can use the asymptotic expansion of the generalized hypergeometric function (App. \ref{app:asymp}) to derive the asymptotics of the number of maximal elements for arbitrary even dimensions, yielding
\begin{equation}
\expec{n}=\frac{2^{\frac{2d-2}{d}}\pi (d-1)}{d\sin\left(\frac{2\pi}{d}\right)\Gamma\left(\frac{2d-2}{d}\right)} \,N^{\frac{d-2}{d}}+...,\quad\text{for}\quad N\rightarrow \infty.
\end{equation} 
From this result one can see that the number of maximal elements scales like $\expec{n}\!\sim\! A/4$, where $A$ is the volume of the boundary of $S_{d-1}(R)$, i.e.
\begin{equation}
A\equiv\Vol\left(\Bcal(S_{d-1}(R))\right)=\frac{\pi^{\frac{d-1}{2}}(d-1)}{\Gamma(\frac{d+1}{2})} R^{d-2}.
\end{equation}
Further, we obtain $\expec{n}\!=\! A/4$ precisely for the following value of the fundamental length scale,\index{Causal sets!fundamental length scale}
\begin{equation}\label{eq:dimension}
l_f=\left[ \frac{16 \left(\frac{\pi d^2}{4}\Gamma\left(\frac{d+1}{2}\right)^2\right)^{\frac{1}{d}}}{d^2\sin\left(\frac{2\pi}{d}\right)\Gamma\left(2\frac{d-1}{d} \right) }  \right]^{\frac{1}{d-2}}
\end{equation}
The analytic continuation of \eqref{eq:dimension} as a function of the
dimension is shown in Fig. \ref{fig:dimension} together with the explicit
values for 2+1 and 4+1 dimensions as determined in the previous sections. One
observes that the expression \eqref{eq:dimension} agrees with these values.
This suggests that \eqref{eq:dimension} also holds for
arbitrary odd dimensions. For $d=2$ the value of \eqref{eq:dimension}
diverges, since the Planck length $l_p\!=\!(\hbar G_N/c^3)^{(d-2)/2}$ is not
well defined in two dimensions. This also reflects the fact that the entropy
bound only holds for dimensions $d\!\geqslant\!3$. For all values
$d\!\geqslant\!3$ the fundamental length scale is of order of the Planck
length.

\subsection{Generalizations to different spatial hypersurfaces}\label{sec:hyp}

In the previous section we have derived the Susskind bound for the case where the spacelike hypersurface was chosen to be a $(d-1)$-dimensional ball in $d$ dimensional Minkowski space-time. Further, from this we determined the fundamental discreteness scale of space-time.  However, it is important to prove that the fundamental discreteness scale so determined yields the same entropy bound for all spacelike hypersurfaces of a certain dimension. 

\begin{figure}[t]
\begin{center}
\includegraphics[width=4in]{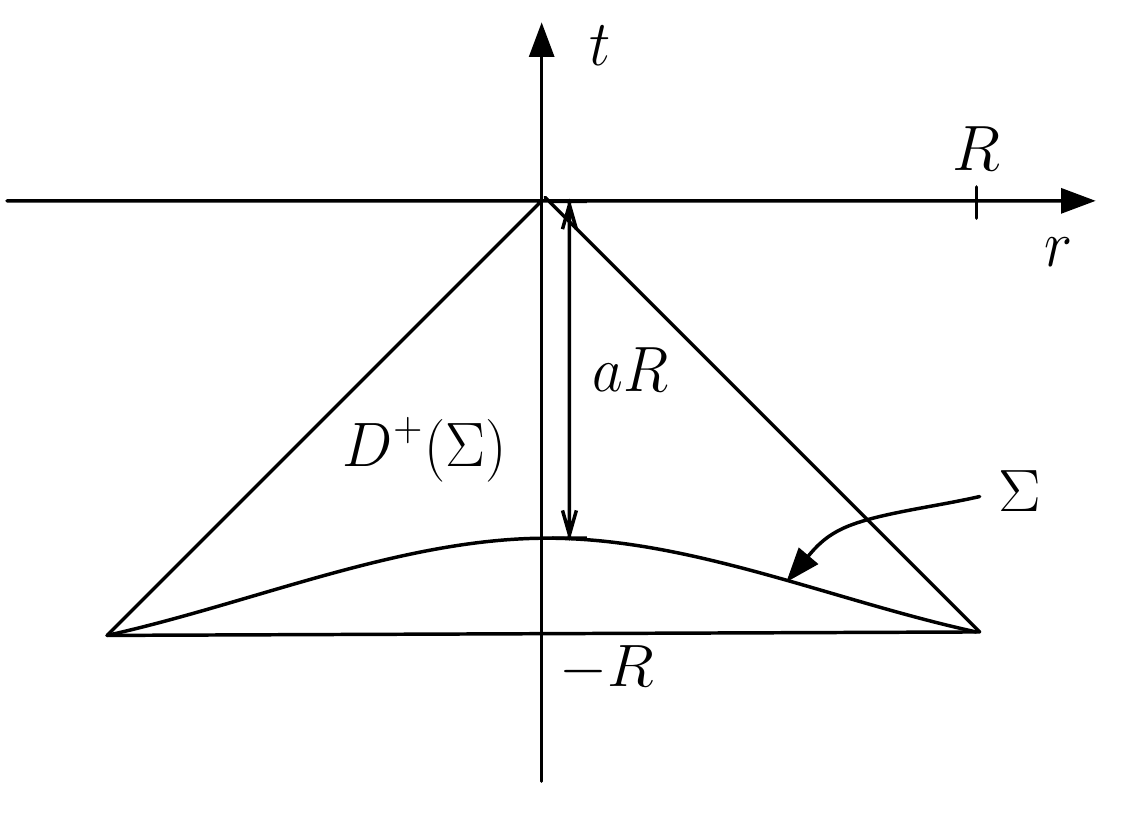}
\caption{Illustration of the different hyperbolic spherically symmetric spatial hypersurfaces $\Scal$ parameterized by $a$, and the domain of dependence $D^+(\Scal)$.}\label{fig:hyp}
\end{center}
\end{figure}

In this section we show that the claim also holds for spacelike
hypersurfaces in Minkowski space-time different from those discussed in the
previous section. We investigate hyperbolic spherically symmetric spacelike
hypersurfaces given by 
\begin{equation}\label{eq:hypdef}
t = -\sqrt{r^2 + a^2(R^2-r^2)},\quad 0 \leqslant a \leqslant 1,
\end{equation}
as shown in Fig.\ \ref{fig:hyp} together with its domain of dependence. For the special case of $a\!=\! 1$ the spacelike hypersurface given by \eqref{eq:hypdef} is equivalent to the $(d-1)$-dimensional ball $S_{d-1}(R)$ for which we have determined an analytic expression for the expected number of maximal elements $\expec{n}$ in the previous section. For other values $0\!<\!a\!<\!1$ one cannot determine the number of maximal elements analytically. In the following we will investigate this problem numerically for the physically most important cases of 2+1 and 3+1 dimensions.

\subsubsection{2+1 dimensions}

For $d=3$ dimensions we use Monte-Carlo methods to numerically obtain the number of maximal elements in the domain of dependence of the spacelike hypersurfaces $\Scal$ defined by \eqref{eq:hypdef} for different values of $a$ as a function total number $N$ of elements in the domain of dependence. Since all these spacelike hypersurfaces $\Scal$ have the same boundary $\Bcal(\Scal)$, it is useful to express $\expec{n}$ as a function of the length of the boundary $A\!=\! 2\pi R$. One can do this by using that $N\!=\! \rho V$, where the volume of the domain of dependence of $\Scal$ is given by
\begin{equation}
V=\frac{2\pi}{3}\frac{a^2}{1+a}R^3.
\end{equation}
Further, we use the value for the fundamental density of space-time as obtained
in Sec. \ref{sec:ball2+1}, i.e.\ $\rho\!=\!(3^{5/2}\Gamma(1/3)^3)/(8192 \pi)$.

\begin{figure}[t]
\begin{center}
\includegraphics[width=5in]{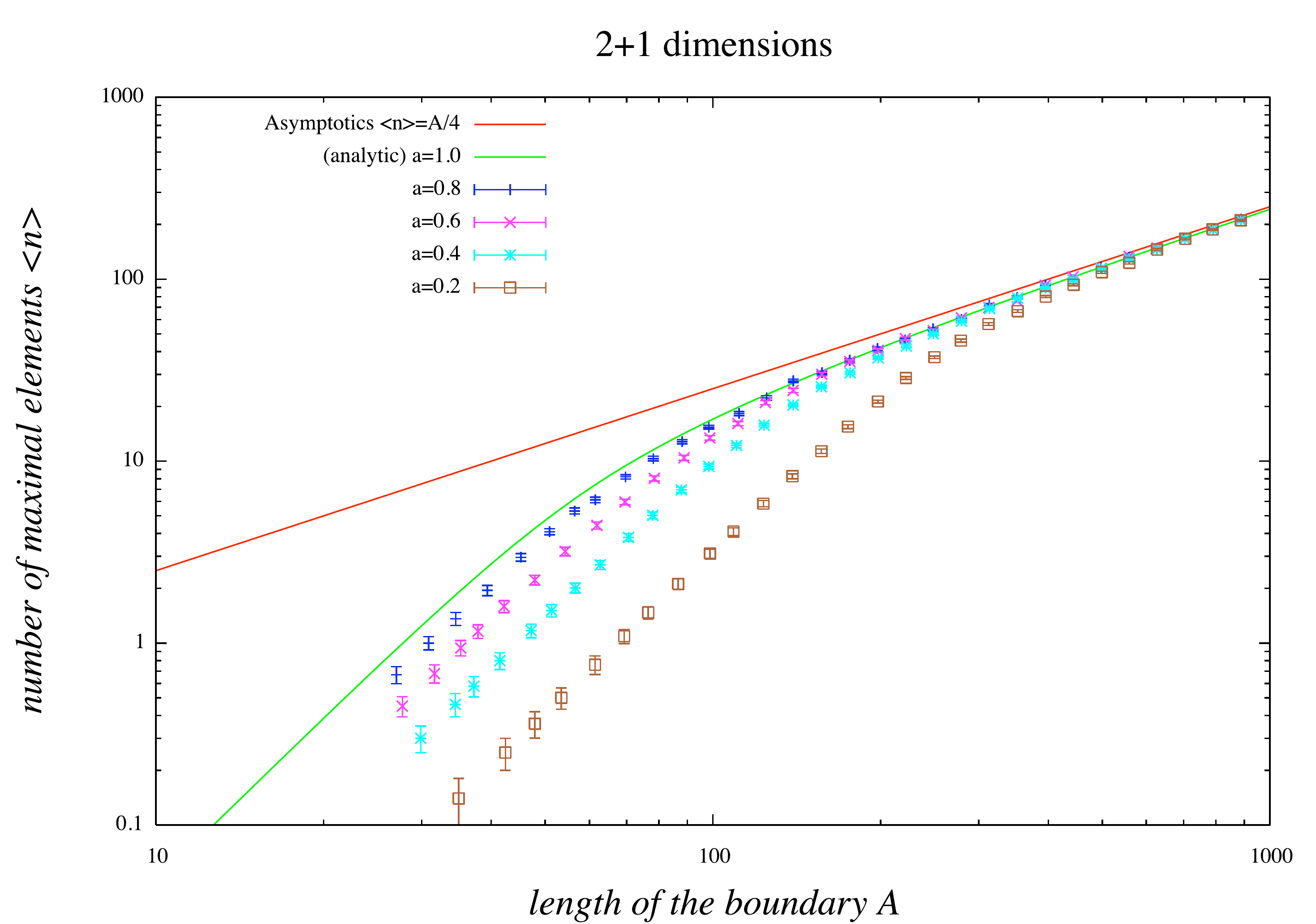}
\caption{Shown is the plot of the expected number of maximal elements $\expec{n}$ in the domain of dependence as a function of the length of the boundary $A$ for different hyperbolic spherical symmetric spacelike hypersurfaces parameterized by $a$ in 2+1 dimensional Minkowski space-time. One sees that all functions approach the same asymptotic $\expec{n}=A/4$.}\label{fig:hyp2+1}
\end{center}
\end{figure}

The results of the simulations are summarized in Fig.\ \ref{fig:hyp2+1}. Shown
is the expected number of maximal elements $\expec{n}$ as a function of the
boundary length $A$. For the special case of $a\!=\!1$ in \eqref{eq:hypdef} we
know that the result is given by \eqref{eq:nmax2+1}, where $N$ is replaced by
$N\!=\!A^3/(2^{13}\Gamma(2/3)^3)$. The asymptotics of this analytic result is
given by $\expec{n}=A/4$ as shown earlier. For different values of
$0\!<\!a\!<\!1$ the simulations show that even though the number of maximal
elements as a function of $A$ differs for small values $A\!\lesssim\!10^3$ in
Planck units, in the expansion for large $A$, corresponding to the continuum
approximation, all functions $\expec{n}$ for different $a$ enter the same
asymptotic expansion $\expec{n}=A/4$, yielding Susskind's entropy bound. In addition, it shows that
the prediction of the fundamental discreteness scale is universal in 2+1
dimensions, at least for all investigated cases of spacelike
hypersurfaces in Minkowski space-time. A generalization to examples in curved
space-time has still to be shown and will be investigated in future work.

\subsubsection{3+1 dimensions}

As in the case of 2+1 dimensions we use Monte-Carlo methods to numerically obtain the number of maximal elements in the domain of dependence of the spacelike hypersurfaces $\Scal$ defined by \eqref{eq:hypdef} for $d=4$ and different values of $a$. Again, the result is expressed as a function of the area of the boundary of $\Scal$, i.e.\ $A\!=\! 4\pi R^2$. This can be done by using the relation $N\!=\! \rho V$ and noticing that the volume of the domain of dependence of $\Scal$ is given by
\begin{equation}
V=R^4 \left[\frac{\pi}{3}-\frac{\pi}{6(1-a^2)^{\frac{3}{2}}}\left(\sqrt{1-a^2}(2-5a^2)+3a^4\log\left( \frac{1+\sqrt{1-a^2}}{a}\right) \right)\right],
\end{equation}
where we use $\rho\!=\!1/6$ for the value of the fundamental density of four dimensional space-time.

\begin{figure}[t]
\begin{center}
\includegraphics[width=5in]{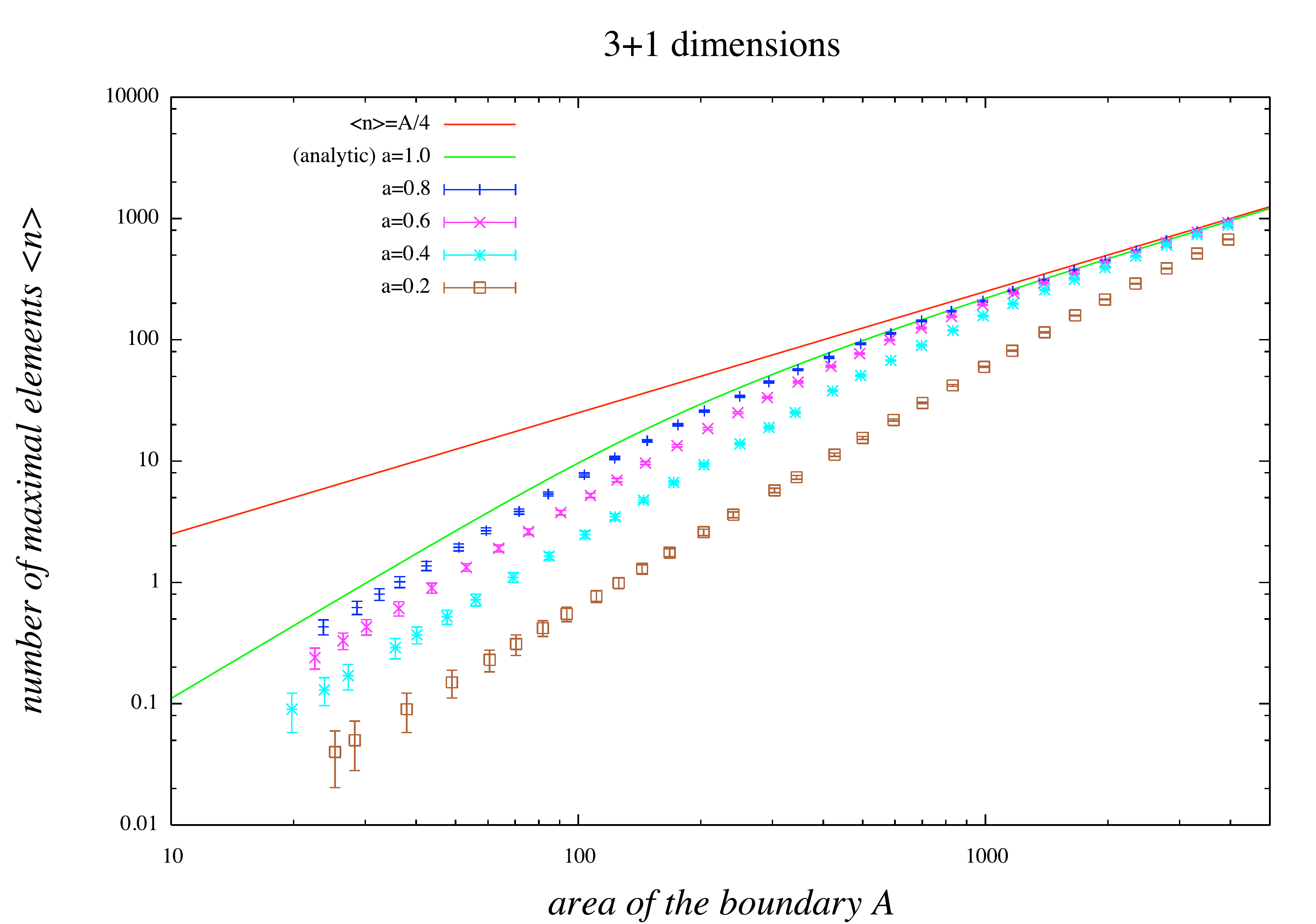}
\caption{Shown is the plot of the expected number of maximal elements $\expec{n}$ as a function of the length of the boundary $A$ for different hyperbolic spherical symmetric spacelike hypersurfaces in 3+1 dimensional Minkowski space-time parameterized by $a$. One sees that all functions approach the same asymptotics $\expec{n}=A/4$.}\label{fig:hyp3+1}
\end{center}
\end{figure}

The numerical and analytical results are shown in
Fig.\ \ref{fig:hyp3+1}. Displayed is the expected number of maximal elements
$\expec{n}$ as a function of the area $A$ of the boundary. For the special
case of $a\!=\!1$ the analytical result was given by \eqref{eq:nofN3+1}, with
$N$ replaced by $N\!=\!A^2/(2 \pi12^2)$. The simulations show that for
different values of $0\!<\!a\!<\!1$ all functions $\expec{n}$ enter the same
asymptotics, $\expec{n}=A/4$, in the continuum limit, giving further evidence
for the claim. For very small values of $a$ the expected number
of maximal elements enters the asymptotic regime only very slowly. However, the upper
value for $A$ of the simulations is still very small ($\sim\!10^{-33}m$). As
in the lower dimensional case the simulations show that the prediction for the
value of the fundamental discreteness scale of four-dimensional space-time,
namely $l_f=\sqrt[4]{6}$, is a universal quantity for this dimension, at least
for all investigated cases of spherically symmetric
spacelike hypersurfaces in Minkowski space-time.

\section{Discussion and outlook}

Using the phenomenological model of causal set theory as introduced in the previous chapter we argued for a bound on the entropy in a spherically symmetric spacelike region from a
counting of potential horizon ``degrees of freedom'' of the fundamental theory, namely
the maximal elements of the future domain of dependence of
the region. It was then shown that, for different spherically
symmetric spacelike regions in Minkowski space-time of arbitrary
dimension, this leads to Susskind's spherical entropy bound. The evidence was given in terms of analytical results
for spatial $(d-1)$-dimensional balls
in $d$-dimensional Minkowski space-time, and in terms of numerical results
obtained from Monte-Carlo simulations for the case of hyperbolic spherically
symmetric hypersurfaces in Minkowski space-time. 

So far results were 
given only in terms
of examples in flat Minkowski space-time. Clearly this is a 
very restricted class and further evidence should be provided in different space-times. Spherically symmetric spacelike regions
in curved space-time such as in Friedmann-Robertson-Walker cosmology will be
investigated in future work. 

Another step would be to formulate the proposal
intrinsically in terms of order invariants, without making explicit reference
to the continuum.  Such a formulation 
may be fruitful in providing a fundamental understanding of Bousso's covariant entropy bound. 

Further work in progress in the direction of entropy bounds from causal sets is the implementation of entropy evaluation through link counting as proposed in \cite{Dou:2003af,Sorkin:1997gi} for black holes, to other situations such as dS space-time.

\part{2D causal dynamical triangulations}\label{part2}

\chapter{Path integrals and quantum gravity \label{Chap:QG}}
	


In Part I of this thesis we introduced causal sets as a simple model of causal random geometry. The remaining three parts of this thesis will be centered around the approach of causal dynamical triangulations. As already discussed in the introduction, one way to confront the non-renormalizability of perturbative quantum gravity in four dimensions is to find a proper non-perturbative definition of the gravitational path integral. Before describing dynamical triangulations as such, we first give a brief introduction to path integrals in general. Specifically, we comment on problems one encounters when formulating the gravitational path integral.    

\section{Random paths and one-dimensional gravity}
\index{Random path}\index{One-dimensional gravity}

Path integrals were first introduced by Dirac and Feynman \cite{Dirac1933,Feynman1948} as a quantization scheme to first quantize physical systems. One of the most instructive examples is the free relativistic particle in $D$-dimensional Minkowski space-time $\mathbb{M}^D$. The amplitude of a particle moving from $x$ to $y$ can be expressed by the so-called propagator,
\index{Relativistic particle}
\beq\label{eq:NRPpropagator}
G_m(x,y)=\int_x^y \cD P(x,y) e^{i S[P(x,y)]}
\eeq  
where the action is simply given by the mass $m$ times the length of the path
\beq \label{eq:NRPaction0}
S[P(x,y)]=m\int_{P(x,y)} dl.
\eeq
The classical equations of motion are derived by choosing a specific parametrization
\beq \label{eq:NRPrep}
x(\xi): [0,1]\to \mathbb{M}^D,\quad x(0)=x,\quad x(1)=y.
\eeq
The action now reads
\beq \label{eq:actionparNRP}
S[P(x,y)]=m\int_0^1 d\xi \sqrt{(\dot{x}^{\mu}(\xi))^2},
\eeq
where we defined $\dot{x}^{\mu}\!:=\! dx^{\mu}/d\xi$. The classical equations of motion can be easily obtained by varying the action
\beq\label{eq:eomNRP}
\frac{\delta S}{\delta x^{\mu}(\xi)}=\frac{d}{d\xi}\frac{\dot{x}^{\mu}}{|\dot{x}|}=0
\eeq
which are simply straight lines from $x$ to $y$ and reparametrizations of it.

\begin{figure}[t]
\begin{center}
\includegraphics[width=5in]{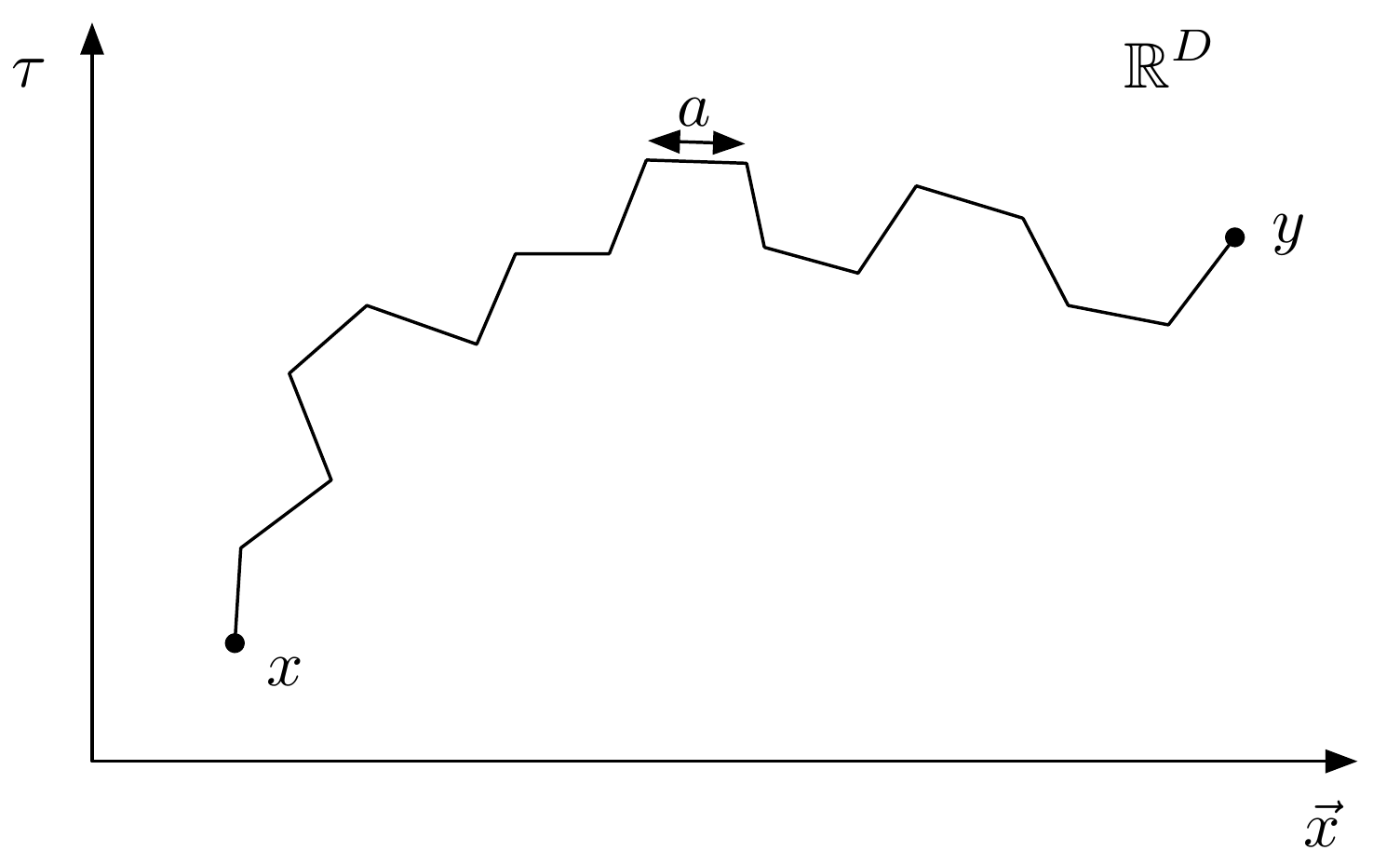}
\caption{Illustration of the path integral for a one-dimensional relativistic quantum mechanical problem, e.g. a propagating particle. One possible path of the configuration space (path space) is drawn. The ``virtual'' particle is propagating from $x$ to $y$ in a piecewise linear path of $n$ steps each of size $a$.}
\label{fig:NRP}
\end{center}
\end{figure}

To be able to define the path integral \eqref{eq:NRPpropagator} one usually performs a Wick rotation by taking time imaginary \index{Wick rotation}
\beq
t\to- i \tau =- i x^{D}.
\eeq
One now deals with the corresponding Euclidean model in $\mathbb{R}^D$. 

One way to define the path integral \eqref{eq:NRPpropagator} is to introduce a ``geometrical'' discretization scheme where one directly discretizes the length of the worldline (see for instance \cite{Ambjorn:1997di}). Therefore it is useful to work with the action \eqref{eq:NRPaction0}, since its definition does not depend on parametrized quantities. Specifically, in the employed discretization each worldline consists of a piecewise linear path, where each edge is of size $a$ (see Fig.\ \ref{fig:NRP}). Here $a$ serves as a reparametrization invariant cut-off which will be taken to zero in the continuum limit. For a piecewise linear path of length $n$ the action simply becomes $S=m_0 a n$, where $m_0$ is the bare mass. Further, the path integral can be written as a sum over all possible paths connecting $x$ and $y$: 
\beq
\int_x^y \cD P(x,y) \to \sum_{n=1}^\infty \int \prod_{i=0}^n d\hat{e}_i\, \delta(a\sum_{i=0}^n \hat{e}_i -(y-x) ).
\eeq  
Here the $a\hat{e}_i$ denote the vectors of the linear parts of the discretized worldline, where the $\hat{e}_i$'s are unit vectors in $\mathbb{R}^D$.
The propagator can now be defined as follows
\beq
G_{m_0}(x,y)= \sum_{n=1}^\infty e^{-m_0 a n} \int \prod_{i=0}^n d\hat{e}_i\, \delta(a\sum_{i=0}^n \hat{e}_i -(y-x) ).
\eeq  
To perform the integration it is helpful to use the Fourier transform
\beq
G_{m_0}(p)= \int dx e^{-i p(x-y)} G_{m_0}(x,y)= \sum_{n=1}^\infty e^{-m_0 a n} \int \prod_{i=0}^n d\hat{e}_i e^{-i a p \hat{e}_i}.
\eeq  
The integral in the above expression can be expanded as 
\beq
\int d\hat{e}\, e^{-i a p \hat{e}}\equiv f(a|p|) = f(0)(1-c^2 (a p)^2 +\mathcal{O}(a^4)),
\eeq
where $f(0)\equiv\Vol(S_{D-1}(1))$ is given in \eqref{volumeSd} and $c$ is an irrelevant constant.
We can now perform the summation leading to
\beq
G_{m_0}(p) =\sum_{n=1}^\infty \left(e^{-m_0 a } f(a |p|) \right)^n = \frac{1}{1-e^{-m_0 a} f(a|p|)}.
\eeq
One can define the continuum limit by taking the cut-off $a$ to zero and at the same time taking the number of steps in the discretized path to infinity. The latter is fulfilled when $\mu\equiv m_0 a$ is taken to its critical value $\mu_c=\log f(0)$ at which $G_{m_0}(p)$ approaches its radius of convergence. Hence, we renormalize the bare mass $m_0$ such that
\beq
\mu-\mu_c= m^2 c^2 a^2
\eeq
where $m$ is the physical mass. Using this renormalization we obtain
\beq
G_{m_0}(p) \xrightarrow[a\to 0]{} \frac{1}{a^2c^2}\frac{1}{p^2+m^2}.
\eeq
After a wave function renormalization the continuum propagator $G^c_m(p)$ becomes the well-known expression for the Feynman propagator\index{Feynman propagator}
\beq \label{eq:NRPKG}
G^c_m(p)=a^2 c^2 G_{m_0}(p)=\frac{1}{p^2+m^2}.
\eeq
In terms of its Fourier transform this is precisely the solution of the field equation for the Green function of a Klein-Gordon field,\index{Klein-Gordon field}
\beq\label{eq:NRPKG2}
(-\prt_y^2+m^2)G_m(x,y)=\delta^{(D)}(y-x).
\eeq

 In the above description we used a discretization scheme where we directly discretized the length of the worldline. A different prescription is to use the parametrized action \eqref{eq:actionparNRP} and to introduce a cut-off on the worldline coordinates (see for instance \cite{Kleinertbook}). However, since the square root in this expression is very difficult to handle, it is useful to look at an alternative action for the free relativistic particle first introduced in \cite{Brink:1976sc},\index{Brink-Di Vecchia-Howe action}
 \beq \label{eq:actionparNRP2}
S[x,g]_{(x,y)}=\int_0^1 d\xi \sqrt{|g(\xi)|}(g^{-1}(\xi)\dot{x}^2+m^2),
\eeq
where $g(\xi)$ is the internal metric of the one-dimensional worldline. The action \eqref{eq:actionparNRP2} can also be viewed as the action of one-dimensional gravity coupled to $D$ scalar fields $x^\mu$. Solving for the equations of motion for $g$ and plugging the result back into the equations of motion for $x$ one finds \eqref{eq:eomNRP}. This shows that the systems desribed by both actions \eqref{eq:actionparNRP} and \eqref{eq:actionparNRP2} are equivalent on the classical level. Whether this is also true on the quantum level remains to be checked. To obtain the propagator we now also have to integrate over the worldline metric $g$,
\beq\label{eq:propNRP2}
G_m(x,y)=\int \cD [g]\int_x^y \cD_g x \,\, e^{i S[x,g]_{(x,y)}},
\eeq 
where $\cD [g]$ refers to the measure of the space of worldline metrics modulo diffeomorphisms, i.e.\ coordinate transformations on the worldline. Let us in the following briefly sketch how to calculate the propagator \eqref{eq:propNRP2} (for details see \cite{Kleinertbook}). The first step is to perform the Wick rotation. One proceeds by defining the integral over $ \cD_g x$ for fixed $g$. Therefore one discretizes the worldline coordinates by forming $N+1$ slices of arbitrary small parameter differences
\beq
\epsilon_n= \xi_{n}-\xi_{n-1}, \quad \xi_0=0,\quad \xi_N=1.
\eeq
 The integration over $ \cD_g x$ and the continuum limit $\epsilon_n\to0$ can now be performed in a similar manner as one usually does for the non-relativistic particle. The remaining integral over $\cD [g]$ can then be done after a suitable gauge fixing. The result is precisely given by \eqref{eq:NRPKG}. This indicates that both actions \eqref{eq:actionparNRP} and \eqref{eq:actionparNRP2} not only lead to the same classical equations of motion, but also to the same dynamics on the quantum level. 
A comparison of the two discretization schemes further shows that the ``geometrical'' discretization employed above is computationally rather simple in comparison to the continuum method. Dynamical triangulations, as will be introduced in the next chapter, is an analogous ``geometrical'' discretization scheme for surfaces and higher-dimensional manifolds.


\section{Random surfaces and strings}
\index{Random surfaces}

\begin{figure}[t]
\begin{center}
\includegraphics[width=5in]{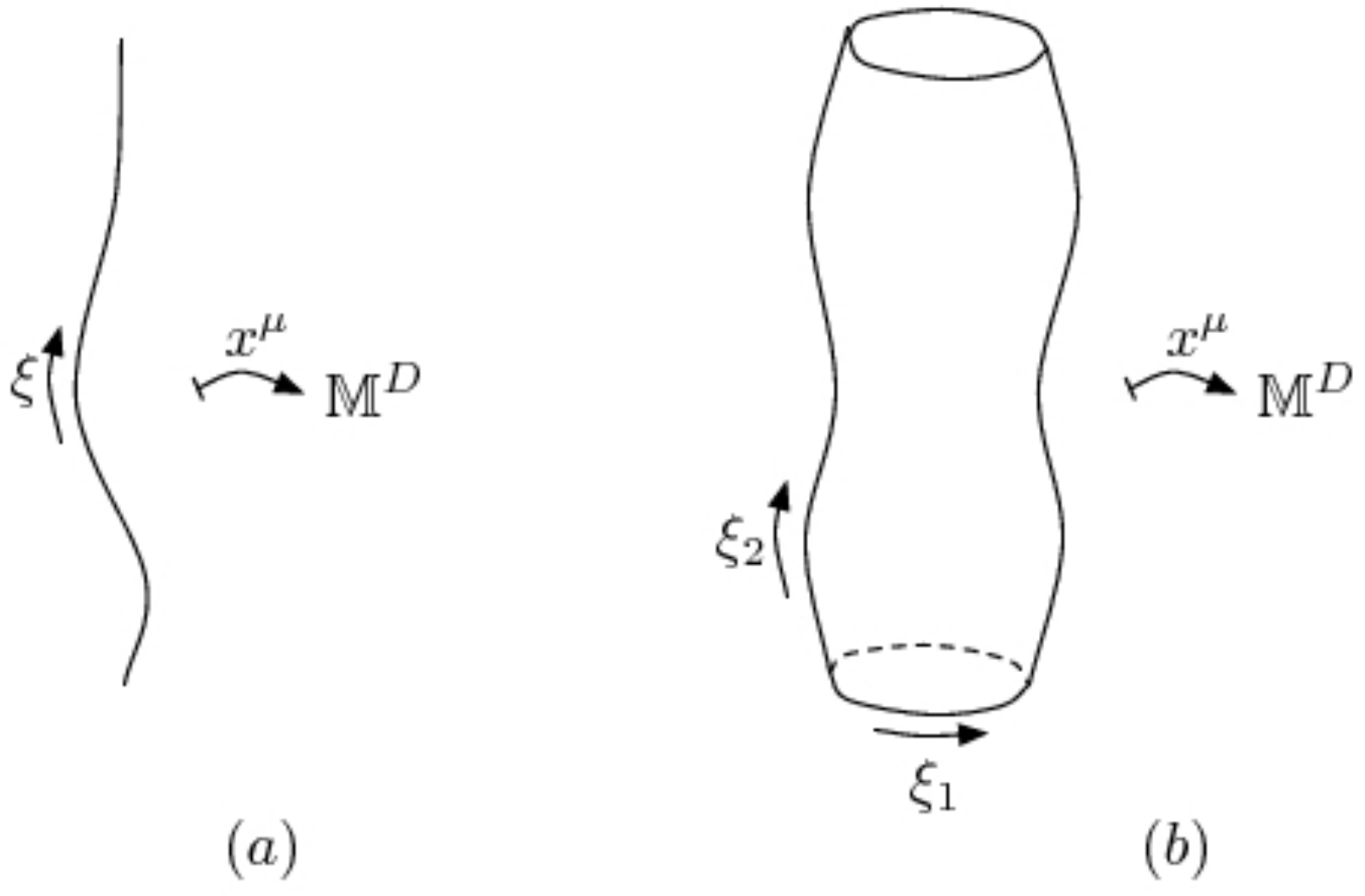}
\caption{(a) A worldline parametrized by $\xi$ is embedded into $\mathbb{M}^D$ by the mapping $x^{\mu}$. (b) Analogous to the worldline a two-dimensional worldsheet of a closed string parametrized by $\xi_1$ and $\xi_2$ is embedded into $\mathbb{M}^D$ by the mapping $x^{\mu}$.}
\label{fig:string}
\end{center}
\end{figure}

The natural generalization to the random paths, as considered in the previous section, are random surfaces or worldsheets of propagating (closed) strings (Fig.\ \ref{fig:string}). In analogy to the relativistic particle we expect those to describe the relativistic string. \index{Relativistic string}
The simplest action for a surface $F$ is given by its area
\beq\label{eq:NGaction}
S[F]= \l \int_F dA(F),
\eeq 
where $\l$ is a (cosmological) constant conjugate to the area. The action \eqref{eq:NGaction} is a direct analog of \eqref{eq:NRPaction0}. The parametrized form of \eqref{eq:NGaction} is called the Nambu-Goto action \cite{Goto1971} and reads \index{Nambu-Goto action}
\beq
S[F]= \l \int_F d^2\xi \sqrt{h}, 
\eeq
where
\beq
h=\det h_{ab}=\det\frac{\prt x^\mu}{\prt\xi^a}\frac{\prt x^\mu}{\prt\xi^b} 
\eeq
is the determinant of the metric $h_{ab}$, $a,b\equ 1,2$, of the embedded worldsheet induced by the mapping $x$. \index{String worldsheet}

As for the relativistic particle we can also write an action in terms of the internal metric $g_{ab}$, $a,b\equ 1,2$, of the worldsheet. This is the so-called Polyakov action \cite{Brink:1976sc}, the analog of \eqref{eq:actionparNRP2},\index{Polyakov action}
\beq\label{eq:Polyakovaction}
S[g,x]=\int_M d^2\xi \sqrt{|g(\xi)|}(g^{ab}h_{ab}+\l),
\eeq
where one should integrate over both $g$ and $x$ in the path integral \cite{Polyakov:1981rd}.
Both \eqref{eq:NGaction} and \eqref{eq:Polyakovaction} are equivalent on the classical level, however, it is not so obvious that this also holds at the quantum level (see for instance \cite{Ambjorn:1997di}).

We will see in the following two sections how \eqref{eq:Polyakovaction} can also be viewed as the action for two-dimensional quantum gravity coupled to $D$ scalar fields. An explicit ``geometrical'' discretization to define the corresponding path integral will be discussed in the next chapter. 

\section{A path integral for quantum gravity}

In the previous section we discussed the Polyakov and Nambu-Goto action as simple generalizations of random paths to random surfaces. 
The natural action for a theory of quantum gravity in $d$-dimensions is the so-called Einstein-Hilbert action, \index{Einstein-Hilbert action}
\beq\label{eq:EHaction}
S_{EH}([g_{ab}]) =  \frac{1}{16\pi G_N}\int_{M} d^d \xi \sqrt{|g(\xi)|} (2\La - R),
\eeq
where $G_N$ is the Newton's constant, $\La$ the cosmological constant, $R$ the scalar curvature of the metric $g_{ab}$ and $g=\det g_{ab}$ the determinant of the metric.
For manifolds with boundaries one also has to include the Gibbons-Hawking-York boundary term \cite{PhysRevLett.28.1082,PhysRevD.15.2752}
\beq\label{eq:GHYaction}
S_{GHY}([g_{ab}]) =  \frac{1}{8\pi G_N}\int_{\partial M} d^{d-1} \xi \sqrt{|h(\xi)|}K,
\eeq
where $h$ and $K$ are the induced metric and curvature on the boundary.
All physical degrees of freedom are encoded in the equivalence class $[g_{ab}]$ which is $g_{ab}$ modulo diffeomorphisms, i.e.\ coordinate transformations.

\index{Gravitational path integral}\index{Partition function}
The expression of the gravitational path integral for the partition function is then formally written as
\beq\label{eq:gravpathint}
\cZ(G_N,\La) =\int_{\Geom(M)} \D[g_{ab}] e^{i S_{EH}([g_{ab}])},
\eeq
where the integration should be taken over the space of all closed geometries, i.e.\ the coset space $\Geom(M)=\mathrm{Metrics}(M)/\Diff(M)$. Generalizations to geometries with boundaries are also possible. 
When trying to define the formal expression \eqref{eq:gravpathint} several problems arise:

First of all, $\D[g_{ab}]$ has to be defined in a covariant way to preserve the diffeomorphism invariance. Unfortunately, to perform further calculations one would then have to gauge fix the field tensors which would give rise to Faddeev-Popov determinants \cite{Faddeev:1980be} whose non-perturbative evaluation is exceedingly difficult.\footnote{See \cite{David:1988hj,Distler:1988jt,Mottola:1995sj} for an evaluation in the setting of two-dimensional Euclidean quantum gravity in the light-cone gauge. In \cite{Dasgupta:2001ue} a calculation for three- and four-dimensional Lorentzian quantum gravity in the proper-time gauge is presented. It is anticipated there that the Faddeev-Popov determinants cancel the divergences coming from the conformal modes of the metric non-perturbatively.}

\index{Wick rotation!quantum gravity}
The second problem is due to the complex nature of the integrand. In quantum field theory this problem is solved by doing the Wick rotation $t\mapsto- i \tau$, where $t$ is the time coordinate in Minkowski space $(\vec{x},t)$. Clearly, a prescription like this does not work out in the gravitational setting, since all components of the metric field tensor depend on time. Further $t\mapsto -i\tau$ is certainly \textit{not} diffeomorphism invariant (as a simple example consider the coordinate transformation $t\mapsto t^2$ for $t>0$). Hence, the question arises: What is the natural generalization of the Wick rotation in the gravitational setting? This problem is often circumvented in the approach of \emph{Euclidean} quantum gravity, where one does an ad hoc substitution
\beq
e^{i S_{EH}([g^{(lor)}_{ab}])}\to e^{- S_{EH}([g^{(eu)}_{ab}])}.
\eeq
In this substitution the path integral over Lorentzian manifolds $M^{(lor)}$ is replaced by an integral over Euclidean manifolds $M^{(eu)}$. Clearly, the path integral now also includes many ``acausal'' manifolds and we will see in the next chapter when discussing two-dimensional Euclidean quantum gravity what consequences this brings with it. The main interest of this thesis is however to analyze ways to define the gravitational path integral in a Lorentzian setting including a proper definition of the Wick rotation.
 
Another problem in the definition of \eqref{eq:gravpathint} is the following. Since we are working in a field theoretical context, some kind of regularization and renormalization will be necessary, and again, has to be formulated in a covariant way.
In quantum field theory the use of lattice methods provides a powerful tool to perform non-perturbative calculations, where the lattice spacing $a$ serves as a cut-off of the theory. An important question to ask at this point is whether or not the theory becomes independent of the cut-off. Consider for example QCD and QED on the lattice. All evidence suggests that QCD in four space-time dimensions is a genuine continuum quantum field theory. By genuine continuum quantum field theory we mean a theory which needs a cut-off at an intermediate step, but whose continuum observables will be \textit{independent} of the cut-off at arbitrarily small scales. In QED the situation seems to be different, one is not able to define a non-trivial theory with the cut-off removed, unless the renormalized coupling $e_{ren}=0$ (trivial QED). Therefore, QED is considered as a low-energy effective theory of a fundamental theory at the Planck scale.

Finally, one could argue to also include a sum over topologies in the path integral \eqref{eq:gravpathint}. However, in higher dimensions such a sum over topologies is quite intractable, since for $d\geq 3$ there does not even exist an obvious classification of topologies in terms of a finite set of parameters.

These are some of the reasons to first look at the simpler case of two dimensions. 

\section{Two-dimensional Euclidean quantum gravity and Liouville theory}
\index{Euclidean quantum gravity}

One of the big advantages of the two-dimensional Einstein-Hilbert action is that the curvature term is a topological invariant. This follows from the Gauss-Bonnet theorem which states that \index{Gauss-Bonnet theorem}
\beq
\int_{M} d^2\xi  \sqrt{|g(\xi)|} R = 4 \pi \chi(M), 
\eeq
where $\chi(M)=2-2h-b$, is the so-called \emph{Euler-characteristic}, $h$ is the genus of the manifold $M$, i.e.\ the number of holes in the surface, and $b$ the number of boundary components of $M$.\index{Euler-characteristic}

Using the Gauss-Bonnet theorem we can write the partition function for two-dimensional Euclidean quantum gravity including a sum over topologies (using a rescaling of the couplings)
\beq \label{eq:QGtopoexp}
\cZ(G_N,\La) =\sum_{h=0}^\infty e^{-\chi(M_h)/G_N } \cZ_{h}(\La),
\eeq
where the partition function for manifolds with fixed genus $h$ is given by
\beq
\cZ_h(\La) =\int_{\Geom(M_h)} \D[g] e^{-S(g,\La)}
\eeq
and
\beq\label{eq:QGactionSh}
S(g,\La)=\La V_g,\quad V_g= \int_{M_h} d^2 \xi \sqrt{|g(\xi)|}. 
\eeq
Here $V_g$ is the volume of the manifold $M_h$. Eq. \eqref{eq:QGtopoexp} is called the \emph{topological expansion}. Note that since the sum is taken over closed surfaces, we have $\chi(M)=2-2h$.\index{Topological expansion}
Often one is only interested in the genus zero contribution, i.e.\ the sum over surfaces with topology of $S^2$. In this case we denote the partition function by $\cZ(\La)\equiv \cZ_0(\La)$.

If the path integral is taken over surfaces with $b$ boundaries the Euler characteristic reads $\chi(M)=2-2h-b$. Further, it is natural to include boundary terms in the action
\beq
S(g,\La,Z_1,...,Z_b)=\La V_g+\sum_{i=1}^b Z_i L_i(g),
\eeq
where $L_i(g)$ is the length of the $i$th boundary loop $\cL_i$ with respect to the metric $g$. Here the $Z_i$ take the role of boundary cosmological constants. One should notice that in two dimensions the intrinsic boundary geometry is completely determined by its lengths $L_i(g)$.
The partition function for the case of boundaries is given thus by
\beq \label{eq:multiloopZ}
W(\La,Z_1,...,Z_b) =\int_{\Geom(M)} \D[g] e^{-S(g,\La,Z_1,...,Z_b)}
\eeq
and is called the multi-loop amplitude. Instead of fixing the boundary cosmological constants $Z_i$ one could have alternatively also fixed the boundary lengths,
\beq\label{eq:multiloopL}
W(\La,L_1,...,L_b) =\int_{\Geom(M)} \D[g] e^{-S(g,\La)} \prod_{i=1}^b \delta(L_i-L_i(g)).
\eeq 
The length space versions of the multi-loop amplitudes are also called the Hartle-Hawking wave functions \cite{Hartle:1983ai}. For the case of one boundary component we also call it the disc function. One observes that \eqref{eq:multiloopL} and \eqref{eq:multiloopZ} are related by Laplace transforms,\index{Hartle-Hawking wave functions}
\beq
W(\La,Z_1,...,Z_b) =\int_0^\infty \prod_{i=1}^b dL_i e^{-Z L_i} W(\La,L_1,...,L_b).
\eeq

So far we discussed only pure two-dimensional Euclidean quantum gravity, i.e.\ without any coupling to matter fields. If one would like to couple $D$ scalar fields $x^\mu$ one should add the following term in the action
\beq
S_M(g,x)=\int_{M_h} d^2 \xi \sqrt{|g(\xi)|} g^{ab} \frac{\prt x^\mu}{\prt\xi^a} \frac{\prt x^\mu}{\prt\xi^b}.
\eeq
It is interesting to notice that $S_M(g,x)+S(g,\La)$ is precisely the Polyakov action \eqref{eq:Polyakovaction}, discussed in the previous section. Hence we see that pure two-dimensional Euclidean quantum gravity corresponds to string theory in zero-dimensional target space, i.e.\ non-critical string theory.

\index{Liouville field theory}
In the following section we explain how to define the path integral for two-dimensional Euclidean quantum gravity using a discretization scheme called dynamical triangulations. Before doing so let us briefly mention a different way of computing the gravitational path integral using continuum methods. This is done by fixing the metric to the conformal gauge\index{Conformal gauge}
\beq
g=e^{\phi} \hat{g},
\eeq
where $\hat{g}$ is the reference metric. One sees that the theory is now expressed in terms of a single field, the so-called Liouville field $\phi$. There are interesting analytical results for Liouville field theory on the quantized level (see for example \cite{Ginsparg:1993is,DiFrancesco:1993nw}). Further, whenever the continuum theory and the discretized theory (as described in the next chapter) can be compared the results agree \cite{Martinec:2003ka}. However, contrary to what one might have expected the discretized theory allows one to compute many quantities which are not accessible from the continuum theory.

\chapter{Two-dimensional dynamical triangulations \label{Chap:dt}}
	In this chapter we introduce dynamical triangulations as a discretization of the Euclidean gravitational path integral. In the special case of two dimensions analytical results can be obtained. In particular, we show how both the Hartle Hawking wave function as well as the gravitational propagator can be calculated explicitly within this approach. The results agree with those of the continuum formulation of two-dimensional Euclidean quantum gravity through Louiville theory as briefly mentioned in the previous chapter.

\section{Geometry from simplices}\label{sec:geometry}

\index{Regge calculus}\index{Simplicial quantum gravity}
The use of discrete approaches to quantum gravity has a long history \cite{Loll:1998aj}. In classical general relativity, the idea of approximating a space-time manifold by a triangulation of space-time goes back to the early work of Regge \cite{Regge:1961px} and was first used in \cite{Rocek:1982fr,Frohlich:1994gc} to give a path integral formulation of gravity. By a triangulation we mean a piecewise linear space-time obtained by a gluing of simplicial building blocks. This one might think of as the natural analogue of the piecewise linear path we used to describe the relativistic particle. In two dimensions these simplicial building blocks are \textit{flat} Euclidean triangles\footnote{In the next chapter, when discussion two-dimensional causal dynamical triangulations, we will consider a similar construction with Minkowskian triangles.}, where flat means isomorphic to a piece of Euclidean space respectively. One could in principle assign a coordinate system to each triangle to recover the metric space $(M,g_{\mu\nu})$, but the strength of this ansatz lies just in the fact that even without the use of coordinates, each geometry is completely described by the set of edges length squared $\{l_i^2\}$ of the simplicial building blocks. This provides us with a regularized parametrization of the space of all geometries $\Geom(M)$ in a diffeomorphism invariant way, and hence, is the first step towards defining the gravitational path integral. 

\begin{figure}[t]
\begin{center}
\includegraphics[width=4.5in]{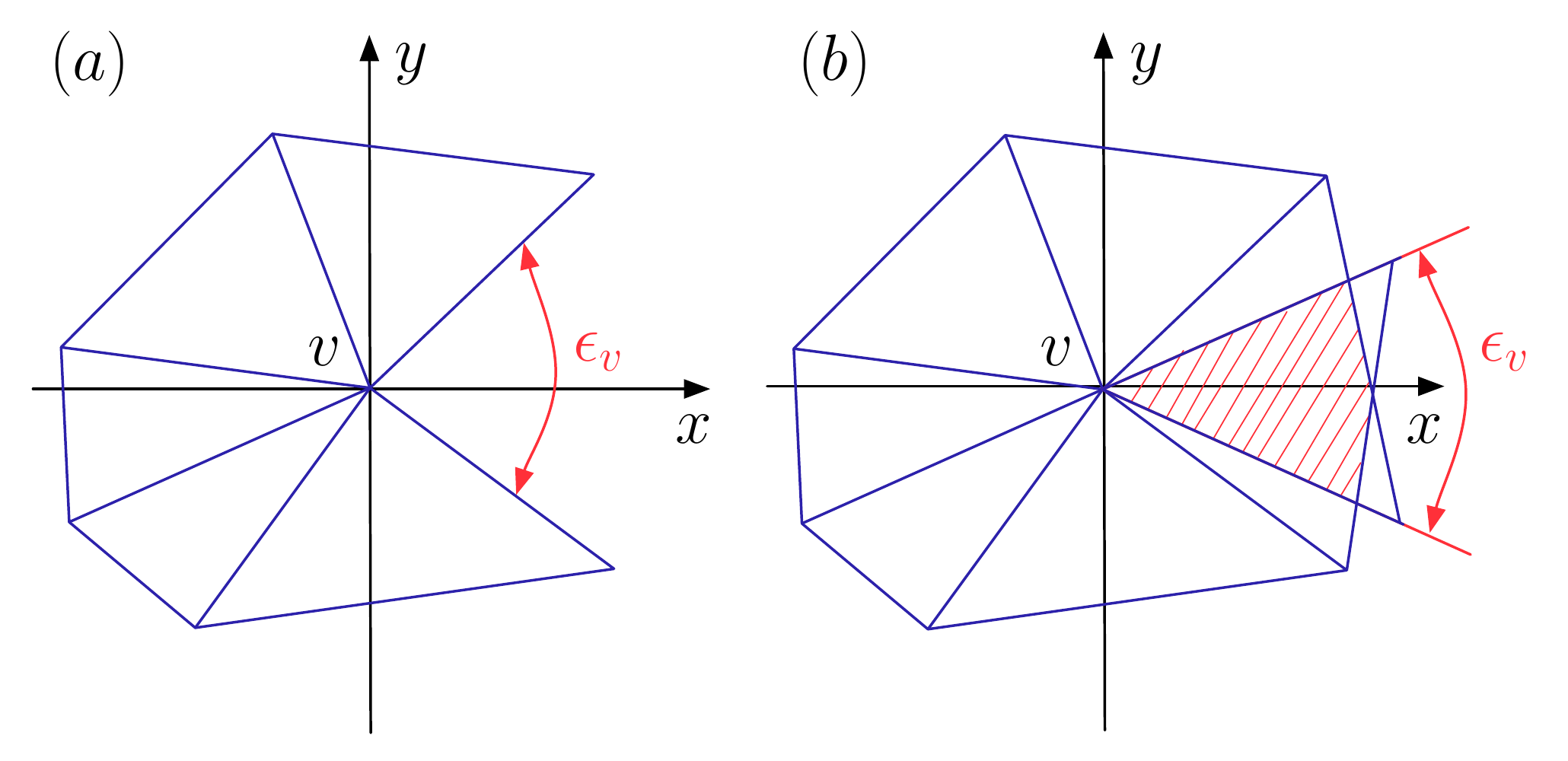}
\caption{Illustration of a positive (a) and negative (b) deficit angle $\epsilon_v$ at a vertex $v$.}
\label{fig:curvature}
\end{center}
\end{figure}

\begin{figure}[t]
\begin{center}
\includegraphics[width=2.5in]{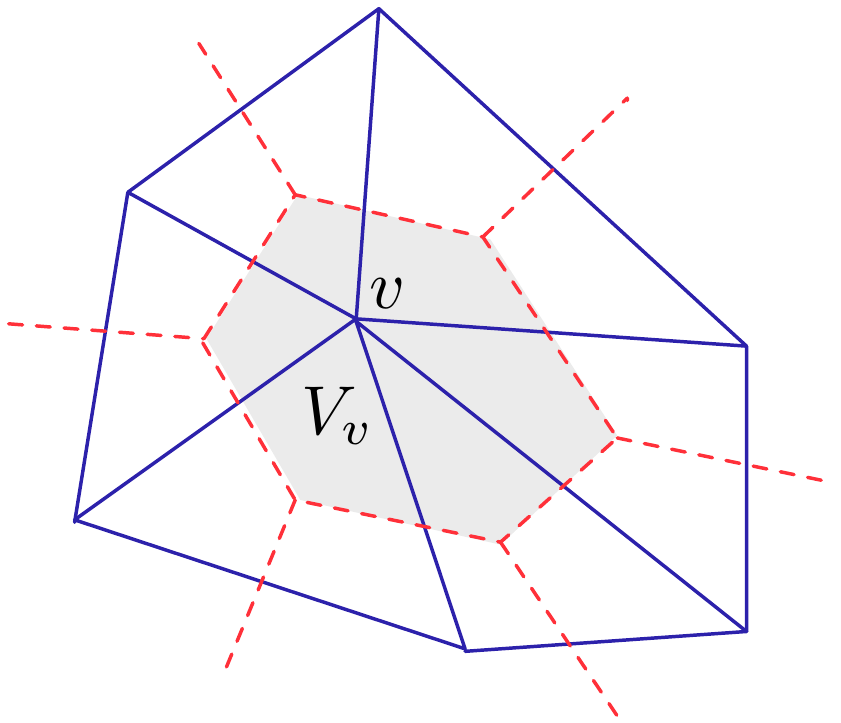}
\caption{Shown is a triangulation and its dual triangulation (dashed line). Highlighted is a vertex $v$ of the triangulation and its dual cell of volume $V_v$.}\index{Triangulation!dual}
\label{fig:curvature2}
\end{center}
\end{figure}

For the further discussion it is essential to understand how a geometry is encoded in the set of edges length squared $\{l_i^2\}$ of the simplicial building blocks of the corresponding triangulation. In the case of two dimensions this can be easily visualized, since the triangulation consists just of triangles. Further, the Riemann scalar curvature $R(x)$ coincides with the Gaussian curvature $K(x)$ up to a factor of $1/2$. There are several ways to reveal curvature of a simplicial geometry. The most convenient method is by parallel transporting a vector around a closed loop. Since all building blocks are flat, a vector parallel transported around a vertex $v$ always comes back to its original orientation unless the angles $\theta_i$ of the surrounding triangles do not add up to $2 \pi$, but differ by the so-called \textit{deficit angle} $\epsilon_v=2\pi-\sum_{i\supset v}\theta_i$ (Figure \ref{fig:curvature}). The Gaussian curvature located at the vertex $v$ is then given by\index{Deficit angle}
\begin{equation}
\label{eq:simpl:gauss}
K_v=\frac{\epsilon_v}{V_v},
\end{equation}
where $V_v$ is the volume associated to the vertex $v$; more precisely, the dual volume of the vertex as shown in Fig.~\ref{fig:curvature2}. Note that the curvature at each vertex takes the form of a conical singularity.
Then one can write the simplicial discretization of the usual curvature and volume terms appearing in the two-dimensional Einstein-Hilbert action,
\begin{eqnarray}
\frac{1}{2}\int_M d^2x \sqrt{|\det g|}\, R(x) & \rightarrow & \sum_{v\in \mathcal{R}} \epsilon_v, \label{eq:simpl:eps}\\
\int_M d^2x \sqrt{|\det g|}  & \rightarrow &  \sum_{v\in \mathcal{R}} V_v ,\label{eq:simpl:A}
\end{eqnarray}
where $\mathcal{R}\equiv \{l_i^2\}$ is a triangulation of the manifold $M$ described by the set of edge lengths squared. From this one can write down the simplicial discretization of the two-dimensional Einstein-Hilbert action, the so-called \emph{Regge action},
\begin{equation}\label{eq:simpl:reggeaction}\index{Regge action}
S_{\mathrm{Regge}}(\mathcal{R})=  \sum_{v\in \mathcal{R}} V_v \left(\lambda-k\,\frac{\epsilon_v}{V_v} \right),
\end{equation}
where $k$ is the inverse Newton's constant and $\lambda$ the cosmological constant. It is then an easy exercise of trigonometry to evaluate the right hand side of \eqref{eq:simpl:reggeaction} in terms of the squared edge lengths.

From this notion of triangulations one can now write the gravitational path integral as the integral over all possible edge lengths $\int\mathcal{D}l$, where each configuration $\mathcal{R}$ is weighted by the corresponding Regge action \eqref{eq:simpl:reggeaction}. In this approach the triangulation is fixed and only the length of the edges is varied. A potential problem one encounters with this ansatz is a possible overcounting of triangulations, due to the fact that one can continuously vary each edge length, as is clear from flat space-time. Further, one still has to introduce a suitable cut-off for the length variable $l$. Among other things, this motivated the approach of ``rigid'' Regge calculus or (Euclidean) dynamical triangulations (DT) where one considers a certain class $\Tri$ of simplicial space-times as an explicit, regularized version of Euclidean geometries, 
where each triangulation $T\in\Tri$ only consists of simplicial building blocks whose edges have all the same edge length squared $L_a^2=a^2$. Here, the geodesic distance $a$ serves as the short-distance cut-off, which will be sent to zero in the continuum limit. 
\index{Rigid Regge calculus}\index{Dynamical triangulations (DT)}\index{Triangulation}

Following early work by Tutte \cite{Tutte1962c,Tutte1962a,Tutte1962b,Tutte1963} and Brezin et al.\ \cite{Brezin:1977sv}, DT have in a quantum gravity context first been introduced as an regularization of the bosonic string \cite{Ambjorn:1985az,Ambjorn:1985dn,David:1985nj,Billoire:1986ab,Kazakov:1985ea}. Further developments have been made in many contributions, for an overview the reader is referred to \cite{Ambjorn:1997di}. 
The motivation for using a discretization of Euclidean geometries instead of Lorentzian geometries is mainly to avoid the need to define a gravitational Wick rotation. One simply starts off with a sum over Euclidean triangulations weighted by the Euclidean (real) Regge action. In the continuum limit this theory of (Euclidean) DT corresponds to two-dimensional Euclidean quantum gravity. It is important to notice that this theory is distinct from two-dimensional Lorentzian quantum gravity which we will introduce in the next chapter.

Putting the above discussion into formulas we can give a definite meaning to the formal continuum path integral of two-dimensional Euclidean quantum gravity as a discrete sum over inequivalent triangulations,
\begin{equation}
\label{eq:simpl:pathsum}
\cZ =\int_{Geom(M)} \D[g_{\mu\nu}] e^{- S_{\mathrm{EH}}[g_{\mu\nu}]} \rightarrow \sum_{T\in\Tri} \frac{1}{C(T)} e^{- S_{\mathrm{Regge}}(T)},
\end{equation}\index{Triangulation!automorphism group}
where $1/C(T)$ is the measure on the space of discrete geometries, with $C(T)=|\mathrm{Aut}(T)|$ the dimension of the automorphism group of the triangulation $T$. 

Now the problem of calculating the gravitational path integral reduces to a statistical physics problem. Here the weight $e^{- S_{\mathrm{Regge}}(T)}$ is the so-called Bolzmann factor and the sum together with the measure factor determines the entropy of configurations with equal weight. In the following section we solve this statistical problem analytically for the two-dimensional model.

\section{The disc function}
\index{Dynamical triangulations (DT)!disc function}

In this section we calculate the disc function or Hartle-Hawking wave function of two-dimensinal Euclidean quantum gravity defined through DT. The disc function is defined as a sum over all triangulations with topology of a disc with fixed boundary of length $l$. In other words the disc function  describes the amplitude for the creation of a universe from nothing \cite{Hartle:1983ai}. Although rather ``global'' it is a simple observable which can be used to calculate other interesting observables.  

\subsection{Discrete solution}\label{subsec:discdiscreteDT}

As we are in two space-time dimensions we have seen that the curvature term in the Einstein-Hilbert action is a purely topological quantity. Hence, if we are interested in computing sums over triangulations of fixed space-time topology, the contribution from the curvature term reduces to an irrelevant normalization constant. 

The Regge action \eqref{eq:simpl:reggeaction} is thus simply given by
\beq
S_{\mathrm{Regge}}(T)=  \l \sum_{v\in T} V_v= \l a^2 N(T),
\eeq
where $\l$ is the bare (dimensionfull) cosmological constant and $N(T)$ is the number of triangles in the triangulation $T$. The factor of $a^2$ comes from the volume of a single triangle where the order one constant has been absorbed into $\l$.

\begin{figure}[t]
\begin{center}
\includegraphics[width=5in]{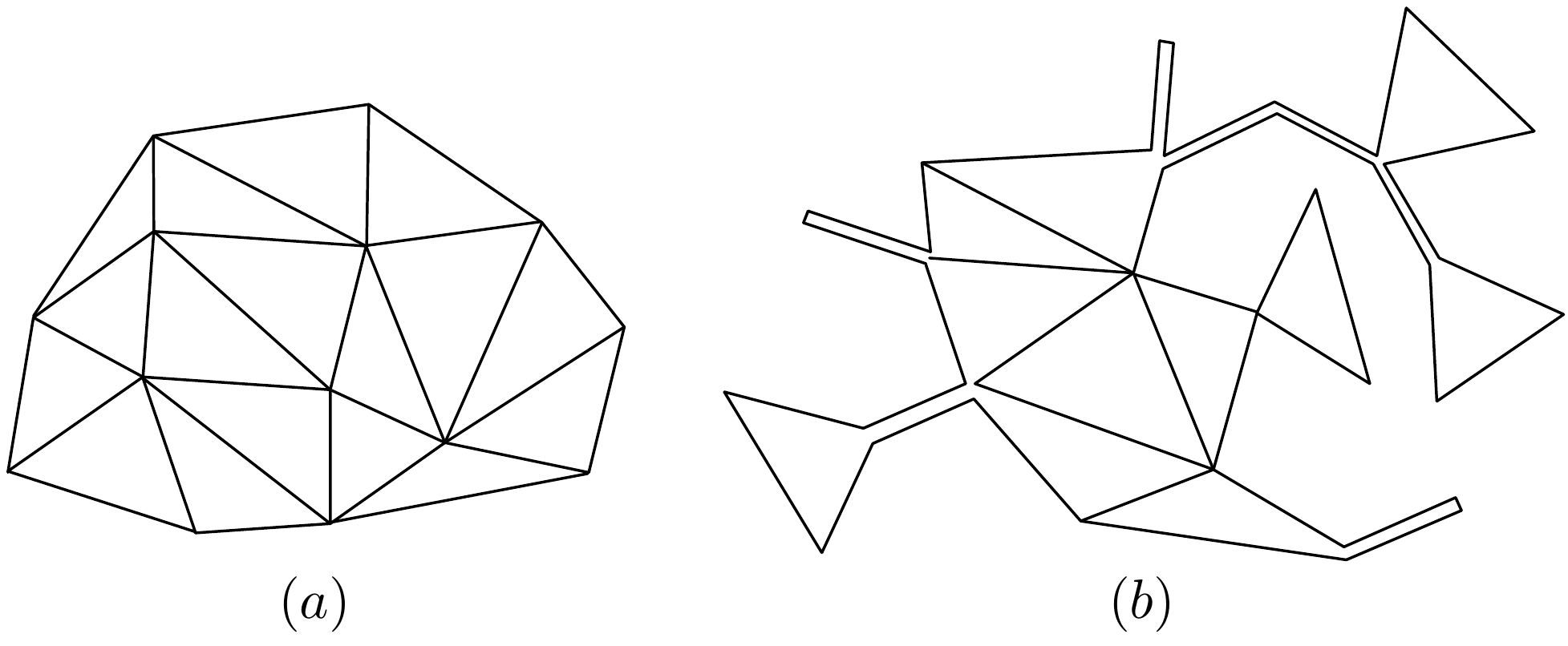}
\caption{Shown is a regular triangulation with topology of the disc $(a)$ and a unrestricted triangulation with topology of a disc $(b)$.}
\label{fig:disc}
\end{center}
\end{figure}

The expression for the disc function $w_l(g)$ now becomes
\beq\label{eq:DTdiscdefi}
w_l (g)=\sum_{T\in\Tri(l)} e^{- \l a^2 N(T)}=\sum_{T\in\Tri(l)} g^{N(T)},
\eeq
where we defined $g=e^{- a^2 \l}$ as the fugacity of a triangle. The sum is taken over all triangulations with boundary length $l$ and a mark on one boundary edge. The mark is introduced for convenience to remove a symmetry factor of $1/l$ in the sum which corresponds to the factor $1/C(T)$ in the above expression. We can further write \eqref{eq:DTdiscdefi} as
\beq
w_l(g)=\sum_{N} w_{l,N} g^N,
\eeq
where $w_{l,N}$ is the number of triangulations with $l$ boundary edges and $N$ triangles. We now see that the problem of determine the disc function reduces to a purely combinatorial problem, namely finding the number $w_{l,N}$ of all possible triangulations with $l$ boundary edges and $N$ triangles. Fig.~\ref{fig:disc} (a) shows an example of one triangulation with $l\equ 9$ boundary edges and $N\equ 16$ triangles. This combinatorial problem was first studied by Tutte in 1962 \cite{Tutte1962c,Tutte1962a,Tutte1962b,Tutte1963}.

\index{Triangulation!non-restricted}
In the following we will not follow the procedure by Tutte but rather count the number of all possible triangulations of a wider class, so-called \emph{non-restricted} triangulations. Non-restricted triangulations are more general than the ones considered above in the sense that they can include double links (see Fig.~\ref{fig:disc} (b)). The double links are mainly included to facilitate the calculation and to be able to map the results to those of matrix models. The use of matrix models will be an important part of this thesis and we will discuss them in detail at a later stage. The justification for being able to include double links is that when taking the continuum limit the contribution coming from the double links will be ``washed away'' as they do not carry a factor of $g$. Further, using the results of Tutte it can be checked explicitly that both regular and non-restricted triangulations lead to the same continuum expression for the disc function (see for example \cite{Ambjorn:1997di}).

When trying to solve a combinatorial problem it is often useful to introduce generating functionals,
\beq \label{generatingDT}
\tilde{w}(g,x)=\sum_{l=0}^\infty w_l(g) x^l.
\eeq
Here $x$ can be understood as the fugacity of a boundary edge or in other terms as being related to a boundary cosmological constant $\l_i$ via $x=e^{-a \l_i}$.

Generating functionals are very useful tools especially since there are helpful to find recursion equations for $\tilde{w}(g,x)$, the so-called loop equations. Let us first make the observation that whenever one removes triangles from a triangulation for each volume decrement of one triangle the generating function is decreased by a factor of $g$ and for each decrement in boundary length it is decreased by a factor of $x$. 

\begin{figure}[t]
\begin{center}
\includegraphics[width=5.5in]{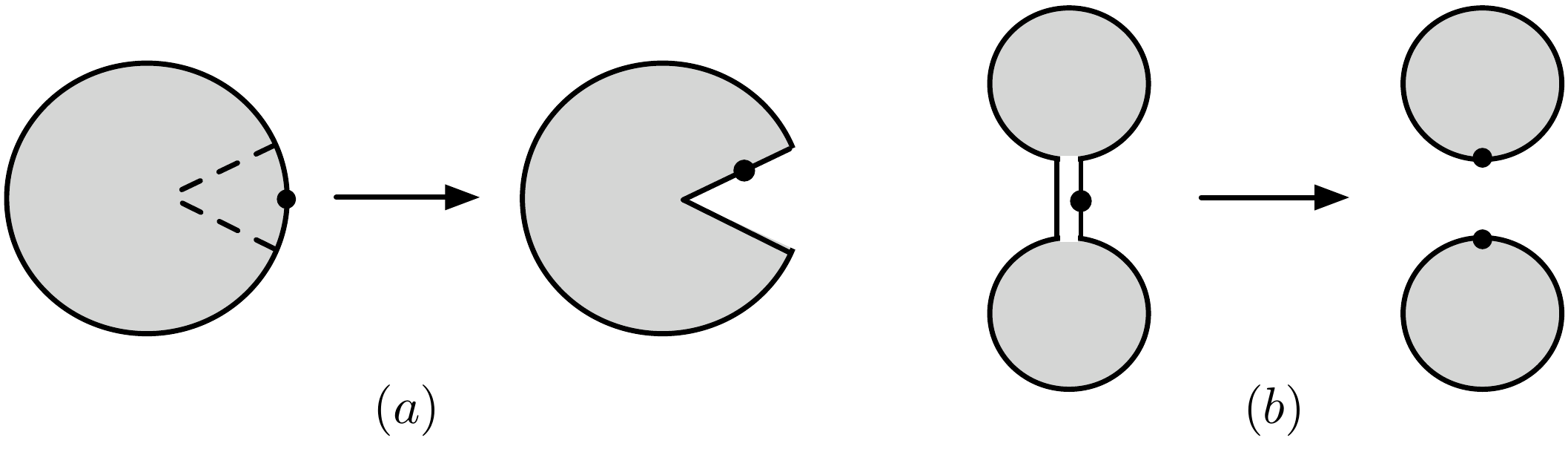}
\caption{Elementary decomposition moves: If the marked edge on the boundary belongs to a triangle this triangle is removed $(a)$, if the mark corresponds to a double link the double link is removed and the triangulation splits into two $(b)$.}
\label{fig:movesDT}
\end{center}
\end{figure}

Starting with a triangulation one can now uniquely decompose it step-by-step to a single dot by the following two elementary moves: Firstly, if the marked edge on the boundary belongs to a triangle this triangle is removed (see Fig.~\ref{fig:movesDT} (a)). Thus one boundary edge is removed and two new edges are created of which by convenience the one further counterclockwise is marked again. In this move the power of $g$ in $\tilde{w}(g,x)$ is decreased by one and the power of $x$ is increased by one. The second move corresponds to the case where the mark belongs to a double link (see Fig.~\ref{fig:movesDT} (b)). In this case the double link is removed and the triangulation splits into two, each with a mark next to where the double link was attached.

\begin{figure}[t]
\begin{center}
\includegraphics[width=4in]{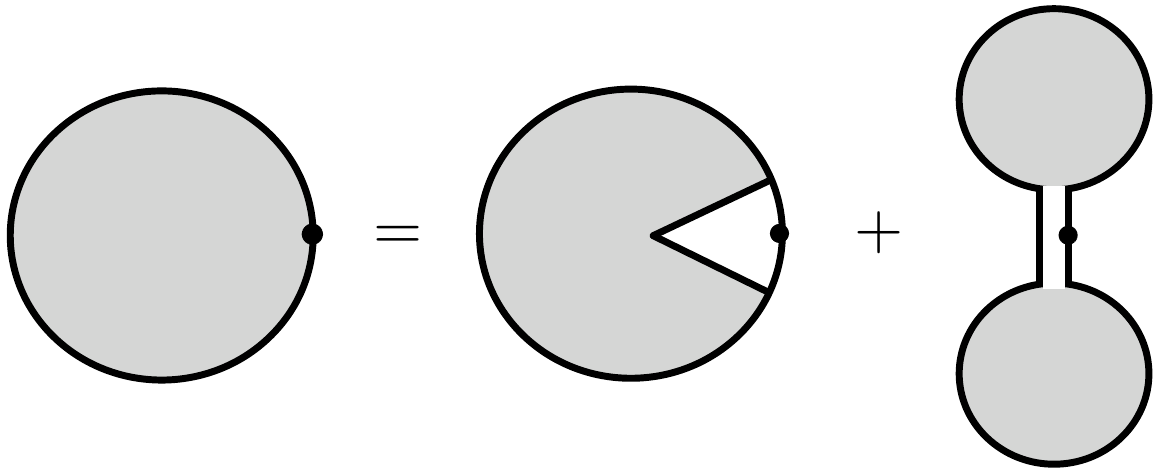}
\caption{Graphical representation of the loop equation \eqref{eq:loop1} for the disc function.}
\label{fig:loopeqDT}
\end{center}
\end{figure}

The step-by-step procedure described above leads to the following recursion relation (see Fig.~\ref{fig:loopeqDT}) \index{Dynamical triangulations (DT)!loop equation}
\beq \label{eq:loop1}
\tilde{w}(g,x)= \frac{g}{x}\left(\tilde{w}(g,x) - w_1(g) x- 1 \right)+ x^2 \tilde{w}^2(x,g),
\eeq
where the first term corresponds to the first move and the last term to the second move. The term $w_1(g) x+ 1$ in the bracket is subtracted to be in accordance with the initial conditions of the the recursion relation. One can now solve for $\tilde{w}(g,x)$, however, the solution is implicit since $w_1(g)$ still has to be determined. 

\begin{figure}[t]
\begin{center}
\includegraphics[width=4in]{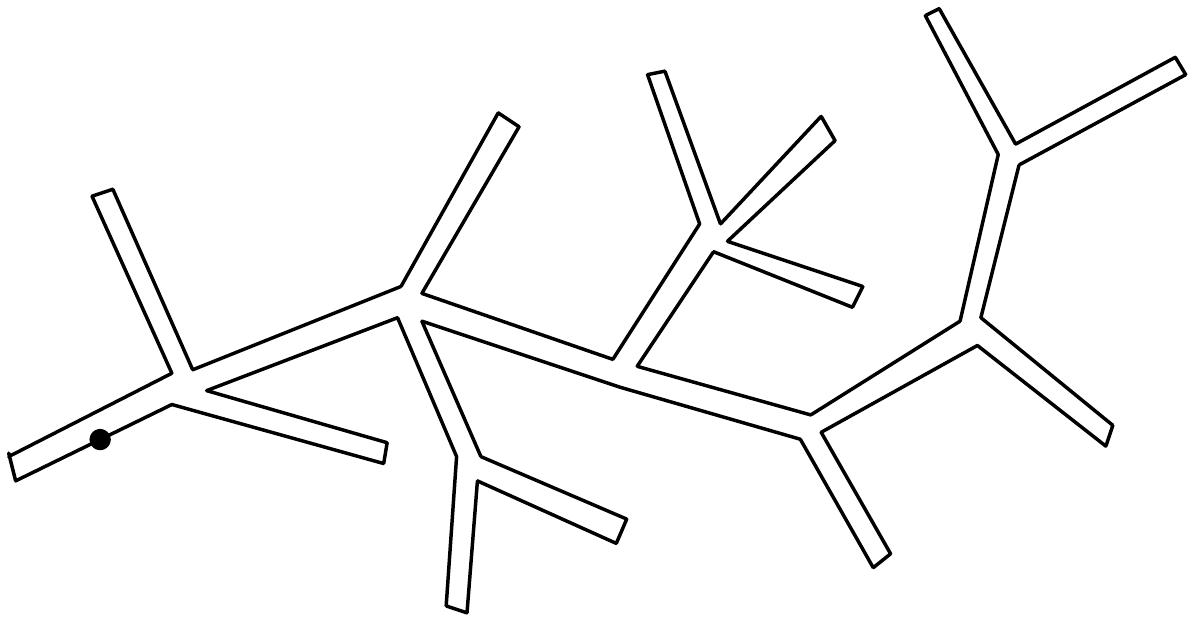}
\caption{A rooted branched polymer created by gluing double links with one marked link.}
\label{fig:BP}
\end{center}
\end{figure}

To be in accordance with the notation adopted in matrix models and also in later parts of this thesis one often rewrites this equation using a slightly different definition of the generating functional\index{Generating functional}
\beq \label{eq:defgenDT}
w(g,z)=\sum_{l=0}^\infty w_l(g) z^{-l-1}=\frac{\tilde{w}(g,1/z)}{z}.
\eeq
Here $z^{-1}$ takes the roles of the fugacity of a boundary edge. The formal solution to the recursion relation \eqref{eq:loop1} reads in terms of the newly defined generating function
\beq\label{eq:disc:sol1}
w(g,z)=\frac{1}{2}\left(V'(z) - \sqrt{(V'(z))^2 -4 Q(z)} \right),
\eeq
where anticipating the discussion of matrix models in later chapters we introduced the notation
\beq
V'(z)=z- g z^2,\quad Q(z)=1-w_1(g) -gz.
\eeq
The sign in front of the square root is determined by the initial condition $w_{0,0}=1$ which leads to the requirement $w(g,z)=1/z + \mathcal{O}(1/z^2)$.
In principle one would now have to find a combinatorial solution for $w_1(g)$, however, there exists a way of determining an explicit solution for $w(g,z)$ by making use of the analytic structure of \eqref{eq:disc:sol1}.
Therefore we first solve \eqref{eq:disc:sol1} for the limiting case of $g=0$, where only double links are present. In this case $w(z)\equiv w(g\!=\!0,z)$ corresponds to the generating function of rooted branched polymers (see Fig.~\ref{fig:BP}). The implicit solution \eqref{eq:disc:sol1} becomes explicit in this special case \index{Branched polymers}
\beq \label{eq:BPgen}
w(z)=\frac{1}{2}\left(z-\sqrt{z^2-4} \right).
\eeq
This is precisely the generating function of the Catalan numbers\index{Catalan numbers}
\beq
w(z)=\sum_{l=0}^\infty \frac{w_{2l}}{z^{2l+1}},\quad w_{2l}=\frac{(2l)!}{(l+1)!l!}.
\eeq
Hence we see that the number of rooted branched polymers with $l$ links is given by the Catalan numbers $w_{2l}$.

Anticipating the general solution for $g \!\neq\! 0 $ it is instructive to analyze the analytic structure of \eqref{eq:BPgen}. The generating function $w(z)$ is analytic in $\mathbb{C}$ with a branch cut on the real axis along the interval $[c_{-}(0),c_{+}(0)]\equiv [-2,2]$. The endpoints of the cut precisely determine the radius of convergence of $w(z)$ which is needed to find the critical points when taking the continuum limit later on. 

The key to finding the solution for the general case of $g \!\neq\! 0 $ is to notice that the analytic structure of $w(g,z)$ as a function of $z$ cannot change discontinuously at $g\!=\!0$. For $g\!=\!0$ the branch cut is determined by the following expression under the square root
\beq
(V'(z))^2 -4 Q(z)=(z-c_{+}(0))(z-c_{-}(0))=(z-2)(z+2)
\eeq
For $g \!\neq\! 0 $ this expression becomes fourth order in $z$,
\bea
(V'(z))^2 -4 Q(z)=\left[ z-(2+2g)+ \mathcal{O}(g^2)\right] \times\nonumber\\
\left[ z+(2-2g)+ \mathcal{O}(g^2)\right]\left[ gz -(1-2g^2)+ \mathcal{O}(g^3)\right]^2.
\eea
Since the analytic structure of $w(g,z)$ should not change in the neighborhood of $g\!=\!0$, this implies that the term under the square root must be of the general form
\beq\label{eq:cpcmeq}
(V'(z))^2 -4 Q(z)=(c(g)-g z)^2(z-c_{+}(g))(z-c_{-}(g)),
\eeq
where we labeled the roots such that $c_{-}(g)\leq c_{+}(g) \leq c(g)$. The solution for the disc function then reads
 \beq \label{eq:DTdiscdiscrete}
w(g,z)=\frac{1}{2}\left(z- g z^2- (c(g)-g z)\sqrt{(z-c_{+}(g))(z-c_{-}(g))} \right).
\eeq
Matching the coefficients of \eqref{eq:cpcmeq} in the expansion in $z$ yields four equations which completely determine $c_{-}(g)$, $c_{+}(g)$, $c(g)$ and $w_1(g)$. This gives the full combinatorial solution for the disc function in DT.

\subsection{Continuum limit}\label{sec:contlimDT}
\index{Dynamical triangulations (DT)!continuum limit}

Having solved the gravitational path integral as a statistical sum over triangulations we now have to perform the continuum limit to extract the continuum physics. This is done by taking the lattice spacing $a$ to zero while at the same time taking the number of triangles $N$ and the discrete boundary length $l$ to infinity, such that physical area $A$ and length $L$ stay finite.

In terms of critical phenomena the continuum limit relates to a fine-tuning of the coupling constants, generally denoted by $g$, to the critical point $g_c$ of the phase transition. This means a divergence of the correlation length $l$ in the scaling limit
\begin{equation}\index{Critical phenomena}
l(g)\sim\frac{1}{|g-g_c|^\nu}\xrightarrow[]{g\rightarrow g_c} \infty,
\end{equation}
while the cut-off $a$ goes to zero as
\begin{equation}
a(g)\sim\ |g-g_c|^\nu \xrightarrow[]{g\rightarrow g_c} 0,
\end{equation}
where $\nu$ is a critical exponent.\index{Critical exponent}\index{Correlation length}
In terms of the cut-off $a$ and the correlation length $l$, the physical correlation length is given by $L\equ l(g)\cdot a(g)$ and the continuum limit is defined as the simultaneous limit
\begin{equation}
a(g)\rightarrow 0,\quad l(g)\rightarrow \infty \quad\text{for fixed}\quad L=l(g)\cdot a(g).
\end{equation}

From standard techniques of renormalization theory, we expect the couplings with positive mass dimension, i.e. the cosmological constant $\lambda$ and the boundary cosmological constant $\lambda_{i}$ to undergo an additive renormalization,\index{Additive renormalization}
\begin{equation}
\lambda=\frac{A_\l}{a^2}+\Lambda,\quad
\lambda_{i}=\frac{A_{\l_i}}{a}+Z,
\end{equation}
where $\Lambda$, $Z$ denote the corresponding renormalized values.
Introducing the critical values \index{Dynamical triangulations (DT)!scaling relations}
\begin{equation}
g_c=e^{-A_{\l}},\quad z_c=e^{A_{\l_i}},
\end{equation}
it follows that
\begin{equation}\label{eq:scalingrelationsDT}
g=g_c \,e^{-a^2\,\Lambda},\quad z=z_c \,e^{a\,Z}.
\end{equation}

The critical point for the coupling $g$ can be extracted from the radius of convergence of $w_1(g)$ at which the number of triangles $N$ diverges. Using the result for $w_1(g)$ which can be derived easily as described below \eqref{eq:DTdiscdiscrete} one finds $g_c=1/\sqrt[4]{432}$. To obtain the critical value for $z$ at which also the boundary length diverges we look at the radius of convergence of $w(g_c,z)$ as a function of $z$. From the analysis of the analytic structure we know that the radius of convergence is precisely given by the endpoints of the branch cut and hence we conclude that $z_c=c_+(g_c)$ (in the following we omit the precise numerical values of the critical couplings, since they are not of essential importance for the discussion).    

Inserting the scaling relations \eqref{eq:scalingrelationsDT} and the critical values into the discrete solution of the disc function \label{eq:DTdiscdiscrete} yields
\beq
w(g,z)= w_{ns} + a^{3/2} (Z-\frac{1}{2}\sL)\sqrt{Z+\sL}+\cO(a^2),
\eeq
where $w_{ns}= V'(z)/2$ is the so-called non-scaling part of the disc function which consists of terms of lower order in the cut-off, more precisely an order one constant and a term proportional to $a Z$. Further, for convenience we rescaled $\La$ and $Z$ by positive multiplicative constants to simplify the result. \index{Non-scaling part}

\index{Wave function renormalization}
Introducing a wave function renormalization of the continuum disc function
\beq \label{eq:DTscalingdisceta}
W_\La(Z)=a^{-3/2} w(g,z)
\eeq
we obtain
\beq \label{eq:DTdisc}
W_\La(Z)= (Z-\frac{1}{2}\sL)\sqrt{Z+\sL},
\eeq
where we dropped the non-scaling part. This is precisely the result obtained from Liouville theory. 

We can now also go back to the physical length $L$. The continuum limit of the defining equation for the generating function \eqref{eq:defgenDT} is precisely the Laplace transform,
\beq
W_\La(Z)=\int_0^\infty dL W_\La(L) e^{-L Z}.
\eeq
Hence performing the inverse Laplace transform
\beq
W_\La(L)=\int_{- i\infty}^{+i\infty} \frac{dZ}{2\pi i} W_\La(Z) e^{L Z}
\eeq
of \eqref{eq:DTdisc} we obtain
\beq \label{eq:discLDT}
W_\La(L)=L^{-5/2}(1+\sL L)e^{-\sL L}.
\eeq
 Further we can also calculate the resulting disc function without a mark on the boundary by simply dividing \eqref{eq:discLDT} by $L$, yielding
 \beq \label{eq:discLDT}
W^{(u)}_\La(L)=L^{-7/2}(1+\sL L)e^{-\sL L},
\eeq
where the $(u)$ stands for ``unmarked''.

\section{Geodesic distance and the two-loop amplitude} \label{sec:DTtwo-loop}

\subsection{Defining geodesic distance and the two-loop function}

\index{Geodesic distance}

In the previous section we calculated the disc function of two-dimensional Euclidean quantum gravity using the framework of dynamical triangulations. This is in a sense a rather ``global'' observable and in the following we are interested in calculating more ``local'' observables. The quantity we calculate is the fixed geodesic distance two-loop amplitude which we also sometimes refer to as the gravitational propagator or simply propagator.

To be able to calculate this quantity we first need to make some definitions of basic concepts. Firstly, we have to give a proper definition of geodesic distance. If we would reintroduce coordinates on the simplicial building blocks one could simply employ the standard continuum definition as used in general relativity. However, we have seen that the strength of this approach lies in the fact that we have a coordinate independent representation of geometries. Hence, we should better look for a geodesic distance definition intrinsic to dynamical triangulations. There are several such definitions. The one we employ here defines the geodesic distance $d(e,e')$ between two links (edges) $e$ and $e'$ in a triangulation as the number of links of the shortest path between the corresponding links of the dual triangulation.
Now we want to extend the above definition for the case of geodesic distance of boundaries of the triangulation, where each boundary consists of a collection of links. In particular, we define the geodesic distance of a boundary loop $\cL$ to a link $e$ as $\min_{e'\in \cL}d(e,e')$. Finally, we say that a boundary loop $\cL_1$ has geodesic distance $t$ to another boundary loop $\cL_2$ if all links $l\in\cL_1$ have a geodesic distance $t$ to $\cL_2$. One should notice that this  definition is not symmetric in $\cL_1$ and $\cL_2$. 

\begin{figure}[t]
\begin{center}
\includegraphics[width=4in]{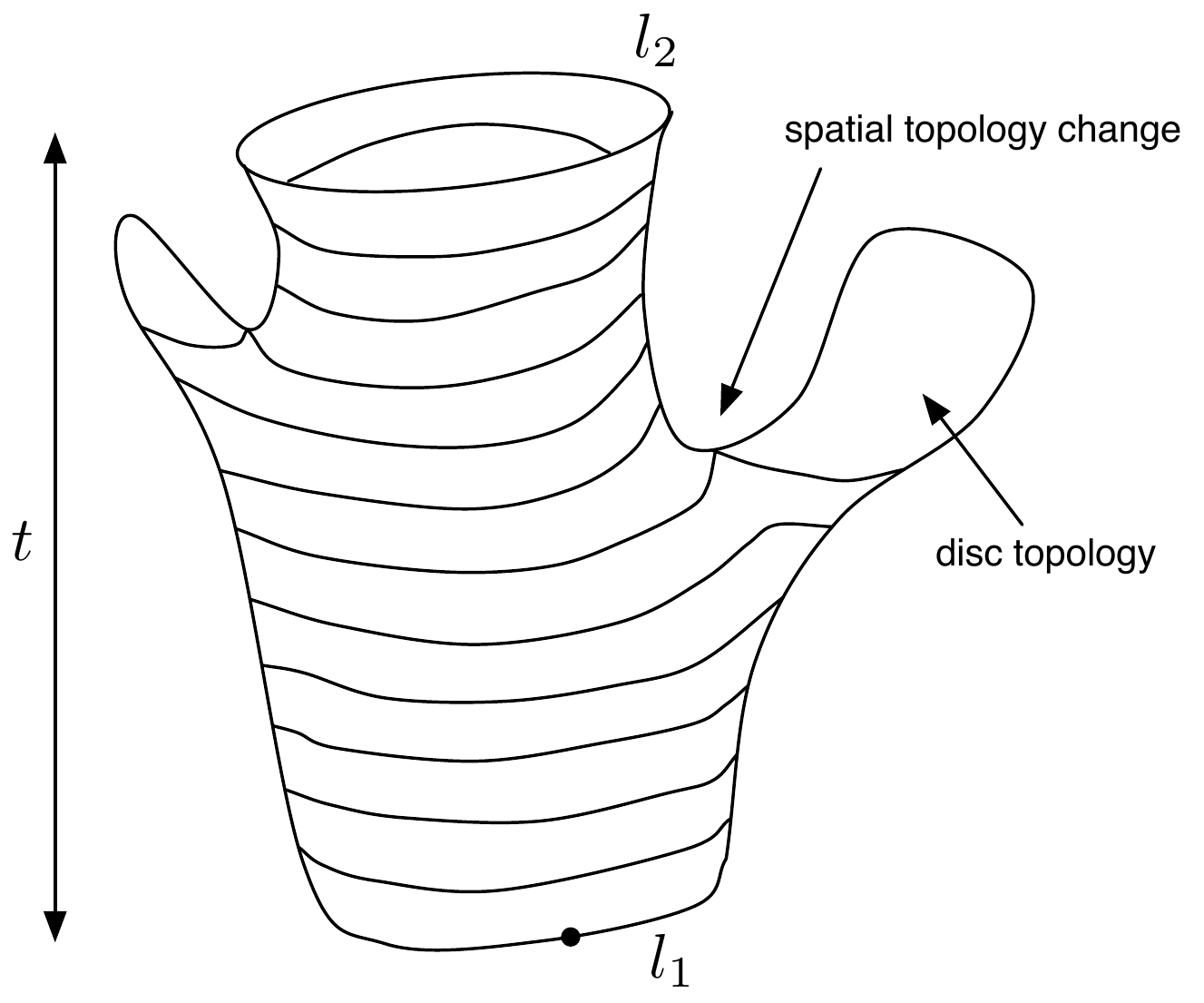}
\caption{Fixed geodesic distance two-loop amplitude $G(l_1,l_2;g;t)$.}
\label{fig:geodesicslicing}
\end{center}
\end{figure}

As an example, Fig.~\ref{fig:geodesicslicing} shows an amplitude with a boundary of length $l_1$ and another boundary of length $l_2$ at geodesic distance $t$ to the first boundary. One observes that the geodesic distance induces a natural foliation or time-slicing for the triangulations by following the sequence of loops of geodesic distance $r=0,1,2,...,t$ to $\cL_1$.\footnote{This notion of time essentially suggests the existence of a Hamiltonian for two-dimensional Euclidean quantum gravity. The formulation of such Hamiltonian dynamics lead to interesting results in the field of string field theory \cite{Ishibashi:1993pc,Ishibashi:1993sv,Ishibashi:1995in,Fukuma:1993tp}. We will not review this work, however, in Chap.\ \ref{Chap:sft} we will propose an analogous string field theory formulation for Lorentzian quantum gravity.} Whereas the space-time topology of the triangulation is fixed to $S_1\times [0,1]$, the topology of the spatial sections changes in general as a function of geodesic distance $r$. We call these \emph{spatial topology changes}. One should notice that spatial topology changes are naturally present in Euclidean dynamical triangulations. However, as we will argue in the next chapter, in a proper theory of Lorentzian quantum gravity the branching points of changing spatial topology relate essentially to points in the continuum geometry of anomalous causal structure. This is one motivation for studying triangulations without spatial topology change as we will do in the next chapter. \index{Spatial topology changes}

We now have all ingredients at hand to define the fixed geodesic two-loop amplitude
\beq \label{eq:defDT2loop}
G(l_1,l_2;g;t)=\sum_{T\in \cT(l_1,l_2,t)} g^{N(T)},
\eeq
where the sum is taken over all triangulations with an initial boundary of length $l_1$ and a final boundary of length $l_2$ at geodesic distance $t$ to the initial boundary. For $t\!=\!0$ the initial and final loop coincide which leads to the following initial condition
\beq \label{eq:initialDT2loop}
G(l_1,l_2;g;t\equ 0)=\delta_{l_1,l_2}.
\eeq
As for the disc function it will be useful to the introduce generating function
\beq
G(z,w;g;t)=\sum_{l_1,l_2=0}^{\infty}G(l_1,l_2;g;t) z^{-l_1-1}  w^{-l_2-1},
\eeq
with the initial condition equivalent to \eqref{eq:initialDT2loop},
\beq \label{eq:initialconditionDT}
G(z,w;g;t\equ 0)=\frac{1}{z w}\frac{1}{z w-1}.
\eeq

Solving \eqref{eq:defDT2loop} is a purely combinatorial problem. In the following we describe two different methods to solve this problem. The first method of solving the geodesic two-loop amplitude is the so-called peeling procedure due to Watabiki \cite{Watabiki:1993ym}. It is very simple and instructive, since it essentially uses the above loop equations and only views them as a ``time-dependent'' process. The disadvantage however is that not all steps of the calculation are rigorous. The second method due to Kawai et al.\ \cite{Kawai:1993cj} uses transfer matrices (see also \cite{Gubser:1993vx,Aoki:1995gi}). Though more complicated it is on the other hand better defined. Further, it introduces important techniques which will be essential to study problems arising in the next chapter.

\subsection{The peeling procedure} \label{sec:peelingDT}
\index{Peeling procedure}

\begin{figure}[t]
\begin{center}
\includegraphics[width=3.5in]{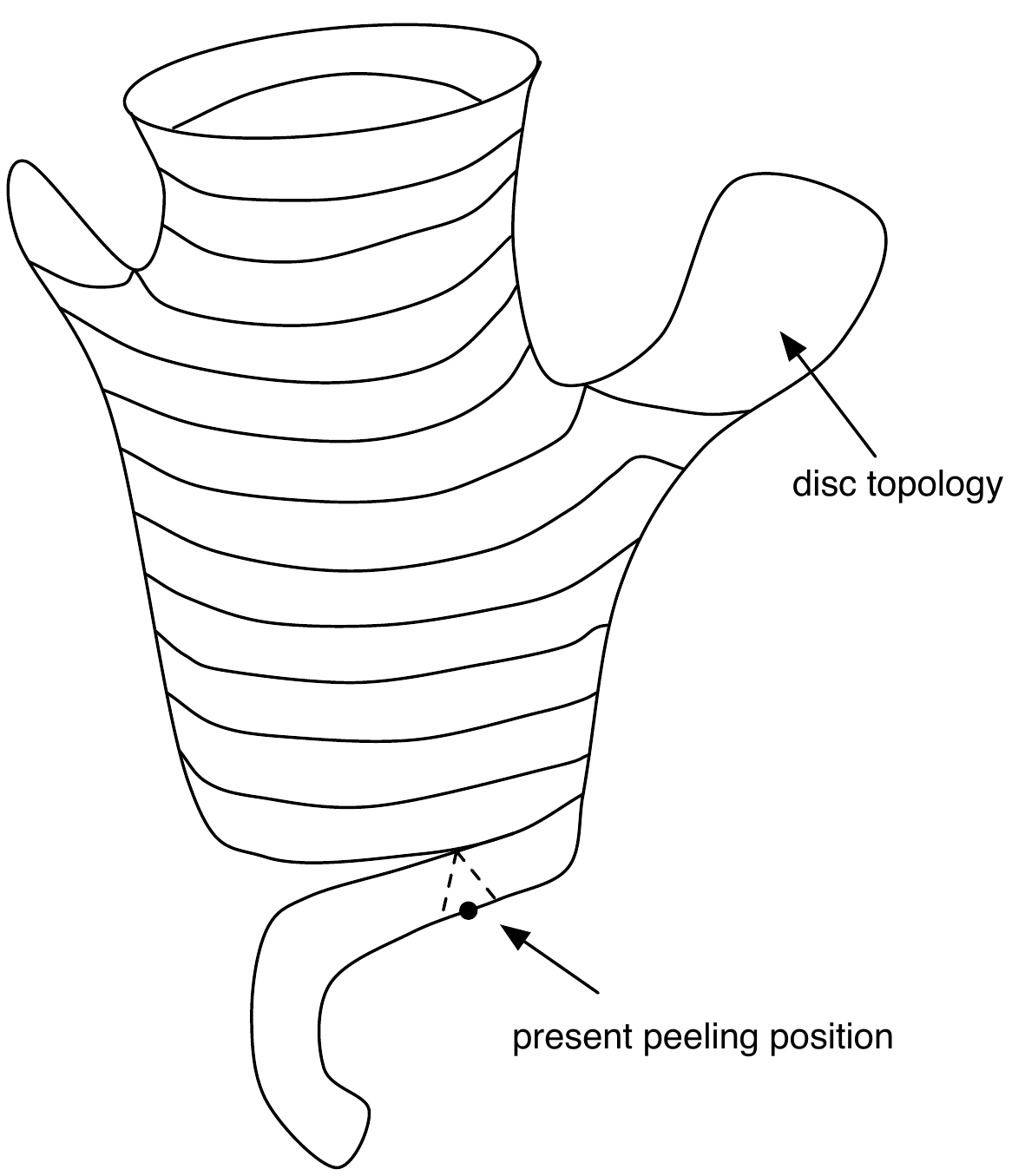}
\caption{Decomposition of the triangulation for the fixed geodesic distance two-loop amplitude by a peeling procedure .}
\label{fig:peeling}
\end{center}
\end{figure}

In Fig.~\ref{fig:geodesicslicing} we illustrate the fixed geodesic distance two-loop amplitude as decomposed in slices of thickness $\Delta r=1$. In the spirit of the loop equation for the disc function, as discussed in the previous section, we can also decompose the triangulation by a peeling procedure, a sequel of elementary decomposition moves (see Fig.~\ref{fig:peeling}). 

\begin{figure}[t]
\begin{center}
\includegraphics[width=6in]{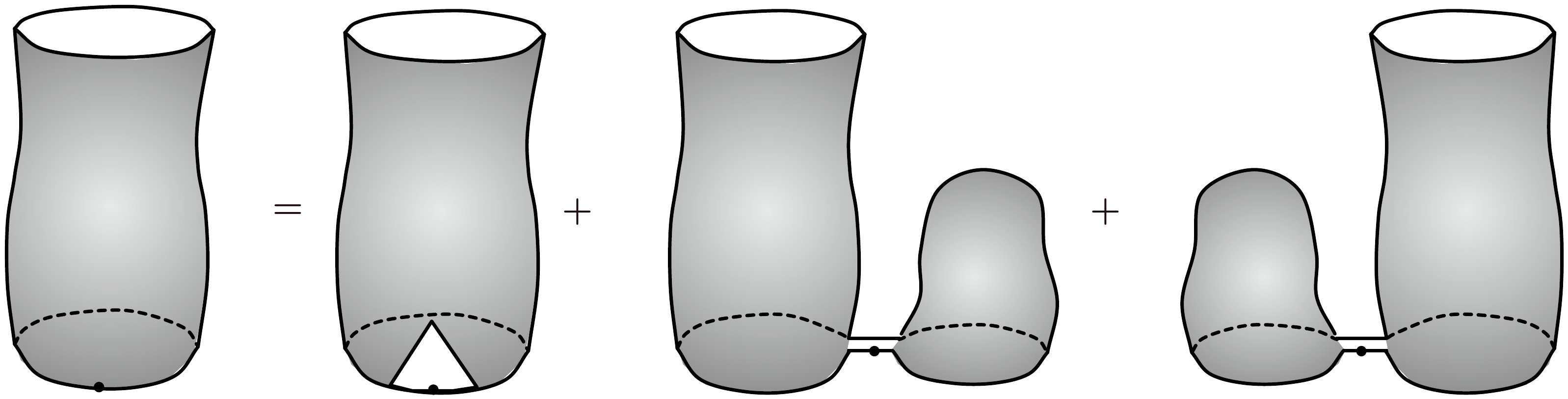}
\caption{The peeling decomposition moves are the same as already used in the loop equation for the disc function (recall Fig.\ \ref{fig:loopeqDT}): If the marked edge on the initial loop belongs to a triangle the triangle is removed. If it belongs to a double link then the double link is removed causing the triangulation to split. The off-splitting part of the triangulation with the final loop has topology of a cylinder, whereas the other part has topology of a disc.}
\label{fig:peelingeq}
\end{center}
\end{figure}

Similar to the loop equation for the disc function we can write down a recursion relation for the geodesic two-loop amplitude (see Fig.\ \ref{fig:peelingeq})
\beq \label{eq:loopeq2loopL}
G'(l_1,l_2;g;t) =g G(l_1+1,l_2;g;t) +2 \sum_{l=0}^\infty w_l(g) G(l_1-l-2,l_2;g;t).
\eeq
Here the two decomposition moves are essentially the same as for the disc function. The factor of two comes from the two possibilities of having the final boundary on either of the two off-splitting triangulations. The prime on the expression on the left-hand-side means that the surfaces contributing to this quantity do not exactly have $\cL_2$ at geodesic distance $t$ from $\cL_1$. 

Rather than giving a precise definition of $G'(l_1,l_2;g;t)$ one makes a proposal which approximates $G'(l_1,l_2;g;t)$ for large $t$. When comparing to the exact transfer matrix calculation in the next section we shall see that this approximation indeed leads to the correct result. The basic idea in approximating $G'(l_1,l_2;g;t)$ is that when applying the elementary decomposition moves $l_1$ times on $G(l_1,l_2;g;t)$ we will on average get $G(l_1,l_2;g;t-1)$, since roughly speaking we peeled off a whole slice. Ignoring the fact that $t$ is an integer we can then view $G'(l_1,l_2;g;t)\!=\!G(l_1,l_2;g;t+1/l_1)$. For large $l_1$ one can Taylor expand 
\beq
G(l_1,l_2;g;t+1/l_1)=G(l_1,l_2;g;t)+1/l_1 \partial_t G(l_1,l_2;g;t)+....
\eeq
 Inserting this into \eqref{eq:loopeq2loopL} gives
\bea\label{eq:DTpeeingsolxx1}
\frac{\partial}{\partial t}G(l_1,l_2;g;t) &=& -l_1 G(l_1,l_2;g;t) +g l_1 G(l_1+1,l_2;g;t)+\non 
&&\,\, + \,\,  2 l_1 \sum_{l=0}^\infty w_l(g) G(l_1-l-2,l_2;g;t),
\eea
or in terms of the generating functions
\beq\label{eq:DTpeeingsolxx2}
\frac{\partial}{\partial t}G(z,w;g;t)=\frac{\partial}{\partial z}\left[ (z-gz^2 -2 w(g,z)) G(z,w;g;t)\right].
\eeq
Plugging in the result for the disc function \eqref{eq:DTdiscdiscrete} we see that the first term in the bracket cancels the non-scaling part of the disc function and we get
\beq \label{eq:geodDTdis}
\frac{\partial}{\partial t}G(z,w;g;t)=-2\frac{\partial}{\partial z}\left[ f(g,z) G(z,w;g;t)\right],
\eeq
where we defined 
\beq
f(g,z)=w(g,z)-\frac{1}{2}(z- g z^2)= \frac{1}{2} (gz-c(g))\sqrt{(z-c_{+}(g))(z-c_{-}(g))}. 
\eeq

Instead of trying to solve the differential equation and then take the continuum limit of the result, we will in the following take the continuum limit directly of the differential equation \eqref{eq:geodDTdis} and its initial condition \eqref{eq:initialconditionDT}. 
Therefore we extend the scaling relations \eqref{eq:scalingrelationsDT} to 
\begin{equation}\label{eq:scalingrelationsDT2}
g=g_c \,e^{-a^2\,\Lambda},\quad z=z_c \,e^{a\,Z},\quad w=z_c \,e^{a\,W},
\end{equation}
using the same critical values as described above. Inserting these relations into the initial conditions \eqref{eq:initialconditionDT} gives \index{Fixed geodesic distance two-loop amplitude} \index{Dynamical triangulations (DT)!propagator}
\beq\label{DT:eqinitial}
G_\La(Z,W;T\equ 0)=\frac{1}{Z+W},
\eeq
where we introduced the following wave function renormalization for the geodesic two-loop amplitude
\beq \label{eq:scalingGDT}
G_\La(Z,W;T)= a G(z,w;g;t).
\eeq
Further, inserting the scaling relations \eqref{eq:scalingrelationsDT2} and \eqref{eq:scalingGDT} into \eqref{eq:geodDTdis} we observe that the geodesic distance should scale as
\beq \label{eq:DTscalingdiscepsilon}
T\sim t \sqrt{a},
\eeq 
yielding the following continuum differential equation for the geodesic distance two-loop amplitude
\beq \label{eq:geodDT}
\frac{\partial}{\partial T}G_\La(Z,W;T)=- \frac{\partial}{\partial Z}\left[ W_\La (Z) G_\La(Z,W;T)\right].
\eeq
One observes that time does not scale canonically as maybe expected. We will see in the next section that this is due to the fractal structure of the quantum geometry.

This partial differential equation can be solved under the above initial condition, yielding 
\beq \label{eq:DTgeodsol}
G_\La(Z,W;T)=\frac{W_\La(\Zbt)}{W_\La(Z)}\frac{1}{Z+W},
\eeq
where $\Zbt$ is the solution to the characteristic equation
\beq \label{eq:DTgeodsolXb}
\dT \Zbt =-W_\La(\Zbt),\quad \Zb(T\equ 0,Z)=Z.
\eeq
Equivalently we can write
\beq
T=\int_{\Zbt}^Z \frac{dX}{W_\La(X)}.
\eeq
One can now insert the solution for the disc function \eqref{eq:DTdisc} into this equation and perform the integral,
\beq
T=-\frac{2}{\a\sqrt[4]{\La}} \left[\atanh\left(\frac{1}{\a\sqrt[4]{\La}}\sqrt{X+\sL}\right) \right]_{X=\Zbt}^{X=Z},
\eeq
where $\a=\sqrt{3/2}$.
Inverting this formula gives
\beq\label{eq:ZbtDT}
\Zbt= \a^2\sL\tanh\left( \frac{\a\sqrt[4]{\La} }{2} T +\atanh\left(\frac{1}{\a\sqrt[4]{\La}}\sqrt{X+\sL}\right) \right)^2-\sL,
\eeq
which together with \eqref{eq:DTgeodsol} yields the explicit solution for $G_\La(Z,W;T)$.

The geodesic two-loop amplitude is a useful quantity from which many other quantities can be calculated. For the discussion of the quantum geometry in the next section we will be especially interested in the dependence on the geodesic distance. For simplicity, we shall analyze this dependence by shrinking both boundaries to zero. The resulting quantity is called the geodesic distance two-point function and was introduced in \cite{Ambjorn:1995dg}. It is much simpler than the full two-loop amplitude, but still contains all essential dependence on the geodesic distance.

Let us first shrink the outer boundary to zero. Performing the inverse Laplace transform we get
\beq
G_\La(Z,L_2;T)=\frac{W_\La(\Zbt)}{W_\La(Z)} e^{-L Z}
\eeq
and hence
\beq\label{eq:GLeq0DT}
G_\La(Z,L_2=0;T)=\frac{W_\La(\Zbt)}{W_\La(Z)}
\eeq
One could have also obtained this result directly from the Laplace transformed quantity \eqref{eq:DTgeodsol} by taking $\lim_{W\to\infty} W^{\beta} G_\La(Z,W;T)$, where the coefficient is $\beta\equ 1$ which lead to a finite expression. The quantity $G_\La(Z,L_2=0;T)$ is in a sense similar to $W_\La(Z)$ with the difference that it has a marked point in the bulk which has a geodesic distance $T$ to the entrance loop. This marked point comes from the boundary which was shrunken to zero. If we would integrate over $T\in[0,\infty]$ this expression should correspond to the disc function with a marked point anywhere in the bulk, given by $-\partial_\La W_\La(Z)$. A quick check reveals that this is indeed true, i.e. that
\beq \label{eq:DTdiscproprel}
-\frac{\partial}{\partial \La}W_\La(Z) \sim \int_0^\infty dT G_\La(Z,L_2=0,T).
\eeq
From \eqref{eq:GLeq0DT} we then get for the two-point function $G_\La(T):=G_\La(L_1\!=\!0,L_2\!=\!0,T)$,\index{Dynamical triangulations (DT)!two-point function}
\beq
G_\La(T)=\lim_{Z\to\infty} Z^\beta \frac{W_\La(\Zbt)}{W_\La(Z)}= W_\La(\Zb(T,Z=\infty)),
\eeq
where we chose $\beta=3/2$ to get a finite expression.\footnote{The different values for $\beta$ when shrinking the initial and final loop to a point (i.e. $3/2$ and $1$) come from the fact that the initial loop was marked and the final loop not.} Using \eqref{eq:ZbtDT} one gets
\beq \label{eq:two-pointDT}
G_\La(T)=(\a\sqrt[4]{\La})^3\frac{\cosh\frac{\a}{2} \sqrt[4]{\La} T }{\sinh^3\frac{\a}{2} \sqrt[4]{\La} T}
\eeq

\subsection{The transfer matrix approach}

In this subsection we want to rederive the continuum result for the fixed geodesic distance two-loop function as obtained by \eqref{eq:DTgeodsol} in the previous subsection using the so-called \emph{transfer matrix} approach. This exact method is more rigorous than the peeling method described in the previous subsection. Further, it employs useful techniques which will be of importance in the forthcoming chapters.

It was argued in \cite{Kawai:1993cj} that any triangulation $\cT(l_1,l_2,t)$ can in principle be uniquely decomposed into triangulations of geodesic distance one, i.e.\ $\cT(l_1,l_2,t\equ1)$, by slicing it in the way described above. This situation is shown in Fig.~\ref{fig:geodesicslicing}. 

\begin{figure}[t]
\begin{center}
\includegraphics[width=3in]{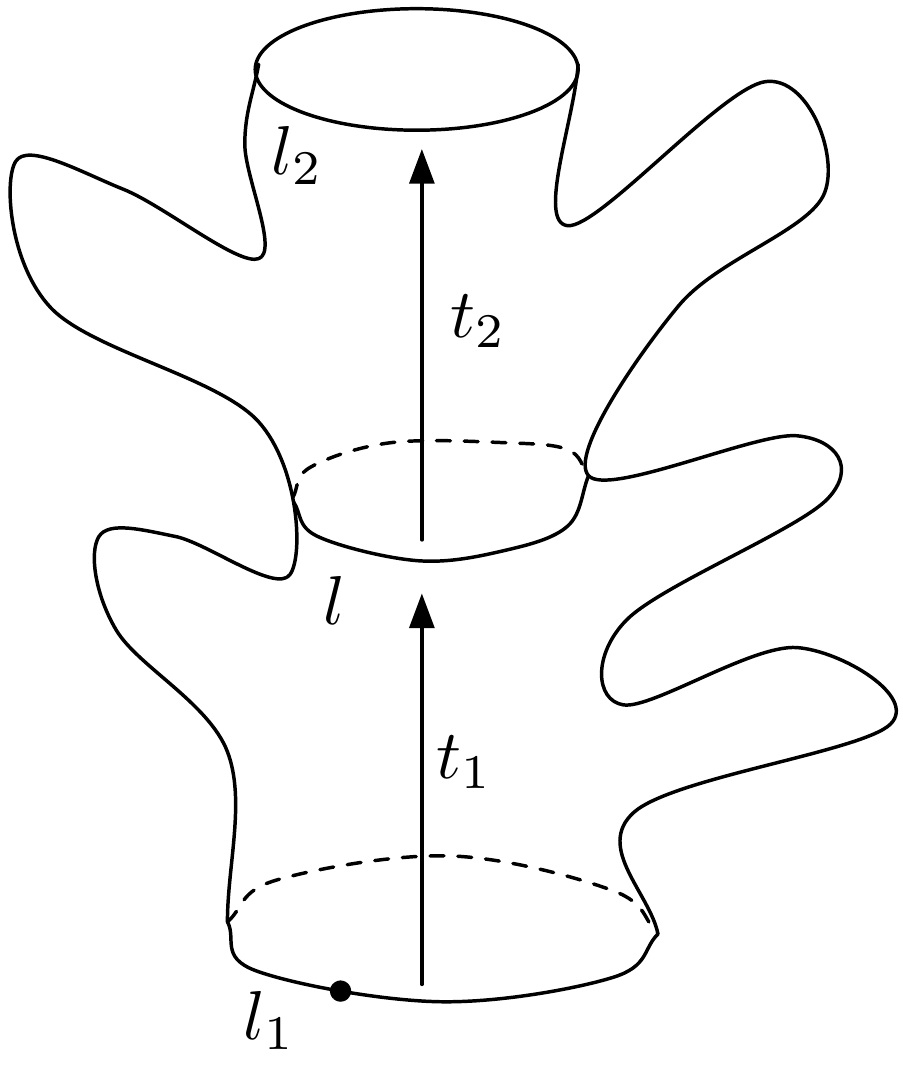}
\caption{Graphical illustration of the composition law \eqref{eq:discrcompositionllmarkedDT} for the fixed geodesic distance two-loop amplitude.}
\label{fig:compositionlaw}
\end{center}
\end{figure}

A nice feature of this construction is that it implies the following semi-group property or composition law for the propagator \index{Semi-group property}\index{Composition law}
\beq \label{eq:discrcompositionllmarkedDT}
G(l_1,l_2;g;t_1+t_2)  = \sum_{l} G(l_1,l;g;t_1) \; G(l,l_2;g;t_2),
\eeq
or in terms of the generating functions
\beq\label{eq:discrcompositionxymarkedDT}
G(z,w;g;t_1+t_2)  = \oint \frac{dx}{2\pi i} G(z,x^{-1};g;t_1)  G(x,w;g;t_2).
\eeq
Fig.~\ref{fig:compositionlaw} gives an graphical interpretation of the composition law.
Setting $t_1=1$ and $t_2=t-1$ one obtains
\beq \label{eq:discrcompositionllmarkedDT}
G(l_1,l_2;g;t)  = \sum_{l} G(l_1,l;g;t\equ 1) \; G(l,l_2;g;t-1).
\eeq
Here $G(l_1,l_2;g;t\equ 1)$ is the so-called one-step propagator which can be seen as the matrix element of the so-called transfer matrix $\cT(g)$,
\beq
G(l_1,l_2;g;t\equ 1)= \bra{l_2} \cT(g)\ket{l_1}.
\eeq
Knowing the transfer matrix one can in principle iterate \eqref{eq:discrcompositionllmarkedDT} $t$-times to obtain the fixed geodesic distance two-loop amplitude \index{Transfer matrix}\index{One-step propagator}
\beq
G(l_1,l_2;g;t)= \bra{l_2} \cT^t(g)\ket{l_1}.
\eeq

If one is however only interested in calculating the continuum expressions there exists a shortcut to obtain $G_\La(L_1,L_2,T)$ directly from the continuum limit of \eqref{eq:discrcompositionllmarkedDT} without having to perform the iteration as we will see in the following.

Our first task is now to calculate the one-step propagator $G(l_1,l_2;g;t\equ 1)$ by combinatorial methods.
So far we have been working with amplitudes which have a mark on the initial boundary. In some cases it is however useful to work with amplitudes without a mark on the boundary. The marked and unmarked propagators are related as follows
\beq\label{eq:markedrelationDT}
G(l_1,l_2;g;t)=l_1 G^{(u)}(l_1,l_2;g;t),
\eeq
where the $(u)$ stands for unmarked. Remember that the marked amplitude has only the initial loop marked and not the final loop. In terms of generating functionals, \eqref{eq:markedrelationDT} reads
\beq\label{eq:markedrelationDTz}
G(z,w;g;t)=-\frac{\prt}{\prt z}z\,G^{(u)}(z,w;g;t).
\eeq
For the unmarked propagator the composition law \eqref{eq:discrcompositionllmarkedDT} becomes
\beq \label{eq:discrcompositionllDT}
G^{(u)}(l_1,l_2;g;t_1+t_2)  = \sum_{l} l\, G^{(u)}(l_1,l;g;t_1) \; G^{(u)}(l,l_2;g;t_2),
\eeq
where the measure factor $l$ in the sum is the direct manifestation of the symmetry factor $C(T)$ in \eqref{eq:simpl:pathsum}.

We will now follow the results of \cite{Kawai:1993cj} to combinatorially determine $G^{(u)}(z,w;g;t)$. It was shown there that every one-step propagator is made out of the following four elementary building blocks:
\begin{eqnarray} 
\raisebox{-10pt}{\includegraphics[height=30pt]{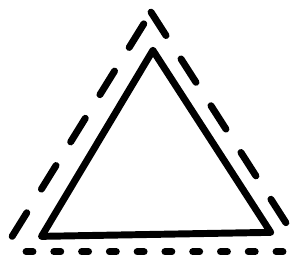}}&=& \frac{g}{z w^2}\\
\raisebox{-18pt}{\includegraphics[height=45pt]{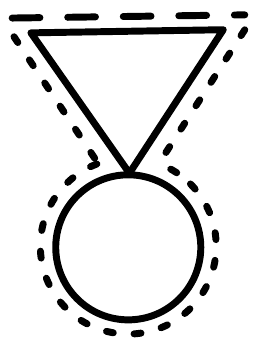}}&=& \frac{g w(g,z)}{z^2 w}\\
\raisebox{-18pt}{\includegraphics[height=45pt]{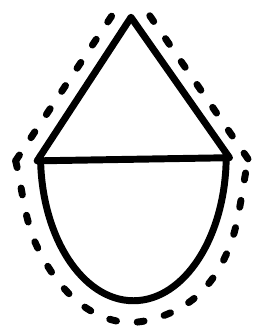}}&=&\frac{g(z w(g,z)-1)}{z^2}\\
\raisebox{-18pt}{\includegraphics[height=45pt]{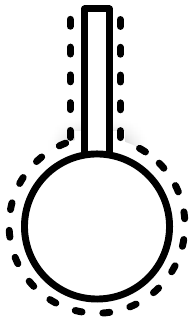}}&=&\frac{w(g,z)}{z^2}.
\end{eqnarray}
Here the dotted lines refer to the initial boundary with weight $1/z$ per edge and the dashed lines refer to final boundary with weight $1/w$ per edge. Again each triangles carries a weight $g$. Further, the blobs represent the DT disc function $w(g,z)$ as derived in Sec.\ \ref{subsec:discdiscreteDT}. 

Now we can write the one-step propagator as the composition of all these terms
\bea\label{eq:resG1DT}
G^{(u)}(z,w;g;1)&=&\sum_{k=1}^\infty \frac{1}{k} \left( 
\raisebox{-5pt}{\includegraphics[height=20pt]{block1}}+
\raisebox{-15pt}{\includegraphics[height=35pt]{block2}}+
\raisebox{-15pt}{\includegraphics[height=35pt]{block3}}+
\raisebox{-15pt}{\includegraphics[height=35pt]{block4}}
\right)^k \non
&=& - \log\left( 1- \frac{g}{z w^2} -\frac{g w(g,z)}{z^2 w} -\frac{g(z w(g,z)-1)}{z^2} - \frac{w(g,z)}{z^2} \right).
\eea\index{Dynamical triangulations (DT)!one-step propagator}
Here the factor of $1/k$ reflects the circular symmetry, i.e.\ $1/C(T)$.

In the following we perform the continuum limit using the same scaling relations as in the previous subsection
\begin{equation}
g=g_c \,e^{-a^2\,\Lambda},\quad z=z_c \,e^{a\,Z},\quad w=z_c \,e^{a\,W},
\end{equation}
with the same critical values. Inserting the scaling relations and the solution for   $G^{(u)}(z,w;g;1)$, i.e.\ \eqref{eq:resG1DT}, into \eqref{eq:markedrelationDTz} we get to lowest order in $a$
\beq \label{eq:expanG1DT}
G(z,w;g;t\equ 1)=\frac{1}{a}\frac{1}{Z+W}- \frac{1}{\sqrt{a}} \frac{\prt}{\prt Z}\left(\frac{W_\La(Z)}{Z+W}\right)+ \cO(1).
\eeq
Hence we introduce the following wave function renormalization for the geodesic two-loop amplitude to match the desired initial condition \eqref{DT:eqinitial},
\beq 
G_\La(Z,W;T)= a G(z,w;g;t).
\eeq
From the second term in the expansion of \eqref{eq:expanG1DT} we see that time has to scale as
\beq
T\sim t \sqrt{a}.
\eeq
Inserting the full scaling relations and the solution \eqref{eq:expanG1DT} for the one step propagator into the composition law \eqref{eq:discrcompositionxymarkedDT} we obtain
\beq\label{eq:geodDT2} 
\frac{\partial}{\partial T}G_\La(Z,W;T)=- \frac{\partial}{\partial Z}\left[ W_\La (Z) G_\La(Z,W;T)\right].
\eeq
This is precisely the differential equation for the fixed geodesic distance two-loop function as derived in the previous subsection, i.e.\ \eqref{eq:geodDT}. Thus, we rederived the previous result obtained from the peeling method using the transfer matrix formalism. The solution to the differential equation \eqref{eq:geodDT2} is again given by \eqref{eq:DTgeodsol}.

\section{Physical observables}

One of the simplest observables in two-dimensional Euclidean quantum gravity is the disc function or Hartle-Hawking wave function which describes the amplitude of creation of a universe from nothing. The result was obtained to be \index{Dynamical triangulations (DT)!disc function}
\beq
W_\La(Z)=(Z-\frac{1}{2}\sqrt{\La})\sqrt{Z+\sqrt{\La}}
\eeq
or
\beq
W_\La(L)=L^{-5/2}(1+\sL L)e^{-\sL L}.
\eeq

Having determined the dependence on the geodesic distance through the two-point function in Sec.\ \ref{sec:peelingDT} we can further calculate the \emph{Hausdorff dimension} $d_H$ of the quantum geometry. This dimension estimator is formally defined as
\beq
\expec{V(T)}\sim T^{d_H}\quad\text{for } T\to\infty.
\eeq

\begin{figure}[t]
\begin{center}
\includegraphics[width=5in]{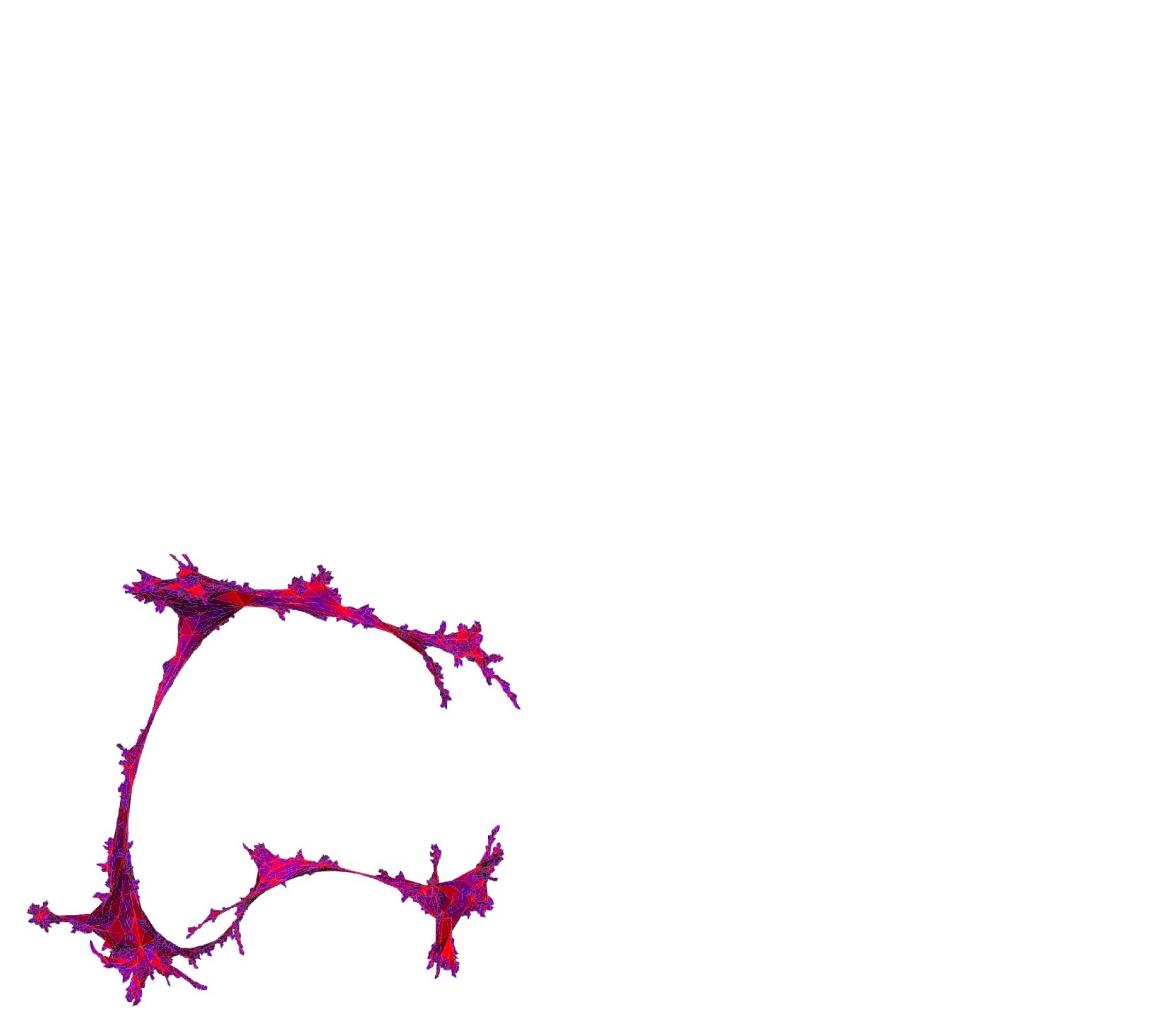}
\caption{A typical quantum geometry of two-dimensional Euclidean quantum gravity \cite{Weigelthesis}.}
\label{fig:universeDT}
\end{center}
\end{figure}

Using the expression for the two-point function, i.e. \eqref{eq:two-pointDT},
\beq
G_\La(T) \sim e^{- \sqrt[4]{\La} T/\a}\quad\text{for } T\to\infty.
\eeq
we obtain
\beq
\expec{V(T)}=-\frac{1}{G_\La(T) }\frac{\partial}{\partial\La}G_\La(T) \sim \frac{T}{\La^{3/4}}.
\eeq
This equation reflects the fact that at large $T$ the quantum geometry looks effectively one-dimensional, i.e. the volume is proportional to the length. Further one can extract from this expression that the average spatial length of the quantum geometry at intermediate $T$ behaves as \index{Dynamical triangulations (DT)!Hausdorff dimension}\index{Hausdorff dimension}
\beq
\expec{L}=\frac{\expec{V(T)}}{T}\sim \frac{1}{\La^{3/4}}.
\eeq
Hence for typical time scales of $T\sim1/\sqrt[4]{\La}$ we have 
\beq
\expec{V(T)}\sim T^{4}\quad\text{for } T \sim \frac{1}{\sqrt[4]{\La}}.
\eeq
Thus we conclude that the Hausdorff dimension of two-dimensional Euclidean quantum gravity is given by $d_H=4$. This result is somehow unexpected since we started off with a theory of two-dimensional triangulations. In fact, this is an indication that the quantum geometry is fractal \cite{Knizhnik:1988ak,Kawai:1993cj,Ambjorn:1995rg,Ambjorn:1997jf}. As we will see explicitly in Sec.\ \ref{sec:CDTtoDT} the quantum geometry is dominated by a proliferation of baby universes at the cut-off scale, meaning that at every point in the geometry there splits off a geometry with disc topology. The infinite number of baby universes causes the increment in the Hausdorff dimension. This fractal structure can be nicely observed in Fig.~\ref{fig:universeDT}, which shows a typical configuration of the path integral obtained from a Monte Carlo simulation \cite{Weigelthesis}.

\section{Discussion and outlook to higher dimensions}

In this chapter we introduced dynamical triangulations as a regularization scheme to define the Euclidean gravitational path integral. In particular, we considered the case of two dimensions in which analytical results can be obtained. We calculated the disc or Hartle-Hawking wave function, the fixed geodesic distance two-loop amplitude as well as the two-point function. Using these quantities we were able to analyze several properties of the quantum geometry. One very interesting observation is the fractal structure of the quantum geometry which manifests itself in a proliferation of baby universes at the cut-off scale. This is related to the unexpected value of the Hausdorff dimension of $d_H\equ 4$. Fig.~\ref{fig:universeDT} illustrates this situation by showing a typical configuration of the path integral obtained from a Monte Carlo simulation \cite{Weigelthesis}.

It is interesting to note how this situation translates to higher dimensions. In an analogous manner to two dimensions one can define the gravitational path integral as a sum over $d$-dimensional simplicial complexes (see for example \cite{Ambjorn:1997di} for a detailed exposition). The corresponding Regge action takes the following simple form \index{Regge action}\index{Dynamical triangulations (DT)!higher-dimensional}
\beq
S_T(k_{d-2},k_d)= k_d N_d(T) - k_{d-2} N_{d-2}(T),
\eeq 
where $N_d(T)$ denotes the number of $d$-simplices and $N_{d-2}(T)$ the number of $(d-2)$-simpices in the triangulation $T$. The analog expression to \eqref{eq:simpl:pathsum} for the partition function is then written as the sum over closed $d$-dimensional triangulations $\cT_d$,
\beq
\cZ(k_{d-2},k_d)=\sum_{T\in \cT_d} \frac{1}{C_T} e^{- S_T(k_{d-2},k_d)}.
\eeq

\begin{figure}[t]
\begin{center}
\includegraphics[width=5in]{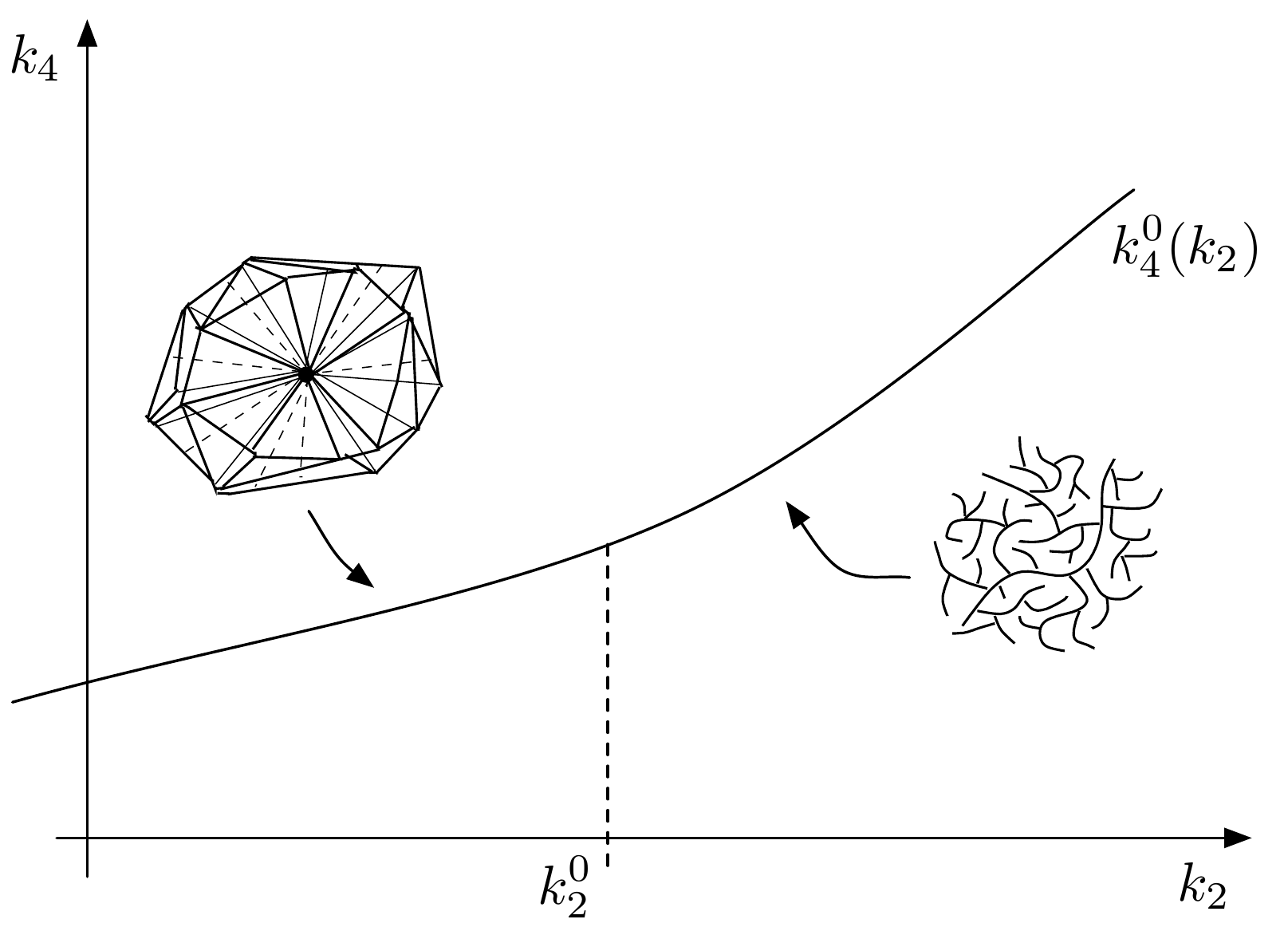}
\caption{The phase diagram of four-dimensional 
Euclidean quantum gravity defined through DT.}
\label{fig:phase}
\end{center}
\end{figure}

Since analytical methods are absent in $d>2$, one has to perform Monte Carlo simulations of the quantum geometry (see for instance \cite{Ambjorn:1997di}). DT in three dimensions were considered in \cite{Ambjorn:1990ge,Agishtein:1990yc,Boulatov:1991hg,Ambjorn:1991rx}. The four-dimensional case was first analyzed in \cite{Ambjorn:1991pq,Agishtein:1991cv,Varsted:1992qs}, followed by further progress in \cite{Ambjorn:1992aw,Agishtein:1992xx,Ambjorn:1993sy,Brugmann:1992js,deBakker:1994zf,Catterall:1994pg}. The results for $d\equ 4$ are summarized by the following schematic phase diagram (Fig.\ \ref{fig:phase}).\footnote{Although we only consider the case of $d\equ 4$ here, the picture 
turns out to be essentially the same in the case of $d\equ 3$.}
One observes that the model possesses an infinite-volume limit everywhere along the 
critical line $k_4^{0}(k_2)$, which fixes the bare
cosmological constant as a function of the inverse Newton
constant $k_2 \sim G_N^{-1}$. 
Along this critical line, there
is a critical point $k_2^{0}$. Below this critical point the geometries generically have a very large
Hausdorff dimension $d_H\approx\infty$. This can be understood in terms of a situation where every point in the quantum geometry is effectively at Planck distance to any other point. We call this the \emph{crumpled phase} of the quantum geometry. Above $k_2^{0}$
we find the opposite situation where
a configuration contributing to the state sum looks likes a branched polymer with Hausdorff dimension $d_H=2$ (see Fig.\ \ref{fig:BP}). This is called the \emph{branched polymer phase} of the quantum geometry.     
\index{Dynamical triangulations (DT)!crumpled phase}
\index{Dynamical triangulations (DT)!branched polymer phase}

One could hope that right at the transition point $k_2^{0}$ genuine extended triangualtions with a finite Hausdorff dimension might be dominating. For this to happen the phase transition at this point should be of second or higher order. Unfortunately, it was found to be of first kind\footnote{Evidence that the phase transition is of first order was given both in three dimensions \cite{Ambjorn:1991rx,Warner:1998mv} as well as in four dimensions \cite{Catterall:1994sf,deBakker:1996zx,Warner:2000pf}.} which leaves us with little hope that a sensible continuum theory of Euclidean quantum gravity can be obtained from DT for $d\!>\!2$.
 
 This unsatisfactory situation is one of the reasons for incorporating causality in DT. We have already seen above that many of the unappealing features of the quantum geometry were related to the proliferation of baby universes at the cut-off scale. In a genuine Lorentzian framework we expect such ``acausal'' configurations to be suppressed and the path integral should be much better behaved. The development of such a genuine Lorentzian framework by so-called causal dynamical triangulations is the subject of the next chapter.

\chapter{Two-dimensional causal dynamical triangulations \label{Chap:cdt}}
	In the previous chapter we introduced the reader to the concept of dynamical triangulations (DT). As we discussed at the end of the chapter, DT seems to be incompatible with a well-behaved continuum limit in three and higher dimensions. This problem is related to the failure of incorporating the Lorentzian signature into the gravitational path integral. In this chapter we see how this issue can be resolved in the framework of Lorentzian quantum gravity defined through causal dynamical triangulations \cite{Ambjorn:1998xu}. The presentation of this chapter closely follows \cite{Ambjorn:1998xu,Zohren:2006tg,Willemthesis}.

\section{Incorporating causality in dynamical triangulations}

\index{Causal dynamical triangulations (CDT)}
Inspired by early ideas of Teitelboim \cite{Teitelboim:1983fh,Teitelboim:1983fk}, we want to construct a gravitational path integral, where we insist in starting from space-times with Lorentzian signature. However, to address the Lorentzian nature of the path integral it is essential to look at geometries that have an intrinsically defined notion of time. In the approach of causal dynamical triangulations (CDT), introduced by Ambj\o rn and Loll in 1998 \cite{Ambjorn:1998xu} for the case of two dimensions\footnote{Further developments for two-dimensional CDT have been made in \cite{Ambjorn:1999nc,Ambjorn:1998fd,Ambjorn:1999fp,DiFrancesco:1999em,DiFrancesco:2000nn,Loll:2005dr,Ambjorn:2006hu,Ambjorn:2007jm,Ambjorn:2008ta,Ambjorn:2008we,Ambjorn:2008jf} which will be partically discussed in more detail in the following chapters. References to results in higher dimensions will be given in the last section of this chapter.}, this problem is addressed by studying piecewise linear geometries that have an entirely time-sliced structure. By virtue of the time-slicing one can globally distinguish timelike and spacelike edges. We will see that this distinction is essential to define a consistent Wick rotation.

\begin{figure}[t]
\begin{center}
\includegraphics[width=4in]{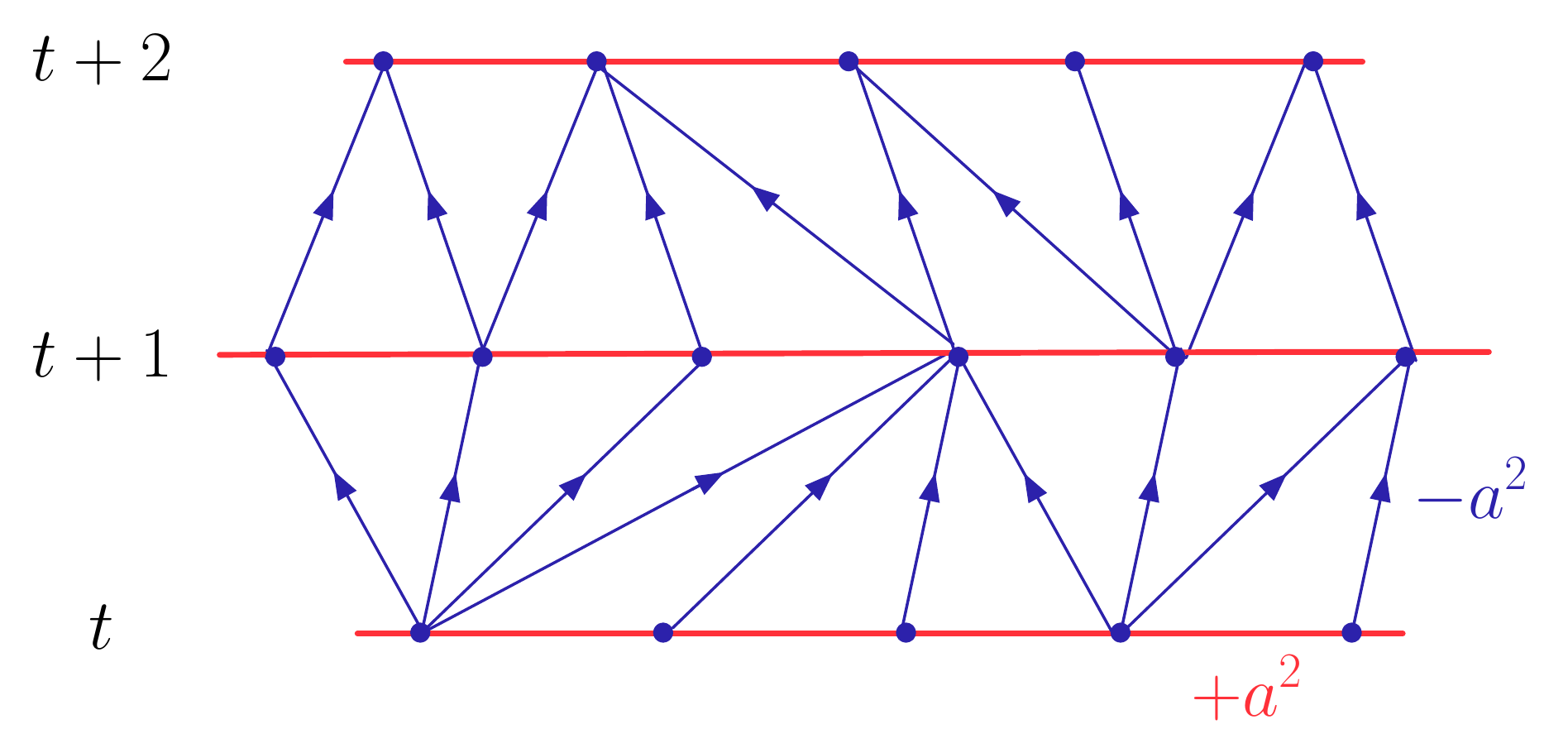}
\caption{Section of a two-dimensional Lorentzian triangulation consisting of spacetime strips of height $\Delta t\equ 1$.
Each spatial slice is periodically identified, such that the simplicial manifold has topology $[0,1]\!\times\! S^1$.
One sees that the lower strip with initial boundary length $l(t)$ and final boundary length $l(t+1)$ consists exactly of $l(t)$ up-pointing triangles and $l(t+1)$ down-pointing triangles.}
\label{fig:triangulation}
\end{center}
\end{figure}

As in the case of DT one uses triangles at cut-off length to discretize the geometries. However, in contrary to DT, CDT uses Minkowskian  instead of Euclidean triangles. These have one spacelike edge of length squared $L_s^2 = + a^2$ and two timelike edges of length squared $L_{t}^2= -a^2$. By construction one chooses triangulations with topology of $S^1 \times [0,1]$. Each triangulation consists of $t$ time slices of height $\Delta t \equ 1$. 
The geometry of an individual time slice is now determined by the composition of
$l(t)$ up-pointing triangles and $l(t+1)$ down-pointing triangles as illustrated in Fig.~\ref{fig:triangulation}. \index{Triangulation!Lorentzian}

In this formulation the time $t$ by definition takes the role of geodesic distance. Further, there are no spatial topology changes with respect to the time-slicing. Hence, the layered structure does not only introduce an intrinsic time-direction that enables us to define a Wick rotation, but it also excludes spatial topology changes which in the Lorentzian picture lead to singular points with ill-defined signature.\footnote{In the next chapter we will relax this constraint and we will show how one can make sense of spatial topology changes in a Lorentzian setting.} 

Since the triangles are simply patches of flat Minkowski space, they are naturally equipped with a local light cone structure. Furthermore, due to the sliced structure of the triangulations which glues spacelike to spacelike and timelike to timelike edges, the triangulation is naturally equipped with a global causal structure. By virtue of this global causal structure it is possible to define a Wick rotation for curved manifolds on the discretized level.
In the triangulation context this is done by performing a Wick rotation on the timelike edges, i.e. $L_t\to i L_t$. The Wick rotation acting on the whole manifold is then defined as
\beq
\mathcal{W}: \cT^{lor}=\{T, L_s^2=a^2, L_{t}^2=-a^2\} \mapsto  \cT^{eu}=\{T, L_s^2=a^2, L_{t}^2=a^2\} .
\eeq \index{Causal dynamical triangulations (CDT)!Wick rotation}\index{Wick rotation!CDT}
As in the continuum case this rotation should be treated with some care and one must show that the Lorentzian action defined with $L_{\tau}^2=-a^2$ can be analytically continued to the Euclidean action defined with $L_{\tau}^2=a^2$.
This is done in appendix \ref{App:Lorentzian}, where we discuss in more detail the Regge action for two-dimensional Lorentzian triangulations. It is shown there that under the above defined Wick rotation the Boltzmann weight for each triangulation indeed transforms as
\beq
\mathcal{W}:\quad  e^{i\,S_{\mathrm{Regge}}(T^{lor}) }\mapsto  e^{-\,S_{\mathrm{Regge}}(T^{eu}) }.
\eeq
We shall see in the forthcoming section that the above defined kinematical structure reduces the computation of the path integral for the fixed geodesic distance correlator or finite-time propagator to a simple statistical mechanics problem.

\section{The discrete solution} \label{sec:The discrete solution}

As for DT we fix the space-time topology of the triangulation to be of the form
$S^1\times[0,1]$. The Regge action is then simply proportional to the volume of the triangulation,
\beq
S_{\mathrm{Regge}}(T) =\tilde{\lambda} \,a^2\, N(T),
\eeq
where $\tilde{\lambda}$ is the bare cosmological constant and $N(T) $ the number of triangles in the triangulation $T$. Note that an order one factor coming from the volume term has been absorbed into $\tilde{\lambda}$. Since all slices have the same geodesic distance, the propagator  or fixed geodesic distance two-loop function can be defined in a simple manner%
\beq
G_{\tilde{\lambda}}^{(u)}(l_1,l_2;t) =\sum_{\TriC(l_1,l_2,t)} \frac{1}{C_T}\:\: e^{i\,\tilde{\lambda}\,a^2\,N(T) },
\eeq
where $\TriC$ denotes the causal triangulations with initial boundary length $l_1$ and final boundary length $l_2$ constructed out of $t$ time-slices. Again,
$C_T$ denotes the volume of the automorphism group of a triangulation. As for DT we add a superscript $(u)$ to denote the propagator with no marked points on the boundary. As discussed above, after  the Wick rotation the discrete sum over complex amplitudes is converted to a genuine statistical model with a real Boltzmann weight,
\beq
G_{\lambda}^{(u)}(l_1,l_2;t) =\sum_{\TriC(l_1,l_2,t)} \frac{1}{C_{T}}\:\: e^{-\lambda\,a^2\,N(T) },
\eeq
where  $\lambda$ and $\tilde{\lambda}$ differ by an order one constant due to the
different volume of Minkowskian and Euclidean triangles (see App.\ \ref{App:Lorentzian}).
As for the fixed geodesic distance two-loop amplitude in the case of DT the propagator satisfies the following semi-group property or composition law (e.g.\ \eqref{eq:discrcompositionllDT}),\index{Semi-group property}\index{Composition law}
\beq \label{eq:discrcompositionllnonmarked}
G_\l^{(u)}(l_1,l_2;t_1+t_2)  = \sum_{l} l\: G^{(u)}_\l(l_1,l;t_1) \; G^{(u)}_\l(l,l_2;t_2),
\eeq
where again the measure factor $l$ in the composition law comes from the periodic boundary conditions. Writing the composition law
for $t_1=1$ we find the following relation for the one-step propagator,
\beq \label{eq:transfermatrixequation}
G^{(u)}_\l(l_1,l_2;t+1)  = \sum_{l} l\: G_\l(l_1,l;1) \; G_\l(l,l_2;t).
\eeq
As for DT we introduce a generating function for $G_\l(l_1,l_2;t) $
\beq
G_\l^{(u)}(x,y;t) \equiv \sum_{k,l} x^k\,y^l \;G_\l^{(u)}(k,l;t),
\eeq
where $x$ and $y$ can again be related to the boundary cosmological constants of
individual triangles,
\beq
x=e^{-\l_ia},~~~~y=e^{-\l_oa}.
\eeq
The analogous fugacity of a triangle is again given by,
\beq
g=e^{-\l a^2}.
\eeq

The total Boltzmann weight of one strip can be determined by noting that for the one-step propagator in CDT
a factor of $g x$ is assigned for triangles that have the spacelike edge on the initial loop and a factor
$g y$ for triangles where the spacelike edge is on the final loop. This relation is much simpler than in the case of DT. The one-step propagator is now easily computed by,

\begin{eqnarray} \label{eq:Gxyg1def}
G^{(u)}(x,y;g;1)  & = & \raisebox{-10pt}{\includegraphics[height=25pt]{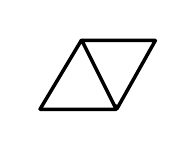}}+
\raisebox{-10pt}{\includegraphics[height=25pt]{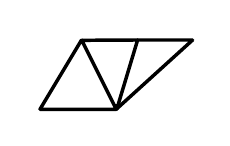}}+
\raisebox{-10pt}{\includegraphics[height=25pt]{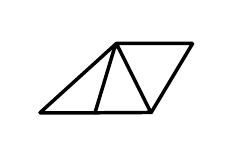}}+... \nonumber\\
 & = &  \sum^{\infty}_{k=1}\frac{1}{k}\left(\sum^{\infty}_{m=1} (gx) ^m\sum^{\infty}_{n=1} (gy) ^n\right) ^k,
\end{eqnarray}
where the factor $1/k$ comes again from dividing by the volume $C(T)$ of the automorphism group for circular triangulations.
Performing the summations in \rf{eq:Gxyg1def} we readily obtain \index{Causal dynamical triangulations (CDT)!one-step propagator}
\beq \label{onesteppropagatorxyZ2}
G^{(u)}(x,y;g;1)  = \sum^{\infty}_{k=1}\frac{1}{k}\left(\frac{gx}{1-gx}\frac{gy}{1-gy} \right) ^k=-\log\left(1-\frac{gx gy}{(1-gx) (1-gy) }\right).
\eeq
From the inverse discrete Laplace transform of \rf{onesteppropagatorxyZ2} it can be seen that the one-step propagator with fixed boundary lengths is given by
\beq \label{eq:onesteppropagatorll}
G^{(u)}(l_1,l_2;g;1)  = g^{l_1+l_2}\frac{1}{l_1+l_2}\binom{l_1+l_2}{l_1},
\eeq
where the division by the volume of the automorphism group now makes its appearance in the factor $1/(l_1+l_2)$. Similar expressions for different boundary conditions can for instance be found in \cite{Zohren:2006tg}.

Instead of obtaining the continuum limit directly of the one-step propagator as done in the previous chapter we want to iteratively compute the finite-time propagator $G(x,y;g;t) $ and then perform the continuum limit. To do so we rewrite the composition law \rf{eq:discrcompositionllnonmarked}
in terms of generating functions and obtain,
\beq\label{eq:discrcompositionxynonmarked}
G^{(u)}(x,y;g;t_1+t_2)  = \oint \oint \frac{dz}{2\pi i}\:\frac{dz'}{2\pi i}\: \frac{1}{\left(1-zz'\right) ^2}\:G(x,z;g;t_1)  G(z',y;g;t_2).
\eeq
Setting $t_2=1$ and performing the contour integration over $z$ one gets
\beq\label{eq:discrcompositionxynonmarkediteration2}
G^{(u)}(x,y;g;t)  = \oint \frac{dz'}{2\pi i z'^2}\: \left[\frac {d}{d z}G^{(u)}(x,z;g;1) \right]_{z=1/z'} G^{(u)}(z',y;g;t-1).
\eeq
Inserting the expression for the one-step propagator, i.e.\ \rf{onesteppropagatorxyZ2}, yields
\beq\label{eq:iterationequationnonmarked}
G^{(u)}(x,y;g;t)  = G^{(u)}\left(\frac{g}{1-gx},y;g;t-1\right) -G^{(u)}\left(g,y;g;t-1\right).
\eeq
This equation can be iterated and the implicit solution written as
\beq\label{eq:discrGxytF}
G^{(u)}(x,y;g;t)  = \log \left(\frac{1}{1-F_t(x)  y}\right) -\log \left(\frac{1}{1-F_{t-1}(g)  y}\right),
\eeq
where $F_t$ is defined iteratively by
\beq \label{eq:iterationF}
F_t(x)  = \frac{g}{1-gF_{t-1}(x) },~~~~F_0(x) =x.
\eeq
The fixed point $F$ of this equation as defined by $F_t(x) =F_{t-1}(x) $ is given by
\beq \label{eq:defF}
F=\frac{1-\sqrt{1-4g^2}}{2g},~~~~g=\frac{1}{F+1/F}.
\eeq
By standard techniques one can use the fixed point to give the explicit solution to the iterative equation
\rf{eq:iterationF}
\beq\label{eq:FxABC}
F_t(x) = \frac{B_t-x C_t}{A_t-x B_t},~~~~F_{t-1}\left({\textstyle{g}}\right) =\frac{B_t}{A_t},
\eeq
where we have defined
\beq \label{eq:defABC}
A_t =1-F^{2t+2},~~~B_t=F-F^{2t+1},~~~C_t=F^2-F^{2t}.
\eeq
The explicit solution for the propagator is now obtained by substituting \rf{eq:FxABC} in \rf{eq:discrGxytF}, yielding
\beq \label{eq:discrGxytZ}\index{Causal dynamical triangulations (CDT)!propagator}
G^{(u)}(x,y;g;t) = -\log\left(1-Z(x,y;g;t) \right),
\eeq
where we have defined
\beq \label{discrZxyt}
Z(x,y;g;t) ={\textstyle \left(1-\frac{A_t C_t}{B_t^2}\right) }\frac{\frac{B_t}{A_t} x \frac{B_t}{A_t} y}{\left(1- \frac{B_t}{A_t} x\right) \left(1- \frac{B_t}{A_t} y\right) }.
\eeq
The combined region of convergence of this result as an expansion in powers $g^k x^{l_1} y^{l_2}$ is
\beq \label{eq:convergencecondition}
|g|< \frac{1}{2}, \quad |x|< 1,\quad|y|<1.
\eeq
As in the case of DT we will use the radius of convergence to obtain the critical points.

\section{The continuum limit} \label{sec:The continuum limit}
\index{Causal dynamical triangulations (CDT)!continuum limit}

 To define the continuum limit of the theory we assume canonical scaling dimensions for
the bulk and boundary cosmological constants as in the case of DT, \index{Causal dynamical triangulations (CDT)!scaling relations}
\beq
g=g_c e^{-a^2 \La/2},\quad x=x_c e^{-a X},\quad y=y_c e^{-a Y}.
\eeq
The critical values for these couplings are again determined by the region of convergence of the propagator, i.e.\ \rf{eq:convergencecondition}, yielding
\beq
g_c = 1/2,~~~~x_c = 1,~~~~y_c=1,
\eeq
One can quickly convince oneself that these are the correct critical vales by looking at the average number of triangles in one strip
\beq
\langle N \rangle =g\frac{d}{d g}G^{(u)}(x,y;g;1)= 1 - \frac{1}{1 - g x} - \frac{1}{1 - g y} + \frac{1}{1 - g (x + y) }.
\eeq
We see that the average number of triangles diverges at the critical point $(g_c,x_c,y_c)\equ(1/2,1,1)$. We hence arrive at the following scaling relations
\beq \label{eq:scalingrelations}
g=\frac{1}{2} e^{-a^2 \La/2},\quad x= e^{-a X},\quad y=e^{-a Y}.
\eeq
The continuum limit of the propagator can now be determined by inserting these scaling relations  into \rf{eq:discrGxytZ}. To get a sensible continuum limit we require time to scale as
\beq
T\sim a t. 
\eeq
which yields the following result \index{Causal dynamical triangulations (CDT)!propagator}
\beq \label{eq:contGXYTZ}
G_{\La}^{(u)}(X,Y,T) = -\log\left(1-Z_{\La}(X,Y,T) \right),
\eeq
where
\beq
Z_{\La}(X,Y,T) = \frac{\La}{\sinh^2\SL T(X+\SL \coth\SL T) (Y+\SL \coth\SL T) }.
\eeq
As in the case of DT the propagator undergoes a wave function renormalization
\beq
G_{\La}^{(u)}(X,Y,T) = a G^{(u)}(x,y;g;t).
\eeq
Again the power of $a$ is uniquely determined by the initial condition coming from the composition law, i.e.\ \rf{eq:discrcompositionxynonmarked}.

The continuum expression for the propagator, where the boundary lengths are fixed instead of the boundary cosmological
constants can now be obtained by an inverse Laplace transformation of \rf{eq:contGXYTZ} with respect to both $X$ and $Y$, \index{Causal dynamical triangulations (CDT)!propagator}
\beq \label{eq:GLT}
G_\La^{(u)}(L_1,L_2;T)  = \frac{e^{- \SL \coth \SL T (L_1+L_2) }}{\sinh \SL T}\; \frac{\sqrt{\La}}{\sqrt{L_1 L_2}}\; \;
I_1\left(\frac{2\sqrt{\La L_1 L_2}}{\sinh \SL T}\right),
\eeq
where $I_1(x) $ is a modified Bessel function of the first kind. \index{Modified Bessel function}
The continuum version of the composition law \rf{eq:discrcompositionxynonmarked} for fixed lengths then reads
\beq \label{eq:compositionlawGLLT}
G_\La^{(u)} (L_1,L_2;T_1+T_2)  = \int_0^\infty d L  \,L\;G_\La^{(u)} (L_1,L;T_1) \,G_\La^{(u)}(L,L_2,T_2).
\eeq
An interesting observation is that in contrast to the Euclidean model the time or geodesic distance scales canonically. We will further comment on this when discussing the physical observables of the model.

\section{Marked amplitudes}\label{Marking the causal propagator}

To better compare to the results of the previous chapter for Euclidean DT it is important to introduce the same convention of a marked edge on the initial boundary in CDT. In terms of the propagator this convention reads,
\beq
G_\l(l_1,l_2;1) = l_1 G_\l^{(u)}(l_1,l_2;1).
\eeq
As before this marking removes the measure factor in the
composition law for the propagator \rf{eq:discrcompositionllnonmarked}
\beq \label{eq:discrcompositionllmarked}
G_\l(l_1,l_2;t_1+t_2)  = \sum_{l} G_\l(l_1,l;t_1) \; G_\l(l,l_2;t_2),
\eeq
which corresponds to the following composition law for the generating function,
\beq\label{eq:discrcompositionxymarked}
G_\l(x,y;t_1+t_2)  = \ointz G_\l(x,z^{-1};t_1)  G_\l(z,y;t_2).
\eeq

The one-step propagator with a mark on the boundary can easily be computed from \rf{onesteppropagatorxyZ2} by taking a derivative with respect to the boundary cosmological constant,
\beq \label{eq:G10xyg1}
G(x,y;g;1)  = x\frac{d}{d x}G^{(u)}(x,y;g;1)  =\frac{g^2 x y}{(1-g x) (1-g(x+y) ) }.
\eeq
Inserting this expression into the composition law \rf{eq:discrcompositionllmarked} gives an iterative equation analogous to
\rf{eq:iterationequationnonmarked} ,
\beq \label{eq:iterationequationmarked}
G(x,y;g;t)  = \frac{gx}{1-gx}\; G\left(\frac{g}{1-gx},y;g;t-1\right).
\eeq
The explicit solution can be obtained from \rf{eq:discrGxytZ} by taking a derivative with respect to the boundary cosmological constant,
\beq
G(x,y;g;t)  = x\frac{d}{d x}G^{(u)}(x,y;g;t) = \frac{F^{2t}(1-F^2)^2 xy}{(A_t-B_t x)(A_t-B_t(x+y)+C_t xy)},
\eeq
where $F$, $A_t$, $B_t$ and $C_t$ were defined in \rf{eq:defF} and \rf{eq:defABC}.

To obtain the continuum marked propagator we could now apply the scaling relations \rf{eq:scalingrelations} to
this expression. Instead, one can also apply the scaling relations
directly to \rf{eq:iterationequationmarked} in the manner we proceeded for DT. Doing so one finds the following differential equation for the continuum marked propagator,
\beq \label{eq:differentialequationGXYT}
\frac{\partial}{\partial T}G_{\La}(X,Y,T) =-\frac{\partial}{\partial X}\left[\hcW(X) G_{\La}(X,Y,T) \right],
\eeq
where
\beq \label{eq:defhcWpure}
\hcW(X) =X^2-\La.
\eeq
The notation $\hcW(X)$ is introduced in analogy to the Euclidean result, however, it should be noted that in the case of CDT $\hcW(X) $ is \emph{not} the disc function.

Solving the differential equation \rf{eq:differentialequationGXYT} with initial condition 
\beq
G_\La(X,Y;T\equ 0)  = \frac{1}{X+Y}.
\eeq
gives
\beq \label{eq:G10XYT}
G_{\La}(X,Y;T) =\frac{\hcW(\bXt ) }{\hcW(X) }\frac{1}{\bXt +Y},
\eeq
where $\bXt$ is the solution to the characteristic equation
\beq \label{eq:characteristicequation}
\frac{d\bXt}{dT}=-\hcW(\bXt),\quad \bX(T\equ 0,X) =X.
\eeq
Proceeding as in Sec.\ \ref{sec:peelingDT} we obtain,
\beq \label{eq:xbar}
\bXt = \SL \;
\frac{(\SL+X) -e^{-2\SL T}(\SL-X) }{(\SL+X) +e^{-2\SL T}(\SL-X) }.
\eeq
Inserting this expression in \rf{eq:G10XYT} gives the explicit solution for the continuum marked propagator. Further, one can check that if one marks the unmarked propagator \rf{eq:contGXYTZ} by taking a derivative with
respect to the initial boundary cosmology constant $X$, the resulting expression precisely coincides with this result. Inverse Laplace transforming \rf{eq:G10XYT} with respect to $X$ and $Y$ yields the continuum marked propagator in the length representation,
\beq \label{eq:GmLT}
G_\La(L_1,L_2;T)  = \frac{e^{- \SL \coth \SL T (L_1+L_2) }}{\sinh \SL T}\; \frac{\sqrt{\La} L_1 }{\sqrt{L_1 L_2}}\; \;
I_1\left(\frac{2\sqrt{\La L_1 L_2}}{\sinh \SL T}\right),
\eeq
 which is simply \rf{eq:GLT} multiplied by $L_1$ as expected.

As in the case of DT we can also calculate the disc function from the propagator. In CDT it can be defined as follows from the finite-time propagator with one boundary shrunken to zero,
\beq \label{eq:WCDTdef}
W_\La (L)= \int_0^\infty dT G_\La(L,L_2=0,T).
\eeq
One notices that this definition differs slightly from the corresponding relation in the Euclidean setting, i.e.\ \eqref{eq:DTdiscproprel}. This is due to the fact that the mark coming from the zero-length boundary is always at latest $T$ and cannot be anywhere in the bulk as in the case of DT. Therefore we do not have to differentiate with respect to $-\partial/\partial\La$. 

Using \rf{eq:GLT} we get
\beq \label{eq:GL0CDT}
G_\La(L,L_2=0,T)=\frac{\La}{\sinh^2 \sL T}e^{-\sL L \coth \sL T}.
\eeq
Inserting this into \rf{eq:WCDTdef} one readily arrives at the simple expression  \index{Causal dynamical triangulations (CDT)!disc function}
\beq\label{eq:discCDT} 
W_\La(L)=e^{-\sqrt{\La} L}
\eeq
or in Laplace transform language
\beq \label{discCDTresult}
W_\La(X)=\frac{1}{X+\sqrt{\La}}.
\eeq
The corresponding result for the disc function without mark reads
\beq
W_\La^{(u)}(L)=\frac{e^{-\sqrt{\La} L}}{L}.
\eeq
From \eqref{eq:GL0CDT} we can also easily compute the two-point function for CDT, i.e.
\beq\label{eq:GTCDT}
G_\La(T)= G_\La(L_1=0,L_2=0,T)= \frac{\La}{\sinh^2 \sL T}.
\eeq
 \index{Causal dynamical triangulations (CDT)!two-point function}

\section{Hamiltonians in causal quantum gravity}\label{sec:Hamiltonians in causal quantum gravity}

As mentioned earlier the concept of time or geodesic distance directly relates to a notion of Hamiltonian dynamics. This can be seen explicitly when viewing the differential equation for the continuum marked propagator \rf{eq:differentialequationGXYT} as a Wick rotated Schr\"{o}dinger equation,
\beq \label{eq:WickSchrodinger}
-\frac{\partial}{\partial T}G_{\La}(X,Y,T) =\hat{H}_X G_{\La}(X,Y,T).
\eeq
Here $\hat{H}_X$ can be interpreted as the \emph{effective quantum Hamiltonian} of the system and is given by \index{Causal dynamical triangulations (CDT)!effective quantum Hamiltonian}
\beq
\hat{H}_X = (X^2-\La) \frac{\partial}{\partial X}+2X.
\eeq
The effective quantum Hamiltonian in length space can be obtained from \rf{eq:WickSchrodinger} by inverse Laplace transformation, yielding
\beq \label{eq:CDTHamiltonian1}
\hat{H}_L = -L\frac{\partial^2}{\partial L^2}+ \La \: L.
\eeq
The corresponding Hamiltonian for the case of unmarked amplitudes reads
\beq \label{eqHu}
\hat{H}^{(u)}_L = -L\frac{\partial^2}{\partial L^2}- 2 \frac{\partial}{\partial L} + \La \: L.
\eeq
These kind of Hamiltonians are analyzed in detail in App.\ \ref{App:Calogero}, where several properties such as the spectrum and eigenfunctions are computed. In particular, it is shown there that the Hamiltonian \rf{eqHu} is self-adjoint with respect to the measure $LdL$. This is also reflected in the corresponding semi-group property, where the integration is taken over the same measure. As we will see later for some comparisons it is useful to transform this Hamiltonian to one which is self-adjoint on a flat measure. This can be done by absorbing the measure in the wave functions corresponding to the initial and final loop, i.e.\ $\psi_n(L) =\tfrac{1}{\sqrt{L}} \varphi_n(L) $. Commuting \rf{eqHu} with $1/\sqrt{L}$ gives\footnote{See App.\ \ref{App:Calogero} for more details.}
\beq \label{eq:flathamiltoniancdt}
\hat{H}^{\scriptscriptstyle{flat}}_L = -L\frac{\partial^2}{\partial L^2}- \frac{\partial}{\partial L} + \frac{1}{4L}+ \La \: L.
\eeq

In the following we present a calculation by Nakayama \cite{Nakayama:1993we} which reproduces the effective quantum Hamiltonian of CDT, i.e.\ \rf{eq:flathamiltoniancdt} from a continuum calculation. More precisely, Nakayama considered the non-local ``induced'' action of two-dimensional quantum gravity in the proper time gauge. The general form of this action was first introduced by Polyakov \cite{Polyakov:1981rd} and reads \index{Induced action}
\beq \label{eq:inducedaction}
S[g]= \int d t d x \sqrt{g} \left( \frac{1}{16} R_g \frac{1}{-\Del_g}R_g +\La \right),
\eeq
where $R$ is the scalar curvature corresponding to the metric $g$, $t$ denotes time and $x$ the
spatial coordinate. In the proper time gauge the metric has the form \index{Proper time gauge}
\beq\label{eq:metricpropertimegauge}
g = \begin{pmatrix} 1
&  0 \\ 0&
\g(t,x) \end{pmatrix}.
\eeq
This is in contrast to the conformal gauge with $g\equ e^{\phi} \hat{g}$. It was shown by Nakayama that in this gauge the classical dynamics is described entirely by the following
effective action
\beq \label{eq:nakayamaaction}
S_\k = \int_0^T d t\left(\frac{(\dot{l}(t))^2 }{4l(t) }  + \La l(t) + \frac{\k}{l(t) }\right),
\eeq
where
\beq \label{eq:l(t) isint}
l(t)  = \frac{1}{\pi}\int d x \sqrt{\g},
\eeq
and where $\k$ is an integration constant coming from the solution for the energy-momentum tensor component $T_{01}=0$. As can be seen from Eq.\ \rf{eq:l(t) isint} $L_{cont}\equiv \pi l(t) $ is precisely the length of a spatial universe at constant $t$, as calculated from the metric
\rf{eq:metricpropertimegauge}. 

One can now quantize the action $S_\k$ for $\k = (m+1) ^2$, where $m\!\in\!\mathbb{N}$ can be interpreted as a winding number. It was further argued in \cite{Nakayama:1993we} that in the quantum theory one should shift $m$ by $-1/2$ leading to $\k = (m+\tfrac{1}{2}) ^2$.
The classical Hamiltonian corresponding to the effective action \rf{eq:nakayamaaction} with $\k = (m+\tfrac{1}{2}) ^2$ then reads
\beq
H_m = \Pi_l l \Pi_l + \left(m+\tfrac{1}{2}\right) ^2\frac{1}{l}+ \La l,
\eeq
where $\Pi_l$ is the canonical momentum conjugate to $l$. To canonically quantize the system we make the standard replacement $\Pi_l \rightarrow -i\tfrac{\partial}{\partial l}$, yielding\footnote{For completeness the spectrum of this Hamiltonian is also derived in App.\ \ref{App:Calogero}.}
\beq \label{eq:quantumhamiltoniannakayama}
\hat{H}_m = -l\frac{\partial^2}{\partial l^2}- \frac{\partial}{\partial l} + \left(m+\tfrac{1}{2}\right) ^2 \frac{1}{l}+ \La \: l.
\eeq
One observes that for $m=0$ this Hamiltonian precisely coincides with the effective quantum Hamiltonian for CDT, i.e.\ \rf{eq:flathamiltoniancdt}. An interpretation for $m\!>\!0$ 
in the context of CDT was given in \cite{DiFrancesco:1999em}. 

Quite remarkably, when computing the propagator,
\beq
G^{(m)}_\La(L_1,L_2;T)=\bra{L_2}e^{-T \hat{H}_m }\ket{L_1},
\eeq
one observes that
\beq
\sum_{m=0}^\infty (-1)^m(2m+1)\int_0^\infty dT G^{(m)}_\La(L_1,L_2;T) =\frac{\sqrt{L_1L_2}}{L_1+L_2}e^{-\sL(L_1+L_2)}
\eeq
precisely coincides with the two-loop amplitude of \emph{Euclidean} DT \cite{Nakayama:1993we}. While a summation over the winding number might seem natural, a complete understanding of this relation is still lacking.

Before moving to the next section let us briefly mention that, since in CDT we are using a lattice regularization, the continuum variables are only defined up to a factor of proportionality which should be fixed by comparing to a continuum calculation. Recall that the length $l$ appearing in Nakayama's Hamiltonian is not the physical length, $L_{cont}$, but $L_{cont}/\pi$. Hence, since both models are described by the same effective quantum Hamiltonian, it is natural to also define the physical length $L_{cont}=\pi L$ for CDT. We will use this assignment in Chap.\ \ref{Chap:emergence} when discussing the emergence of a semiclassical background in the context of CDT.

\section{Physical observables: Comparing CDT and DT}\label{sec:CDT:observables}

As in the case of DT one of the simplest observables is the disc function or Hartle-Hawking wave function which describes the amplitude of creation of a universe from nothing. The result is given by \eqref{discCDTresult}, i.e \index{Causal dynamical triangulations (CDT)!disc function}
\beq
W_\La(X)=\frac{1}{X+\sqrt{\La}}.
\eeq
The corresponding Euclidean expression \rf{eq:DTdisc}, here labeled with a superscript $(eu)$ to distinguish from the analogous CDT expressions, is
\beq
W_\La^{(eu)}(X)=(X-\frac{1}{2}\sqrt{\La})\sqrt{X+\sqrt{\La}}.
\eeq


Another interesting quantity to calculate is the Hausdorff dimension $d_H$ of the quantum geometry formally defined as
\beq
\expec{V(T)}\sim T^{d_H}\quad\text{for } T\to\infty.
\eeq
We saw that in Euclidean DT we had a fractal dimension of $d_H=4$ which was related to the non-canonical scaling of time. We already observed above that in CDT time scales canonically and we therefore expect the Hausdorff dimension to be $d_H=2$. In the following we will check this by a simple calculation.

From \rf{eq:GTCDT} we have
\beq
G_\La(T) \sim e^{-2 \sqrt{\La} T}\quad\text{for } T\to\infty.
\eeq
From this we obtain
\beq
\expec{V(T)}=-\frac{1}{G_\La(T) }\frac{\partial}{\partial\La}G_\La(T) \sim \frac{T}{\sqrt{\La}}.
\eeq
This equation reflects the fact that at large $T$ the quantum geometry looks effectively one-dimensional. Further we see from this expression that the average spatial length of the quantum geometry at intermediate $T$ behaves as
\beq
\expec{L}=\frac{\expec{V(T)}}{T}\sim \frac{1}{\sqrt{\La}}.
\eeq
Hence for typical time scales of $T\sim1/\sqrt{\La}$ we have 
\beq
\expec{V(T)}\sim T^{2}\quad\text{for } T \sim \frac{1}{\sqrt{\La}}
\eeq
which yields a Hausdorff dimension of $d_H=2$. In this sense the quantum geometry of CDT is much better behaved as a model of two-dimensional quantum gravity in contrary to DT which has a fractal dimension of $d_H=4$.  \index{Causal dynamical triangulations (CDT)!Hausdorff dimension}

\begin{figure}[t]
\begin{center}
\includegraphics[width=2.5in]{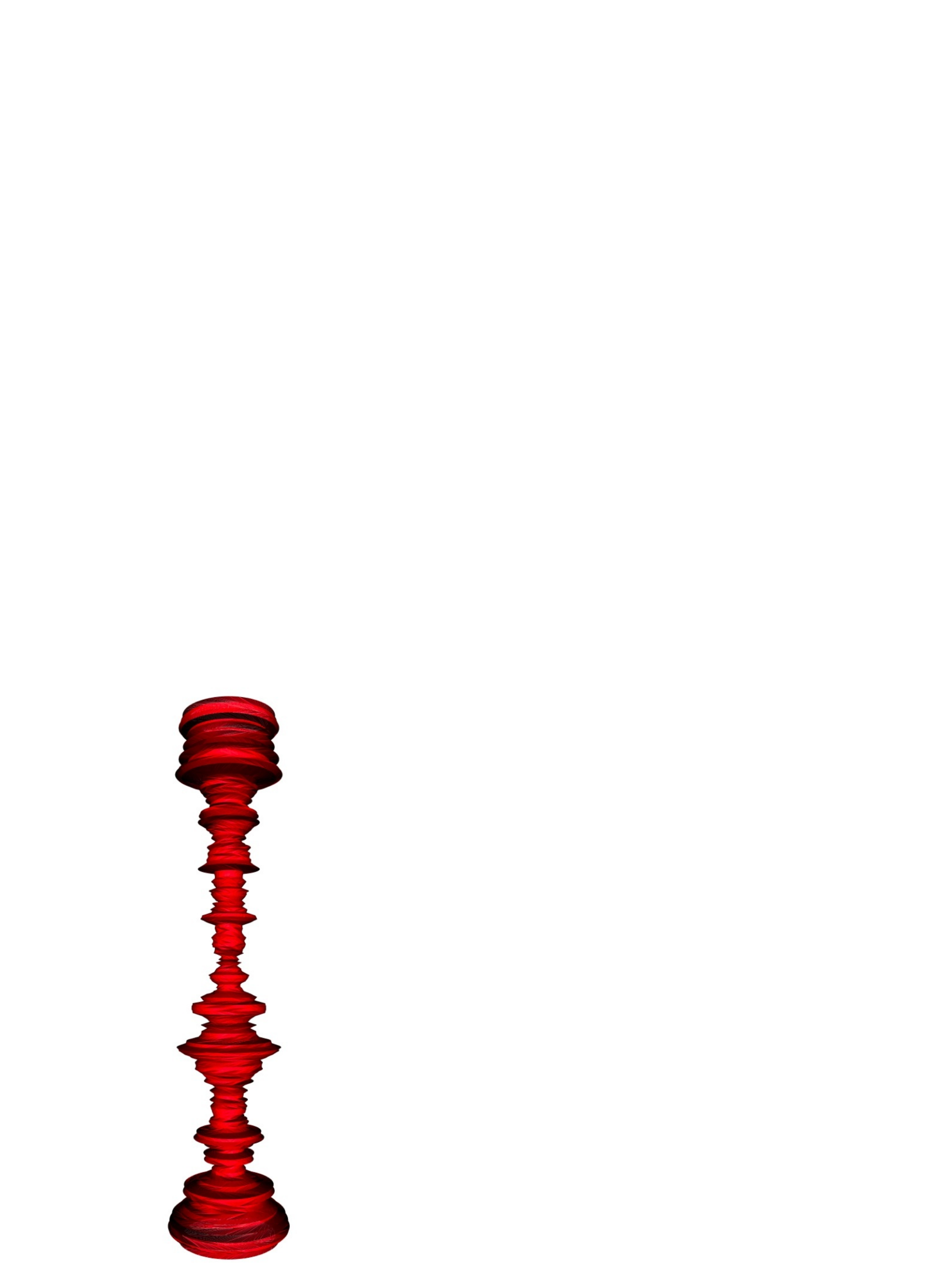}
\caption{A typical two-dimensional Lorentzian space-time. The compactified direction shows the spatial hypersurfaces of length $\expec{L}$ and the vertical axis labels proper time $T$. Technically, the picture was generated by a Monte-Carlo simulation, where a total volume of $N=18816$ triangles and a total proper time of $t=168$ steps was used. Further, initial and final boundary has been identified.}
\label{fig:CDTuniverse}
\end{center}
\end{figure}

To get a better picture of the quantum geometry of CDT it is instructive to look at a typical space-time geometry (Fig.~\ref{fig:CDTuniverse}). We see that in contrast to the Euclidean model there are no spatial topology changes. Further, as we have already seen above the average spatial length of the quantum geometry behaves as 
\beq
\expec{L} \sim \frac{1}{\sqrt{\La}}.
\eeq
A simple calculation as will be presented in Chap.\ \ref{Chap:emergence} reveals that also the fluctuations scale as 
\beq
\expec{\Delta L}\equiv \sqrt{\expec{L^2}-\expec{L}^2}\sim \expec{L} \sim \frac{1}{\sqrt{\La}}.
\eeq
Hence we see that the two-dimensional quantum geometry is purely governed by quantum fluctuations and there does not exist a sensible semiclassical background geometry. We will see in Chap.\ \ref{Chap:emergence} how this situation changes when one of the boundaries is taken to infinity and the geometry becomes non-compact.

\section{Summary and outlook to higher dimensions}\label{Sec:CDToutlook}

In this chapter we introduced two-dimensional Lorentzian quantum gravity defined through CDT.

The triangulations used in CDT have a global time-slicing without spatial topology changes. This is unlike DT where spatial topology changes are naturally present. The global notion of time enables us to define a gravitational Wick rotation. We solved the combinatorial problem and performed the continuum limit. As a result we saw that two-dimensional CDT and DT are distinct theories. Whereas the continuum dynamics of DT is completely dominated by spatial topology changes leading to a fractal dimension of $d_H\equ4$, the situation is different in CDT where we have $d_H\equ2$. In this sense two-dimensional CDT is arguably better suited as a model of two-dimensional quantum gravity than DT. Before discussing several relations between both theories in the next chapter, let us first comment on the results for higher dimensions.

In contrast to the disappointing situation for DT in higher dimensions, as described in the previous chapter, CDT lead to very interesting results for both $d\equ 3$ \cite{threed1,Ambjorn:2001cv,Ambjorn:2001br,threed2,Ambjorn:2003ct,Ambjorn:2003sr,Benedetti:2007pp} as well as $d\equ 4$ \cite{Ambjorn:2004qm,Ambjorn:2004pw,Ambjorn:2005db,Ambjorn:2005qt,Ambjorn:2007jv,Ambjorn:2008wc} (see also \cite{Ambjorn:2008sw,Ambjorn:2006jf,Ambjorn:2005qt,Ambjorn:2005jj} for a general overview). \index{Causal dynamical triangulations (CDT)!higher dimensional}

Let use briefly mention some of the numerical results obtained from Monte Carlo simulations in 3+1 dimensions. A very important non-trivial test for every non-perturbative formulation of quantum gravity is whether it can reproduce a sensible classical limit at macroscopic scales.
The numerical results indicate that the scaling behavior of the spatial volume as a function of space-time volume is that of a four-dimensional universe at large scales, a first indication of sensible classical behavior \cite{Ambjorn:2004qm}. Moreover, after integrating out all dynamical variables apart from the spatial volume as a function of proper time, one can derive the scale factor whose dynamics is described by the simplest minisuperspace model used in quantum cosmology \cite{Ambjorn:2004pw,Ambjorn:2005qt,Ambjorn:2007jv,Ambjorn:2008wc}. In Chap.\ \ref{Chap:emergence} we will describe an analogous model in two dimensions where a semi-classical background emerges from quantum fluctuations when the geometries become non-compact.

Having passed the first consistency checks regarding the \textit{macro}scopical structure of space-time it is very interesting what predictions one can make for the quantum nature of the \textit{micro}structure of space-time. One important observable which has been measured is the spectral dimension of space-time which is the dimension a diffusion process would feel on the space-time ensemble. Surprisingly, this quantity depends on the scale at which it is measured. More precisely, one observes a dimensional reduction from four at large scales to two at small scales within measurement accuracy \cite{Ambjorn:2005db}. This gives an indication that non-perturbative quantum gravity defined through CDT provides an effective ultraviolet cut-off through a dynamical dimensional reduction of space-time and might therefore be non-perturbatively renormalizable. The origin of the renormalizability could be a nontrivial fixed point scenario as described by Weinberg \cite{Weinberg:1980gg}. It is interesting to note that similar results were also obtained in  the exact renormalization group flow method for Euclidean quantum gravity in the continuum \cite{Souma:1999at,Reuter:2002kd,Reuter:2001ag,Litim:2003vp,Lauscher:2005qz}. \index{Causal dynamical triangulations (CDT)!dimensional reduction}\index{Causal dynamical triangulations (CDT)!spectral dimension}

An interesting observation at this point is that such an effective ultraviolet cut-off through a dynamical dimensional reduction of space-time provides us with a non-perturbative mechanism which regulates the theory. To come back to the discussion in Part I of this thesis, this means in particular that it is priori not necessary to introduce a fundamental cut-off such as the fundamental discreteness scale by hand.

\chapter{Relating Euclidean and causal dynamical triangulations \label{Chap:Relating}}
	
In the previous two chapters we introduced two-dimensional Euclidean quantum gravity defined through DT and two-dimensional Lorentzian quantum gravity defined through CDT. Even though both theories have some physically very distinct features such a the Hausdorff dimension it is nevertheless possible to establish certain relations between them.
In particular, following \cite{Ambjorn:1998xu,Ambjorn:1999fp}, we describe how CDT and DT can be related by respectively introducing or ``integrating out'' baby universes, i.e.\ spatial topology changes.

\section{Introducing spatial topology changes: From CDT to DT}\label{sec:CDTtoDT}

In this section we discuss the incorporation of spatial topology changes into the discretized framework of CDT and show how this model relates to DT in the continuum limit. Although there exist several ways of implementing spatial topology changes on the discrete level, they all lead to the same continuum description \cite{Ambjorn:1998xu}. 

\begin{figure}[t]
\begin{center}
\includegraphics[width=4in]{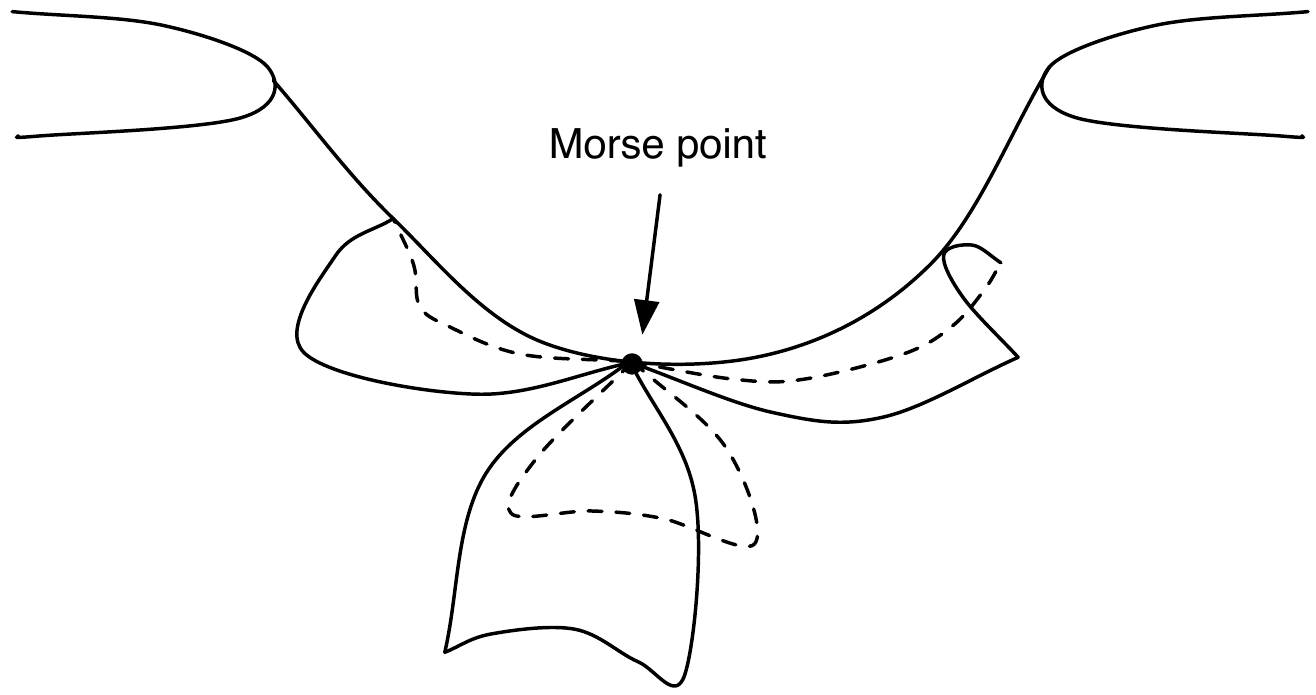}
\caption{Illustration of the double light cone causal structure at a Morse point in an up-side down trousers geometry.}
\label{fig:morsepoint}
\end{center}
\end{figure}

Before constructing the model let us first discuss some Lorentzian aspects of spatial topology changes. While in the Euclidean model the presence of baby universes was natural, their appearance in the Lorentzian model is far from obvious. Even though the metrics are Wick rotated from Lorentzian to Euclidean signature, one in general still has to perform an inverse Wick rotation. However, this is not so easy for geometries with baby universes, since these geometries do not admit a Lorentzian metric everywhere, as we already mentioned above. 
This is related to the fact that it is not possible to find
a non-vanishing vector field everywhere. However, since the geometry is compact there are only finitely many isolated points where this is not possible. These are so-called Morse points \cite{Dowker:1997hj,Borde:1999md,Dowker:1999cp,Dowker:1999wu,Dowker:2002hm}. If one imagines the splitting of a spatial universe into two, there is a Morse point precisely at
the moment where the spatial topology is that of a ``figure eight''. Embedded in $\mathbb{R}^3$ the two-dimensional geometry looks like an up-side down trousers and the Morse point precisely corresponds to the saddle point. Contrary to a generic point on the manifold, such a Morse point does not have a unique timelike vector field perpendicular to its spatial slice. In fact it has
two future, and two past light cones which is referred to as a double light cone structure \cite{Dowker:2002hm} and the resulting geometry is called causally discontinuous. Such a double light cone structure is illustrated in Fig.~\ref{fig:morsepoint} for the case of the up-side down trousers. One observes that both legs of the trousers carry a future light cone belonging to the Morse point at the splitting, while the mother universe carries the two past light cones.
\index{Morse point}\index{Baby universe} \index{Spatial topology changes}\index{Trousers geometry}\index{Double light cone}

Since there are indications that the Einstein-Hilbert action develops complex valued singularities at these Morse points \cite{Louko:1995jw}, it is not a priori clear that the Wick rotation as employed in CDT is valid at these points. However, since we want to relate our model to Euclidean DT, we can confine ourselves to the Euclidean sector of CDT. In Chap.\ \ref{Chap:cap} we present a generalized model of CDT which includes spatial topology changes, and where one in principle would like to Wick rotate back to the Lorentzian sector. However, in this model there is a coupling constant associated to every splitting which might be used to absorb possible divergencies. 

\begin{figure}[t]
\begin{center}
\includegraphics[width=6in]{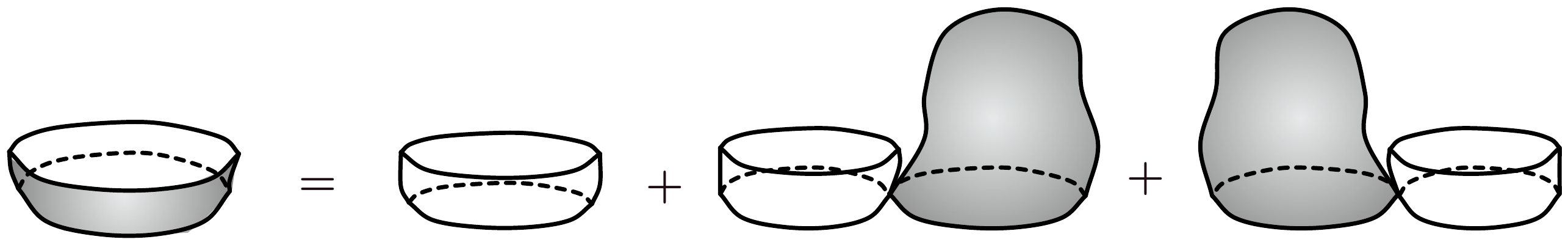}
\caption{Illustration of the one-step propagator with an off-splitting ``baby universe''.}
\label{fig:dressedonesteppropagator}
\end{center}
\end{figure}

Let us now, after this short digression, focus on the explicit construction of the discretized CDT model with spatial topology changes. The first step is to generalize the one-step propagator by allowing the initial loop to have topology of a ``figure eight''. One possibility to do so is to non-locally identify two points of an initial spatial universe with topology $S^1$. For a boundary of length $l_1$ this so-called pinching leads to a combinatorial factor of $l_1$ in the full one-step
propagator. A baby universe is then created by assigning a disc function to one of the loops of the ``figure eight'' and the initial loop of a ``bare'' one-step propagator to the other one
(see Fig.~\ref{fig:dressedonesteppropagator}).
Putting the above relation into formulas, we see that the new, or ``dressed'', one-step propagator is related to the old, or  ``bare'', one-step propagator as follows
\beq \label{eq:onsteppropagatordressed}
G_\l(l_1,l_2;t\equ 1)  = G_\l^{(b) }(l_1,l_2;t\equ 1) + 2 \sum_{l=1}^{l_1-1} l_1 w(l_1\mi l,g) \,G_\l^{(b) }(l,l_2;t\equ 1),
\eeq
where $w(l,g)$ denotes the so far undetermined disc function and the factor of two comes again from the two different ways of attaching the disc function. One observes the analogy to the peeling equation for the fixed geodesic distance two-loop amplitude in DT.

It is readily checked that the ``dressed'' one-step propagator obeys the usual semi-group property, i.e. \rf{eq:discrcompositionllmarked},
\beq
G_\l(l_1,l_2;t_1+t_2)  = \sum_l G_\l(l_1,l;t_1) G_\l(l,l_2;t_2).
\eeq
Making the standard choice $t_1\equ 1$ and $t_2\equ t-1$ to relate to the one-step propagator we obtain
\beq\label{eq:discrcompositionlldressed}
G_\l(l_1,l_2;t)  = \sum_l G_\l(l_1,l;1) \;G_\l(l,l_2;t\mi 1),
\eeq
or in terms of generating functions 
\bea
{G(x,y;g;t)  =}~~~~~~~~~~~~~~~~~~~~~~~~~~~~~~~~~~~~~~~~~~~~~~~~~~~~~~~
~~~~~~~~~~~~~~~~~~~~~~
 & &  \nonumber \\
\ointz  \left[G_\l^{(b) }(x,z^{-1};1) \pl 2 x \frac{\prt}{\prt x}
\Bigl( w(g,x)  G_\l^{(b) }(x,z^{-1};1) \Bigr)  \right]  G(z,y;g;t \mi 1).
\label{eq:GxygtGdressed}
\eea
Inserting the explicit solution for the ``bare'' one-step propagator $G^{(b)}(x,z;g;1) $,
i.e.\ \rf{eq:G10xyg1}, as derived in the previous chapter, we get
\beq 
G(x,y;g;t)  = \Bigl[1+2 x\frac{\prt w(g,x) }{\prt x}+ 2 x w(g,x) \frac{\prt}{\prt x} \Bigr]
\, \frac{gx}{1-gx} \, G\Bigl( \frac{g}{1\mi gx},y;g;t\mi 1\Bigr) \label{eq:Gxygtdressed}.
\eeq
At this point neither the dressed disc amplitude $w(g,x)$ nor the propagator $G(x,y;g;t) $ are known. In particular, $w(g,x)$ is \emph{not} fixed to be the Euclidean disc function. In the following we will show how one can use scaling arguments to uniquely determine the continuum propagator $G_\La(X,Y;T) $ and the continuum disc amplitude $W_\La(X)$ from this equation. Let us assume the standard scaling relations for the boundary and bulk cosmological constants as used in DT and the ``bare'' CDT model,
\beq \label{eq:relscaling1}
g=\frac{1}{2}e^{-a^2 \La},\quad x=e^{-a X},\quad y=e^{-a Y}.
\eeq
Inserting the scaling relations into \rf{eq:Gxygtdressed} and setting $t\equ0$ we obtain the initial condition
\beq
G_\La(X,Y;T\equ0)=\frac{1}{X+Y} 
\eeq
where we introduced the following wave function renormalization
\beq\label{top8x}
G_\La(X,Y;T)=aG_\l(x,y;t) 
\eeq
or in terms of the lengths variables
\beq
G_\La(L_1,L_2;T)= a^{-1} G_\l(l_1,l_2,t).
\eeq

\begin{figure}[t]
\begin{center}
\includegraphics[width=1.4in]{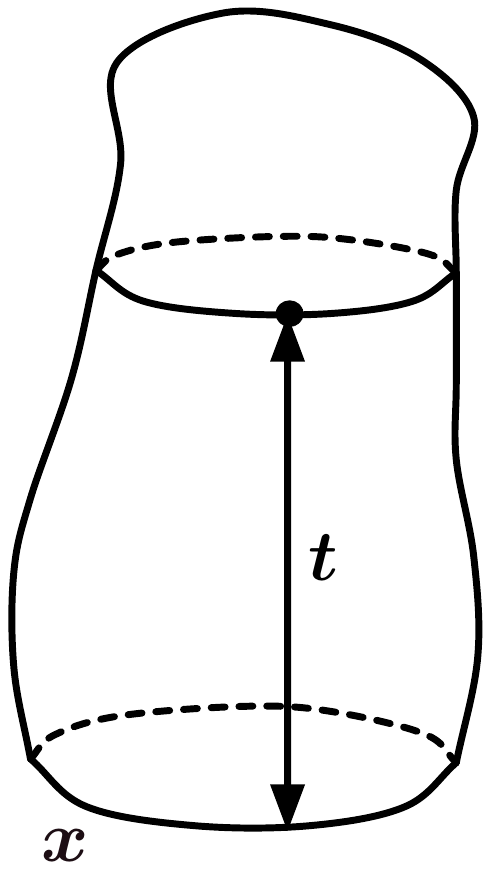}
\caption{Decomposition of the disc function with a marked point in the bulk into a propagator with arbitrary time and a disc function with a mark on the boundary.}
\label{fig:discfunctiondecomposition}
\end{center}
\end{figure}

Extracting the scaling of the disc function is more difficult, since it actually depends on the scaling of time. To do so Ambj{\o}rn and Loll used the following combinatorial identity
\beq\label{an1}
g\ \frac{\prt w(g,x) }{\prt g} = \sum_t \sum_l G(x,l;g;t)  \, l\, w(l,g),
\eeq
which in terms of the generating functions reads
\beq\label{an2}
g\ \frac{\prt w(g,x) }{\prt g}=\sum_t \ointz G(x,z^{-1};g;t) \; \frac{\prt w(g,z) }{\prt z}.
\eeq
Fig.~\ref{fig:discfunctiondecomposition} illustrates this relation: If we introduce a mark in the bulk by taking a derivative with respect to the fugacity $g$, the disc function can be decomposed into a propagator and a smaller disc function which has this mark on its boundary. Further, one sums over all $t$ to ensure that the mark can be anywhere in the bulk. 

Recall from Chap.\ \ref{Chap:dt} and Chap.\ \ref{Chap:cdt} that the disc function for DT scales as
\beq
w^{(eu)}(g,x) = w_{ns} + a^{3/2}W^{(eu)}_\La(X) 
\eeq
while the disc function for CDT scales as
\beq
w^{(b)}(g,x) = a^{-1}W^{(b)}_\La(X). 
\eeq
Hence, we make the following general ansatz for the scaling relation which includes both cases 
\beq\label{an3}
w(g,x) = w_{ns} + a^{\eta}W_\La(X)  + \mbox{less singular terms}.
\eeq
Further, the general scaling ansatz for the time variable reads
\beq \label{eq:relscaling2}
T = a^{\ep}t,~~~~~\ep >0.
\eeq
which describes both the non-canonical scaling of time for DT, i.e.\ $\ep\equ1/2$ as well as
the canonical scaling of time for CDT, i.e.\ $\ep=1$.
Below we show that by allowing the branching into baby universes to contribute in the continuum limit one is led to a non-canonical scaling of time as for DT.

Inserting the scaling relations \rf{eq:relscaling1}, \rf{an3} and \rf{eq:relscaling2} into the combinatorial identity \rf{an2} we obtain
\bea
&&\frac{\prt w_{ns}}{\prt g}- 2a^{\eta-2}\frac{\prt W_\La(X) }{\prt \La} = \nonumber \\
&&\hspace{.6cm} \frac{1}{a^{\ep}}\int d T \int dZ\;  G_\La(X,-Z;T) \bigg[ \frac{\prt w_{ns}}{\prt z} -a^{\eta-1}\frac{1}{z_{c}} \frac{\prt W_\La(Z) }{\prt Z}\bigg],
\label{an4}
\eea
where we have $(x,g) =(x_c,g_c) $ in the non-singular part.

From the last equation and the requirement that $\epsilon >0$ it follows that there are
only two consistent choices for $\eta$:
\begin{itemize}
\item{{\it Scaling 1: }$\eta<0$}

In this range the non-scaling part does not survive the continuum limit and Eq.\ \rf{an4} becomes
\beq\label{an5c}
a^{\eta-2} \frac{\prt W_\La(X) }{\prt \La} = \frac{a^{\eta-1}}{2 a^{\ep}}
  \int d T \int dZ\;  G_\La(X,-Z;T)  \;\frac{1}{z_{c}}
  \frac{\prt W_\La(Z) }{\prt Z}.
\eeq
Hence, we conclude that $\ep=1$ as in the case of CDT. We observe that whenever the non-scaling part of the disc function is not present in the continuum result time must scale canonically.

\item{{\it Scaling 2: }$1<\eta<2$}

In this range Eq.\ \rf{an4} splits into two equations
\beq\label{an5}
-a^{\eta-2} \frac{\prt W_\La(X) }{\prt \La} = \frac{1}{2 a^{\ep}}\,\frac{\prt w_{ns}}{\prt z}\bigg|_{z=x_c}\;\int d T \int dZ\;  G_\La(X,-Z;T),
\eeq
and
\beq\label{an6}
\frac{\prt w_{ns}}{\prt g}\bigg|_{g=g_c} = -\frac{a^{\eta-1}}{a^{\ep}} \int d T \int dZ\;  G_\La(X,-Z;T) \;\frac{1}{z_{c}} \frac{\prt W_\La(Z) }{\prt Z}.
\eeq
From the first equation we see that $a^{\eta-2}= 1/a^{\ep}$ while from the second equation we have $ a^{\eta-1}/a^{\ep}=1$. Combining these requirements we get $\ep= 1/2$ and $\eta=3/2$
which precisely correspond to the scaling relations of DT, i.e \rf{eq:DTscalingdiscepsilon} and \rf{eq:DTscalingdisceta}. Further, we observe that using $\ep= 1/2$ and $\eta=3/2$ Eq.\ \rf{an5} reads
\beq\label{an5a}
-\frac{\prt W_\La(X) }{\prt \La} \sim \int_0^\infty dT G_\La(X,L_2=0,T).
\eeq
which is nothing but the relation between the disc function and the propagator as obtained in DT, i.e.\ \rf{eq:DTdiscproprel}. In addition for $\ep= 1/2$ and $\eta=3/2$ Eq.\ \rf{an6} becomes
\beq\label{an7}
\int d T \int dZ\;  G_\La(X,-Z;T) \;\frac{\prt W_\La(Z) }{\prt Z} = \mbox{const},
\eeq
where the constant is related to the non-scaling part of the disc function and does not play any role in the continuum theory.
\end{itemize}

\begin{figure}[t]
\begin{center}
\includegraphics[width=5.5in]{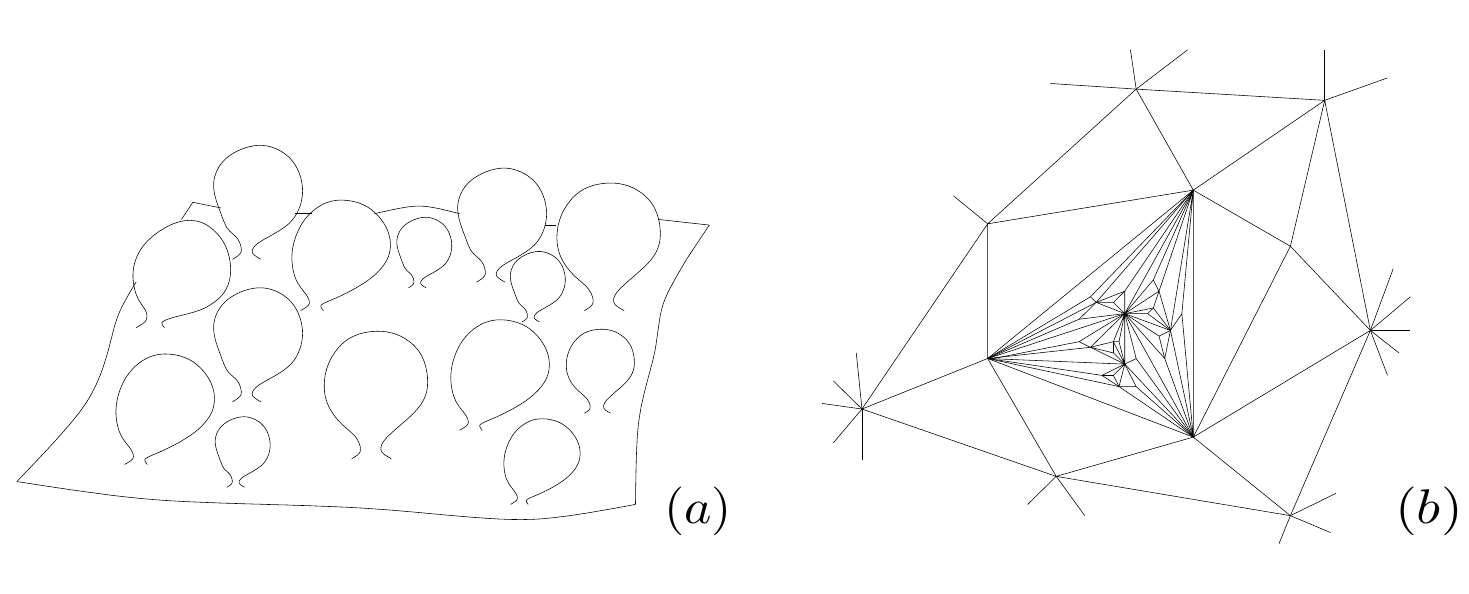}
\caption{$(a)$ At every point in the quantum geometry there is an infinitesimal baby universe. $(b)$ When trying to draw the baby universes on a plane one is forced to draw smaller and smaller triangles reflecting the fractal structure of the geometry.} \index{Dynamical triangulations (DT)!fractal structure}
\label{fig:mini}
\end{center}
\end{figure}

The relation \rf{an5a} has an intersting interpretation in terms of baby universes, namely, it shows that at any mark in the bulk there is a baby universe whose boundary is of length of the cut-off. Further, since the mark can be anywhere, we conclude that
\emph{near every point of the quantum geometry there is a baby universe with cut-off size boundary.} This situation is illustrated in Fig.~\ref{fig:mini} (a). Mapping the picture into the plane, as shown in Fig.~\ref{fig:mini} (b), clearly visualizes the fractal structure of the quantum geometry. Another indication for the importance of geometries with infinitesimal boundaries comes from the Laplace transform of \rf{an7}, i.e.\
\beq
\int d T\int dL\;  G_\La(L,L';T) \;L'\;W_\La(L')  \sim \delta(L).
\eeq
Essentially, this equation shows that the distribution of geometries is peaked around universes that have infinitesimal boundary length. 

Using the above scaling relations one can now analyze the scaling limit of \rf{eq:Gxygtdressed} to obtain a differential equation for the dressed propagator. In order for the scaling limit of the equation to exist, the critical values $x_c,\, g_c$ and $w_{ns}$ must satisfy two relations which can be straightforwardly determined from \rf{eq:Gxygtdressed}. The remaining continuum differential equation equation reads
\bea
a^\ep\frac{\prt}{\prt T}\, G_\La(X,Y;T) & =&-a \, \frac{\prt}{\prt X} \Bigl[ (X^2-\La)  G_\La(X,Y;T) \Bigr] \nn
&&-2 a^{\eta-1}\frac{\prt}{\prt X} \Bigl[W_\La(X)  G_\La(X,Y;T) \Bigr].
\label{top13}
\eea
The first term on the right-hand-side of \rf{top13} is related to the bare CDT model, while the second term corresponds to the creation of baby universes as is also present in DT. In particular, we saw in the previous chapter that the first term was related to the effective quantum Hamiltonian of CDT and hence represents the kinetic term of the more general model \eqref{top13}.

In the case where $\eta \leq 0$ (more precisely $\eta \leq 1$) the last term in \rf{top13} diverges. This reflects the fact that the bare model is incompatible with spatial topology changes. However, dropping the divergent term, i.e.\ excluding spatial topology changes we have already seen that
for $\eta \leq 0$ we need to have that $\ep=1$ leaving us with the continuum differential equation for the propagator of the bare CDT model, i.e.\ \rf{eq:differentialequationGXYT}, as a \emph{unique} solution
\beq \label{eq:difequationGXYTpure}
\frac{\prt}{\prt T}\, G_\La(X,Y;T)  = - \frac{\prt}{\prt X} \Bigl[ (X^2-\La)  G_\La(X,Y;T) \Bigr].
\eeq

For the second possible scaling, where $ 1 < \eta < 2 $, the last term on the right-hand-side of \rf{top13} will dominate over the first while still being compatible with the time derivative term. Hence, the kinetic term of the model does \emph{not} survive the continuum limit and \emph{the dynamics is entirely governed by the off-spliting baby universes}, or as Ambj{\o}rn and Loll express it, {\em once we allow for the creation of baby universes, this process will completely dominate the continuum limit.} We have seen above that the only scaling consistent with $ 1 < \eta < 2 $ is $(\eta,\ep) =(3/2,1/2) $. Inserting
the corresponding scaling relation into \rf{top13} we obtain the following differential equation
\beq\label{top16}
\frac{\prt}{\prt T}\, G_\La(X,Y;T)  = -2\frac{\prt}{\prt X} \Bigl[W_\La(X)  G_\La(X,Y;T) \Bigr].
\eeq
This is precisely of the form of the differential equation \rf{eq:geodDT} for the propagator of DT. However, at this stage $W_\La(X)$ and with that $G_\La(X,Y;T)$ are still undetermined.
In the following we want to show how this equation combined with Eq.\ \rf{an5a}, completely determines the continuum disc function $W_\La(X)$ to be that of DT.
Integrating \rf{top16} with respect to $T$ and using the initial condition
\beq
G_\La(X,Y;T\equ 0) =\frac{1}{X+Y}
\eeq
i.e. in Laplace transform language
\beq\label{top18a}
G_\La(X,L_2\equ 0;T\equ 0) =1,
\eeq
we obtain
\beq\label{top18}
-1 = \frac{\prt}{\prt X}\bigg[ W_\La(X)  \frac{\prt}{\prt\La} W_\La(X)\bigg].
\eeq
From dimensional analysis one can easily see that $W_\La(X)$ must have mass dimension $-3/2$ and hence can be written as $W_\La^2(X)  = X^3 F(\SL/X)$. This condition implies the following general form for the solution of the disc function
\beq\label{top19}
W_\La(X)  = \sqrt{-2\La X + b^2 X^3+ c^2 \La^{3/2}}.
\eeq
We can further fix one of the two constants by noting that $W_\La(X)$ is not allowed to have any singularities or cuts for $\Re(X)\!>\!0$. This requirement is essentially the same as saying that 
the inverse Laplace transform of \rf{top19}, i.e.\ $W_\La(L)$, should be bounded for
$L\to \infty$. This completely fixes the analytic structure of the disc function, yielding 
\beq\label{top20}
W_\La(X)  = b \Big(X-\frac{\sqrt{2}}{b\,\sqrt{3}} \,\SL\Big)
\sqrt{X+ \frac{2\sqrt{2}}{b\,\sqrt{3}}\SL}.
\eeq
Upon rescaling of the boundary and bulk cosmological constants we can absorb the constant $b$ and obtain the disc function of DT, i.e.\ \rf{eq:DTdisc}, 
\beq\label{top21}
W_\La (X)  = (X -\oh \SL )  \sqrt{X+\SL}.
\eeq

\section{Integrating out baby universes: From DT to CDT} \label{sec:integrationout}
\index{Integrating out baby universes}

In the previous section we showed how to obtain DT from CDT by introducing spatial topology changes, i.e.~by allowing baby universes to appear. In the following we want to show the opposite, namely how to obtain CDT from DT by ``integrating out'' the baby universes \cite{Ambjorn:1999fp}. Since the baby universes completely dominate the continuum dynamics of DT such a relationship has to be established on the discrete level.

\begin{figure}[t]
\begin{center}
\includegraphics[width=3.5in]{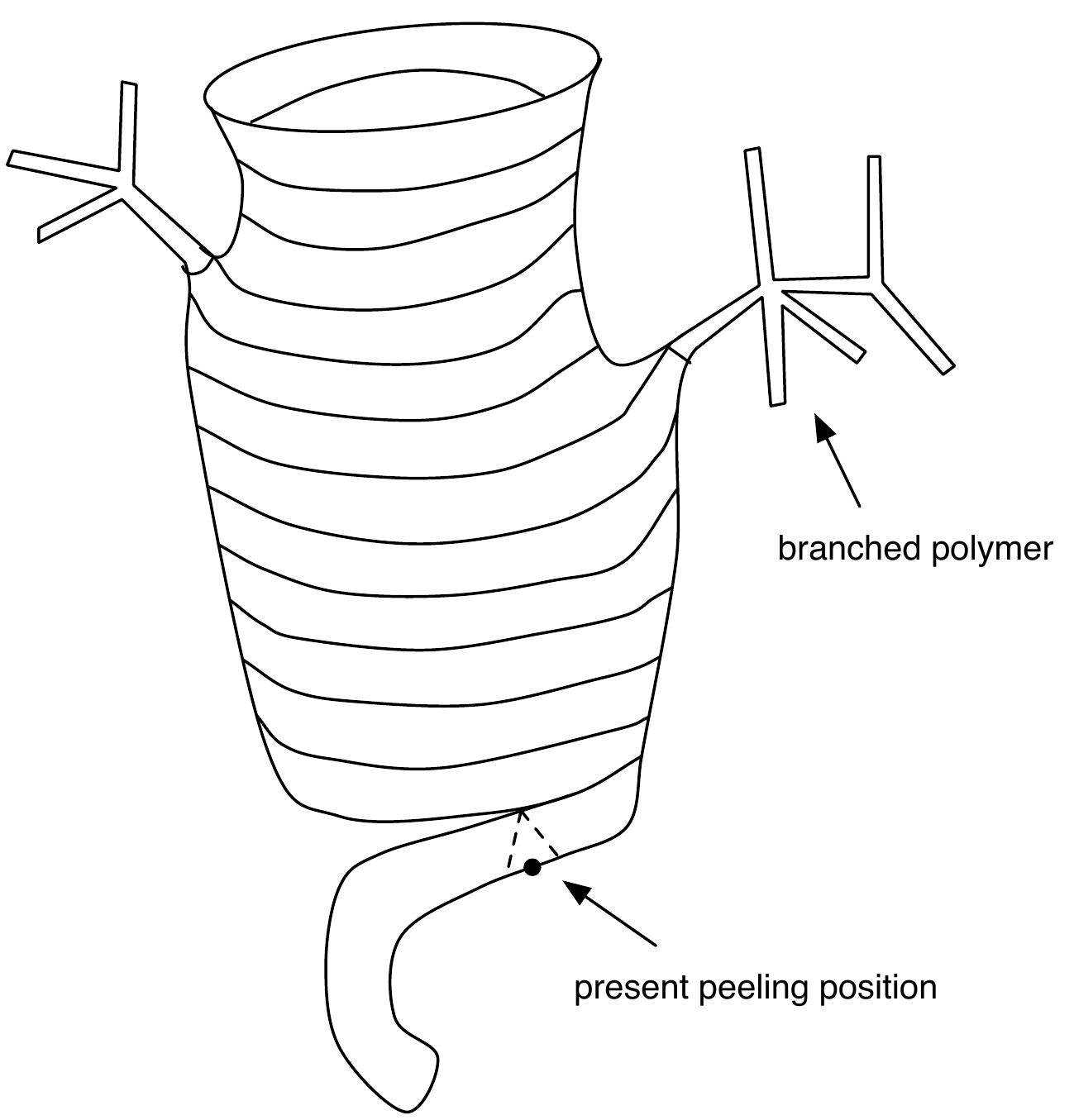}
\caption{Illustration of the peeling relation \eqref{eq:PDErelatingCDT}. In each step of the decomposition either a triangle is removed or a double link which results into an off-splitting of a branched polymer.}
\label{fig:peelingBP}
\end{center}
\end{figure}

We remind the reader of the following differential equation for the discrete fixed geodesic distance two-loop function in DT, i.e.\ \eqref{eq:geodDTdis}, which we obtained from the peeling procedure,
\beq 
\frac{\partial}{\partial t}G^{(eu)}(z,w;g;t)=-2\frac{\partial}{\partial z}\left[ f^{(eu)}(g,z) G^{(eu)}(z,w;g;t)\right],
\eeq
where we defined 
\beq\label{eq:defoffDT}
f^{(eu)}(g,z)=-\frac{1}{2}(z- g z^2)+w^{(eu)}(g,z). 
\eeq
The initial condition of this differential equation was given in \eqref{eq:initialconditionDT}, i.e.
\beq\label{eq:initialconditionDTrepeat}
G(z,w;g;0)=\frac{1}{z w}\frac{1}{z w-1}.
\eeq
Here the first term in \eqref{eq:defoffDT} corresponded to the first decomposition move, e.g.\ removing a triangle, while the second term corresponded to the second move, where a double link was removed and a baby universe (disc function) splits off. A first idea for integrating out the baby universes would therefore be to drop the second term in \eqref{eq:defoffDT}. However, as we are using unrestricted triangulations (recall Fig.~\eqref{fig:disc} (b)) also double links are present. Hence, there could still be off-splitting branched polymers which correspond to zero volume disc functions, i.e.\ \eqref{eq:BPgen},
\beq
w^{(bp)}(z)\equiv w^{(eu)}(g\equ 0,z)=\frac{1}{2}\left(z-\sqrt{z^2-4} \right).
\eeq
Thus we integrate out the (non-zero volume) baby universes by simply replacing 
the term $w^{(eu)}(g,z)$ in \eqref{eq:defoffDT} by $w^{(bp)}(z)$, yielding
\beq \label{eq:PDErelatingCDT}
\frac{\partial}{\partial t}G(z,w;g;t)=-2\frac{\partial}{\partial z}\left[ f(g,z) G(z,w;g;t)\right],
\eeq
with 
\beq\label{eq:defoffCDT}
f(g,z)=-\frac{1}{2}(z- g z^2)+w^{(bp)}(z),
\eeq
subject to the same initial conditions \eqref{eq:initialconditionDTrepeat}. Fig.~\ref{fig:peelingBP} illustrates the peeling equation \eqref{eq:PDErelatingCDT}.

We now want to show how to obtain the continuum CDT propagator by taking a suitable continuum limit of \eqref{eq:PDErelatingCDT}. 
It is natural to use the standard scaling relations for the boundary and bulk cosmological constants which were applicable for both DT and CDT, 
\begin{equation}\label{eq:relatingscalingrel}
g=g_c \,e^{-a^2\,\Lambda},\quad z=z_c \,e^{a\,Z},\quad w=z_c \,e^{a\,W},
\end{equation}
As before the demanded continuum initial condition
\beq
G_\La(Z,W;T\equ 0)=\frac{1}{Z+W},
\eeq
already fixed the wave function renormalization of the propagator, i.e.
\beq 
G_\La(Z,W;T)= a G(z,w;g;t).
\eeq 
Further, we note that to obtain a non-trivial continuum dynamics we demand that $f(g,z)\sim\cO(a^2)$. This requirement completely fixed the scaling relations: Firstly, form $f(g,z)\sim\cO(a^2)$ and the scaling relations \eqref{eq:relatingscalingrel} it follows that
\beq
T\sim a t
\eeq 
as it is the case for CDT. Secondly, the requirement that the order one and order $a$ part of $f(g,z)$ should vanish determines
\beq
g_c=\frac{1}{4},\quad z_c=1. 
\eeq
Inserting the complete scaling relations into \eqref{eq:PDErelatingCDT} yields (under rescaling of the couplings)
\beq
\frac{\partial}{\partial T}G_{\La}(Z,W,T) =-\frac{\partial}{\partial Z}\left[(Z^2-\La) G_{\La}(Z,W,T) \right]
\eeq
with the above initial conditions. This differential equation precisely coincides with the differential equation for the propagator of CDT, i.e. \eqref{eq:differentialequationGXYT}. Hence we showed how to obtain CDT from DT by integrating out baby universes.

\section{The renormalization relation}
\index{Causal dynamical triangulations (CDT)!renormalization relation to DT}
\index{Dynamical triangulations (DT)!renormalization relation to CDT}

As an interesting relation between DT and CDT, it was shown in \cite{Ambjorn:1999fp} that it is possible to ``renormalize'' the couplings and time of DT in such a way that the differential equation for the DT propagator is matched to the one of the CDT propagator. This ``renormalization'' relation can be established both in the discrete as well as in the continuum \cite{Ambjorn:1999fp}. We will in the following only derive the continuum relation. 

Let us for the remainder of this chapter label all Euclidean quantities with a tilde. The differential equation for the Euclidean propagator thus reads,
\beq\label{top16new}
\frac{\prt}{\prt \tT}\, \tG_\tL(\tX,\tY;\tT)  = -2\frac{\prt}{\prt \tX} \Bigl[\tW_\tL(\tX)  \tG_\tL(\tX,\tY;\tT) \Bigr],
\eeq
where
\beq\label{top16newS}
\tW_\tL (\tX)  = (\tX -\oh \tsL )  \sqrt{\tX+\tsL}
\eeq
was the Euclidean disc function.
The corresponding equation in the framework of CDT reads
\beq\label{top16new2}
\frac{\prt}{\prt T}\, G_\La(X,Y;T)  = -\frac{\prt}{\prt X} \Bigl[\hat{W}_\La(X)  G_\La(X,Y;T) \Bigr]
\eeq
with
\beq\label{top16newS2}
\hat{W}_\La(X) =X^2-\La.
\eeq
Here the hat (not to be confused with the tilde) should remind ourself that $\hat{W}_\La(X)$ is not the disc function of CDT but rather related to the effective quantum Hamiltonian of the system.

To relate \eqref{top16new} and \eqref{top16new2} we make the following ansatz for the renormalization relations of the couplings and time
\beq \label{renormalization1}
\tX=\tX(X,\La),\quad \tT= b\,\La^{1/4} \, T,\quad \tL=\tL(\La),
\eeq
where $b$ is a constant and the factor of $\La^{1/4}$ in front of the $T$ has been chosen for dimensional reasons. To map the differential equation \eqref{top16new} for DT into the differential equation \eqref{top16new2} for CDT using the relations \eqref{renormalization1}, we must have the following wave function renormalizations
\beq
G_\La(X,Y;T)= \left( \frac{\partial X}{\partial \tX} \right)^{-1} \tG_{\tL(\La)}(\tX(X,\La),\tY(Y,\La); b\, \La^{1/4}\, T) 
\eeq
and
\beq
\hat{W}_\La(X) = 2\, b \, \La^{1/4}  \left( \frac{\partial X}{\partial \tX} \right) \tW_{\tL(\La)} (\tX(X,\La)). 
\eeq
Integrating the second relation gives
\beq\label{renormalization2}
  2 \,  b\, \La^{1/4} \int \frac{dX}{\hat{W}_\La(X)} =  \int \frac{\tX}{ \tW_\tL(\tX)}+k.
 \eeq
Since we require that $\tX\to\infty$ should imply $X\to\infty$ the integration constant must be $k\equ0$.
Inserting \eqref{top16newS} and \eqref{top16newS2} into \eqref{renormalization2} with $k\equ0$ and performing the integrals gives
\beq 
\frac{b}{\sqrt[4]{\La}} \log\left( \frac{X+\sL}{X-\sL} \right)= \sqrt{\frac{2}{3}}\,\frac{1}{\sqrt[4]{\tL}}
\log\left( \frac{\sqrt{\frac{2}{3}} \tL^{1/4} \sqrt{\tX +\sqrt{\tL}} +\sqrt{\tL} }{ \sqrt{\frac{2}{3}} \tL^{1/4} \sqrt{\tX +\sqrt{\tL}} -\sqrt{\tL}} \right)
\eeq
To further simplify this expression we perform a rescaling of both $X$ and $\sL$ to choose
\beq\label{renormalizationrelation1}
\sL=\frac{3}{2}b^2 \sqrt{\tL}.
\eeq
Using this relation \eqref{renormalizationrelation1} becomes
\beq\label{renormalization3}
\frac{X}{\sL}=\sqrt{\frac{2}{3}} \frac{\sqrt{\tX+\sqrt{\tL}}}{\tL^{1/4}}.
\eeq
Together with
\beq
 T=\frac{\tT}{b\,\La^{1/4}}
\eeq
these equations determine the whole set of renormalization relations. 

These renormalization relations map the differential equation of the DT propagator \eqref{top16new} to the corresponding differential equation of the CDT propagator \eqref{top16new2}. Further, it shows that the process of ``integrating out'' baby universes which we discussed in the previous section can be understood as a renormalization of the cosmological constants and the time variable combined with a dressing of the propagator. 

It is interesting to notice that the relation \rf{renormalization3} is similar to one encountered in
regularized bosonic string theory in dimensions 
$d\geq 2$ \cite{Durhuus:1984in,Ambjorn:1985az,Ambjorn:1987wu}, where the world sheet degenerates into so-called branches polymer. The two-point function of these branched polymers is related to
the ordinary two-point function of the free relativistic particle
by chopping off (i.e.\ integrating out) the branches, just leaving
for each branched polymer connecting two points in target space
one {\it path} connecting the two points. The mass-parameter
of the particle is then related to the corresponding parameter
in the partition function for the branched polymers as
$X/\SL$ to $\tX/\sqrt{\tL}$ in \rf{renormalization3}. 

\section{Discussion and outlook}

In this chapter we explained several relations between two-dimensional Euclidean quantum gravity defined through DT and two-dimensional Lorentzian quantum gravity defined through CDT.

It was shown that when introducing spatial topology changes into the framework of CDT one arrives at a continuum differential equation for the gravitational propagator which only allows for two possible scalings. In the first scaling time scales canonically and spatial topology changes, i.e.\ off-splitting baby universes are forbidden. This scaling leads back to the `bare'' model of CDT without spatial topology changes as described in the previous chapter. However, in the second scaling where time scales non-canonical spatial topology changes are possible and the model naturally leads to DT. Further, we observed how in this scaling the baby universes completely dominate the continuum limit, explaining the fractal structure of two-dimensional Euclidean quantum gravity.

In the Sec.\ \ref{sec:integrationout} of this chapter we showed the opposite relation, namely, when integrating out the baby universes in DT one recovers CDT. This was done by introducing a new peeling equation similar to the one used to calculated the fixed geodesic distance two-loop function in Sec.\ \ref{sec:peelingDT}, where outgrowing baby universes are replaced by branched polymers, i.e.\ zero-volume baby universes. 

In the last section of this chapter we showed how the process of ``integrating out'' baby universes could be understood as a renormalization of the cosmological constants and the time variable combined with a dressing of the propagator. 

In summary, we showed many relations between CDT and DT by respectively introducing or ``integrating out'' baby universes. An unsatisfactory situation at this point is that we were in a sense not able to ``regularize'' the spatial topology changes. Either baby universes are absent as in CDT or the average number of baby universes is infinite as in DT which completely dominates the continuum dynamics. In Chap.\ \ref{Chap:cap} we show how one can naturally introduce a coupling which regulates spatial topology changes, leading to a finite number of baby universes in the continuum. Further, this formulation lead to the discovery of many other relations between CDT and DT within the framework of matrix models as will be discussed in Part IV.

\chapter{The emergence of background geometry in two dimensions \label{Chap:emergence}}
	In the previous chapters we studied both Euclidean as well as causal dynamical triangulations with compact geometries. In particular, we saw that the quantum geometry in both cases does not posses a semiclassical background, but is rather dominated entirely by quantum fluctuations. In this chapter we show how this situation changes when one studies the case of non-compact space-times. Before moving to our prime interest of non-compact space-times in CDT \cite{Ambjorn:2006hu,Ambjorn:2006tt} let us first review some results of the Euclidean model \cite{aagk}. \index{Non-compact space-times}

\section{Two-dimensional Euclidean quantum gravity with non-compact space-time}

In this thesis we have so far only considered two-dimensional Euclidean and Lorentzian quantum gravity with compact space-time. The study of two-dimensional Euclidean quantum gravity
with non-compact space-time was initiated by the Zamolodchikovs (ZZ) in the context of Liouville theory \cite{zz}. In particular, they
showed how to use conformal bootstrap and the cluster-decomposition
properties to quantize Liouville theory on the pseudo-sphere, i.e.\ the Poincar\'e disk.
\index{Poincar\'e disk}\index{Pseudo-sphere}

\index{Dynamical triangulations (DT)!non-compact space-times}

It was later shown by Martinec \cite{Martinec:2003ka} and Seiberg et al.\ \cite{Seiberg:2003nm,Kutasov:2004fg}
how the work of the Zamolodchikovs fitted into the framework of
non-critical string theory, where the ZZ-theory could
be reinterpreted as special branes, now called ZZ-branes.
Particularly, they found that the ZZ-brane of two-dimensional Euclidean gravity
was associated with the zero of of the disc function, \index{ZZ-brane}
\beq\label{0.0}
W^{(eu)}_\La(X) = (X -\oh \sqrt{\La})\sqrt{X + \sqrt{\La}}.
\eeq
In \cite{aagk,Ambjorn:2007iq,Ambjorn:2007xe} it was shown how this could be understood
in terms of worldsheet geometry, i.e.\ from a two-dimensional quantum gravity
point of view. Let us briefly describe the main idea. We can use the fixed geodesic distance two-loop function, i.e.\ \eqref{eq:DTgeodsol}, to compute the average length of the final boundary,
\beq\label{eq:emergence1xxx}
\la L(T)\ra_{X,Y,T} = -\frac{1}{G_\La (X,Y;T)} \, 
\frac{\prt G_\La (X,Y;T)}{\prt Y} =  \frac{1}{\bX(T,X)+Y}.
\eeq 
Further, we see from this that the average length of the boundary of a disc with a marked point in the bulk at geodesic distance $T$ to the boundary is given by 
\beq
\expec{L(T)}_{Y,T}=\frac{1}{\bX(T)+Y}
\eeq
 with $\bX(T)\equiv \bX(T,X\equ\infty)$. Viewing $\bX(T)$ as a ``running'' boundary cosmological constant with scale $T$, we see that both geodesic distance and boundary length diverge simultaneously if we choose $Y\equ-\bX(\infty)\equ-\sqrt{\La}/2$ which was precisely determined by $0=W_\La(\bX(\infty))$ (recall Eq.\ \eqref{eq:DTgeodsolXb}). Hence, the geodesic distance from a generic point on the disk to the boundary diverges and in this
way effectively creates a non-compact space-time. 
\index{Dynamical triangulations (DT)!running boundary cosmological constant}

In this chapter we show that the same phenomenon occurs in the framework of two-dimensional Lorentzian quantum gravity defined by CDT. 

\section{The emergence of background geometry}

Recall from Sec.\ \ref{Marking the causal propagator} that the propagator of CDT is given by
\beq\label{2.a3}
G_\La (X,Y;T) = \frac{\bX^2(T,X)-\La}{X^2-\La} \; \frac{1}{\bX(T,X)+Y},
\eeq
where $\bX(T,X)$ is the solution of the characteristic equation
\beq\label{2.3}
\frac{d \bX}{d T} = -(\bX^2-\La),~~~\bX(0,X)=X.
\eeq
The explicit solution to this equation was given by \rf{eq:xbar} which can be rewritten as 
\beq\label{2.4}
\bX(t,X)= \SL \coth \SL(t+t_0),~~~~X=\SL \coth \SL \,t_0.
\eeq\index{Causal dynamical triangulations (CDT)!running boundary cosmological
constant}
As described above we can view $\bX(T)$ as a ``running'' boundary cosmological
constant, with $T$ being the scale. For $X > - \SL$ we have
$\bX(T) \to \SL$ for $T \to \infty$, i.e.\ $\SL$ can be seen as a ``fixed point'', or in other words a zero of the ``$\b$-function'' $-(\bX^2-\La)$ in Eq.\ \rf{2.3}.\footnote{Note that here $\hat{W}_\La=\bX^2-\La$ is not the disc function of CDT, as it was for the analogous term in DT.}

Using the semi-group property of the marked propagator
\beq\label{2.a6}
G_\La (X,Y;T_1+T_2) = \int_0^\infty d L \;
G_\La (X,L;T_1)\,G(L,Y,T_2).
\eeq
we can now calculate the expectation value of the length
of the spatial slice at intermediate proper time $t \in [0,T]$:
\beq\label{2.a7}
\la L(t)\ra_{X,Y,T} =
\frac{1}{G_\La (X,Y;T)} \int_0^\infty d L\;
G_\La (X,L;t) \;L\; G_\La (L,Y;T-t).
\eeq
Since the classical action is trivial, there is no reason to expect $\la L(t) \ra$ to
have a classical limit. We have already seen in Sec.\ \ref{sec:Hamiltonians in causal quantum gravity} and Sec.\ \ref{sec:CDT:observables} that for a generic situation the system is purely governed by quantum fluctuations. Except for boundary 
effects these quantum fluctuations are determined by the ground state of the effective quantum Hamiltonian $\hat{H}_{X}$ as introduced in Sec.\ \ref{sec:Hamiltonians in causal quantum gravity}. More precisely, an explicit calculation using \rf{2.a7} shows that in the situation 
where $X$ and $Y$ are larger than $\SL$ and where $T \gg 1/\SL$ the system has forgotten everything about the boundaries and the expectation value of $L(t)$ is, up to 
corrections of order $e^{-2\SL t}$ or $e^{-2\SL (T-t)}$, determined 
by the ground state of the effective Hamiltonian $\hat{H}_{X}$.

We will now study the situation where a non-compact 
space-time is obtained as a limit of the compact space-time in the same manner as described 
in the previous subsection for the case of DT. Hence we want to take $T \to \infty$ and at the same time also take the length of the outer boundary at time $T$ to infinity. 
Following the same reasoning as above, we see that the only choice of the boundary cosmological constant $Y$ independent of $T$ where the length  $\la L(T)\ra_{X,Y,T}$ goes
to infinity for $T \to \infty$ is the ``fixed point'' $Y\equ \mi \SL$, since
\beq\label{2.a8}
\la L(T)\ra_{X,Y,T} = -\frac{1}{G_\La (X,Y;T)} \, 
\frac{\prt G_\La (X,Y;T)}{\prt Y} =  \frac{1}{\bX(T,X)+Y}
\eeq 
which differs from the analogous DT expression, i.e.\ \rf{eq:emergence1xxx}, only in the definition of $\bX(T,X)$.

Inserting $Y \equ \mi \SL$ into \rf{2.a7} one obtains in the 
limit $T \to \infty$:
\beq\label{2.a9}
\la L(t) \ra_{X} = \frac{1}{\SL} \; \sinh (2\SL(t+t_0(X))),
\eeq
where $t_0(X)$ is inversely defined through Eq.\ \rf{2.4}.

At this point we want to remind the reader that starting from a lattice
regularization and taking the continuum limit the length $L$ is only determined 
up to a constant of proportionality which should be fixed by 
comparing with a continuum effective
action. In Sec.\ \ref{sec:Hamiltonians in causal quantum gravity} we made such a comparison with a continuum calculation by Nakayama and concluded that $L$ has to be identified with $L_{cont}/\pi$. We thus have
\beq\label{2.a10}
L_{cont}(t) \equiv \pi \la L(t)\ra_X = \frac{\pi}{\SL} \; 
\sinh (2\SL(t+t_0(X))).
\eeq
We now consider the line element of the classical surface where the intrinsic geometry is defined
by proper time $t$ and spatial length $L_{cont}(t)$ at constant $t$
\beq\label{2.a11}
d s^2 = d t^2 + \frac{L_{cont}^2}{4\pi^2}\; d \th^2 =
d t^2 + \frac{\sinh^2 (2\SL (t+t_0(X)))}{4 \La} \;d \th^2.
\eeq
Here $t \ge 0$ and $t_0(X)$ is a function of the boundary cosmological
constant $X$ at the boundary corresponding to $t \equ 0$ (see \rf{2.4}).
A remarkable observation about Eq. \rf{2.a11} is that the surfaces
for different boundary cosmological constants $X$ can be viewed as
sections of the same surface, namely the Poincar\'e disk with curvature $R= -8\La$,
since $t$ can be continued to $t=\mi t_0 $. The whole Poincar\'e disk is
then obtained in the limit $X \to \infty$, where the initial boundary is shrunken to a point and $t_0\equ 0$.

\section{The classical effective action}
\index{Causal dynamical triangulations (CDT)!classical effective action}

Even though the classical action of our theory is trivial, we discussed in Sec.\ \ref{sec:Hamiltonians in causal quantum gravity} that the effective quantum Hamiltonian of CDT can be derived from the following classical effective action
\beq\label{eq:classeff}
S_\k = \int_0^T d t \left(\frac{(\dot{l}(t))^2 }{4l(t) }  +
\La l(t) + \frac{\k}{l}\right).
\eeq
With respect to this it is interesting to see in how far the results of this chapter relate to the classical solutions of this effective action. 
Solving the equations of motion of \rf{eq:classeff} gives 
\bea
l(t) =& \frac{\sqrt{\k}}{\SL} \; \sinh 2\SL t,~~~~~~~~~~&\k>0~~
\mbox{elliptic case},
\label{3.6a}\\
l(t) =& \frac{\sqrt{-\k}}{\SL} \;
\cosh 2\SL t,~~~~~~~&\k<0~~\mbox{hyperbolic case},
\label{3.6b}\\
l(t) =& e^{2\SL t},~~~~~~~~~~~~~~~~~~~~~~~&\k=0 ~~\mbox{parabolic case}.
\label{3.6c}
\eea
One observes that all solutions correspond to cylinders with constant negative curvature $-8 \La$. In the elliptic case, there is
a conical singularity at $t =0$ unless $\k = 1$ for which
the geometry is regular at $t=0$ and
corresponds precisely to the Poincar\'{e} disc as described by \rf{2.a11}. However, $\k = 1$ is exactly the value for which Nakayama's Hamiltonian corresponds to the effective quantum Hamiltonian of CDT. Concluding, we see that upon the required identification $L_{cont}(t)\equ\pi l(t)$ the classical solution of Nakayama's effective action coincides with \rf{2.a10}.

\section{Quantum fluctuations}

In the considerations above we fixed the
outer boundary cosmological constant to a specific value, namely the ``fixed point'' of the running boundary cosmological constant, rather than fixing the outer length
of the boundary to a specific value. Since the boundary length and the boundary cosmological constant are conjugate variables, one pays the price that the
fluctuations of the boundary size are large. In particular, it can be checked from \rf{2.a8} that the fluctuations are of
order of the average length of the boundary itself, as was also true for the compact case
\beq\label{5.1emerg}
\la L^2(T) \ra_{X,Y;T} - \la L(T) \ra^2_{X,Y;T} =
-\frac{\prt \la L(T) \ra_{X,Y;T}}{\prt Y} =  \la L(T) \ra^2_{X,Y;T}.
\eeq
The same relation also holds for DT as was shown by Ambj{\o}rn et al.\ \cite{aagk}. These large fluctuations are not only present at the outer boundary as described by \rf{5.1emerg}, but also at $t< T$. From this point of view
it is even more remarkable that the emergent semiclassical background has such a nice interpretation in terms of the Poincar\'e disc. 

In the following we want to analyze the situation, where we by hand fix the boundary lengths $L_1$ and $L_2$ instead of the boundary cosmological constants. This is done in the Hartle-Hawking Euclidean path integral when the geometries $[g]$ are
fixed at the boundaries \cite{Hartle:1983ai}. Recall from the previous chapters that the boundary geometry is complete specified by its length and that amplitudes with fixed boundary cosmological constants can be related to amplitudes with fixed boundary lengths by a Laplace transformation. Let us in the following analyze the temporal correlations between two spatial slices at intermediate times $t$ and $t+\Delta$ with $0< t \leq t+\Del < T$. For simplicity we only consider the situation where the entrance loop is shrunken to a point, i.e.\ $L_1\equ0$.
The temporal correlations are determined by the so-called connected loop-loop correlator
\beq\label{5.a1}
\la L(t)L(t+\Del)\ra^{(c)}_{L_2,T} \equiv
\la L(t+\Del)L(t)\ra_{L_2,T}-\la L(t)\ra \la L(t+\Del)\ra_{L_2,T}.
\eeq
Using the semi-group property \rf{2.a6} and the length version of the propagator \rf{2.a3} one obtains 
\bea\label{5.a2}
\la L(t)L(t\pl\Del)\ra^{(c)}_{L_2,T} &=&
\frac{2}{\La}
\frac{\sinh^2 \SL t \sinh^2 \SL (T\mi (t\pl\Del))}{\sinh^2 \SL T}+
 \\
&&
\frac{2L_2}{\SL}
\frac{\sinh^2 \SL t \sinh\SL (t\pl\Del)
\sinh\SL(T \mi(t\pl\Del))}{\sinh^3 \SL T}.\no
\eea
In comparison the average length is given by
\beq\label{5.b2}
\la L(t)\ra_{L_2,T}=
\frac{2}{\SL} \frac{\sinh \SL t \sinh \SL (T\mi t)}{\sinh \SL T}+
L_2\frac{\sinh^2 \SL t}{\sinh^2 \SL T}.
\eeq
For fixed $L_2$ and $T \to \infty$, i.e.\ in the situation of compact geometries, we obtain
\beq\label{5.a3}
\la L(t)L(t+\Del)\ra^{(c)}_{L_2} =
\frac{1}{2\La} \; e^{-2\SL \Del } \left(1-e^{-2\SL t} \right)^2
\eeq
and 
\beq\label{5.b3}
\la L(t)\ra_{L_2}=\frac{1}{\SL}\left( 1-e^{-2\SL t}\right). 
\eeq
These equations precisely agree with the picture presented above, namely that except for small $t$, where we are close to the initial boundary we have $\la L(t)\ra_{L_2}\equ 1/\SL$, i.e.\ the dynamics is determined by the ground state of the effective quantum Hamiltonian. Further, the quantum 
fluctuations are
\beq
(\Del L(t))^2 = \la L(t)L(t)\ra^{(c)} \equiv \la L^2(t)\ra -\la L(t)\ra^2 \sim
\la L(t)\ra^2 \sim \frac{1}{\sL}.
\eeq
Thus the quantum geometry is entirely governed by quantum fluctuations as is illustrated in Fig.\ \ref{fig:CDTuniverse}. The time correlation between two slices
$L(t)$ and $L(t+\Del)$ is also dictated by the same scale $1/\SL$. Particularly, we observe that temporal correlations between ``elementary'' spatial elements of size $1/\SL$,
separated in time by $\Del$ fall off exponentially as $e^{-2\SL \Del}$. This is 
precisely what one would expect from the two-dimensional Einstein-Hilbert action with boundary terms, i.e.
\beq
S[g] = \La \int \int d x d t \sqrt{g(x,t)} +
X \oint d l_1 + Y \oint d l_2.
\eeq
If we force $T$ to be large and choose a generic $Y$ at which $\la L_2(T)\ra$ is not 
large, the quantum geometry will look like a thin tube which expect close to the boundaries is ``classically'' of zero spatial extension, but due to quantum fluctuations has average spatial length $1/\SL$.

Let us now repeat the above analysis for the case of non-compact geometries. Recall that only when the boundary cosmological constant was chosen to $Y = -\SL$ one obtains a non-compact geometry in the limit $T\to \infty$. To implement this ``critical value'' of $Y$ in a situation where $L_2$ is fixed instead of $Y$ we fix $L_2(T)$ to the average value \rf{2.a8} for 
$Y\equ \mi \SL$, i.e.\ 
\beq\label{5.2emerg}
L_2(T) = 
\la L(T) \ra_{X,Y= -\SL;T} = \frac{1}{\SL} \; \frac{1}{\coth \SL T -1}.
\eeq
From \rf{5.a2} and \rf{5.b2} one obtains in the limit $T \to \infty$ the following expression for the average length at intermediate time $t$
\beq\label{5.3} 
\la L(t) \ra = \frac{1}{\SL} \; \sinh 2\SL t
\eeq
which agrees with \rf{2.a9} as expected. Further, the loop-loop correlator equates to 
\beq\label{looploopnc}
\la L(t+\Del)L(t)\ra^{(c)}=
\frac{2}{\La}\;\sinh^2 \SL t= \frac{1}{\SL} \left( \la L(t)\ra 
-\frac{1}{\SL} \left(1-e^{-2\SL t}\right)\right).
\eeq

In contrast to the analogue expression for compact geometries, one observes that \rf{looploopnc} is independent of $\Del$.
In particular, using $\Del \equ 0$ we see that the quantum fluctuations for  $t \gg 1/\SL$ are given by
\beq\label{5.5emerg}
(\Del L(t))^2 \equiv \la L^2(t)\ra -\la L(t)\ra^2 \sim
\frac{1}{\SL} \la L(t)\ra.
\eeq
As in the case of compact geometries we can give an interpretation of \rf{5.5emerg} in terms of correlations between ``elementary'' spatial elements of size $1/\SL$. More precisely, we can view the curve of length $L(t)$ as consisting of $N(t) \approx \SL L(t) \approx e^{2\SL t} $ independently fluctuating ``elementary'' spatial elements of size $1/\SL$, each with a fluctuation of size $1/\SL$, i.e.\ each corresponding to a section of compact geometry. Thus the fluctuation $\Del L(t)$ of the total spatial slice $L(t)$ are of order
$\sqrt{N(t)}/\sL$, i.e.\
\beq\label{5.a5}
\frac{\Del L(t)}{\la L(t)\ra} \sim \frac{1}{\sqrt{\SL \la L(t)\ra}}
\sim e^{-\SL t}.
\eeq
Thus the fluctuations of $L(t)$ around $\la L(t)\ra$ are small for
$t \gg 1/\SL$. In the same way we can understand the independence of the loop-loop correlator of $\Del$ as the combined result, where $L(t+\Del)$
is growing exponentially in length with $e^{2\SL \Del}$ and
the correlation between  ``elementary'' spatial elements of $L(t)$ and $L(t+\Del)$
is exponentially decreasing with $e^{-2\SL \Del}$, causing a correlation constant in $\Del$.

\section{Discussion and outlook}

In this chapter we described the transition form compact to non-compact quantum geometries within the framework of two-dimensional CDT. In particular, the resulting quantum geometry
can be viewed as the Poincar\'e disk dressed with quantum fluctuations. Remarkably, it was shown that the quantum fluctuations of $L(t)$ are small compared to the average value of $L(t)$. In contrast to the analogous construction in the context of DT, this enables us to view the average geometry as a true semiclassical background.

The main idea underlying this construction is similar to the appearance of $ZZ$-branes when described from a worldsheet perspective, i.e.\ form a two-dimensional quantum gravity point of view \cite{aagk}. In DT the
non-compactness arose when the running boundary cosmological
constant ${\bX(T)}$ approached its fixed point described by $W^{(eu)}_\La(\bX(\infty)) \equ 0$,
i.e.\ to $\bX(\infty) \equ \sqrt{\tL}/2$ (see \rf{0.0}). In the case of CDT the transition occurs in exactly the same manner, namely when the running boundary cosmological constant $\bX(T)$ goes to $\SL$ for $T \to \infty$ which is determined by $\hat{W}_\La(\bX(\infty)) \equ 0$. However, in the case of CDT $\hat{W}_\La(X)$ is not the disc function, but is related to the effective quantum Hamiltonian of the system. It is interesting to observe that both processes are essentially the same, since both fixed points can be related by the mapping between the coupling constants of both theories as derived in the previous chapter, i.e.\ \rf{renormalization3},
\beq\label{6.1}
\frac{X}{\sL}=\sqrt{\frac{2}{3}} \frac{\sqrt{\tX+\sqrt{\tL}}}{\tL^{1/4}},
\eeq
where the Euclidean couplings are labeled with a tilde again.
From this expression we can see that
$X \to \SL$ corresponds precisely to $\tilde{X} \to \sqrt{\tL}/2$.

Another interesting observation is that the geometries described by \rf{2.a11}, for different initial boundary conditions, are all sections of the same surface, namely the Poincar\'e disc. Hence, they obey the Euclidean Hartle-Hawking no-boundary condition.\index{Hartle-Hawking no-boundary condition}
This is particularly surprising, since before rotating to the Euclidean sector we initially started with a path integral over entirely Lorentzian geometries. It would be interesting to investigate whether this also holds in higher dimensional CDT. So far, the computer simulations, as briefly described in Sec.\ \ref{Sec:CDToutlook}, seem to be compatible with this condition.

\part{Third quantization of 2D causal dynamical triangulations}

\chapter{A field theoretic perspective of spatial topology changes \label{Chap:cap}}
	
In the second part of this thesis we introduced causal dynamical triangulations from a two-dimensional quantum gravity point of view in which the two-dimensional geometries were interpreted as a toy model for four-dimensional space-time. Taking the string theoretic point of view we see the surfaces as worldsheets of propagating strings. Using this interpretation one is naturally lead to the idea of formulating a third-quantization of the theory, a so-called string field theory, in which strings can be annihilated, and created from the vacuum. Models of string field theory (SFT) have been intensively studied in the framework of the Euclidean model \cite{Ishibashi:1993pc,Ikehara:1994vx,Ishibashi:1995in,Ikehara:1994xs,Watabiki:1993ym,Ambjorn:1996ne}. In this part of the thesis we develop an analogous formulation also for the Lorentzian theory. While in this chapter we compute certain ``Feynman diagrams'' of the model \cite{Ambjorn:2007jm}, we will in the next chapter introduce the whole string field theoretic framework.   

\section{Taming spatial topology changes}\label{sec:tamingtop}

In Chap.\ \ref{Chap:Relating} we described how to introduce spatial topology changes into CDT. 
We saw that when
viewing them as a purely
geometric process\footnote{By this we mean that each distinct geometry (distinct in the
sense of Euclidean geometry) appears with equal
weight in the sum over two-dimensional geometries.}, one obtains 
Euclidean quantum gravity. In particular, we recall from Sec.\ \ref{sec:CDTtoDT} that there are two possible scaling relations entering the differential equation for the propagator \eqref{top13}, i.e.
\bea
a^\ep\frac{\prt}{\prt T}\, G_\La(X,Y;T) & =&-a \, \frac{\prt}{\prt X} \Bigl[ (X^2-\La)  G_\La(X,Y;T) \Bigr] \nn
&&-2 a^{\eta-1}\frac{\prt}{\prt X} \Bigl[W_\La(X)  G_\La(X,Y;T) \Bigr].
\label{top13new}
\eea
In the scaling described by 
\bea\label{2.51}
w(g,x) &\xrightarrow[a\to 0]{}& a^{\eta}\, W_\La(X),~~~~\eta <0, \\
t &\xrightarrow[a\to 0]{}&  T/a^\ep,~~~~\ep =1.
\label{2.51a}
\eea
the creation of baby universes is forbidden, leading back to the ``bare'' CDT model. Further, in the scaling with
\bea\label{2.52}
w(g,x) &\xrightarrow[a\to 0]{}& w_{ns} +a^{\eta}\, W_\La(X), 
~~~~\eta=3/2\\
t& \xrightarrow[a\to 0]{}&  T/a^\ep,~~~~~\ep=1/2,
\label{2.52a}
\eea
the baby universes completely dominate the continuum limit, leading to Euclidean quantum gravity described by DT.

 \begin{figure}[t]
\begin{center}
\includegraphics[width=3in]{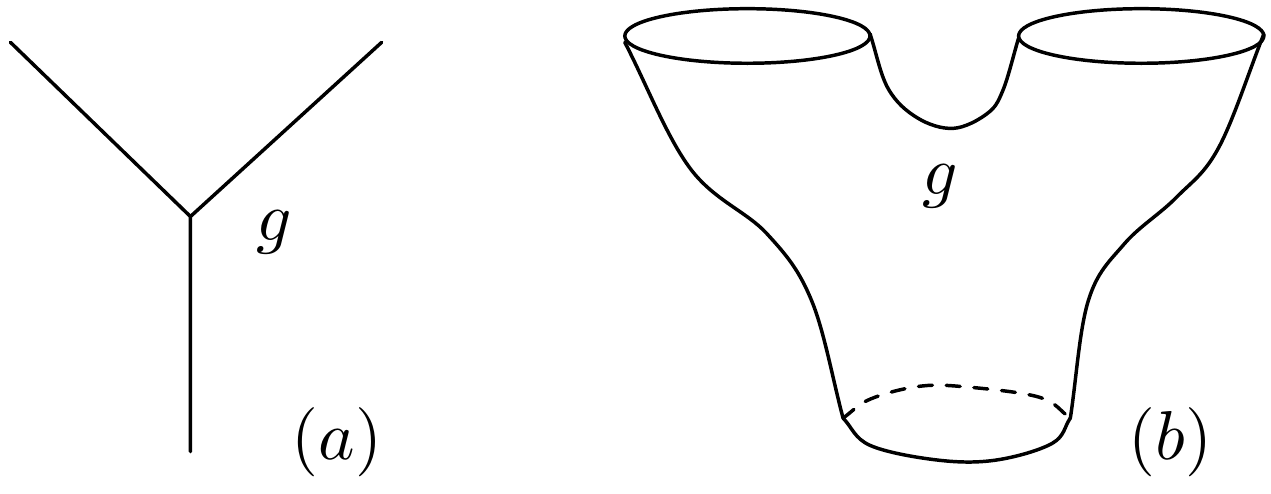}
\caption{Feynman rules: $(a)$ shows a vertex of $\varphi^3$-theory which is assigned a coupling $g$. $(b)$ shows the analogous interaction term for a splitting string. Similar to $\varphi^3$-theory we assign a string coupling $g$ to this interaction.}
\label{fig:feynman}
\end{center}
\end{figure}

\begin{figure}[t]
\centerline{\scalebox{0.5}{\rotatebox{0}{\includegraphics{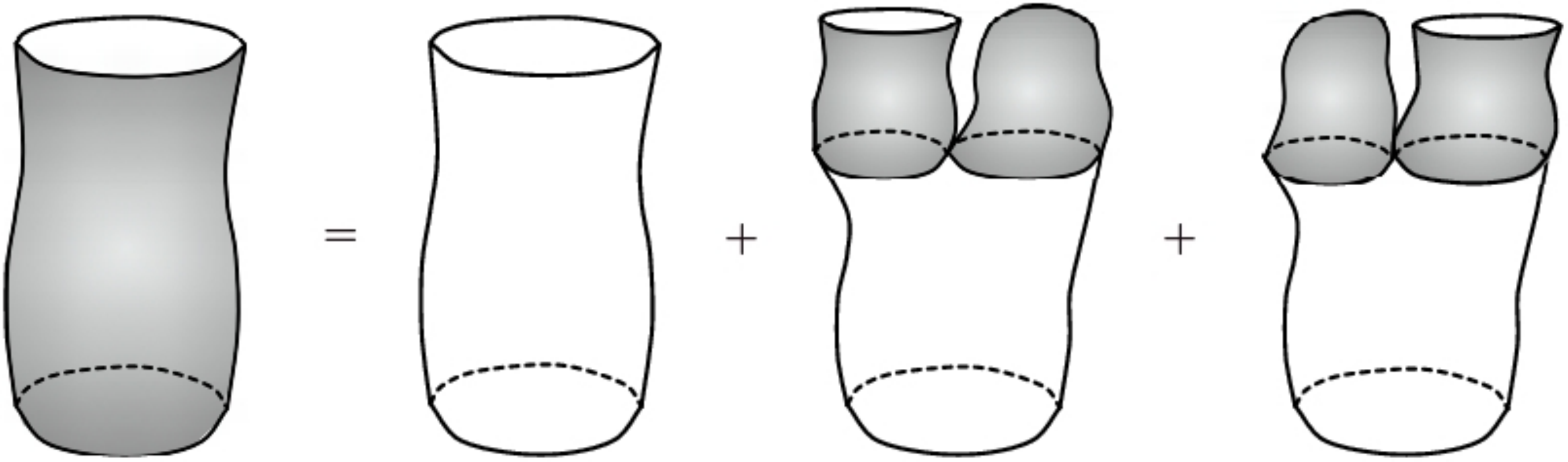}}}}
\caption{
In all four graphs, the geodesic distance from the final to the initial 
loop is given by $T$. Differentiating
with respect to $T$ leads to eq.\ \rf{2.55}. Shaded parts of graphs represent
the full, $g_s$-dependent propagator and disc amplitude, and non-shaded 
parts the CDT propagator.}
\label{fig2}
\end{figure}

Taking a string field theoretic perspective, however, one can view each spatial topology change as a vertex of a Feynman diagram of the corresponding string field theory (see Fig.~\ref{fig:feynman}). In this spirit it is natural to assign a string coupling $g$ to this process. This is in analogy to the coupling $g$ in front of the interaction term of zero-dimensional $\varphi^3$-theory, described by the action \index{$\varphi^3$-theory}
\beq \label{eq:capphi3}
S[\varphi] = \frac{1}{2}\varphi^2 - \frac{g}{3!} \varphi^3.
\eeq
 While in the Euclidean model such a coupling is absent\footnote{This is also reflected in the Hamiltonian of the string field theory \cite{Ishibashi:1993pc}.} we will see in the following that in the case of CDT it can be used to regulate the number of baby universes.     
From renormalization arguments one expects the coupling $g$ also to scale, i.e.\ to be a non-constant function $g \equ g(a)$ of the cut-off $a$. A geometric interpretation of this
assignment will be given in the discussion at the end of this chapter. 
Since we are 
interested in a theory  which smoothly recovers CDT in the limit as $g \to 0$, it 
is natural to assume that $\eta = - 1$,  as in CDT. Consequently, the 
only way to obtain a non-trivial consistent equation is 
to assume that $g$ scales to zero with the cut-off $a$ according to \index{Causal dynamical triangulations (CDT)!string coupling constant}
\beq\label{2.54}
g = g_s a^3,
\eeq
where $g_s$ is a coupling constant of mass dimension three, which is 
kept constant when $a \to 0$ \cite{Ambjorn:2007jm}. With this choice, \eqref{top13new} is turned into \index{Causal dynamical triangulations (CDT)!generalized model} \index{Generalized CDT}
\beq\label{2.55}
 \frac{\prt}{\prt T} G_{\La,g_s}(X,Y;T) = 
- \frac{\prt}{\prt X} \Big[\Big((X^2-\La)+2 g_s\;W_{\La,g_s}(X)\Big) 
G_{\La,g_s}(X,Y;T)\Big].
\eeq
The graphical representation of \rf{2.55} 
is shown in Fig.\ \ref{fig2}.
Differentiating the integral equation corresponding to this 
figure with respect to the time $T$ one obtains \rf{2.55}.
The disc amplitude $W_{\La,g_s}(X)$ is at this stage unknown. 

Note that one could in principle have considered an a priori more general branching process, 
where more than one baby universe is allowed to sprout at any given time step $T$. 
However, one observes from the scaling relation \rf{2.54} that the corresponding extra terms in 
relation \eqref{top13new} would be suppressed by higher orders of $a$ and therefore play no role in the continuum limit.

In the next section we show that the quantum geometry, in the sense defined 
above, together with the requirement of recovering standard CDT in the
limit as $g_s\rightarrow 0$,
uniquely determines the disc amplitude and thus 
$G_{\La,g_s}(X,Y;T)$.

\section{The disc amplitude}\label{disc}

\begin{figure}[t]
\centerline{\scalebox{0.55}{\rotatebox{0}{\includegraphics{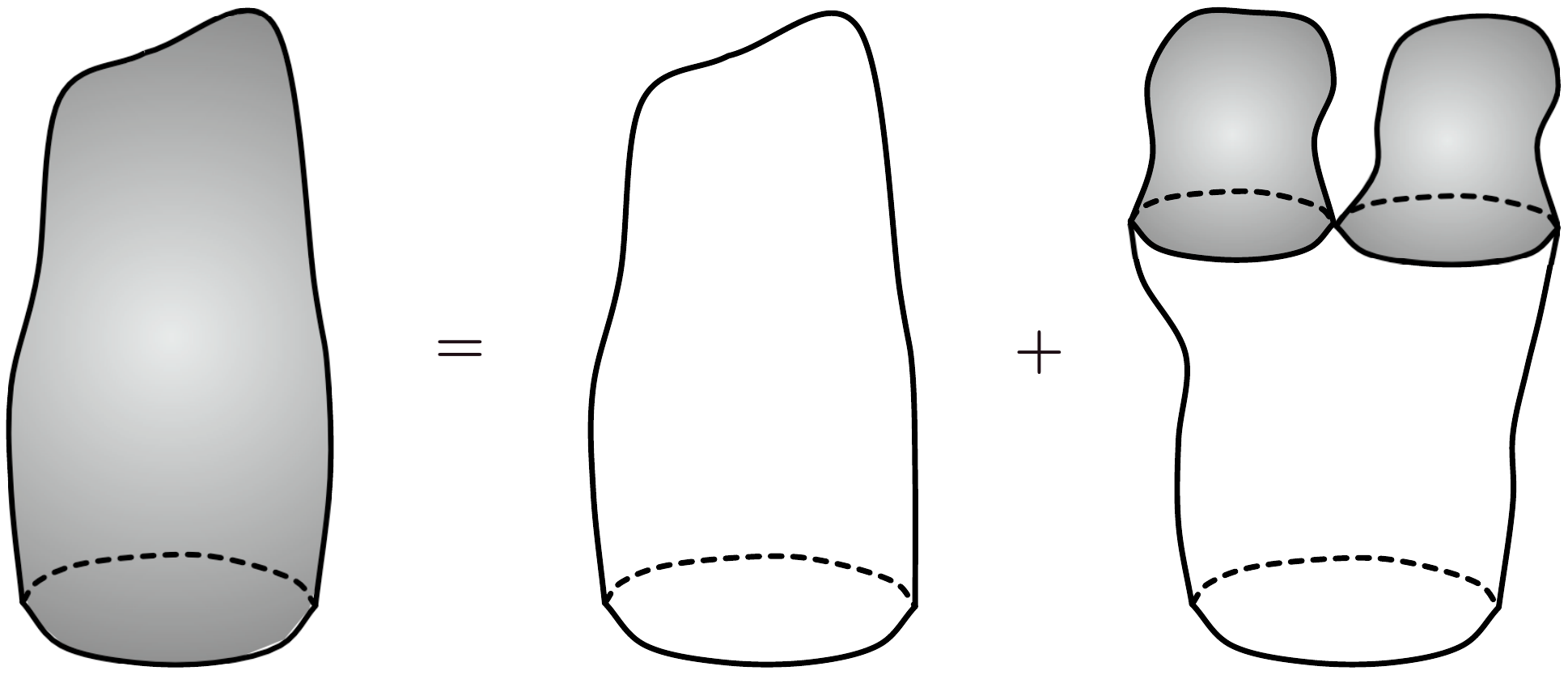}}}}
\caption{Graphical illustration of eq.\ \rf{3.2}. Shaded
parts represent the full disc amplitude, unshaded parts the CDT disc
amplitude and the CDT propagator. 
}
\label{fig3}
\end{figure}

The disc amplitude of CDT was calculated in Sec. \ref{Marking the causal propagator}, i.e. \eqref{eq:discCDT}.
In \eqref{eq:WCDTdef} it was determined directly by integrating $G_\La(L_1,L_2=0;T)$
over all times. This decomposition is unique, since by assumption 
$T$ is a global time and no baby universes can be created. In Sec.\ \ref{sec:integrationout}
it was shown that it could also be obtained from Euclidean quantum gravity
(DT) by peeling off baby universes in a systematic way.
By either method we found
\beq\label{3.1}
W_\La (X) = \dfrac{1}{X+\sL}
\eeq
for the disc amplitude as function of the boundary cosmological constant $X$.
In the present, generalized case we allow for baby universes, leading to a
graphical representation of the decomposition of the disc amplitude as 
shown in Fig.\ \ref{fig3}.
It translates into the equation 
\bea
\!\!\!\!\!\! W_{\La,g_s} (X) &=& W_{\La,g_s} ^{(b)}(X) +\non
\!\!\!\!\!\!\!\!\! &+&\!\! \!g_s\! \int\limits_0^\infty d T\!\! \int\limits_0^\infty d L_1 d L_2  \;
(L_1+L_2) G^{(b)}_{\La,g_s}  (X,L_1+L_2;T) W_{\La,g_s} (L_1)W_{\La,g_s} (L_2)\label{3.2}
\eea
for the full propagator $W_{\La,g_s}(X)$, where we have introduced 
a superscript $(b)$ to indicate the ``bare'' CDT amplitudes, that is,
\beq
W_{\La,g_s} ^{(b)}(X)\equiv W_{\La,g_s=0} (X)=W_\La(X),\label{relabel}
\eeq 
and similarly for $G^{(b)}_{\La,g_s}$, quantities which were defined in eqs. \rf{3.1} and \rf{2.a3}
respectively.
The integrations in \rf{3.2} can be performed, yielding
\beq\label{3.3}
W_{\La,g_s}(X) = \dfrac{1}{X+\sL} +\dfrac{g_s}{X^2-\La}\Big( W_{\La,g_s}^2(\sL)-W_{\La,g_s}^2(X)\Big).
\eeq
Solving for $W_{\La,g_s}(X)$ we find
\beq\label{3.4}
W_{\La,g_s}(X) = \frac{-(X^2-\La) + \hat{W}_{\La,g_s}(X)}{2g_s},
\eeq
where we have defined
\beq
\hat{W}_{\La,g_s}(X) = \sqrt{(X^2-\La)^2 +4g_s\Big(g_s W^2_{\La,g_s}(\sL) + X-\sL\Big)}.
\label{3.4a}
\eeq
The sign of the square root is fixed by the requirement that 
$W_{\La,g_s}(X) \to W_\La(X)$ for $g_s \to 0$, and $W_{\La,g_s}(X)$ is determined
up to the value $W_{\La,g_s}(\sL)$. We will now show that this value is also 
determined by consistency requirements of the quantum geometry. If we insert the 
solution \rf{3.4} into \rf{2.55} we obtain
\beq\label{3.5}
 \frac{\prt}{\prt T} G_{\La,g_s}(X,Y;T) = 
- \frac{\prt}{\prt X} \Big[\hat{W}_{\La,g_s}(X)\, G_{\La,g_s}(X,Y;T)\Big].
\eeq
In analogy with \rf{eq:G10XYT} and \rf{eq:characteristicequation}, this is solved by
\beq\label{3.6}
G_{\La,g_s} (X,Y;T) = \frac{\hat{W}_{\La,g_s}(\bX(T,X))}{\hat{W}_{\La,g_s}(X)} \; \frac{1}{\bX(T,X)+Y},
\eeq
where $\bX(T,X)$ is the solution of the characteristic equation for \rf{3.5}, 
\beq\label{3.7}
\frac{d \bX}{d T} = -\hat{W}_{\La,g_s}(\bX),~~~\bX(T\equ0,X)=X,
\eeq
such that
\beq\label{3.8}
T = \int^X_{\bX(T)} \dfrac{d Y}{\hat{W}_{\La,g_s}(Y)}.
\eeq
Physically, we require that $T$ can take values from 0 to $\infty$, as opposed to just in a
finite interval. From expression \rf{3.8} for $T$ this is 
only possible if the polynomial under the square root in the defining equation 
\rf{3.4} has a double zero, which fixes the function $\hat{W}_{\La,g_s}(X)$ to
\beq\label{3.9}
\hat{W}_{\La,g_s}(X) = (X-C)\sqrt{(X+C)^2-2g_s/C},
\eeq
where 
\beq\label{3.9a}
C = U\sL, ~~~U^3-U+\dfrac{g_s}{\La^{3/2}}=0.
\eeq
In order to have a physically acceptable $W_{\La,g_s}(X)$, 
one has to choose the solution to the third-order equation which is closest to 1. 
Quite remarkably, one can also derive \rf{3.9} from \rf{3.4} by demanding that 
the inverse Laplace transform $W_{\La,g_s}(L)$ falls off exponentially for large $L$.
In this region $W_{\La,g_s}(X)$ equals $W^{(b)}_{\La,g_s}(X)$ plus a convergent 
power series in the
dimensionless coupling constant $g_s/\La^{3/2}$.

One can check the consistency of the quantum geometry by using 
\rf{3.6} in
\beq\label{eq:decompdiscnew}
-\frac{\partial}{\partial \La}W_{\La,g_s} (X) =\int_0^\infty dT\int_0^\infty dL G_{\La,g_s}(X,L;T)\, L\, W_{\La,g_s} (L)
\eeq
which comes form the decomposition of a disc function with a mark in the bulk (see Fig.\ \ref{fig:discfunctiondecomposition}), i.e.\ the continuum version of \eqref{an1}.
Integrating \eqref{eq:decompdiscnew} gives
\beq\label{3.10}
\dfrac{\prt W_{\La,g_s} (X)}{\prt \La} = \dfrac{W_{\La,g_s} (X) - W_{\La,g_s} (C)}{\hat{W}_{\La,g_s}(X)},
\eeq
which is indeed satisfied by the solution \rf{3.4}.

\section{The loop-loop amplitude and the consistency condition}\label{sec:consitencyCap}

We mentioned earlier that the propagator can be regarded as a 
building block for other, more conventional ``observables" in 
two-dimensional quantum gravity. One of the most beautiful illustrations of this
and at the same time a non-trivial example of what we have called quantum 
geometry is the calculation in two-dimensional Euclidean quantum gravity of the loop-loop 
amplitude from the fixed geodesic distance two-loop amplitude \cite{Aoki:1995gi}. The full loop-loop amplitude is obtained by summing 
over all Euclidean two-dimensional geometries with two boundaries, without any particular 
restriction on the boundaries' mutual position. This amplitude 
was first calculated using matrix model techniques (for cylinder topology) \cite{Ambjorn:1990ji}. 

\begin{figure}[t]
\centerline{\scalebox{0.55}{\rotatebox{0}{\includegraphics{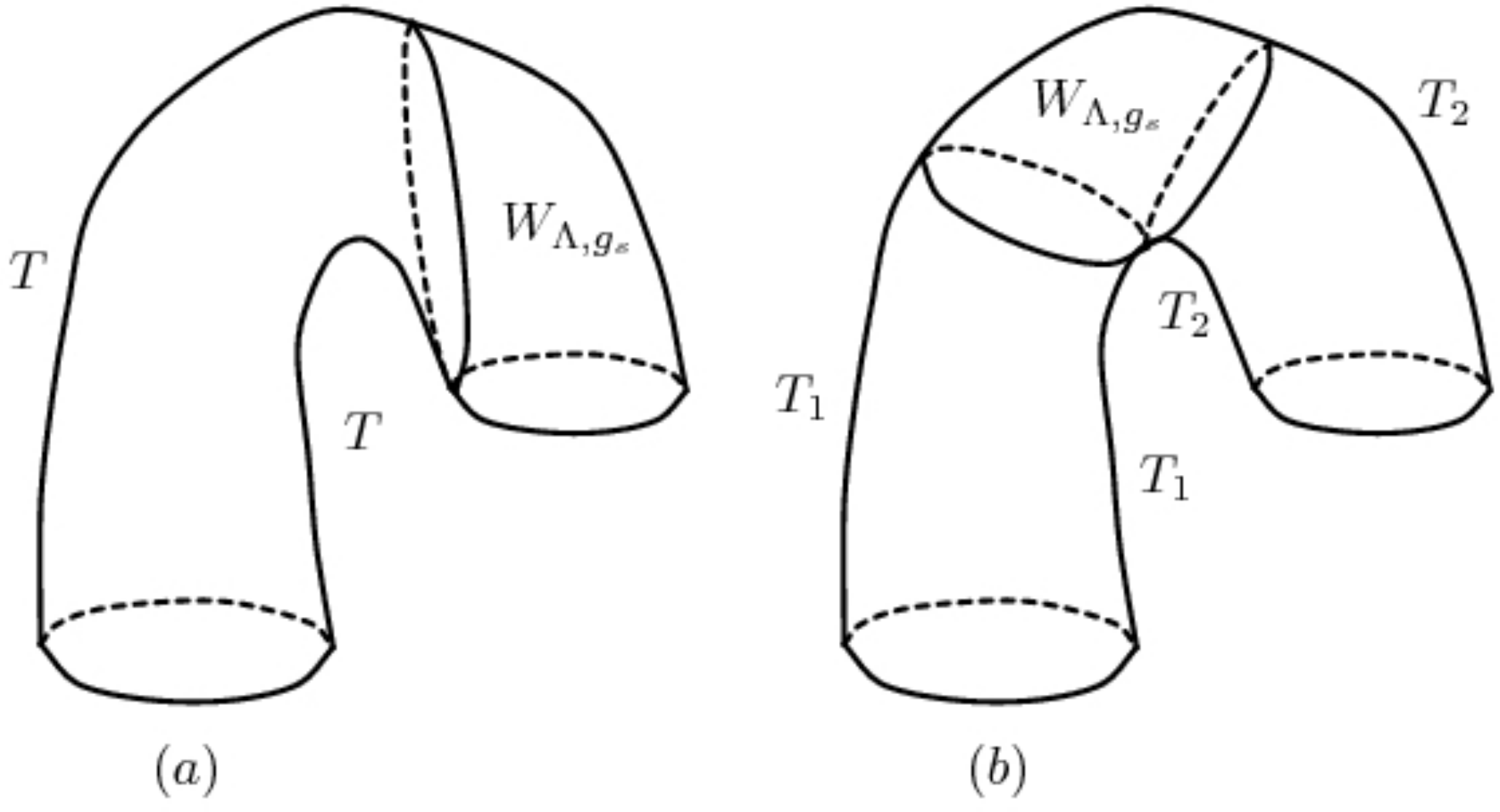}}}}
\caption{ Two different ways of decomposing
the loop-loop amplitude into proper-time propagators and a disc amplitude.
Two points touch in the disc amplitude $W$, pinching the boundary to a figure-8, 
which combinatorially 
implies a substitution $W_{\La,g_s} (L) \to L \,W_{\La,g_s}(L)$ in the formulas. The time variables
are related by $T_1+T_2 = T$.
}
\label{fig4}
\end{figure}

To appreciate the underlying construction, consider a given geometry of cylindrical
topology.
Its two boundaries will be separated by a geodesic distance $t$, in the sense of
minimal distance of any point on the final loop to the initial loop. It follows that we can 
consider the geometry as composed of a cylinder where the entire final loop (i.e.\ {\it each}
of its points) has a distance $t$ from the initial one and a ``cap" related to the
disc amplitude, as illustrated in Fig.\ \ref{fig4} (a). 
One can now obtain the
loop-loop amplitude by integrating over all $T$ and all gluings of the cap
(we refer to \cite{Aoki:1995gi} for details). An intriguing aspect of the 
construction is that the decomposition of a given geometry into cylinders
and caps is not unique. One can choose another decomposition
consisting of two cylinders of length $T_1$ and $T_2$, with $T_1+T_2 = T$,
joined by a cap, as illustrated in Fig.\ \ref{fig4} (b). 
As shown in \cite{Aoki:1995gi}, the end result is indeed independent of this decomposition.

The whole construction can be repeated for our new, generalized CDT model, in this 
way {\it defining} a loop-loop amplitude. More precisely, although an exact equality of
amplitudes corresponding to different
decompositions like those depicted in Fig.\ \ref{fig4} (a) and (b) is not 
immediately obvious at
the level of the triangulations of the discretized theory\footnote{because of the 
different arrangements of the proper-time slicings}, the continuum ansatz \rf{4.1} 
below is self-consistent, in the sense that it leads to a non-trivial symmetric 
expression for the amplitude with a well-defined $g_s\rightarrow 0$ limit. 
The algebra is similar to that of \cite{Aoki:1995gi}.

We will denote the loop-loop amplitude by $W_{\La,g_s}(X,Y)$, and its
Laplace transform by $W_{\La,g_s}(L_1,L_2)$. 
The integral equation corresponding
to Fig.\ \ref{fig4} (a) is given by 
\beq\label{4.1}
W_{\La,g_s}(L_1,L_2) = \int_0^\infty d T \int_0^\infty d L \; 
G_{\La,g_s}(L_1,L;T) L W_{\La,g_s}(L+L_2).
\eeq
Laplace-transforming eq.\ \rf{4.1}, the integrals can be performed
using  eqs.\ \rf{3.6}-\rf{3.9}. After some non-trivial algebra one obtains \index{Causal dynamical triangulations (CDT)!loop-loop amplitude}
\beq\label{4.2}
W_{\La,g_s}(X,Y) = \dfrac{1}{f(X)f(Y)}\dfrac{1}{4g_s}
\left( \dfrac{[(X+C)+(Y+C)]^2}{(f(X)+f(Y))^2}-1\right),
\eeq
where we are using the notation
\beq\label{4.3}
f(X) = \sqrt{(X+C)^2-2g_s/C} = \hat{W}_{\La,g_s}(X)/(X-C).
\eeq
In the limit $g_s \to 0$ one finds
\beq\label{4.4}
W^{(b)}_{\La,g_s\equ0}(X,Y) = \dfrac{1}{2\sL (X+\sL)^2(Y+\sL)^2},
\eeq
a result which could of course also have been obtained directly from
\rf{4.1} using \rf{eq:G10XYT}, \rf{eq:characteristicequation} and \rf{3.1}.
We note that the corresponding expression in the case of Euclidean two-dimensional quantum gravity is
given by \index{Dynamical triangulations (DT)!loop-loop amplitude}
\beq\label{4.5}
 W^{(eu)}_\La(X,Y) = \dfrac{1}{2h(X)h(Y)(h(X)+h(Y))^2},~~~h(X)=\sqrt{X+\sL},
\eeq
which can be obtained from expressions similar to \rf{3.6}-\rf{3.9},
only with $\hat{W}_{\La,g_s}(X)$ replaced by the Euclidean disc function
\beq\label{4.6}
W_\La^{(eu)}(X) = (X-\sL/2)\; h(X).
\eeq
We observe a structural similarity between \rf{4.2} and \rf{4.6}, with the function $f(X)$ having the 
same relation to $\hat{W}_{\La,g_s}(X)$ as $h(X)$ has to $W^{(eu)}_\La(X)$.

Let us now prove the consistency condition that the two composition of the two-loop as shown in Fig.~\ref{fig4} lead to the same result. Decomposing as in Fig.~\ref{fig4} (a) leads to
\beq\label{ds23}
W_{\La,g_s}(L_1,L_2;T) = \int_0^\infty d L \; 
G_{\La,g_s}(L_1,L;T)  \, L\,W_{\La,g_s}(L+L_2). 
\eeq
while the decompositions of Fig.~\ref{fig4} (b) reads
\beq\label{ds24}
W_{\La,g_s}(L_1,L_2;T) = \int_0^\infty d L \int_0^\infty dL'\; 
G_{\La,g_s}(L_1,L;T_1) \,L \, W_{\La,g_s}(L+L')  \, L_2 G_{\La,g_s}(L',L_2;T-T_1).
\eeq
Thus we arrive at the consistency condition:
\beq\label{ds25}
0= \frac{\prt}{\prt T_1} \;\int_0^\infty dL \int_0^\infty dL'\; 
G_{\La,g_s}(L_1,L;T_1)\,L\,  W_{\La,g_s}(L+L') \, L_2  G_{\La,g_s}(L',L_2;T-T_1).
\eeq
This condition was checked for the Euclidean model in \cite{Ishibashi:1993pc}.
Remarkably, it is also satisfied in the our case.
After a Laplace transformation eq.\ \rf{ds25} reads:
\beq\label{ds26}
0= \frac{\prt^2}{\prt X \prt Y} \; \frac{\prt}{\prt T_1} \; 
\int_{-i\infty+c}^{i\infty +c} 
\frac{dZ}{2\pi i}\; 
G_{\La,g_s}(X,-Z;T_1)  W_{\La,g_s}(Z) G_{\La,g_s}(Y,-Z;T-T_1).
\eeq
Using the explicit form of $G_{\La,g_s}(X,Y;T)$, eq.\ \rf{3.6}, we 
can perform the $Z$-integration in eq.\ \rf{ds26}. Using 
\rf{3.3} we can express $W_{\La,g_s}(X)$ in terms of 
$\hat{W}_{\La,g_s}(X)$ given by \rf{3.2} and finally using \rf{3.7}
we can differentiate after $T_1$. The result is\footnote{The corresponding 
equation in the case of non-critical string theory is 
$$
0= \frac{\prt^2}{\prt X \prt Y} \; \Big(\bar{X}(T_1,X)-\bar{Y}(T-T_1,Y)\Big).
$$
Again one obtains the remarkable result that the amplitude $W_\La^{(eu)}(X,Y;T)$
is independent of the subdivision of $T=T_1+T_2$ as shown in Fig.\ \ref{fig4}.
In addition we have in the non-critical SFT setting the additional 
consistency test that $\int_0^\infty dT W_\La^{(eu)}(X,Y;T) = W_\La^{(eu)}(X,Y)$, where 
$W_\La^{(eu)}(X,Y)$ is the so-called universal loop-loop correlator calculated
from matrix model \cite{Ambjorn:1990ji,Ambjorn:1990wg}, i.e. \rf{4.5}. This was verified in \cite{Ishibashi:1993pc}.}
\beq\label{ds27}
0= \frac{\prt^2}{\prt X \prt Y} \; 
\Big(\bar{X}^2(T_1,X)-\bar{Y}^2(T-T_1,Y)\Big),
\eeq
which is satisfied.

The existence of well-defined, symmetric expressions for the unrestricted loop-loop
amplitudes in our generalized CDT model (at genus 0) and thus in standard two-dimensional
CDT, formulas \rf{4.2} and \rf{4.4}, gives strong support to the claims that (i) the proper-time
propagator does indeed encode the complete information on the quantum-gravitational system,
and (ii) following the arguments given in \cite{Aoki:1995gi} concerning the decomposition invariance of
the loop-loop amplitude (c.f. Fig.\ \ref{fig4}), the continuum theory is diffeomorphism-invariant.

\section{Discussion and outlook}

The generalized CDT model of two-dimensional quantum gravity we have defined in this
chapter is a perturbative deformation of the original model in the sense
that it has a convergent power expansion of the form
\beq\label{5.1}
W_{\La,g_s}(X) = \sum_{n=0}^\infty c_n (X,\La) \left( \dfrac{g_s}{\La^{3/2}}\right)^n
\eeq 
in the dimensionless 
coupling constant $g_s/\La^{3/2}$ \cite{Ambjorn:2007jm}. This implies in particular that 
the average number $\langle n\rangle$ of  ``causality violations" in a 
two-dimensional universe
described by this model is finite, a property already observed in previous two-dimensional models with
topology change \cite{Loll:2003rn,Loll:2003yu,Loll:2005dr,Loll:2006gq}.
The expectation value of the number $n$ of 
branchings can be computed according to
\beq\label{5.2}
\la n\ra = \dfrac{g_s}{W_{\La,g_s}(X)}\dfrac{d W_{\La,g_s}(X) }{d g_s},
\eeq
which is finite as long as we are in the range of convergence of $W_{\La,g_s}(X)$.
As already mentioned, this coincides precisely with the range where the
function $W_{\l,g_s}(X)$ behaves in a physically acceptable way, namely, 
$W_{\La,g_s}(L)$
goes to zero exponentially in terms of the length $L$ of the boundary loop. 
The same is true for the other functions considered, namely, $G_{\La,g_s}(L_1,L_2;T)$ 
and $G_{\La,g_s}(L_1,L_2)$.

The behaviour \rf{5.2} should be contrasted with that in two-dimensional Euclidean quantum gravity,
and is reflected in the different scaling behaviours \rf{2.51a} and \rf{2.52a} for the time $t$. 
These scaling 
relations show that the effective continuum ``time unit" in Euclidean quantum gravity is
much longer than in CDT, giving rise to infinitely many causality violations for a 
typical space-time history which appears in the path integral  
when the cut-off $a$ is taken to zero. This phenomenon 
was discovered in the seminal paper \cite{Kawai:1993cj}.

As we will see in the next chapter, the calculations presented here should 
be seen as pertaining to the genus-0 sector of a generalized CDT model, which also 
includes a sum over space-time topologies. Although we have not given a precise
definition of the higher-genus amplitudes in this chapter, one would expect them to be finite
order by order. If the handles are as scarce as are the baby universes in the genus-0 
amplitudes, it might even be that the sum over all genera is uniquely defined. Whether 
or not this is so will clearly also depend on the combinatorics of allowed handle configurations.

In the context of higher-genus amplitudes, it is natural to associate each handle with a
``string coupling constant", because one may think of it as a process where (one-dimensional) 
space splits and joins again, albeit as a function of an intrinsic proper time, rather than the
time of any embedding space. An explicit calculation reveals that in the generalized CDT model
this process is related with a coupling constant $g_s^2$ (see next chapter), which one may think of
as two separate factors of $g_s$, associated with the splitting and joining respectively.

How does the disc amplitude fit into this picture?
From a purely Euclidean point of view all graphs appearing in
Fig.\ \ref{fig3} have the fixed topology of a disc.
However, from a Lorentzian point of view, which comes with a notion of time, it is
clear that the branching of a baby universe is associated with a change of
the {\it spatial} topology, a
singular process in a Lorentzian space-time \cite{Louko:1995jw}.
One way of keeping track of this in a Wick-rotated, Euclidean picture is
as follows. Since 
each time a baby universe branches off it also has to end somewhere,  
we may think of marking the resulting ``tip" with a puncture. From a
gravitational viewpoint, each new puncture corresponds to a topology
change and receives a weight $1/G_N$, where $G_N$ is Newton's constant,
because it will lead to a change by precisely this amount in the two-dimensional 
(Euclidean) Einstein-Hilbert action
\beq\label{5.3}
S_{EH} = -\dfrac{1}{2\pi G_N} \int d^2\xi \sqrt{g} R.
\eeq
Identifying the dimensionless coupling constant 
in eq.\ \rf{top13new} with $g(a)= e^{-1/G_N(a)}$, one can introduce a {\it renormalized}
gravitational coupling constant by \index{Causal dynamical triangulations (CDT)!double scaling limit}
\beq\label{5.4}
\dfrac{1}{G_N^{ren}} = \dfrac{1}{G_N(a)}+\dfrac{3}{2}\ln \La a^2.
\eeq
This implies that the {\it bare} gravitational coupling constant $G_N(a)$ goes to
zero like $1/|\ln a^3|$
when the cut-off vanishes, $a \to 0$, in such a way that the product $\e^{1/G_N^{ren}}/\La^{3/2}$ is 
independent of the cut-off $a$. We can now identify 
\beq\label{5.5}
e^{-1/G_N^{ren}} = g_s/\La^{3/2}
\eeq
as the genuine coupling parameter in which we expand.

This renormalization of the gravitational (or string) coupling constant is reminiscent 
of the famous double-scaling limit in non-critical string
theory\footnote{It is called the double-scaling limit since  
from the point of view of the discretized theory it involves a
simultanous renormalization of the cosmological constant $\La$ and 
the gravitational coupling constant $G_N$. In this article we have already
performed the renormalization of the cosmological constant. For details
on this in the context of CDT we refer to \cite{Ambjorn:1998xu}.}. 
In that case one also has $g_s \propto e^{-1/G_N^{ren}}$, the only difference
being that relation \rf{5.4} is changed to \index{Dynamical triangulations (DT)!double scaling limit}
\beq\label{5.6}
\dfrac{1}{G_N^{ren}} = \dfrac{1}{G_N(a)}+\dfrac{5}{4}\ln \La a^2,
\eeq
whence the partition function of non-critical string theory appears  
precisely as a function  of the dimensionless coupling 
constant $g_s /\La^{5/4}$.

\chapter{A causal string field theory \label{Chap:sft}}
	In the previous chapter we introduced a generalization of CDT by incorporating spatial topology changes regularized by a coupling $g_s$. Using certain Feynman rules for propagation and splitting of strings (spatial universes) we were able to solve the genus zero disc function to all orders in $g_s$. In this chapter we embed these results in a broader framework of a causal string field theory \cite{Ambjorn:2008ta}. 

\section{The string field theory framework}\label{sft}
\index{String field theory}
\index{Third-quantization of gravity}

In quantum field theory particles can be created and annihilated 
if the process does not violate any conservation law of the
theory. In string field theories one operates in the 
same way with operators which can create and annihilate strings. 
From the two-dimensional quantum gravity point of view we thus have a 
third-quantization of gravity: one-dimensional universes can 
be created and destroyed. In \cite{Ishibashi:1993pc,Ikehara:1994vx,Ishibashi:1995in,Ikehara:1994xs,Watabiki:1993ym,Ambjorn:1996ne} such a formalism was
developed for non-critical strings (or two-dimensional Euclidean quantum 
gravity). We will follow the formalism developed there closely
and develop a string field theory or third quantization 
for CDT which will allow us in principle to calculate any
amplitude involving creation and annihilation of universes \cite{Ambjorn:2008ta}.

The starting point is the assumption of a vacuum from 
which universes can be created. We denote this state $\vac$ and
define creation and annihilation operators:
\beq\label{s1} 
[\Psi(L),\Psi^\dg(L')]=L\del(L-L'),~~~\Psi(L)\vac = \cav \Psi^\dg(L) =0. 
\eeq
This assignment corresponds to working with spatial universes where
a point has been marked. This avoids putting in certain combinatorial 
factors by hand when gluing together universes. The operators 
$\Psi(L)$ and $\Psi^\dg(L)$ will be assigned 
dimensions ${\rm dim}\,[\Psi]={\rm dim}\,[\Psi^\dg] =0$.

We could alternatively have chosen creation and annihilation
operators which create and annihilate universes without such
a mark. Then instead of \rf{s1} one would have 
\beq\label{s1a} 
[\Psi(L),\Psi^\dg(L')]=L^{-1}\del(L-L'),~~~\Psi(L)\vac = \cav \Psi^\dg(L) =0, 
\eeq
and the dimensional assignment would be 
${\rm dim}\,[\Psi]=1$ and ${\rm dim}\,[\Psi^\dg] =1$. One could even
let $\Psi^\dg$ create marked universes and $\Psi$ annihilate unmarked
universes if one just compensates for missing combinatorial factors by hand.
Here we will use the assignment \rf{s1}.

Let us write the differential equation for the finite time propagator derived in \eqref{eq:differentialequationGXYT} using the boundary length rather than the boundary cosmological constant as variable
\beq\label{s2}
\frac{\prt}{\prt T} G^{(m)}_\La(L_1,L_2;T) =  
L_1 \Big(\frac{\prt^2}{\prt L_1^2}-\La\Big) G^{(m)}_\La(L_1,L_2;T),
\eeq 
For convenience we have in \rf{s2} also marked the exit-loop $L_2$ in order
to have symmetry between loops at the initial time and the loop at the  final time, 
i.e.\ $G^{(m)}_\La(L_1,L_2;T)= L_2 G_\La(L_1,L_2;T)$, where $G_\La(L_1,L_2;T)$ was given by \eqref{eq:GmLT}.

We can now also write 
\beq\label{s3}
G^{(m)}_\La(L_1,L_2;T) = \la L_2| e^{-T H_0(L) } |L_1\ra, ~~~~
H_0(L) =- L \frac{\prt^2}{\prt L^2}+\La L,
\eeq
where $H_0(L)$ was derived in \eqref{eq:CDTHamiltonian1}.\footnote{We changed the notation from $\hat{H}_L$ in \eqref{eq:CDTHamiltonian1} to $H_0(L)$ in \eqref{s3} to indicate that it is the first-quantized Hamiltonian.} Associated with the spatial universe we have a Hilbert space on the
positive half-line, and a corresponding scalar product making $H_0(L)$
self-adjoint (see App.\ \ref{App:Calogero} for more details)
\beq\label{s4}
\la \psi_1 |\psi_2\ra = \int \frac{dL}{L} \; \psi_1^* (L) \psi_2(L).
\eeq
 
The introduction of the operators $\Psi(L)$ and $\Psi^\dg(L)$ in \rf{s1}
can be thought of as analogous to the standard second quantization 
in many-body theory. The single particle Hamiltonian becomes in our 
case the ``single universe'' Hamiltonian $H_0$. It has eigenfunctions
$\psi_n(L)$ with corresponding eigenvalues $E_n= 2n\sL$, $n=1,2,\ldots$:
\beq\label{s4a}
\psi_n(L) = L\, e^{-\sL L} p_{n-1}(L),~~~~~
H_0(L)\psi_n(L)= E_n \psi_n(L),
\eeq
where $p_{n-1}(L)$ is a polynomial of order $n\mi 1$.\footnote{The precise form of $p_{n-1}(L)$ is not of relevance here, but it can be readily obtained from the general formulas in App.\ \ref{App:Calogero}.} 
We now introduce creation and
annihilation operators $a_n^\dg$ and $a_n$ corresponding to these states,
acting on the Fock-vacuum $\vac$ and satisfying $[a_n,a^\dg_m]=\del_{n,m}$. 
We define
\beq\label{s5}
\Psi(L) = \sum_n a_n \psi_n(L),~~~~\Psi^\dg(L) = \sum_n a_n^\dg \psi^*_n(L),
\eeq
and from the orthonormality of the eigenfunctions with respect to 
the measure $dL/L$ we recover \rf{s1}. The ``second-quantized'' Hamiltonian is
\beq\label{s6}
\hH_0 = \int_0^\infty \frac{dL}{L}\; \Psi^\dg (L) H_0(L) \Psi(L),
\eeq
and the propagator $G^{(m)}_\La (L_1,L_2;T)$ is now obtained as
\beq\label{s7}
G^{(m)}_\La (L_1,L_2;T)= \cav \Psi(L_2) e^{-T \hH_0} \Psi^\dg(L_1) \vac.
\eeq

\begin{figure}[t]
\centerline{\scalebox{0.70}{{\includegraphics{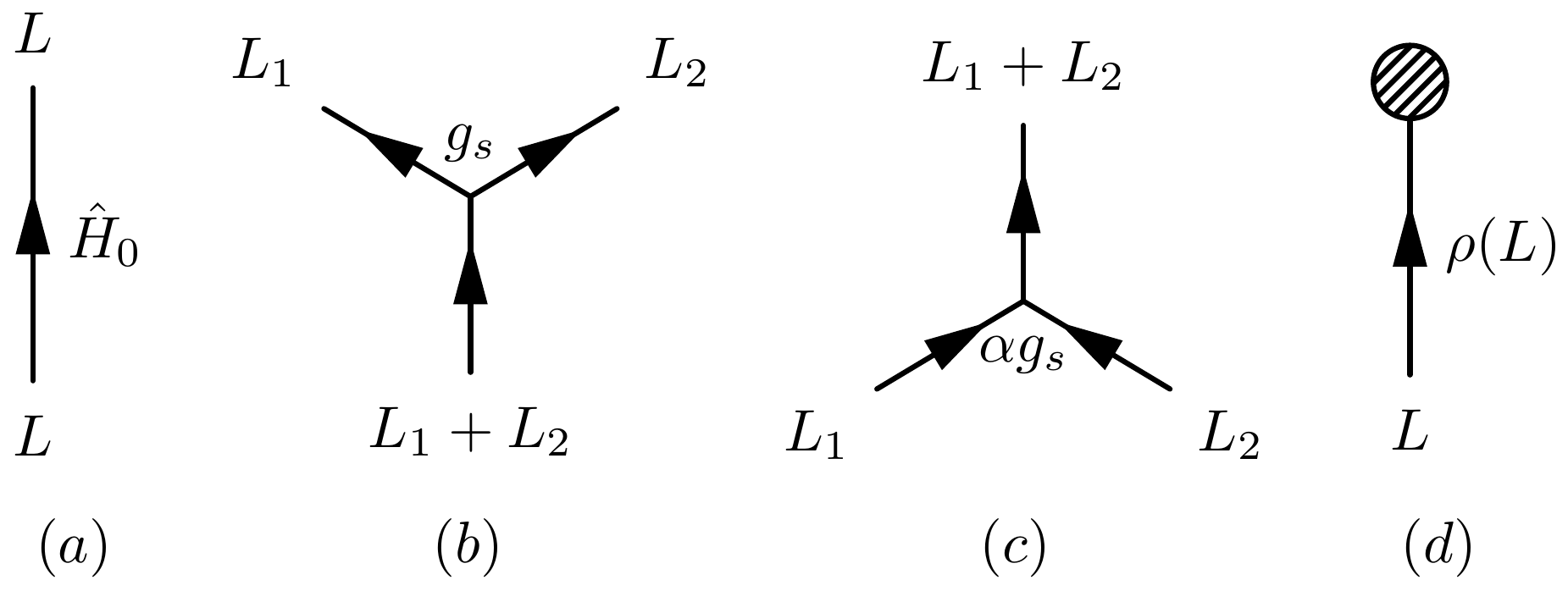}}}}
\caption{ The elementary terms of the string field 
theory Hamiltonian \rf{s8}: (a) the single spatial universe propagator, 
(b) the term corresponding to splitting into two spatial universes, 
(c) the term corresponding to the merging of two spatial universes 
and (d) the tadpole term.    
}
\label{figinter}
\end{figure}

While this is  trivial, the advantage of the formalism
is that it automatically takes care of symmetry factors (like in the 
many-body applications in statistical field theory) both when many 
spatial universes are at play and when they are 
joining and splitting. We can follow
\cite{Ishibashi:1993pc} and define the following Hamiltonian, describing the 
interaction between spatial universes: \index{Causal dynamical triangulations (CDT)!string field theory}
\bea
\hH = \hH_0\!\!\!\!\! \!\!\!&&-~ g_s \int dL_1 \int dL_2 \Psi^\dg(L_1)\Psi^\dg(L_2)\Psi(L_1+L_2)
\non && -\, \a \,g_s\int dL_1 \int dL_2 \Psi^\dg(L_1+L_2)\Psi(L_2)\Psi(L_1)
-\int \frac{dL}{L} \; \rho(L) \Psi(L), \label{s8}
\eea
where the different terms of the Hamiltonian are illustrated in 
Fig.\ \ref{figinter}.
Here $g_s$ is the coupling constant we have already 
encountered in Chap.\ \ref{Chap:cap}
of mass dimension 3. The factor $\a$ is just inserted to be 
able the identify the action of the two $g_s$-terms in \rf{s8} when
expanding in powers of $g_s$. We will think of $\a=1$ unless explicitly
stated differently. Note that the sign of all the interaction
terms in \rf{s8} is negative. This reflects that we want these 
terms to represent the insertion of new geometric structures
compared to the ``free'' propagation generated by $\hH_0$.
These structures should thus appear with positive weight when
we expand $e^{-T \hH}$.   
$\hH$ is hermitian except for the presence of the tadpole term.
It tells us that universes can vanish, but not be created from nothing.
The meaning of the two interaction terms is as follows: the first term
replaces a universe of length $L_1+L_2$ with two
universes of length $L_1$ and $L_2$. This is precisely the 
process shown in Fig.\ \ref{fig2}. The second term represents
the opposite process where two spatial universes merge into one,
i.e.\  the time-reversed picture. The coupling constant $g_s$ seems to play the role of a string coupling constant: one factor $g_s$ for 
splitting spatial universes, one factor $g_s$ for merging spatial
universes and thus a factor $g_s^2$ when the 
space-time topology changes.

In a certain way the appearance of a tadpole term is more 
natural in the CDT framework than in the original Euclidean 
framework in \cite{Ishibashi:1993pc}. Recall the discussion in
Chap.\ \ref{Chap:cap}: In a Lorentzian setting there is no regular
disc geometry, it has to end in a ``puncture''. The tadpole 
term is a formal realization of this process. Also recall that 
we could associate this process with a gravitational 
coupling constant, in this way linking it to $g_s$. It 
was done by observing that to each splitting off of a
baby universe we have a puncture where the baby universe ends.
Thus the coupling constant $g_s$ related to the splitting of 
spatial universes could be identified with the vanishing of 
universes. This shift can be made explicit in our 
string field Hamiltonian $\hH$ in \rf{s8}. In \rf{s8} the 
coupling constant $g_s$ is associated with splitting and joining
of spatial universes. No coupling constant is associated with 
the tadpole term, i.e.\ the vanishing of a spatial universe.
However we can redefine $\Psi$ and $\Psi^\dg$:
\beq\label{r1}
\bar{\Psi} = g_s \, \Psi,~~~~\bar{\Psi}^\dg = \frac{1}{g_s} \, \Psi^\dg.
\eeq
With this substitution the coupling constant $g_s$ is shifted 
from the splitting term  to the tadpole term, i.e.\ precisely 
the shift mentioned above. In addition the term associated
with the joining of spatial universes will have the coupling 
constant $g_s^2$, which matches a change in topology.

Finally, let us identify the real coupling constant appearing 
in \rf{s8}: let us measure everything in terms of $1/\sL$ which
is the natural length scale of our universe. Introducing the 
dimensionless length variable $\tLe = L\sL$, the dimensionless 
boundary cosmological constant $\tX = X /\sL$, the dimensionless 
time variable $\tT = T\sL$, the dimensionless tadpole density 
$\trho(\tLe) = \rho(L)/\sL $, the dimensionless coupling constant 
$\tg_s = g_s/\La^{3/2}$ (already introduced in eq.\ \rf{5.1})
  and finally the dimensionless Hamiltonian $\tH = \hH/\sL$,
we can write
\beq\label{s8b}
\hH_0 = \sL \, \tH_0,~~~ 
\tH_0 = \int \frac{d\tLe}{\tLe}  \tPsi^\dg(\tLe) H_0(\tLe)\tPsi(\tLe),
\eeq
where $\tPsi(\tLe)=\Psi(L)$ and $\tPsi^\dg(\tLe)=\Psi^\dg(L)$ 
satisfy the same commutation relation
as $\Psi(L), \Psi^\dg(L)$ when expressed in terms of $\tLe$, 
and $\hH = \sL \tH$, where   
\bea 
\tH = \tH_0 \!\!\!\!\! \!\!\!&&-~
\tg_s \int d\tLe_1 \int d\tLe_2 \tPsi^\dg(\tLe_1)\tPsi^\dg(\tLe_2)\tPsi(\tLe_1+\tLe_2)
\non 
&& - 
\a \tg_s \int d\tLe_1 \int d\tLe_2 \tPsi^\dg(\tLe_1+\tLe_2)\tPsi(\tLe_2)\tPsi(\tLe_1)
-\int \frac{d\tLe}{\tLe} \; \trho(\tLe) \tPsi(\tLe). \label{s8a}
\eea
From this equation it is clear that the real coupling constant in the 
theory is the dimensionless $\tg_s$, precisely the "double scaling" 
coupling constant which already appeared in the calculation of $W_{\La,g_s}(X)$
and $G_{\La,g_s}(X,Y;T)$ (i.e.\ \rf{5.1}). 
From the discussion above we also observe that
the expansion parameter for topology change of space-time is $\tg_s^2= g_s^2/\La^3$.

In principle we can now calculate the process where we start
out with $m$ spatial universes at time 0 and end with $n$ universes
at time $T$, represented as
\beq\label{s100}
G_{\La,g_s}(L_1,..,L_m;L'_1,..,L'_n;T) =
\cav \Psi(L'_1)\ldots \Psi(L'_n) \; 
e^{-T\hH}\Psi^\dg(L_1)\ldots \Psi^\dg(L_m)\vac.
\eeq

\section{The genus zero limit}\label{alpha}

\subsection{The disc amplitude}

Let us consider the simplest amplitude: a single 
spatial universe which disappears in the vacuum. This is 
precisely the disc-amplitude considered in the previous chapter.
The topology of space-time was not allowed to change in the calculation
in Chap.\ \ref{Chap:cap}. We can incorporate that in the SFT-picture by 
choosing $\a = 0$ in \rf{s8}, i.e.\ the genus zero limit. The disc-amplitude can then be expressed as:
\beq\label{s9}
W_{\La,g_s}(L) =\lim_{T\to \infty}W_{\La,g_s}(L,T) = 
\lim_{T\to \infty} \cav \,e^{-T \hH(\a=0)} \Psi^\dg(L)\vac.
\eeq
It will describe all possible ways in which a spatial loop can develop 
in time and disappear in the vacuum without changing the topology 
of space-time. Note that the tadpole term in \rf{s8} is needed 
in order that the amplitude \rf{s9} is different from zero since 
the state $|L\ra = \psi^\dg (L) \vac$ is orthogonal to the vacuum state
$\vac$. We note that if $\a=0$ we have
\bea\label{s9a}
\!\!\!\!\lefteqn{\cav \,\e^{-T \hH(\a=0)}\Psi^\dg(L_1)\cdots \Psi^\dg(L_m)\vac =}
~~~~~\\ 
&&\!\!\!\!\!\!\!\cav \,\e^{-T \hH(\a=0)}\Psi^\dg(L_1)\vac
\cav \,e^{-T \hH(\a=0)}\Psi^\dg(L_2)\vac\cav\cdots\vac
\cav \,e^{-T \hH(\a=0)}\Psi^\dg(L_m)\vac,
\nonumber
\eea
this factorization being a consequence of the fact that if we 
start out with $m$ spatial universes there is no way they can merge 
at any time if $\a=0$ (it is easy to prove \rf{s9a} using the algebra
of the $\Psi$'s).

Following \cite{Ishibashi:1993pc} we obtain an equation for 
$W_{\La,g_s}(L)$ by differentiating 
\rf{s9} with respect to $T$ and using $\hH \vac =0$:
\beq\label{s10}
0= \lim_{T\to \infty}\frac{\prt }{\prt T}W_{\La,g_s}(L,T) = 
\lim_{T\to \infty} \cav e^{-T \hH(\a=0)}[\hH(\a=0), \Psi^\dg(L)]\vac.
\eeq
The commutator can readily be calculated and after a Laplace transformation
eq.\ \rf{s10} reads
\beq\label{s11}
\frac{\prt}{\prt X}\left((X^2-\La)W_{\La,g_s}(X)  + 
g_s W_{\La,g_s}^2(X)\right) = \rho(X),
\eeq
where the last term on the left-hand-side of \rf{s11} is a consequence 
of the factorization \rf{s9a}.

Eq.\ \rf{s11} has the generalized CDT solution \rf{3.9}-\rf{3.9a}
discussed in Chap.\ \ref{Chap:cap} if 
\beq\label{s12}
\rho(X)=1,~~~{\rm i.e.}~~~\rho(L) = \del(L),
\eeq
which is a reasonable physical requirement: the spatial 
universe can only vanish in the vacuum when the length of 
the universe goes to zero. 

\subsection{Inclusive amplitudes}\label{inclusive}

Above we have understood how to reproduce the generalized CDT disc amplitude 
$W_{\La,g_s}(X)$ as the connected amplitude arising in SFT in
the limit $\a=0$. We now want to understand how
to reproduce the proper-time propagator $G^{(m)}_{\La,g_s}(X,Y;T)$ 
in the context of SFT. We have an ``entrance'' loop at time $T=0$ and
an ``exit'' loop at time $T$. However, the propagator $G^{(m)}_{\La,g_s}(X,Y;T)$
allows baby-universes to split off and propagate further than time 
$T$ if they only vanish into the vacuum eventually (see Fig. \ref{fig2}).

We can reproduce this result in the $\a=0$ limit of SFT by introducing
the ``inclusive'' Hamiltonian \cite{Ishibashi:1993pc}. Since we are working 
in the $\a=0$ limit, we only have universes branching, not merging,
during the time evolution, and all the branching universes except one have
to vanish in the vacuum. The branching process is dictated
by the term
\beq\label{ds19}
g_s \int dL_1 \int dL_2 \Psi^\dg(L_1) \Psi^\dg(L_2) \Psi(L_1+L_2)
\eeq
in the Hamiltonian $\hH$, eq.\ \rf{s8}. Once the branching has 
occurred, only one of the two universes can connect to the exit loop
at time $T$, the other universe has to continue until it eventually
vanishes in the vacuum, a process which is allowed to  occur a
time later than $T$.  This scenario is captured by the replacement
\beq\label{ds20}
\Psi^\dg(L_1) \Psi^\dg(L_2) \to W_{\La,g_s}(L_1)\Psi^\dg(L_2)+
\Psi^\dg(L_1)W_{\La,g_s}(L_2)
 \eeq
in eq.\ \rf{ds19}. Thus we arrive at the following ``inclusive Hamiltonian''
\beq\label{ds21}
\hH_{incl} = \int \frac{dL}{L} \Psi^\dg(L) H_0(L) \Psi(L) -2g_s 
\int dL_1\int dL_2 \; W_{\La,g_s}(L_1)\Psi^\dg(L_2) \Psi(L_1+L_2),
\eeq
and we obtain the following representation of $G^{(m)}_{\La,g_s}(L_1,L_2;T)$
\beq\label{ds22}
G^{(m)}_{\La,g_s}(L_1,L_2;T) = \cav \Psi(L_2) \,e^{-T \hH_{incl}} \Psi^\dg(L_1)\vac.
\eeq
Differentiating eq.\ \rf{ds22} with respect to $T$, commuting $\hH_{incl}$ to the
right to $\vac$ (and using $\hH_{incl}\vac =0$) 
one obtains after a Laplace transformation and removal of the mark on the final boundary \rf{2.55}.
Thus the generalized CDT proper-time propagator also has a simple 
SFT description.

\section{Dyson-Schwinger equations}\label{ds}
\index{Dyson-Schwinger equations}

The disc amplitude is one of a set of functions for which 
it is possible to derive Dyson-Schwinger equations (DSE).  
Here we consider a more general class of functions. 
Define the generating function:
\beq\label{ds1}
Z(J;T)= \cav e^{-T \hH} \; e^{\int dL \, J(L) \Psi^\dg(L)}\vac.
\eeq
We have 
\beq\label{ds2}
\cav \e^{-T \hH} \;\Psi^\dg(L_1)\cdots \Psi^\dg(L_n)\vac = 
\left.\frac{\del^n Z(J;T)}{\del J(L_1)\cdots \del J(L_n)}\right|_{J=0}.
\eeq
We have already seen that in the case where the coupling constant  
$\a =0$ we had factorization:
\beq\label{ds2a}
Z(J,T;\a=0) = e^{\int dL \,J(L)W_{\La,g_s}(L,T)}
\eeq
where $W_{\La,g_s}(L,T)$ denotes the disk amplitude where the universe 
decays into the vacuum before or 
at time $T$, and where $W_{\La,g_s}(L,T\equ\infty)$ 
was the disc  amplitude we already calculated.
 
Following \cite{Ishibashi:1993pc} we can obtain the DSE in the same 
way as for the disc amplitude, the only difference being that when 
the constant $\a$ is no longer zero these equations do not close but
connect various amplitudes of more complicated topology. However,
the equations can be solved iteratively. We denote
\beq\label{ds2b}
Z(J) \equiv \lim_{T\to \infty} Z(J;T),
\eeq
$Z(J)$ being the generating functional for universes that disappear in the 
vacuum. We now have
\beq\label{ds3}
0= \lim_{T\to \infty}\; \frac{\prt}{\prt T}\; 
\cav e^{-T\hH} \; e^{\int dL \, J(L) \Psi^\dg(L)}\vac =
\lim_{T\to \infty} \cav e^{-T\hH} \; 
\hH\;e^{\int dL \, J(L) \Psi^\dg(L)}\vac.
\eeq
Commuting the $\Psi(L)$'s in $\hH$ past the source term effectively replaces
these operators by $L \,J(L)$, after which they can be moved to the left of 
any $\Psi^\dg (L)$ and outside  $\cav$. 
After that the remaining $\Psi^\dg(L)$'s in $\hH$ can
be replaced by $\del/\del J(L)$ and also moved outside
$\cav$, leaving us with a integro-differential operator acting on $Z(J)$:
\beq\label{ds4}
0= \int_0^\infty dL \, J(L) \,O \left(L,J,\frac{\del}{\del J}\right)Z(J)
\eeq
where
\bea\label{ds5}
O \left(L,J,\frac{\del}{\del J}\right)&=& H_0(L) \frac{\del}{\del J(L)} -
\del (L)  \\
&& -g_s L \int_0^L dL'\frac{\del^2}{\del J(L')\del J(L-L')}
-\a g_s L\int_0^\infty dL' L'J(L')\frac{\del}{\del J(L+L')}\nonumber
\eea

$Z(J,T)$ is a generating functional 
which also includes totally disconnected universes which 
never ``interact'' with each other. It is of more interest 
to restrict ourselves to the study of ``connected universes'', 
i.e.\ universes where space-time is connected. The generating 
functional for connected universes is obtained in the standard
way from field theory by taking the logarithm of $Z(J,T)$. Thus 
we write:
\beq\label{ds6}
F(J,T) = \log Z(J,T),
\eeq
and we have
\beq\label{ds7}
\cav \e^{-T\hH} \Psi^\dg(L_1)\cdots \Psi^\dg(L_n)\vac_{con} =
\left.\frac{\del^n F(J,T)}{\del J(L_1) \cdots \del J(L_n)}\right|_{J=0}, 
\eeq
and we can readily transfer the DSE \rf{ds4}-\rf{ds5} into an equation 
for the connected functional
\beq\label{ds8}
F(J) = \lim_{t\to \infty} F(J,T).
\eeq
From \rf{ds4}-\rf{ds5} we obtain
\bea
0= \int_0^\infty dL \, J(L)
\left\{  H_0(L)\, \frac{\del F(J)}{\del J(L)} - \delta(L) 
 -g_s L \int_0^L dL'\;\frac{\del^2 F(J)}{\del J(L')\del J(L-L')} \right. 
\nonumber\\
\left. -g_s L \int_0^L dL'
\frac{\del F(J)}{\del J(L')}\frac{\del F(J)}{\del J(L-L')}
-\a g_s L\int_0^\infty dL' L'J(L')\frac{\del F(J)}{\del J(L+L')}\right\}.
\label{ds9}
\eea
From \rf{ds9} one obtains the DSE by differentiating 
\rf{ds9} after $J(l)$ a number of times and then taking $J(L)=0$.

\section{Application of the DSE}\label{application}

Let us introduce the notation 
\beq\label{ds10}
W(L_1,\ldots,L_n) \equiv 
\left.\frac{\del^n F(J)}{\del J(L_1) \cdots \del J(L_n)}\right|_{J=0}
\eeq
as well as the Laplace transform 
\beq\label{ds11}
W(X_1,\ldots,X_n) \equiv \int_0^\infty dL_1
\cdots\int_0^\infty dL_n \; \e^{-X_1L_1-\cdots -X_nL_n} 
W(L_1,\ldots,L_n)
\eeq
for the higher loop amplitudes.

Let us differentiate eq.\ \rf{ds9} after $J(L)$ one, two and three
times, then take $J(L)=0$
and Laplace transform the obtained equations. We obtain the 
following three equations (where $H_0(X)f(X) = \prt_X [(X^2-\La) f(X)]$):
\bea\label{ds13}
0&=&H_0(X)W(X) -1 +
g_s \prt_X \Big(  W(X,X) +  W(X)W(X)\Big),\\
&& ~ \nonumber\\
0&=&(H_0(X)+H_0(Y))W(X,Y) +g_s\prt_X W(X,X,Y)+
g_s\prt_Y  W(X,Y,Y) \non
&& +2g_s\left(\prt_X [W(X)W(X,Y)] \pl \prt_Y[ W(Y) W(X,Y)]\right) \non
&&+2\a g_s\prt_X\prt_Y \Big(\frac{W(X)\mi W(Y)}{X-Y}\Big)\label{ds15}\\
&& ~ \nonumber\\
0&=&(H_0(X)+H_0(Y) +H_0(Z))W(X,Y,Z) \non
&&+g_s\prt_X W(X,X,Y,Z)+
g_s\prt_Y  W(X,Y,Y,Z) + g_s\prt_Z  W(X,Y,Z,Z) \nonumber\\
&& +2g_s\prt_X [W(X)W(X,Y,Z)] + 2g_s \prt_Y[ W(Y) W(X,Y,Z)]
\nonumber\\
&&+2g_s \prt_Z[ W(Z) W(X,Y,Z)]+2g_s\prt_X [W(X,Y)W(X,Z)] \nonumber\\
&&+ 2g_s \prt_Y[ W(X,Y) W(Y,Z)]+2g_s \prt_Z[ W(X,Z) W(Y,Z)]\non
&&+2\a g_s\left(\prt_X\prt_Y \frac{W(X,Z) \mi W(Y,Z)}{X-Y} 
\pl\prt_X\prt_Z \frac{W(X,Y)\mi W(Y,Z)}{X-Z}\right.\non
&&\left. +\prt_Y\prt_Z \frac{W(X,Y) \mi W(X,Z)}{Y-Z}\right)\label{dsxyz}
\eea
The structure of the DSE should now be clear.

We can solve the DSE iteratively. For this purpose let us introduce 
the expansion of $W(X_1,\ldots,X_n)$
in terms of the coupling constants $g_s$ and $\a$:
\beq\label{ds12}
W(X_1,\ldots,X_n) = \sum_{k=n-1}^\infty \a^k\sum_{m=k-1}^\infty g_s^m \; 
W(X_1,\ldots,X_n;m,k).
\eeq
The amplitude $W(X_1,\ldots,X_n)$ starts with the 
power $(\a g_s)^{n-1}$ since we have to perform $n$ mergings 
during the time evolution in order to create a connected geometry
if we begin with $n$ separated spatial loops. Thus one can 
find the lowest order contribution to $W(X_1)$ from \rf{ds13}, use that
to find the lowest order contribution to $W(X_1,X_2)$ from \rf{ds15}
and use this again in \rf{dsxyz} which involves $W(X_1,X_2,X_3)$, etc. 
Returning to eq.\ \rf{ds13}
we can use the lowest order expression for $W(X_1,X_2)$ to find the 
next order correction to $W(X_1)$, use this and the lowest order
correction for $W(X_1,X_2,X_3)$ to find the next order correction
to $W(X_1,X_2)$, etc. 

Two remarks are in order: firstly, the integration constants which come 
by integrating \rf{ds13}-\rf{dsxyz} and the corresponding higher
order equations are uniquely fixed by the requirement that the 
correlation functions fall off for the lengths $L_i \to \infty$,
i.e.\ the requirement that the Laplace transformed amplitude 
$W(X_1,\ldots,X_n)$ is analytic for $X_i > 0$. Secondly, the 
expressions obtained for $W(X_1,\ldots,X_n)$ can of course be 
obtained directly from a diagrammatic expansion, using the
interaction rules shown in Fig.\ \ref{figinter},
the propagation being defined by $\hH_0$, and then 
integrating in a suitable way over the times $T_i$ involved. 
  
\index{Causal dynamical triangulations (CDT)!genus one disc function}
\index{Causal dynamical triangulations (CDT)!genus two disc function}
Let us just list the first few orders:
\bea\label{g1}
 W(X;0,0)\!\!\!\!& =& \!\!\!\!\frac{1}{X+\sL},\\
W(X;1,0)\!\!\!\!&=& \!\!\!\!\frac{X+3\sL}{\La (X+\sL)^3}. 
\label{g2}\\
W(X,Y;1,1)\!\!\!\!&=&\!\!\!\! \frac{1}{2\sL (X+\sL)^2(Y+\sL)^2}.
\label{g3}\\
W(X,Y,Z;2,2)\!\!\!\!& =&\!\!\!\! 
\frac{7\La^{\frac{3}{2}}+5\La(X+Y+Z)+3\sL(XY+XZ+YZ) + XYZ}{4\La^{\frac{3}{2}} 
(\sL+X)^3(\sL+Y)^3(\sL+Z)^3}\label{g4} 
\eea
These amplitudes involve no change in space-time topology.
The genus one and genus two amplitudes to lowest order in $g_s$,
i.e.\ without any additional baby universes, become, using the results
\rf{g1}-\rf{g4} in the iteration as described above,
\bea\label{ds18}
W(X;2,1) &=& \frac{15\La^{\frac{3}{2}}+11\l X+
5\sL X^2+X^3}{32\La^{\frac{5}{2}}(\sL +X)^5}.
\\
W(X;3,2)&=&\frac{1}{2048\La^{\frac{11}{2}}(\sL \pl X)^9}
\left(11319\La^{\frac{7}{2}} 
\pl 19951\La^3 X \pl 21555\La^{\frac{5}{2}} X^2 \pl \right.\nonumber\\ 
&&\left.  16955 \La^2 X^3 \pl 9765\La^{\frac{3}{2}} X^4 \pl 3885 \La X^5 
\pl 945 \sL X^6 \pl 105 X^7\right). \label{ds19a}
\eea
In terms of diagrams this genus 2 amplitude corresponds to the
following three diagrams (integrated suitably over the times $T_i$):
\beq\nonumber
W(X;3,2)\,\,=\,\,
\raisebox{-35pt}{\includegraphics[height=80pt]{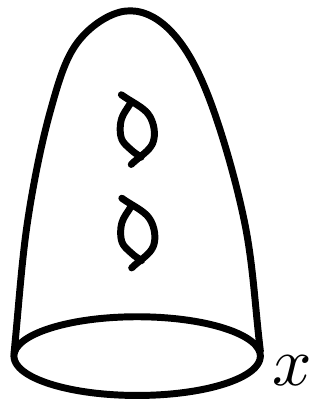}}+\,\,
\raisebox{-35pt}{\includegraphics[height=80pt]{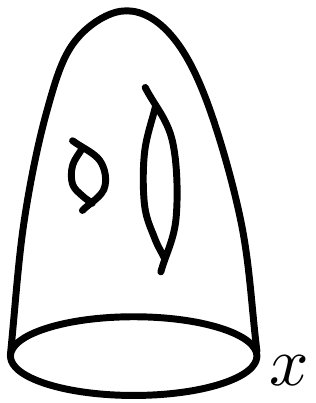}}+\,\,
\raisebox{-35pt}{\includegraphics[height=80pt]{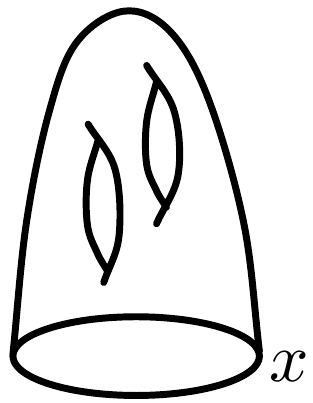}}.
\eeq

As mentioned above the amplitude $W(X_1,\ldots,X_n)$ starts with the 
power $(\a g_s)^{n-1}$ coming from merging the
$n$ disconnected spatial universes. The rest of the powers
of $\a g_s$ will result in a topology change of the resulting, connected 
worldsheet. From a Euclidean point of view it is thus more appropriate
to reorganize the series as follows
\bea\label{ds12a}
&&W(X_1,\ldots,X_n) = (\a g_s)^{n-1}
\sum_{h=0}^\infty (\a g_s^2)^h W_h(X_1,\ldots,X_n)\\
&&W_h(X_1,\ldots,X_n) = \sum_{j=0}^\infty g_s^j 
 W(X_1,\ldots,X_n;n-1+2h +j,n-1+h)\label{ds12b}
\eea
and aim for a topological expansion in $\a g_s^2$, at each order
solving for all possible baby-universe creations which at some 
point will vanish into the vacuum. Thus $W_h(X_1,\ldots,X_n)$ 
will be a function of $g_s$ although we do not write it explicitly.
The DSE allow us to obtain the topological expansion 
iteratively, much the same way we already did as a power expansion 
in $g_s$. 

Since we have $W(X,Y) = O(\a)$ this term does not contribute 
to the lowest order and from the DSE \rf{ds13} we 
obtain a closed equation for $W_0(X)$:
\beq\label{ds13a}
H_0(X)W_0(X;g_s) + g_s \prt_X W_0^2(X;g_s)=1,
\eeq
This equation is of course just eq.\ \rf{s10} and we have 
\beq\label{ds13b}
W_0(X) = W_{\La,g_s}(X).
\eeq

Knowing $W_0(X)$ allows us to obtain $W_0(X,Y)$ from \rf{ds15}
since $W(X,Y,Z)$ is of order $O(\a^2)$. Thus the 3-loop term does 
not contribute to the lowest $\a$ order of eq.\ \rf{ds15}, which is $O(\a)$,
and we have to the lowest order:
\bea
\Big( H_0(X)\pl 2g \prt_x W_0(X)\pl H_0(Y)\pl 2g_s \prt_Y W_0(Y)\Big)W_0(X,Y) = \nonumber \\= 
\mi 2 \prt_X\prt_Y \Big(
\frac{W_0(X) \mi W_0(Y)}{X-Y}\Big)\label{ds15a}
\eea
Thus $W_0(X,Y)$ is entirely determined by the knowledge of $W_0(X)$.
We note that using the definition \rf{3.3} we can simplify \rf{ds15a}:
\bea
\frac{\prt}{\prt X} \Big(\hWg(X)W_0(X,Y)\Big) +
\frac{\prt}{\prt Y} \Big(\hWg(Y)W_0(X,Y)\Big) =\nonumber \\
=- \frac{1}{g_s} \frac{\prt^2}{\prt X\prt Y} \Big(
\frac{\hWg(X) - \hWg(Y)}{X-Y}\Big)\label{ds15b}
\eea
The solution $W_0(X,Y)$ can readily be found from eq.\ \rf{ds15b}, yielding
\beq\label{ds16}
W_0(X,Y)= \frac{1}{f(X)f(Y)}\frac{1}{4g_s} 
\left(\frac{[(X+C)+(Y+C)]^2}{[f(X)+f(Y)]^2} -1\right),
\eeq
where 
\beq\label{ds16a}
f(x) = \sqrt{(x+C)^2-2g_s/C} = \hWg (X)/(X-C).
\eeq
and
\beq
C = U\sL, ~~~U^3-U+\dfrac{g_s}{\La^{3/2}}=0.
\eeq
as in Sec.\ \ref{disc}.

In fact this solution was already found in Sec.\ \ref{sec:consitencyCap}
since we by definition have
\beq\label{ds15c} 
W_0(X,Y) = \int_0^\infty dT \;W_{\La,g_s}(X,Y;T),
\eeq
where $W_{\La,g_s}(X,Y;T)$ is the Laplace transform of the loop-loop function
$W_{\La,g_s}(L_1,L_2;T)$ defined in \rf{ds24} with $T_1=T/2$.
When expanded to lowest order in $g_s$ we reproduce \rf{g3}.

It should now be clear that one can iterate the DSE in 
a systematic way as a power series in the number of handles
of the world sheet exactly the same way as we iterated the DSE 
as a function of the coupling constant $g_s$:
As an instructive example we calculate the genus one amplitude $W_1(X)$.
We expand
\bea\label{dse1}
W(X) &=& W_0(X) +\a g_s^2 W_1(X) + \a^2g_s^4 W_2(X)+\cdots\\
W(X,Y) &=& \a g_s W_0(X,Y) + \a^2 g_s^3 W_1(X,Y)+ \cdots \nonumber
\eea 
While \rf{ds13a} was the 0th order in $\a$ of eq.\ \rf{ds13},
the 1st order reads
\beq\label{dse2}
\frac{\prt}{\prt X} \left( \hWg(X) W_1(X) + W_0(X,X) \right) =0,
\eeq
where $W_0(X,X)$ is given by \rf{ds16}. The integration constant
is fixed by the requirement that $W_1(x)$ is analytic for
$X > 0$, i.e.\ that $W_1(L)$ falls of as $L \to \infty$. We then obtain
\beq\label{dse3}
W_1(X) = \frac{W_0(C,C) -W_0(X,X)}{\hWg(X)} = 
\frac{(X+3C)(X^2+2CX+5C^2-4g_s/C)}{4(2C^3-g_s)((X+C)^2-2g_s/C)^{5/2}}.
\eeq
It is seen that if we expand $W_1(X)$ in powers of $g_s$, the 
first term will reproduce \rf{ds18}.

\section{Discussion}\label{discuss}

We have developed here a string-field theory based on the CDT quantization
of two-dimensional quantum gravity. It shares many properties with
the original non-critical SFT, and the whole formalism was borrowed
from non-critical SFT. Yet, it is different and in some ways simpler.
The tad-pole term is simpler. It is simply $\rho(L)=\del (L)$, telling 
us that universes can only disappear in the vacuum if they have zero
spatial ``volume''. This is in accordance with the interaction
between spatial universes, which preserves the total length (the 
total spatial volume). 
In non-critical SFT the evolution in proper time  results in
a process where the original spatial universe at proper time $T\equ 0$
spawn an infinity of (infinitesimal) baby universe during the 
time evolution. This is linked to the fact that the 
proper time in non-critical SFT has the anomalous length dimension 
1/2. In our new CDT-based SFT the situation is different. The proper
time $T$ has the canonical dimension 1, and the number of 
baby universes created during the time evolution is finite (see \eqref{5.2}).

As discussed in Chap.\ \ref{Chap:cap} it is not possible to 
connect the non-critical SFT and the CDT-based SFT by a 
simple analytic continuation in the coupling constant $g_s$,
not even in the $\a=0$ limit.
As already shown in Chap.\ \ref{Chap:Relating}, starting out with a discretized,
regularized version of the theory, the  Euclidean theory 
(quantum Liouville theory) is obtained if the ``bare'' dimensionless
coupling constant $g$ is of order one. However, the relation between
the bare coupling constant and the dimensionful continuum coupling
constant $g_s$ used here and in the previous chapter is, as mentioned in Sec.\ \ref{sec:tamingtop}, 
\beq\label{dis1}
g = g_s a^3
\eeq
As discussed in Sec.\ \ref{sec:tamingtop} the generalized
CDT continuum limit corresponds to $g_s$ fixed, $a \to 0$, and thus 
to $g(a) \to 0$. The fact that $g(a)$ goes to zero in  
CDT SFT is of course related to the finite number of baby universes
generated in this theory. In contrast we have an infinite number
of baby universes generated in non-critical SFT where $g$ is 
of order one.

However, there is clearly a deep connection between the Euclidean 
theory and the CDT theory awaiting to be fully understood. In
Sec.\ \ref{sec:integrationout} it was shown that if one integrate out the 
``excessive spawn'' of baby universes in Euclidean two-dimensional quantum gravity
one recovers the CDT theory and the mapping between the dimensionless
variables $X/\sL$ in the two theories was found, i.e.\ \eqref{renormalization3}. This mapping was
later discovered by Seiberg et al.\ \cite{Seiberg:2003nm} as the uniformization 
map from the algebraic surface representing the ``semiclassical'' 
non-critical string to the complex plane. The singular points
of the algebraic surface corresponds to ZZ-branes 
where there is a transition from compact to non-compact topology \cite{aagk,Ambjorn:2007iq,Ambjorn:2007xe}.
These singular points are mapped to points in 
the complex plane where one has a similar transition from 
compact to non-compact geometry in CDT context as discussed in Chap.\ \ref{Chap:emergence}.
 
  \index{c=1 barrier}
It is interesting to try to generalize the present CDT-based SFT to include
the coupling to matter. In particular, the presence of a $c=1$ 
barrier can then be addressed. Since the $c=1$ barrier partly 
can be understood as the result of an excessive creation of 
baby universes, such that the two-dimensional worldsheet is torn apart 
\cite{Ambjorn:1987wu,Ambjorn:1992rp} it is clear that the CDT theory may behave differently 
at $c=1$. Numerical simulations
hint that there still might be a barrier for large $c$ \cite{Ambjorn:1999yv},
but nothing is presently known with certainty
and CDT-based SFT may provide a useful analytic tool. 
Work in this direction is in progress.

Equally interesting is the possibility of performing a summation 
over wordsheets of all genera. Again, since the double-scaling
limit in CDT-based SFT is different from the double-scaling 
limit in non-critical 
string theory and the penalty for creation of a higher genus surface
is larger in the sense mentioned above, viewing the creation 
a higher genus worldsheet as a successive creation and annihilation
of a baby universe, one could hope the result of such a summation 
is better behaved and less ambiguous than was the case in non-critical
string theory. Work in this direction is also in progress.

\part{Matrix models for 2D causal dynamical triangulations}

\chapter{Matrix models for two-dimensional Euclidean quantum gravity\label{Chap:matrix}}
	
In the previous chapter we computed higher-loop and higher-genus amplitudes in the framework of a SFT for the continuum model of generalized CDT. An earlier formulation for DT was introduced by Kawai et al.\ \cite{Ishibashi:1993pc} as discussed above. If one would like to perform similar calculations in the discrete framework of DT one could either formulate a discrete STF as has been done in \cite{Watabiki:1993ym} or make use of so-called matrix models and loop equations derived from them. 

In this chapter we want to introduce the use of matrix models to solve combinatorial counting problems of DT on the discretized level. Those techniques are very powerful, since they allow for many generalizations especially in the context of matter couplings.

\index{Matrix models}
\section{Counting triangulations with matrix models}

In the following we want to give a short introduction to how certain matrix integrals can be used as generating functionals for random triangulation. The basic idea underlying this construction goes back to the seminal work by 't Hooft \cite{Hooft:1973jz} on the large-N limit of QCD, followed by work on the saddle point approximation \cite{Brezin:1977sv,Bessis:1980ss}. For further references the reader is referred to several reviews \cite{DiFrancesco:1993nw,DiFrancesco:2004qj,Ginsparg:1993is,Ambjorn:1997di}. 

\begin{figure}[t]
\begin{center}
\includegraphics[width=2.7in]{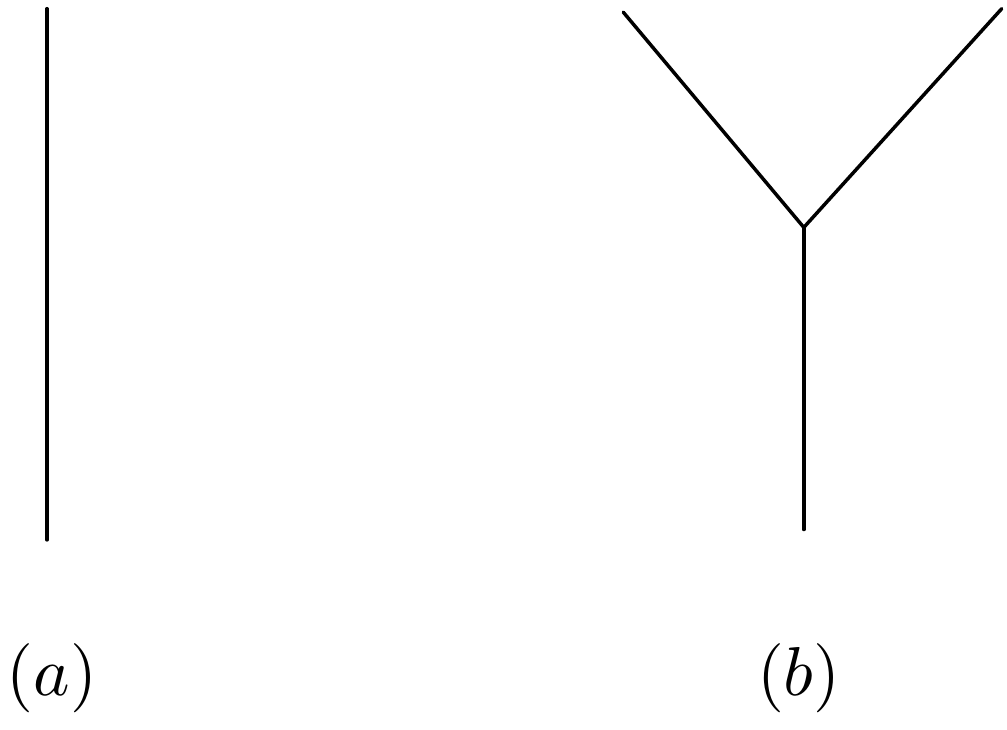}
\caption{Building blocks for Feynman graphs: (a) the scalar propagator and (b) the scalar three-point vertex.}
\label{notfatgraph}
\end{center}
\end{figure}

\begin{figure}[t]
\begin{center}
\includegraphics[width=3in]{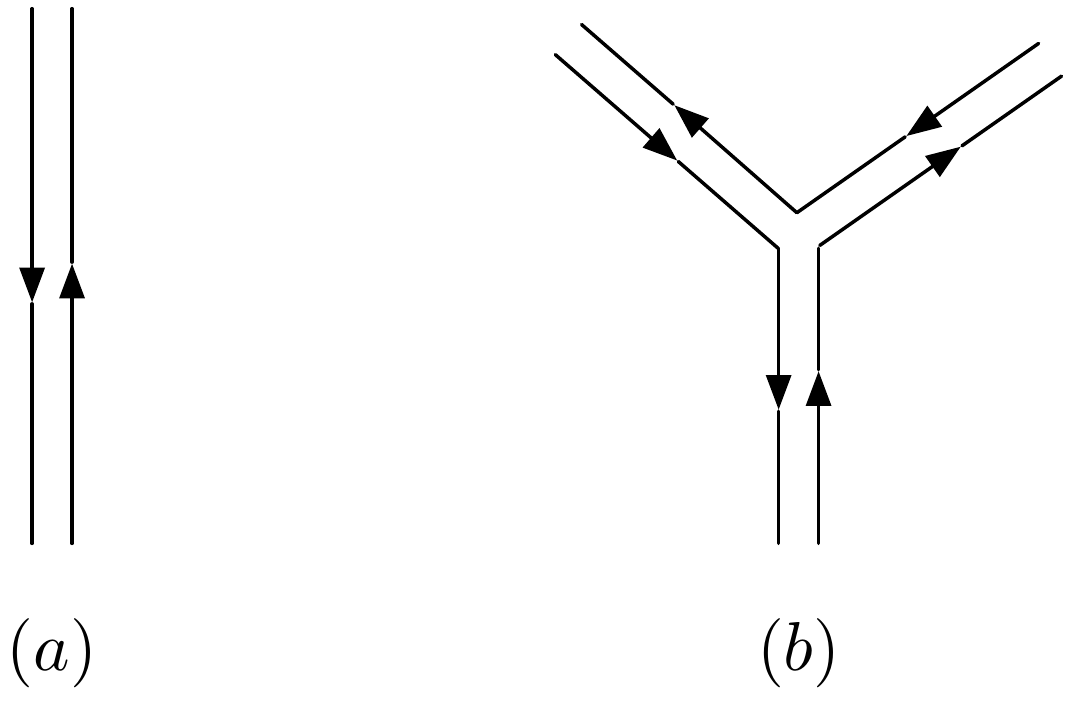}
\caption{Building blocks for fat-graphs: (a) the Hermitian matrix propagator and (b) the Hermitian matrix three-point vertex.}
\label{fatgraph}
\end{center}
\end{figure}

\index{$\varphi^3$-theory}
Before entering the discussion of matrix integrals let us first briefly recall the Feynman expansion of zero-dimensional $\varphi^3$ field theory whose action we already mentioned in \eqref{eq:capphi3},
\beq\label{phi3PI}
Z(g)=\int_{-\infty}^\infty \frac{d\varphi}{\sqrt{2\pi}} e^{-\frac{1}{2}\varphi^2+\frac{g}{3!}\varphi^3} = \sum_{n=0}^\infty I_n(g)
\eeq
with
\beq \label{matrixphi3}
I_n(g)=\int_{-\infty}^\infty \frac{d\varphi}{\sqrt{2\pi}} e^{-\frac{1}{2}\varphi^2} 
\frac{1}{n!}\left( \frac{g}{3!}\varphi^3 \right)^n.
\eeq
The integral \eqref{phi3PI} is formal since it is clearly not convergent until some analytic continuation is performed.
In terms of Wick contractions the integral \eqref{matrixphi3} counts all possible ways of connecting $n$ three-valent vertices (Fig.\ \ref{notfatgraph} (b)) by pairings, represented by the propagator $\expec{\varphi\varphi}$ (Fig.\ \ref{notfatgraph} (a)). These Feynman diagrams already look very much alike the dual triangulations in DT. However, to represent Riemannian geometry more structure is needed. This structure is provided when replacing the usual Feynman graphs by so-called ``fat-graphs'' as shown in Fig.\ \ref{fatgraph} In terms of the integral expression \eqref{matrixphi3} this means that we replace the field $\varphi$ by traces over $N\times N$ Hermitian matrices $\phi=(\phi_{\a\beta})$, i.e. \index{Fat-graphs}
\beq\label{eq:capphi32}
Z(g,N)=\int d\phi e^{-N\left( \frac{1}{2}\tr\phi^2-\frac{g}{3}\tr\phi^3 \right)} = \sum_{n=0}^\infty I_n(g,N)
\eeq
with
\beq  \label{matrixphi32}
I_n(g,N)=\int d\phi e^{- \frac{N}{2}\tr\phi^2} \frac{1}{n!}\left( \frac{Ng}{3}\tr\phi^3 \right)^n.
\eeq
Here the integration measure is defined using the matrix elements 
\beq
d\phi=\prod_{\a\leq\beta} d \Re \phi_{\a\beta} \prod_{\a<\beta} d \Im \phi_{\a\beta}.
\eeq
The propagators become now oriented links (see Fig.\ \ref{fatgraph} (a))
\beq
\expec{\phi_{\a\beta}\phi_{\a'\beta'}}=\delta_{\a,\beta}\delta_{\a',\beta'}
\eeq
and the three-valent vertices become also oriented (Fig.\ \ref{fatgraph}(b)).
From this we see that the integral \eqref{matrixphi32} accounts for all closed, possibly disconnected, unrestricted triangulations with $n$ triangles (recall Fig.\ \ref{fig:disc}). Since each vertex comes with a factor of $gN$ and each propagator with a factor $1/N$ one can convince oneself that the contribution of a given triangulation with $n$ triangles carries a factor 
\beq \label{matrixweight}
C^{-1}(T) g^n N^{\chi(T)},
\eeq
where the Euler characteristic of the triangulation is given by $\chi(T)=2-2h$, with $h$ being the genus of the triangulation.\footnote{The relation \eqref{matrixweight} is also valid for the case of triangulations with boundaries, where the Euler characteristic becomes $\chi(T)=2-2h-b$ with $b$ being the number of boundary components.} Further, it is noted that the factor $1/n!$ in \eqref{matrixphi32} combines with the number of permutation of the vertices yielding the factor $1/C(T)$ which already appeared in \eqref{eq:simpl:pathsum}. To make contact with dynamical triangulations we see that we have to assign $g$ with the bare dimensionless cosmological constant and $N$ with the bare dimensionless Newton's constant $G_N$ as follows\footnote{The relation for $g$ is the same as used in previous chapters, only that we sometimes written it in terms of the bare dimensionfull cosmological constant, i.e\ $g=e^{-a^2\l}$.}:
\beq
g=e^{-\l},\quad N=e^{1/G_N}.
\eeq
Eq. \eqref{matrixweight} now corresponds to the Bolzmann weight of the triangulation. We conclude that the partition function for dynamical triangulations is given by
\beq
\cZ(g,N)=\frac{\log Z(g,N)}{\log Z(0,N)},
\eeq
where the logarithm is introduced to precisely account for the number of \emph{connected} (non-restricted) triangulations. Here $Z(g,N)$ as introduced in \eqref{eq:capphi32} is commonly written as  \index{Dynamical triangulations (DT)!matrix model}\index{Matrix models!DT}
\beq\label{matrixactionDT}
Z(g,N)=\int d\phi e^{-N \, \tr V(\phi)},\quad V(\phi) = \frac{1}{2} \phi^2 - \frac{g}{3} \phi^3, 
\eeq
where $V(\phi)$ is called the potential of the matrix model.
From \eqref{matrixweight} we see that $\cZ(g,N)$ has the expected topological expansion
\beq
\cZ(g,N)= \sum_{h=0}^{\infty} N^{2-2h} \cZ_{h}(g).
\eeq
In the so-called planar or large-$N$ limit one takes $N\to\infty$ and obtains the genus zero partition function \index{Large-$N$ limit}\index{Planar limit}
\beq
 \cZ_{0}(g)= N^{-2} \cZ(g,N) +\cO(1/N^2).
\eeq

We have seen how to represent the partition function of DT as an integral over Hermitian matrices. One of the strengths of using matrix models is that it allows for many generalizations. One possible generalization is to replace the potential $V(\phi)$ in \eqref{matrixactionDT} by
\beq \label{matrixgeneraldef}
Z(g_1,...,N)=\int d\phi e^{-N \, \tr V(\phi)},\quad V(\phi) = \frac{1}{2} \phi^2 - \sum_{j=1}^{\infty}\frac{g_j}{j} \phi^j.
\eeq
This integral counts for closed, possibly disconnected, general non-restricted triangulations which include $j$-gons, each weighted by the corresponding coupling $g_j$. 
Generalizing \eqref{matrixweight} to arbitrary potential gives
\beq\label{matrixweight2}
C^{-1}(T) N^{\chi(T)}  \prod_{j=1}^{\infty} g_j^{N(j)},
\eeq
where $N(j)$ is the number of $j$-gons in the triangulation.

Having established the connection between matrix models and dynamical triangulations we now want to show how to actually calculate certain quantities, such as the disc function, within this framework. 

\section{Loop equations from matrix models} \label{sec:matrixDTloopeq}
\index{Matrix models!loop equations}

Let us for the following consider the matrix model with general potential as described in \eqref{matrixgeneraldef}. We abbreviate the set of couplings by $\gu\equ(g_1,g_2,g_3,...)$.

Denoting the expectation value by 
\beq\label{xx3}
\expec{ O_1(\phi)\cdots O_n(\phi) } = 
\frac{\int \d \phi \; O_1(\phi)\cdots O_n(\phi)\;\e^{-N \tr V(\phi)}}{\int 
\d \phi \; \e^{-N \tr V(\phi)}}
\eeq
we see that
\beq
w_l(\gu,N)=\frac{1}{N}\expec{\tr\phi^l}
\eeq
accounts for the disc function for general non-restriucted triangulations with boundary length $l$ and arbitrary genus.\footnote{The factor of $1/N$ is introduced to cancel the factor $N^{2-b}$, $b\equ 1$ coming from the Euler characteristic. This is done to make the leading order in the large-$N$ limit of order one.} 
We can now also introduce the generating function
\beq
w(\gu,N,z)=\sum_{i=0}^\infty w_l(\gu,N) \, z^{-l-1}=\frac{1}{N}\expec{\tr\frac{1}{z-\phi}}
\eeq
It is sometimes useful to further express the disc function in terms of the density of eigenvalues $\rho(\l)$ defined as \index{Matrix models!density of eigenvalues}
\beq
\rho(\l)= \expec{\sum_{i=1}^N \delta(\l-\l_i)},
\eeq
where $\l_i$, $i\equ 1,2,..,N$ are the eigenvalues of the matrix $\phi$. We can now write the disc function as
\beq \label{matrixdensitydisc}
w_l(\gu,N)=\int_{-\infty}^\infty d\l \rho(\l) \l^l,\quad w(\gu,N,z)= \int_{-\infty}^\infty d\l \frac{\rho(\l)}{ z-\l}.
\eeq
In the limit $N\to\infty$ the density of eigenvalues has finite support $[c_-,c_+]$ on the real axis. This implies that the analytic continuation of $w(\gu,N,z)$ is well defined on $\mathbb{C}\slash [c_-,c_+]$ and we can invert \eqref{matrixdensitydisc} as follows
\beq \label{matrixdensitydisc2}
2\pi i \rho(\l)=\lim_{\epsilon\to 0}\left(w(\gu,N,\l-i\epsilon)-w(\gu,N,\l+i\epsilon) \right). 
\eeq

\begin{figure}[t]
\begin{center}
\includegraphics[width=4.5in]{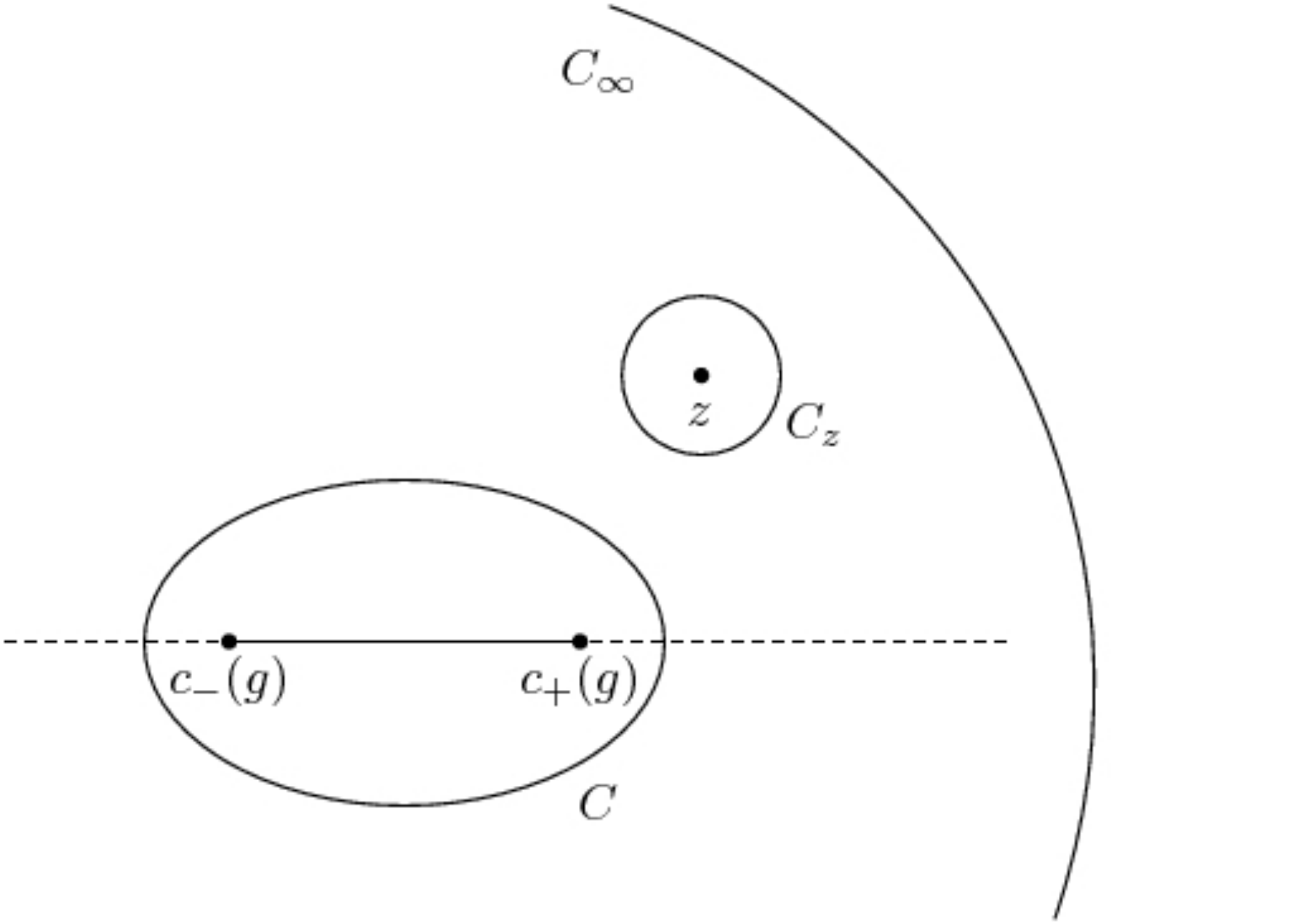}
\caption{The contour $C$ encloses counterclockwise the cut $[c_-,c_+]$ but not $z$.}
\label{contour}
\end{center}
\end{figure}

To derive the loop equations for the disc function from the matrix model \cite{David:1990ge,Ambjorn:1990ji,Ambjorn:1990wg} we will make an infinitesimal change in the field variable of the functional integral. This is a standard method employed in field theories. Let us consider the following infinitesimal field redefinition
\beq
\delta: \phi\to\phi+\epsilon \frac{1}{z-\phi},
\eeq
where $\epsilon>0$ is an infinitesimal parameter. In the limit $N\to\infty$ this field redefinition is well defined for $z\in\mathbb{C}\slash [c_-,c_+]$.  
Under this field redefinition the measure and the potential transform to first order in $\epsilon$ as
\beq
d\phi\to d\phi \left( 1+\epsilon \, \tr\frac{1}{z-\phi} \tr\frac{1}{z-\phi} \right)
\eeq 
and 
\beq
\tr V(\phi) \to \tr V(\phi)+\epsilon\, \tr\left(\frac{1}{z-\phi} V'(\phi)  \right).
\eeq
Using that $\log Z(g,N)$ as given in \eqref{matrixgeneraldef} should be invariant under the transformation $\delta$ we obtain the following identity
\beq \label{matrixloop1}
\delta \log Z(\gu,N)=   Z^{-1}(\gu,N) \int d\phi \left\{ \left(\tr\frac{1}{z-\phi}\right)^2   - N \tr\left(\frac{1}{z-\phi} V'(\phi)  \right) \right\} e^{-N \, \tr V(\phi)}=0.
\eeq
The first term in the brackets is by definition (see \eqref{matrixconnectedexpec})
\beq\label{matrixloop2}
\expec{\left(\tr\frac{1}{z-\phi}\right)^2} = N^2 w^2(\gu,N,z) + w(\gu,N,z,z),
\eeq
where $w(\gu,N,z_1,z_2)$ is the two-loop amplitude, i.e.\ the sum over all triangulations with two-boundaries with boundary cosmological constants $z_1$ and $z_2$. 
The second term can be evaluated using the density of eigenvalues,
\beq\label{matrixloop3}
\frac{1}{N}\expec{\tr \frac{V'(\phi)}{z-\phi}}=\int_{-\infty}^\infty d\l \rho(\l) \frac{V'(\l)}{z-\l}
= \oint_C \frac{d w}{2\pi i}  \frac{V'(w)}{z-w} w(\gu,N,z), 
\eeq
where the first equality follows from \eqref{matrixdensitydisc} and the second equality from \eqref{matrixdensitydisc2}. The contour $C$ encloses counterclockwise the cut $[c_-,c_+]$ of $w(\gu,N,z)$ but not $z$ as shown in Fig.\ \ref{contour}. As discussed above \eqref{matrixdensitydisc} for this relation to hold we must have $N\to\infty$ which implies finiteness of $[c_-,c_+]$. Inserting \eqref{matrixloop2} and \eqref{matrixloop3} into \eqref{matrixloop1} yields the loop equation 
\beq\label{matrixloop4}
 \oint_C \frac{d w}{2\pi i}  \frac{V'(w)}{z-w} w(\gu,N,z)= w^2(\gu,N,z) + \frac{1}{N^2}w(\gu,N,z,z).
\eeq

Since we are only interested in the genus zero disc function we can take the large-$N$ limit which allows us to drop the last term from \eqref{matrixloop4}, yielding
\beq\label{matrixloop5}
 \oint_C \frac{d w}{2\pi i}  \frac{V'(w)}{z-w} w(\gu,z)= w^2(\gu,z).
\eeq
The solution to \eqref{matrixloop5} reads
\beq\label{matrixloopsol}
w(\gu,z)=\frac{1}{2}\left( V'(z) -M(\gu,z)\sqrt{ (z-c_+(\gu))(z-c_-(\gu))} \right),
\eeq
where $M(z)$ is a polynomial whose degree is one less than $V'(z)$. Further the density of eigenvalues reads
\beq\label{matrixloopsoleigen}
\rho(\l)=\frac{1}{2\pi} M(\gu,\l)\sqrt{ (c_+(\gu)-\l)(\l-c_-(\gu))}.
\eeq
As for the loop equations of DT, the solution for the disc function \eqref{matrixloopsol} is completely determined by the requirement that $w(\gu,z)=1/z+\cO(1/z^2)$. In particular, one can derive the following equations by imposing this requirement for \eqref{matrixloopsol} and using \eqref{matrixloop5} (see for example \cite{Ambjorn:1997di} for more details):
\beq\label{matrixloopcond1}
M(\gu,z)=\oint_{C_\infty} \frac{dw}{2\pi i} \frac{V'(w)}{w-z} \frac{1}{\sqrt{(w-c_+)(w-c_-)}}
\eeq 
and
\beq\label{matrixloopcond2}
M_{-1}(\gu,c_-,c_+)=2,\quad M_{0}(\gu,c_-,c_+)=0,
\eeq
where we defined
\beq\label{matrixloopcond3}
M_{k}(\gu,c_-,c_+)=\oint_{C} \frac{dw}{2\pi i}  \frac{V'(w)}{(w-c_+)^{k+1/2}(w-c_-)^{1/2}}.
\eeq
Here the first equation \eqref{matrixloopcond1} determines the polynomial $M(\gu,z)$ as a function of $c_-$, $c_+$ and the couplings $\gu$, while the second equation \eqref{matrixloopcond2} yields two algebraic equations which determine $c_-(\gu)$ and $c_+(\gu)$ as functions of the couplings $\gu$.

Let us now look at two specific cases of the potential. For 
\beq
V(\phi)=\frac{1}{2}\phi^2
\eeq
we expect to obtain the generating function for rooted branched polymers. Inserting $V'(z)$ into  
\eqref{matrixloopsol} and using the conditions \eqref{matrixloopcond1}-\eqref{matrixloopcond2} for the requirement that $w(z)=1/z+\cO(1/z^2)$ we obtain
\beq\label{matrixloopsolBP}
w(z)=\frac{1}{2}\left( z -\sqrt{ (z-2)(z+2)} \right),
\eeq
which is precisely \eqref{eq:BPgen}. Further, using \eqref{matrixloopsoleigen} one gets for the density of eigenvalues
\beq
\rho(\l)=\frac{1}{2\pi} \sqrt{ (z-2)(z+2)}
\eeq
which is Wigner's semicircle law for the eigenvalue distribution of Hermitian matrices \cite{Wigner:1,Wigner:2}.

For the potential 
\beq \label{matrixpotentialDT}
V(\phi)=\frac{1}{2}\phi^2 -\frac{g}{3}\phi^3
\eeq
we expect, according to the considerations made above, to obtain DT. Inserting \eqref{matrixpotentialDT} into \eqref{matrixloopsol} we get using \eqref{matrixloopcond1}-\eqref{matrixloopcond2}
\beq
w(g,z)=\frac{1}{2}\left( z-gz^2 -(gz-c(g))\sqrt{ (z-c_+(g))(z-c_-(g))} \right)
\eeq
which coincides with \eqref{eq:DTdiscdiscrete}, where $c(g)$, $c_-(g)$ and $c_+(g)$ are determined by the same algebraic equations as in $\eqref{eq:DTdiscdiscrete}$.

We have seen how matrix models can be used to count generalized triangulations. Further, we showed how we can obtain DT from matrix models for the specific potential $V(\phi)=\phi^2/2-g\phi^3/3$. In particular, we computed the disc function. In the following section we want to extend our results by calculating higher-loop amplitudes. 

\section{The loop insertion operator and higher loop amplitudes} \label{sec:matrixhigherloop}

\index{Loop insertion operator}\index{Higher loop amplitudes}
Even though in Chap.\ \ref{Chap:dt} we only calculated the one-loop amplitude for DT (disc function), we already encountered higher loop amplitudes in this thesis, namely the two-loop amplitude for DT in the continuum, i.e.\ \eqref{4.5}. 

Matrix model higher-loop amplitudes, i.e.\ amplitudes for generalized triangulations with $b$ boundaries of lengths $l_1,l_2,...,l_b$, can be defined as follows
\beq
w_{l_1,\ldots,l_b}(\gu) = N^{b-2} 
\left\la (\tr \phi^{l_1} )\cdots (\tr \phi^{l_b} )
\right\ra_{c},
\eeq
where $\expec{...}_{c}$ denotes the connected expectation value defined as
\beq \label{matrixconnectedexpec}
\expec{ O_1(\phi)\cdots O_n(\phi) }_{c} = \expec{ O_1(\phi)\cdots O_n(\phi) }- \expec{ O_1(\phi)}\cdots\expec{ O_n(\phi) }. 
\eeq
Equivalently one can also use the generating functional for the higher-loop amplitudes
\beq
w(\gu,z_1,\ldots,z_b) = N^{b-2} 
\left\la (\tr \frac{1}{z_1-\phi})\cdots (\tr \frac{1}{z_b-\phi})
\right\ra_{c}.
\eeq

To calculate the higher loop amplitudes we introduce the so-called loop insertion operator \cite{Ambjorn:1990ji,Ambjorn:1992gw}
\beq\label{matrixloopinsertion}
\frac{d}{d V(z)} = \sum_{j=1}^\infty \frac{j}{z^{j+1}} \; 
\frac{d }{d g_j}. 
\eeq
For a given potential with fixed coupling constants $g_j^0$ one uses 
these relations in the following way. Assume that $g_j$ can vary, 
one now acts with the loop insertion operator $\d/\d V(z_2)$ say on the disc function $w(\gu,z_1)$, and then set $g_j=g_j^0$ again. Each term in the summation over $j$ has the effect that it removes a $j$-sided polygon\footnote{We showed above that each triangulation comes with a weight proportional to $\prod_{i=1}^{\infty} g_i^{N(i)}$, 
where $N(i)$ is the number of $i$-gons in the triangulation. Acting with $\d/\d g_j$ therefore reduces the number of $j$-gons by one. Further it brings down a factor of $N(j)$ which precisely accounts for the number of possibilities to choose a specific $j$-gon.}, leaving a marked boundary of length $j$ to which the new boundary cosmological constant is assigned. 

We now obtain the higher-loop amplitudes by successive action of the loop insertion operator on the disc function
\beq\label{matrixhigherloop}
w(\gu,z_1,\ldots,z_b) = \frac{\d^{n-1}}{\d V(z_b)\cdots \d V(z_2)} \; w(\gu,z_1).
\eeq

Inserting the solution for the disc function \eqref{matrixloopsol} into \eqref{matrixhigherloop} one obtains for $b\equ 2$ the two-loop amplitude for general potential  \cite{David:1990ge,Ambjorn:1990ji,Ambjorn:1990wg}, 
\beq
w(\gu,z_1,z_2)= \frac{1}{2} \frac{1}{(z_1-z_2)^2}\left( 
\frac{z_1z_2- \oh (c_- +c_+)(z_1+z_2)+c_-c_+}{\sqrt{(z_1-c_-)(z_1-c_+)}
\sqrt{(z_2-c_-)(z_2-c_+)}} 
-1\right).
\eeq
The dependence on the potential is implicit through $c_-(\gu)$ and $c_+(\gu)$. For the case of DT this expression can be used to derive the continuum two-loop amplitude \eqref{4.5}.

We will make further use of the loop insertion operator in Chap.\ \ref{Chap:contmatrix}, where we will look at higher genus contributions. It is clear from the loop equation \eqref{matrixloop4} and the genus expansion
\beq
w(\gu,N,z)=\sum_{h=0}^\infty N^{-2 h} w_h(\gu,z)
\eeq
that to calculate the higher genus disc function higher-loop amplitudes are needed. This is similar to the situation when calculating higher genus amplitudes from the SFT in Chap.\ \ref{Chap:sft}. We will see in Chap.\ \ref{Chap:contmatrix} that this is not a coincidence, since the CDT-based SFT can be described by a matrix model.

\section{Discussion and outlook}

In this chapter we introduced matrix models to facilitate combinatorial counting problems in DT. Matrix model techniques are very powerful to compute many quantities on the discrete level and to study further generalizations of those.

We explicitly saw how the disc function and higher-loop functions can be computed within this context. This construction can also be extended to higher genus contributions of the disc function and higher-loop amplitudes as we also commented on above (see, for instance, \cite{Ambjorn:1992gw} or 
the more recent papers \cite{Eynard:2004mh,Chekhov:2005rr}). 

\index{Minimal models}
Further, we saw that matrix model techniques generalize easily to arbitrary potentials. This interesting fact allows one to study simple matter models, so-called \emph{minimal models}, coupled to the quantum geometry (see \cite{Staudacher:1989fy} for the simplest example). In the continuum limit of the pure DT matrix model the first moment of the potential was tuned to zero while the second moment was non-zero.  This situation corresponds to the so-called \emph{2nd multi-critical hypersurface} in the space of coupling constants. Being able to introduce higher terms in the potential we can generalize this situation by going to the $m$th multi-critical hypersurface, where the first $k$ moments, $k<m$ of the potential are zero. It was seen that those minimal models correspond to so-called $(2,2m-1)$-conformal field theories coupled to two-dimensional Euclidean quantum gravity. 

The so-called two-matrix model allows for further applications with respect to matter coupling \cite{Mehta:1981xt,Kazakov:1986hu,Boulatov:1986sb,Gross:1989ni,Staudacher:1993xy,Douglas:1994av,Carroll:1995qd}. This model is used to study the Ising model coupled to two-dimensional Euclidean quantum gravity, where the two matrices are used to describe the triangles with up-pointing and down-pointing spins respectively. We will further comment on this in the final discussion of this thesis.

\chapter{A loop equation and matrix model for CDT\label{Chap:loop}}
	
Matrix model techniques have been a useful tool in the field of DT. However, until recently \cite{Ambjorn:2008jf,CDTmatrix,CDTmatrix2} no matrix model for two-dimensional CDT has been found. In this chapter we introduce a newly discovered matrix model for discretized CDT \cite{CDTmatrix,CDTmatrix2}. After motivating geometrical loop equations for the ``bare'' CDT model in the next section, we see that to make the link to matrix models it is important to include (weighted) spatial topology changes. In particular, we present a matrix model derivation of the generalized CDT disc function as derived in Chap.\ \ref{Chap:cap} purely from continuum methods.

\section{Geometrical loop equations for CDT}

\index{Causal dynamical triangulations (CDT)!loop equations}
In this section we compute the generating function for 
a set $w_{l,N}$  of triangulations which are somewhat similar to 
the original set of causal triangulations and which lead 
to the same continuum physics: Let  $N$
denote the number of triangles and $l$ the number of 
links at the boundary (which has one marked link), and assume
the topology is that of the disk and denote the generating function
$\tilde{w}(g,x)$:
\beq
\tilde{w}(g,x)=\sum^{\infty}_{l,N=0} w_{l,N} g^N x^l=\sum_{l=0}^\infty w_l (g) x^l.
\eeq
Here we adopted the definition of the generating function as used in \eqref{generatingDT}, where $g$ is the fugacity of a triangle and $x$ is the fugacity of a boundary edge. The triangulations considered here do not have the strict time-sliced structure as those considered in Chap.\ \ref{Chap:cdt} (e.g.\ see Fig.\ \ref{fig:triangulation}), but have a spiral structure, since they are derived using loop equations, i.e.\ through a peeling procedure (see Fig.\ \ref{fig:peelingBP}). Nevertheless, we will see that this construction leads to the same continuum expressions as obtained in Chap.\ \ref{Chap:cdt}. This is similar to the derivation of the fixed geodesic distance two-loop function for DT considered in Sec.\ \ref{sec:DTtwo-loop}, where both the transfer matrix method and the peeling procedure lead to the same continuum results.    

\begin{figure}[t]
\begin{center}
\includegraphics[width=6in]{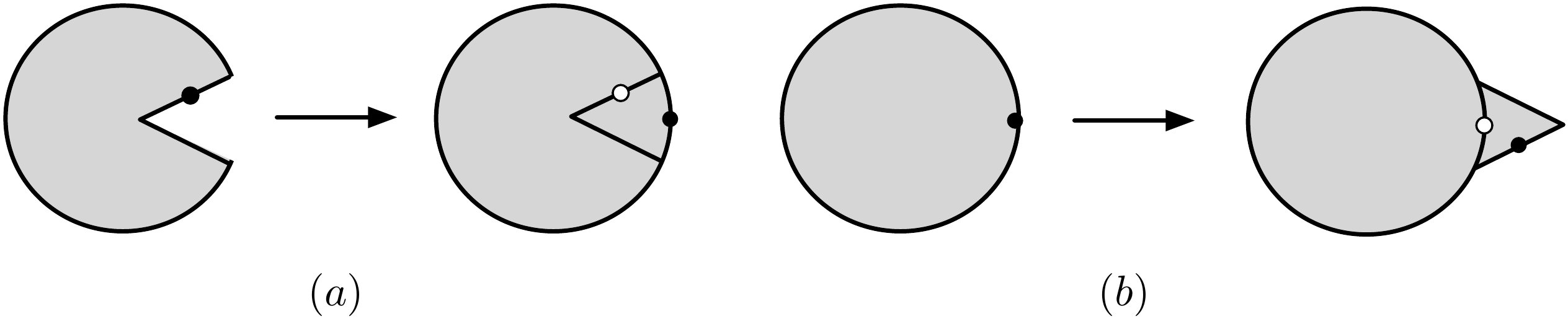}
\caption{Illustration of the two composition moves to add a triangle. The white dot on the right-hand-side shows the position of the mark before the triangle was added whereas the black dot shows the mark after the triangle was added.}
\label{fig:movesCDT1}
\end{center}
\end{figure}
\begin{figure}[t]
\begin{center}
\includegraphics[width=5in]{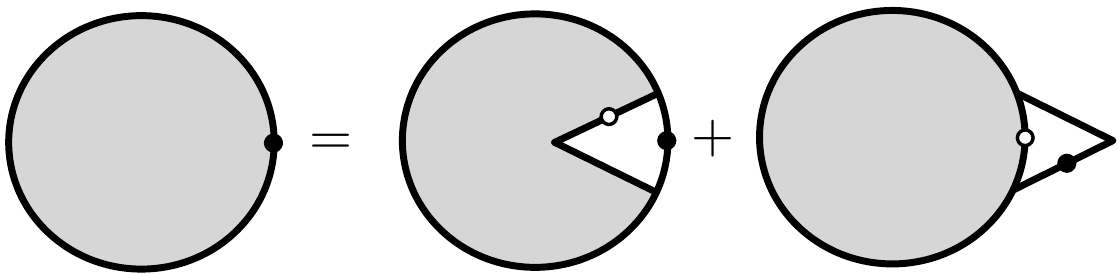}
\caption{Graphical representation of the loop equation \eqref{eq:le1}.}
\label{figloopeq1}
\end{center}
\end{figure}

The triangulations can be generated by recursively adding triangles. 
In our model there are two possible moves.
Firstly, one can glue two edges of the additional triangle to the 
triangulation, one to the marked edge and the other one next to it in 
the clockwise direction (Fig.\ \ref{fig:movesCDT1} (a)). 
Secondly, one can add a triangle by 
simply gluing one of its edges to the marked edge of the triangulation and assigning the new mark to the new edge further clockwise (Fig.\ \ref{fig:movesCDT1} (b)).  
Together, the two moves give the following generating equation for 
large $n$ and $l$ (see Fig.\ \ref{figloopeq1}),
\beq \label{eq:le1}
\tilde{w}(g,x)= 1+ \frac{g}{x}\left(\tilde{w}(g,x) - w_1(g) x- 1 \right) + g x \tilde{w}(g,x).
\eeq
The one in front and the subtraction of $ w_1(g) x+1$ are related to the initial conditions.

Equation \eqref{eq:le1} is a simple linear equation and the solution is given by
\beq \label{eq:phires1}
\tilde{w}(g,x)= g \left(\frac{1 - (1/g  - w_1(g)) ~ x }{g x^{2} - x + g  }\right),
\eeq
where as in previous situations $w_1(g)$ can be determined by demanding that the singularity structure of \eqref{eq:phires1} does not change discontinuously near $g\equ 0$. The poles of \eqref{eq:phires1}  are located at
\beq
x_{\pm} = \frac{1 \pm \sqrt{1-4 g^2}}{2 g},~~~~~~~1/g - x_- = x_+.
\eeq
Since the expansion of $w_1(g)$ needs to be a power series, we have that  $w_1\equ x_-$. Hence the disc function is given by the following simple expression
\beq
\tilde{w}(g,x)= \frac{1}{1-w_1(g)\, x} ,~~~~~~~w_l(g)= w_1(g)^l.
\eeq
Using the same scaling relations as in the transfer matrix formalism of CDT,
\beq \label{eq:pureCDTscaling}
g=\frac{1}{2}e^{- a^2 \La /2},\quad x=e^{-a X},
\eeq
we reproduce the continuum disc amplitude of CDT, e.g.\ \eqref{discCDTresult},
\beq
W_{\La}(X) = \frac{1}{X+\sqrt{\La}},~~~~~~~W_{\La}(L)=e^{-\sqrt{\La}L}.
\eeq

\section{A matrix model for generalized 2D causal quantum gravity}

\begin{figure}[t]
\begin{center}
\includegraphics[width=5in]{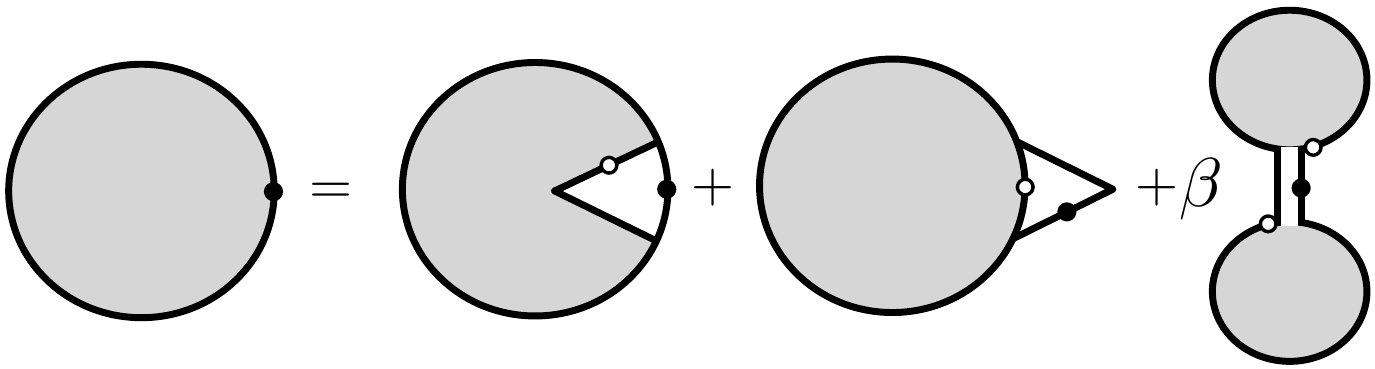}
\caption{Graphical representation of the loop equation \eqref{eq:le2}.}
\label{figloopeq2}
\end{center}
\end{figure}

To include weighted spatial topology changes, as has been done in Chap.\ \ref{Chap:cap} in the continuum, we reintroduce the quadratic term in the loop equation \eqref{eq:le1}
\bea
\tilde{w}_\beta(g,x)= 1+ \frac{g}{x}\left(\tilde{w}_\beta(g,x) - w_1(g,\beta) x- 1 \right) + g x \tilde{w}_\beta(g,x)
      + \beta x^2 \tilde{w}_\beta(g,x)^2, \label{eq:le2}
\eea
where $\beta$ is a coupling constant that determines the rate of the spatial topology fluctuations (see Fig.\ \ref{figloopeq2}).

To conform with matrix model conventions it is useful to introduce the following notation
\beq
w_\beta(g,z)= \frac{\tilde{w}_\beta(g,x=1/z)}{z},
\eeq
which has already been employed in \eqref{eq:defgenDT}.

With these conventions the loop equation is given by 
\beq \label{eq:genloopmat}
\beta w_\beta(g,z)^2  -V' (z)w_\beta(g,z) +  Q_\beta(g,z) = 0,
\eeq
where
\beq \label{eq:vdef}
V(z) = -g z + \tfrac{1}{2} z^2 - \tfrac{1}{3} g z^3,~~V'(z)= -g  +  z - g z^2,
\eeq
and
\beq \label{eq:qdef}
Q_\beta(g,z) = 1  - g (w_1(g,\beta) + z).
\eeq

The solution of the loop equation \eqref{eq:genloopmat}  is of the following form  
\beq
w_\beta(g,z) = \frac{1}{2 \beta} \left( V'(z) - \sqrt{V'(z)^2-4 \beta Q_\beta(g,z)}\right).
\eeq
At this stage the solution of the disc function is still implicit since it depends on $w_1(g,\beta)$ through \eqref{eq:qdef}. Demanding the solution to have only one cut in the complex $z$ plane, as discussed in Sec.\ \ref{sec:matrixDTloopeq}, gives the explicit solution
\bea \label{eq:wdiscrgen}
w_\beta(g,z) = \frac{1}{2 \beta} \left( -g  +  z - g z^2 +(g z-c)\sqrt{ (z - c_+)(z - c_-)}\right),
\eea
where $c$ is the solution of a third order polynomial,
\beq \label{eq:polyc}
2 c^3 - 3 c^2 +\left(2 g^2 + 1\right) c =  g^2 (1-2 \beta ),
\eeq
and
\beq
c_{\pm}=\frac{1-c\pm\sqrt{2} \sqrt{(1-c) c-g^2}}{g}.
\eeq

Written in the form of  \eqref{eq:genloopmat} it is easily seen from the considerations in Sec.\ \ref{sec:matrixDTloopeq} that this equation corresponds
to the loop equation of a simple matrix model, \index{Causal dynamical triangulations (CDT)!matrix model}\index{Matrix models!CDT}
\beq \label{eq:discmat}
Z_{\beta}(g,N)=\int d\phi  \exp\left(\! -\frac{N}{\beta}  \tr V(\phi)  \right),
\eeq 
where $\phi$ is a $N \times N$ Hermitian matrix and the functional form of the potential $V(\phi)$ is given by
\beq 
V(\phi) = - g \phi + \frac{1}{2} \phi^2 - \frac{1}{3} g \phi^3.
\eeq

We see that in comparison with the matrix model of DT, i.e.\ \eqref{matrixpotentialDT}, we have an additional linear term in the potential. For later convenience we also took the factor $1/\beta$ out of the potential, but alternatively one could have included it in the potential to be in accordance with the conventions used in \eqref{matrixgeneraldef}.

\section{Taking the continuum limit}

In the continuum limit of DT, as described in Sec.\ \ref{sec:contlimDT}, or equivalently of the matrix model for DT, the critical value for the boundary cosmological constant $z$ coincides with the critical value of $c_+$ only, i.e.\ $z_c=c_+(g_c)$.  As a result, this standard limit has the peculiar feature that it leaves a non-scaling term as a memory of the discrete theory, since 
\beq
w^{(eu)}(g,z)= w_{ns} + a^{3/2} W^{(eu)}_{\La}(Z) +\cO(a^2),
\eeq
where $w_{ns}=(z-gz^2)/2$ is non-scaling part and $W^{(eu)}_{\La}(X)$ the continuum disc function of DT
\beq
W^{(eu)}_{\La}(Z) =  \left(Z-\frac{\sL}{2} \right) \sqrt{Z+\sqrt{ \La}}
\eeq 

In our specific model the matrix potential is such that the critical points of $c_+(g_c)$ and $c_-(g_c)$ coincide. This leads us naturally to the universality class of two-dimensional CDT implying the same scaling relations as before \eqref{eq:pureCDTscaling}, provided one also scales the coupling constant $\beta$ in exactly the same way as described in \eqref{2.54},
\bea \label{eq:genCDTscaling}
\beta =\frac{1}{2}g_s a^3,~~~~~~~\,c =\frac{1}{2}e^{ a C}.
\eea
Contrary to the standard continuum limit of the one matrix model as employed by DT our new continuum limit is free from leading non-scaling contributions. Inserting the scaling relations of the bare model \eqref{eq:pureCDTscaling}, i.e.
\beq \label{eq:pureCDTscaling2}
g=\frac{1}{2}e^{- a^2 \La /2},\quad z=e^{a Z},
\eeq
 and \eqref{eq:genCDTscaling} into \eqref{eq:wdiscrgen} yields
\beq
w_{\beta}(g,z) =  \frac{1}{a}  \:W_{\La,g_s}(Z) + \cO(a^0),
\eeq
where the continuum disc function $W_{\La,g_s}(Z)$ is given by
\bea \label{eq:loopWsol}
W_{\La,g_s}(Z) = \frac{1}{2 g_s} \left( -(Z^2-\La) + (Z-C)\sqrt{(Z + C)^2-2g_s/C}\right).
\eea
Inserting the scaling relations \eqref{eq:pureCDTscaling2} and \eqref{eq:genCDTscaling} into the algebraic equation for $c$ \eqref{eq:polyc} further gives 
\beq \label{eq:loopCsol}
C^3 - \La C + g_s=0.
\eeq
We see that \eqref{eq:loopWsol} and \eqref{eq:loopWsol} agree precisely with the solution for the generalized CDT model as derived in Chap.\ \ref{Chap:cap} purely in a continuum framework, i.e.\ \eqref{3.9}-\eqref{3.9a}.

\section{The finite time propagator}

Having derived the disc function of generalized CDT from a matrix model we can now also implement the decomposition moves of the above loop equation into an corresponding peeling equation for the finite time propagator. This is in analogy to the calculation of the fixed geodesic distance two-loop function in DT which has been described in Sec.\ \ref{sec:peelingDT}.

Recall from Sec.\ \ref{sec:peelingDT} that in the peeling procedure it is natural to interpret the loop equation as a time dependent process \cite{Watabiki:1993ym,Ambjorn:1999fp,Arnsdorf:2001wh}, where each decomposition move on a boundary of length $l$ is seen as a ``$\nicefrac{1}{l}$-th'' part of a time step. Following \cite{CDTmatrix2} we can obtain an equation for the finite time propagator similar to \eqref{eq:DTpeeingsolxx1} and \eqref{eq:DTpeeingsolxx2} for DT. For large $l_1$ we have
\bea
\frac{\partial}{\partial t}G_\beta(l_1,l_2;g;t) &=& g l_1 G_\beta(l_1-1,l_2;g;t)-l_1 G_\beta(l_1,l_2;g;t) +g l_1 G_\beta(l_1+1,l_2;g;t)+\non 
&&\,\, + \,\,  2 \beta l_1 \sum_{l=0}^\infty w_l(g,\beta) G_\beta(l_1-l-2,l_2;g;t).
\eea
Here the first term on the right-hand-side is new in comparison with \eqref{eq:DTpeeingsolxx1} and it precisely corresponds to the new move in the loop equation. In terms of the generating function we have
\beq
\frac{\partial}{\partial t} G_\beta(z,w;g;t)=\frac{\partial}{\partial z}\left[ (-g+z-gz^2 -2 \beta w_\beta(g,z)) G_\beta(z,w;g;t)\right].
\eeq

Using the scaling relations defined above, i.e.\ \eqref{eq:pureCDTscaling2} and \eqref{eq:genCDTscaling}, and letting time scale as
\beq
T\sim t\, a
\eeq
we obtain 
\beq\label{3.5newxxx}
 \frac{\prt}{\prt T} G_{\La,g_s}(X,Y;T) = 
- \frac{\prt}{\prt X} \Big[\left( X^2 -\La + 2 g_s W_{\La,g_s}(X) \right)\, G_{\La,g_s}(X,Y;T)\Big].
\eeq
This is precisely the result obtained in Chap.\ \ref{Chap:cap}. However, whereas in Chap.\ \ref{Chap:cap} we were generalizing CDT within a continuum framework, we now found an underlying discrete theory reproducing those results.

\section{Discussion and outlook}

In this chapter we developed a loop equation and a matrix model for generalized CDT \cite{CDTmatrix,CDTmatrix2}. For the continuum limit of this matrix model it is important that both endpoints of the cut are taken to the same critical point, i.e. $c_+(g_c)\equ c_-(g_c)\equ z_c$. To accomplish this it is essential that the matrix model has a linear term in the potential with the same coupling as the cubic term. We gave an interpretation of this linear term in terms of a new move in the loop equation.

In the next chapter we will see an intriguing aspect of this new matrix model continuum limit, namely, that it does not require $N$ to scale with the cutoff in the so-called double scaling limit \cite{Douglas:1989ve,Brezin:1990rb,Gross:1989vs}. Moreover the continuum 
results are also described by a matrix model.  
This matrix model has of course both an expansion in the coupling
constants $g_s$ and $\La$, as well as a large-$N$ 
expansion in powers of $1/N^2$, which reorganizes the power expansions 
in $g_s$ and $\La$ in convergent ``subseries''. By comparing 
with the generalized causal dynamical triangulation model the 
powers of $N^{-2h+2}$ in the large $N$ expansion can be identified with
the continuum causal dynamical surfaces of genus $h$ 
\cite{Ambjorn:2008jf}. In this sense our new continuum limit leads to a picture that is much closer in spirit to the original idea by 't Hooft \cite{Hooft:1973jz} for QCD.

An advantage of having a matrix model description is its computational strength. Until now, only the transfer matrix formalism has been available for analytical computations in CDT. 
Here we extended the available tools and presented new, more powerful, matrix model and loop equation methods. Importantly, these new methods allow one to analytically study simple matter models coupled to two-dimensional causal quantum gravity, such as minimal models or the Ising model. Several of these models are currently under investigation. 

It is interesting that two-dimensional CDT without spatial topology change is not described by a matrix model. Pure CDT is only recovered when taking the limit $g_s\to 0$ of the loop equations for generalized CDT. There exists a rather complicated matrix model description for the one-step propagator of three-dimensional CDT \cite{Ambjorn:2001br,Ambjorn:2003ct}. It seems quite plausible that if one allows for branching baby universes weighted by a coupling $g_s$ the matrix model could greatly simplify.

\chapter{A continuum matrix model for CDT\label{Chap:contmatrix}}
	In the previous chapter we found a matrix model for two-dimensional causal quantum gravity defined through CDT. In this chapter we show that not only the discrete, but also the continuum model of CDT can be described by a matrix model \cite{Ambjorn:2008jf}. This is a very interesting result, since in the DT context matrix models have only been used on the discretized level and not in the continuum. 

\section{Mapping the DSE}

One interesting observation is that the \emph{continuum} expression of the disc function for generalized CDT \eqref{3.9}, as derived in Chap.\ \ref{Chap:cap}, is very similar to the expression of the disc function obtained from matrix models, i.e.\ \eqref{matrixloopsol}. Writing \eqref{3.9} in the following way it becomes more obvious:
\bea \label{eq:loopWsol}
W_{\La,g_s}(X) = \frac{1}{2 g_s} \left( -(X^2-\La) + (X-C)\sqrt{(X - C_-)(X-C_+)}\right).
\eea
where the constants $C$, $C_\pm$ are determined by
\beq\label{ds16amm}
C^3-\La C +g_s=0,~~~C_\pm = -C \pm \sqrt{2g_s/C}.
\eeq
This analogy can be pushed much further. Introducing the notation
\beq\label{Vmm}
V'(X) = \frac{1}{g_s} (\La -X^2),~~~~
V(X)= \frac{1}{g_s} \Big(\La X -\frac{1}{3} X^3\Big),
\eeq
we can rewrite the DSE from the CDT-based SFT, i.e. \eqref{ds13}-\eqref{dsxyz}, in the following way
\bea\label{ds13mm}
0&=&\prt_X\Big(-V'(X)W(X)+  W^2(X) +\a W(X,X) \Big) -\frac{1}{g_s}, \\
&& ~ \nonumber\\
0&=&\prt_X \Big([-V'(X)+2W(X)]w(X,Y)+\a W(X,X,Y)\Big)+ \nonumber\\
&&\prt_Y\Big([-V'(Y)+2W(Y)]W(X,Y)+\a W(X,Y,Y) \Big)+\label{ds15mm}\\ 
&& +2 \prt_X\prt_Y \Big(\frac{W(X)\mi W(Y)}{X-Y}\Big), \nonumber\\
&& ~ \nonumber\\
0&=&\prt_X \Big([-V'(X)+2W(X)]W(X,Y,Z) + \a W(X,X,Y,Z)\Big)+ \nonumber\\
&&  \prt_Y \Big([-V'(Y)+2W(Y)]W(X,Y,Z) + \a W(X,Y,Y,Z)\Big)+ \nonumber\\
&&  \prt_Z \Big([-V'(Z)+2W(Z)]W(X,Y,Z) + \a W(X,Y,Z,Z)\Big)+ 
\label{dsxyzmm} \\
&&2\prt_X [W(X,Y)W(X,Z)] + 2 \prt_Y[ W(X,Y) W(Y,Z)]+
\nonumber\\
&&2 \prt_Z[ W(X,Z) W(Y,Z)]+ 2 \left(\prt_X\prt_Y \frac{W(X,Z) \mi W(Y,Z)}{X-Y}+\right. 
\nonumber\\
&&\left.\prt_X\prt_Z \frac{W(X,Y)\mi W(Y,Z)}{X-Z}
\pl\prt_Y\prt_Z \frac{W(X,Y) \mi W(X,Z)}{Y-Z}\right).
\nonumber
\eea
Let us introduce the expansion\footnote{Note that both $W$ and $W_h$ are 
still $g_s$-dependent, although we
do not write the dependence explicitly here.}
\beq\label{ds12amm}
W(X_1,\ldots,X_n) = 
\sum_{h=0}^\infty \a^{h} W_h(X_1,\ldots,X_n).
\eeq
As shown in Chap.\ \ref{Chap:sft}, $h$ can be interpreted
as the number of handles of the worldsheet, and 
the equations above can be solved iteratively in $h$.
More precisely, the equations at order $\a^0$ allow us 
to determine $W_0(x)$, $W_0(x,y)$, ..., and similarly
the equations at general order $\a^{h}$ determine $W_h(x)$,
$W_h(x,y)$, etc. For example, one finds
\beq\label{ds3.9mm}
W_0(X) = \oh\Big(V'(X)+\frac{1}{g_s} (X-C)\sqrt{(X-C_-)(X-C_+)}\Big),
\eeq
\beq\label{ds16mm}
W_0(X,Y)= \frac{1}{2} \frac{1}{(X-Y)^2}\left( 
\frac{XY- \oh (C_- +C_+)(X+Y)+C_-C_+}{\sqrt{(X-C_-)(X-C_+)}
\sqrt{(Y-C_-)(Y-C_+)}} 
-1\right),
\eeq
where the constants $C$ and $C_\pm$ are defined as in \eqref{ds16amm} above.

Writing the amplitudes in this fashion leads one to the surprising
realization that $W_0(X)$ and $W_0(X,Y)$ indeed
coincide with the large-$N$ limit of the 
disc function and two-loop function
of a Hermitian matrix model 
\beq
Z_{g_s}(\La,N)=\int d\Phi e^{ - N \tr V(\Phi)}
\eeq
with potential \index{Matrix models!CDT}\index{Causal dynamical triangulations (CDT)!matrix model}
\beq\label{xx1}
V(\Phi) = \frac{\La}{g_s} \Phi - \frac{1}{3g_s} \Phi^3\, .
\eeq
This is a potentially exciting result, because so far no 
standard formulation in terms of
matrix models has been found for a continuum model of quantum gravity. We will in the following section show a more general result which identifies the Dyson-Schwinger 
equations derived above with the loop equations of a Hermitian
matrix model with the cubic potential \rf{xx1}.

\section{Matrix loop equations}

In Sec.\ \ref{sec:matrixhigherloop} we explained how in a matrix model context loop equations for higher-loop amplitudes can be derived from the loop equation of the disc function by applying the loop insertion operator sufficiently many times, as summarized in \eqref{matrixhigherloop}.
In order to compare with the Dyson-Schwinger
equations of our string field theory, we
differentiate the equations obtained with respect to $X$. For the potential \rf{xx1} we arrive at the following equations
\bea\label{yy4}
0&=&{\prt_X} \Big(-V'(X)W(X) +W^2(X)\
+\frac{1}{N^2}W(X,X)\Big )-\frac{1}{g_s},
 \\
0&=& {\prt_x}\Big( [-V'(X)+2W(X)] W(X,Y)
 +\frac{1}{N^2} W(X,X,Y)\Big)+ \nonumber\\
&& + \prt_{X}\prt_Y
\Big(\frac{W(X)-W(Y)}{X-Y}\Big),\label{yy5}\\
&& \nonumber\\
0&=& {\prt_X}\Big( [-V'(X)+2W(X)] W(X,Y,Z)
+\frac{1}{N^2} W(X,X,Y,Z) \Big)
+ \nonumber\\
&&2\prt_X\Big(W(X,Z)W(X,Y)\Big)+ \label{yy6}\\
&&{\prt_X \prt_Y}\Big( \frac{W(X,Z)-W(Y,Z)}{X-Y}\Big)+{\prt_X \prt_Z} 
\Big(\frac{W(X,Y)-W(Z,Y)}{X-Z}\Big).\nonumber
\eea
Using that $W(X_1,\ldots,X_n)$ is a symmetric function of its arguments,
we see that eqs.\ \rf{yy4}-\rf{yy6} lead to exactly the same 
coupled equations for $W$ as do \rf{ds13mm}-\rf{dsxyzmm} {\it if} we identify
\beq\label{N}
\a = \frac{1}{N^2}.
\eeq
In this case
the discussion surrounding the expansion \rf{ds12amm} is 
nothing but the standard discussion of the large-$N$ expansion  
\beq\label{yy7}
W(X_1,\ldots,X_n) = \sum_{h=0}^\infty N^{-2h} 
\;W_h(X_1,\ldots,X_n)
\eeq 
of the multi-loop correlators (see \cite{Ambjorn:1992gw} or 
for more recent developments \cite{Eynard:2004mh,Chekhov:2005rr}). The iterative solution
of these loop equations is uniquely determined by
$W_0(X)$ provided that $W(X_1,\ldots,X_n)$ is 
analytic in those $X_i$ that do not belong to the cut of the matrix model.
 
\section{Relation to the discrete matrix model}

In the previous chapter we saw how the \emph{discretized} model of CDT could be described by a matrix model with the following partition function and action
\beq\label{cdtmatrixactiond}
Z_{\beta}^{(d)}(g,N)=\int d\phi e^{ - S^{(d)}[\phi]},\quad S^{(d)}[\phi]=\frac{N}{\beta} \tr\left( - g \phi + \frac{1}{2} \phi^2 - \frac{1}{3} g \phi^3 \right).
\eeq
In comparison, in this chapter, we obtained a matrix model for the \emph{continuum} model of CDT, described by
\beq\label{cdtmatrixactionc}
Z_{g_s}^{(c)}(\La,N)=\int d\Phi e^{ - S^{(c)}[\Phi]},\quad S^{(c)}[\Phi]=\frac{N}{g_s}\tr\left( \La \Phi - \frac{1}{3} \Phi^3 \right).
\eeq
It is a very surprising result that both matrix models can be related by taking the continuum limit on the level of the action. Inserting the standard scaling relations
\beq
\beta =\frac{1}{2}g_s a^3,\quad g=\frac{1}{2}e^{- a^2 \La /2}
\eeq
together with a scaling of the matrices
\beq
\phi = e^{a \Phi}=\sum_{n=0}^\infty \frac{1}{n!} (a \Phi)^n, \quad  \Phi^0=1_{N\times N}
\eeq
into the action \eqref{cdtmatrixactiond} of the matrix model for the discrete CDT model gives 
\beq
S^{(d)}[\phi] = \frac{N}{g_s} \tr\left (   \La \Phi - \frac{1}{3} \Phi^3  \right) + const. + \cO(a).
\eeq
Hence we see that the matrix model for continuum CDT follows from the one for the discrete model by simply taking the continuum limit on the level of the action.

This situation is very different to the one encountered in the matrix model for DT. Due to the absence of the coupling $\beta$ one has to perform a \emph{double scaling limit} in which $N$ scales as \index{Causal dynamical triangulations (CDT)!double scaling limit}
\beq
N(a)\equiv e^{1/G_N(a)} = \La^{-5/4} a^{-5/2}
\eeq
such that the dimensionless Newton's constant 
\beq
\frac{1}{G_N^{ren}} =\frac{1}{G_N(a)} + \frac{5}{4} \log ( \La a^{2})=\log( N(a) a^{5/2})
\eeq
remains finite as $a\to 0$ and $N\to\infty$. In contrary, in our continuum CDT matrix model $N$ does not have to scale with the cut-off.

\section{Discussion and Outlook}

Let us consider the matrix model corresponding to the potential
\rf{xx1}. We can perform a simple change of variables 
$\Phi \to -\Phi -\sL$ in the matrix integral
to obtain a standard matrix integral
\beq\label{zz1}
Z_{g_s}(m,N) = \int d \Phi \; \e^{-N \tr V(\Phi)},
\eeq
where the new potential (up to an irrelevant constant term) is given by
\beq\label{zz2}
V(\Phi) = \frac{1}{g_s}\; \Big(\oh m \Phi^2 +\frac{1}{3}{\Phi^3}\Big),~~~m=2\sqrt{\La}.
\eeq \index{Dijkgraaf-Vafa correspondence}
It is amusing to note that the matrix integral \rf{zz1} 
is precisely the kind of matrix integral considered  
in the so-called Dijkgraaf-Vafa correspondence \cite{Dijkgraaf:2002fc,Dijkgraaf:2002dh}, where in this case $V(\Phi)$ is the 
tree-level superpotential of the adjoint 
chiral field $\Phi$, which breaks the 
supersymmetry of the unitary gauge theory from $\cN = 2$ to $\cN = 1$.   
If one demands that this tree-level potential correspond to a 
renormalizable theory, its form is
essentially unique, and precisely of the form \rf{zz2} originally 
used by Dijkgraaf and Vafa, with $g_s$ a 
dimension-three coupling constant coming from topological string theory 
and in the DV-correspondence related to the glueball superfield condensate in
the gauge theory. 

In the ``old'' matrix model representation of non-critical strings
and two-dimensional gravity one had to perform a fine-tuning of the coupling 
constants in order to obtain a continuum string or quantum 
gravity theory. This implemented the 
gluing of triangles (or, more generally, of squares, pentagons, etc.)
which served as a regularization of the worldsheet. The fine-tuning
of the coupling constants reflected the fact that the link length of the 
triangles (the lattice spacing of the dynamical lattice)
was taken to zero in the continuum limit. 
The situation here is different. Although
CDT can be constructively defined as the continuum limit of a dynamical
lattice, we have in the present work been dealing only
with the associated continuum theory.
Thus in our case the matrix model with the potential \rf{xx1} (or \rf{zz1}) 
already describes a {\it continuum} theory of  two-dimensional quantum gravity. Its 
coupling constants can be viewed as continuum coupling constants
and the role of $N$ is exactly as in the original context of 
QCD, namely, to {\it reorganize} the expansion in the coupling 
constant $g_s$. 
't Hooft's large-$N$ expansion of QCD is a reorganization of the perturbative
series in the Yang-Mills coupling $g_{\rm{YM}}$, with $1/N$ taking the role
of a new expansion parameter. In this framework,
after the coefficient of the term $1/N^{2h}$ of some 
observable has been calculated as function of the 't Hooft coupling 
$ g_H^2= g_{\rm{YM}}^2 N$, one must take $N=3$ for $SU(3)$, say.
The situation in CDT string field theory is entirely analogous: starting from a perturbative expansion
in the ``string coupling constant'' $g_s$ (in fact, in the dimensionless
coupling constant $g_s/\La^{3/2}$, as described in previous chapters),
we can reorganize it
as a topological expansion in the genus of the worldsheet by 
introducing the expansion parameter $\a$. 
For the
multi-loop correlators this expansion is exactly the large-$N$ expansion
of the matrix model \rf{xx1} and the coefficients, the 
functions $W_h(X_1,\ldots,X_n)$, are exactly the multi-loop 
correlators for genus-$h$ worldsheets of the CDT string field theory
with $\a=1$.

As a ``bonus" for our treatment of generalized (and therefore slightly 
causality-violating) geometries, we also obtain a matrix formulation
of the original two-dimensional CDT model as discussed in Chap.\ \ref{Chap:cdt}, where 
the spatial universe 
was {\it not} allowed to split. Working out the limit as $g_s \to 0$
of the various expressions derived above, we see that 
this model corresponds to the large-$N$ limit of the matrix model where
the coupling constants go to infinity, but at the same time the cut
shrinks to a point in such a way that the resolvent 
(or disk amplitude) survives, that is,
\beq\label{diskmm}
W_0(X) \to \frac{1}{X +\sL} = W_\La^{(cdt)}(X).
\eeq     
The existence of a matrix model describing the algebraic structure of the DSE leads automatically to the existence of Virasoro-like operators $L_n$, $n \geq -1$
\cite{David:1990ge,Ambjorn:1990ji}, which can be related to redefinitions of the time variable $T$ in 
the string field theory. This 
line of reasoning has already been pursued by Ishibashi, Kawai and collaborators in the
context of non-critical string field theory. It would be interesting to perform the same analysis 
in the CDT model and show that  reparametrization under the change of time-variable
will reappear in a natural way in the model via the operators $L_n$. The results should
be simpler and more transparent than the corresponding results in non-critical string field 
theory since we have a non-trivial free Hamiltonian $H_0$ in the CDT model.


\chapter{Summary and conclusions \label{Chap:Summary}}

In this thesis we discussed the incorporation of causality in models of random geometry. 

In the first part of this thesis we introduced causal sets as a simple and instructive model of causal random geometry. Using a phenomenological model where causal sets are faithfully embedded in a fixed continuum space-time, we proposed a measure for the maximal entropy of spherically symmetric spacelike space-time regions. Using this entropy measure, a bound for the entropy contained in such a region was derived from a counting of potential ``degrees of
freedom'' associated with the Cauchy horizon of its future domain of
dependence. For different spherically symmetric spacelike regions in Minkowski
spacetime of arbitrary dimension, we showed that this proposal leads, in the
continuum approximation, to Susskind's and Bekenstein's well-known spherical entropy bound up to a numerical factor. It was interesting to observe how the interplay between causality and discreteness leads to holography and finite entropy in the continuum approximation.

In the remaining parts of this thesis we focused on the causal dynamical triangulations (CDT) approach to quantum gravity. 

In Chap.\ \ref{Chap:QG} we gave a brief introduction to path integrals. After a short exposition of the path integral for the relativistic particle and the relativistic string, we discussed problems one encounters when defining a path integral for quantum gravity.

In Chap.\ \ref{Chap:dt} we described dynamical triangulations (DT) as a non-perturbative definition of the path integral for two-dimensional Euclidean quantum gravity. We showed an analytical solution and discussed the resulting quantum geometries. An intriguing aspect is the fractal structure of the two-dimensional geometries and the related Hausdorff dimension $d_H\equ 4$. This behavior can be explained in terms of an excessive off-splitting of baby universes. We commented on how this proliferation of spatial topology changes leads to problems when trying to obtain a sensible continuum limit of DT in dimensions higher than two.

In Chap.\ \ref{Chap:cdt} we introduced two-dimensional Lorentzian quantum gravity which can be defined through causal dynamical triangulations (CDT). The triangulations used in CDT have a global time-slicing without spatial topology changes, i.e.\ off-splitting baby universes. This is in contrast to DT where spatial topology changes are naturally present. The global notion of time enables us to define a gravitational Wick rotation. We described how to solve the combinatorial problem analytically using transfer matrix techniques and performed the continuum limit. As a result one observes that two-dimensional CDT and DT are distinct theories. Whereas the continuum dynamics of DT is completely dominated by spatial topology changes leading to a fractal dimension of $d_H\equ4$, in CDT one has $d_H\equ2$. We briefly commented on results in higher dimensions, where the causality constraint is essential to obtain a sensible continuum limit.

In Chap.\ \ref{Chap:Relating} we explained several relations between both models. In particular, we showed how CDT and DT can be related by respectively introducing or ``integrating out'' baby universes, i.e.\ spatial topology changes. 

In Chap.\ \ref{Chap:emergence} we analyzed two-dimensional CDT with non-compact space-times. In this situation a semi-classical background emerges from quantum fluctuations yielding a simple model of two-dimensional quantum cosmology. 

In the following part of this thesis we introduced a generalized CDT model which was formulated as a third quantization of two-dimensional CDT, i.e.\ a model in which spatial universes or strings can be annihilated, and created from the vacuum. In Chap.\ \ref{Chap:cap} we showed how by introducing a coupling constant $g_s$ to the process of off-splitting baby-universes one can non-perturbatively solve for the disc function with an arbitrary number of baby-universes. In Chap.\ \ref{Chap:sft} this result was embedded in the framework of a string field theory (SFT) model of CDT. As an application of the resulting Dyson-Schwinger equations (DSE) of the SFT we calculated amplitudes of higher genus. 

One of the biggest achievements of this thesis is the development of a matrix model for two-dimensional causal quantum gravity defined through CDT. After a short introduction to matrix model techniques employed in DT in Chap.\ \ref{Chap:matrix}, we formulated a matrix model to solve combinatorial problems of the discretized CDT model in Chap.\ \ref{Chap:loop}. In Chap.\ \ref{Chap:contmatrix} we showed that also the generalized CDT model in the continuum is described by a matrix model. Interestingly, both matrix models can be related by taking a continuum limit already on the level of the action. 
Subsequently, the continuum theory also has a large-$N$ expansion in which
terms with powers $N^{-2h+2}$ can be identified with continuum causal surfaces of genus $h$. In this sense, the role of $N$ is exactly as in the original context of 't Hooft's large-$N$ expansion of QCD, namely, to reorganize the expansion in the coupling constant $g_s$. 

One advantage of having a matrix model is the computational strength. Until now, only transfer matrix techniques have been available for analytical computations in CDT. 
Here we extend the available tools by introducing powerful matrix model techniques for CDT. 

One possible application is to lift the analytical methods to higher-dimensional CDT. In fact, there already exists a matrix model for three-dimensional CDT \cite{Ambjorn:2001br,Ambjorn:2003ct}. However, this model is rather complicated which is partially related to the fact that it does not allow for off-splitting baby universes. From the discussion above, it seems quite plausible that if one allows for the outgrowth of baby universes weighted by a coupling $g_s$ the matrix model could greatly simplify. 

Another important application comes from the fact that the new matrix model allows one in principle to analytically study simple matter models coupled to two-dimensional causal quantum gravity, such as minimal models or the Ising model. Several of these models are currently under investigation.

\index{Ising model}
With respect to applying the matrix model to couple the Ising model to CDT there is a very interesting feature, namely, that the triangulations appearing in CDT are much more regular than the triangulations appearing in DT. In particular, we observed that a generic triangulation appearing in DT has a fractal dimension $d_H\equ 4$, while a generic triangulation appearing in CDT has a Hausdorff dimension $d_H\equ 2$. If one couples the Ising model to DT the critical exponents will change from the Onsager-values (see for instance \cite{Montroll1953,McCoy1973,Domb1996}) to the so-called KPZ-values \cite{Kazakov:1986hu,Boulatov:1986sb}.\index{Onsager-values}\index{KPZ-values}
This change can be traced to the very fractal structure of generic triangulations in DT. However, we have seen that generic triangulations in CDT do not have this fractal structure and we expect the critical exponents to be much more similar to those of the flat case. Notably, it has been shown in numerical simulations that the critical exponents for the Ising model coupled to CDT are precisely given by the Onsager-values \cite{Ambjorn:1999gi,Ambjorn:1999yv,Ambjorn:2008jg}. By generalizing the one-matrix model of CDT, as presented in this thesis, to a two-matrix model one should be able to prove analytically that the critical exponents are indeed given by the Onsager-values. In fact, if possible, this would be by far the simplest derivation of the Onsager exponents from an Ising-lattice model. Work in this direction is in progress.

We hope to have convinced the reader that the incorporation of causality in models of random geometry leads to interesting models of causal quantum gravity and to fascinating new results. As described above there are still many interesting challenges and we will hopefully report on them soon.

\appendix

\chapter{Annexes: Causal sets \label{Chap:app}}
\section{Basic definitions regarding the causal structure}
\label{app:causal}

In this appendix we state some of the basic concepts regarding the causal structure of continuum spacetimes. For further references the reader is referred to \cite{Wald:1984rg,largescale}, where we follow the conventions of \cite{Wald:1984rg}.

Let $(\M,g)$ be a time orientable spacetime. We define a differentiable curve
$\lambda(t)$ to be a \emph{future directed timelike curve} if at each point
$p\elem\lambda$ the tangent $t^{a}$ is a future directed timelike
vector. Further, $\lambda(t)$ is called a \emph{future directed causal curve}
if at each $p\elem\lambda$ the tangent $t^{a}$ is a future directed timelike
or null vector.

Using this definition one can define the chronological future (past) and causal future (past) of a spacetime event or a set of spacetime events.\index{Chronological future}
The \emph{chronological future} of $p\elem\M$, denoted by $I^+(p)$, is defined as
\begin{equation}
\label{eq:cfofp}
I^+(p)=\left\{q\in\M \left|\text{ $\exists$ future directed timelike curve $\lambda(t)$ s.t. $\lambda(0)=p$ and $\lambda(1)=q$} \right.\right\}
\end{equation} 
For any subset $\Scal\subset\M$ we thus define
\begin{equation}
\label{eq:cfofS}
I^+(\Scal)=\bigcup_{p\in\Scal}I^+(p)
\end{equation}
In analogy to $I^+(p)$ and $I^+(\Scal)$ we can also define the \emph{chronological past} $I^-(p)$ and $I^-(\Scal)$ by simply replacing ``future'' by ``past'' in \eqref{eq:cfofp}.\index{Chronological past}

In analogy to the chronological future we can also define the \emph{causal future} of a point $p\elem\M$, denoted by $J^+(p)$, by replacing ``future directed timelike curve'' by ``future directed causal curve'' in \eqref{eq:cfofp}. The definition of $J^+(\Scal)$, $J^-(p)$ and $J^-(\Scal)$ then follow accordingly.\index{Causal future}\index{Causal past}

Very important for the formulation of the entropy bound stated in Sec.\ \ref{sec:conjecture} is the concept of the domain of dependence of a spacetime region.

Let $\Scal$ be any spacelike hypersurface of $\M$.  We define the \emph{future domain of dependence} of $\Scal$, denoted by $D^+(\Scal)$, by
\begin{equation}\label{eq:domain}
D^+(\Scal)=\left\{p\in\M \left|\text{every past inextendible causal curve through $p$ intersects $\Scal$} \right.\right\}.
\end{equation}
Analogously, one can define the\emph{ past domain of dependence} of $\Scal$, denoted by $D^-(\Scal)$, by simply replacing ``past'' by ``future'' in \eqref{eq:domain}. \index{Future domain of dependence} \index{Past domain of dependence}

Using the definitions above one can then define the future and past Cauchy horizon.\index{Cauchy horizon}\index{Cauchy horizon!future}\index{Cauchy horizon!past}
Let $\Scal$ be any spacelike hypersurface of $\M$. The \emph{future Cauchy horizon} of $\Scal$, denoted by $H^+(\Scal)$, is defined as
\begin{equation}\label{eq:cauchy}
H^+(\Scal)=\overline{D^+(\Scal)} \setminus I^-(D^+(\Scal)),
\end{equation}
where $\overline{D^+(\Scal)}$ is the closure of  $D^+(\Scal)$. The \emph{past Cauchy horizon}, $H^-(\Scal)$, can be defined in analogy to \eqref{eq:cauchy}. The different concepts defined in this subsection are illustrated in Fig.\ \ref{fig:domain}.

\section{Volume of the causal region between two events}
\label{app:volume}

In this appendix we calculate the volume of the intersection of the causal
future of an event $p$ with the causal past of another event $q\elem J^+(p)$ in Minkowski spacetime $\mathbb{M}^d$, i.e.\ the Alexandrov region $J^+(p)\bigcap J^-(q)$.
This volume depends only on the proper time $\tau$ between $p$ and $q$ and hence we set $\Vol_d(\tau)\!\equiv\!\Vol(J^+(p)\bigcap J^-(q))$. \index{Alexandrov region}

Note that the volume of a $d$-dimensional ball with radius $r$, $S_d(r)$, is given by
\begin{equation}\label{volumeSd}
\Vol(S_d(r))=\frac{\pi^{\frac{d}{2}} r^d}{\Gamma(\frac{d}{2}+1)}=:C_d\, r^d, \quad \text{with}\quad C_d=\frac{\pi^{\frac{d}{2}}}{\Gamma(\frac{d}{2}+1)}.
\end{equation}
Using this we obtain
\begin{eqnarray}
\Vol_d(\tau)&=&2\int_0^\frac{\tau}{2}dt\, \Vol(S_{d-1}(t))\nonumber\\
&=&\frac{\pi^{\frac{d-1}{2}}}{2^{d-1} d\, \Gamma(\frac{d+1}{2})}\tau^d=:D_d\, \tau^d,\quad\text{with}\quad D_d=\frac{C_{d-1}}{2^{d-1}d}.
\end{eqnarray}
Specifically we get $\Vol_2(\tau)=\frac{1}{2}\tau^2$, $\Vol_3(\tau)=\frac{\pi}{12}\tau^3$ and $\Vol_4(\tau)=\frac{\pi}{24}\tau^4$.

\section{Asymptotics of the generalized hypergeometric function}\index{Generalized hypergeometric function!asymptotic expansion}
\label{app:asymp}
In this appendix we obtain the asymptotic expansion of the generalized hypergeometric function as given in \eqref{eq:maxgenerald} to first order. 

We use the general expression for the asymptotic expansion of the generalized hypergeometric function ${}_pF_q$ with $p\!=\!q$ which can be derived from \cite{luke}
\begin{eqnarray}
&&\!\!\!\!\!\!\!\!\!\! {}_pF_p(a_1,...,a_p;b_1,...,b_p;z)  =  \left(\prod_{j=1}^p\frac{\Gamma(b_j)}{\Gamma(a_j)}\right) e^z z^\gamma \left(1+\mathcal{O}\left(\frac{1}{z}\right)\right) +\nonumber\\
&+& \left(\prod_{j=1}^p\frac{\Gamma(b_j)}{\Gamma(a_j)}\right) \sum_{k=1}^p \frac{\Gamma(a_k)\prod_{\substack{j=1\\ j\neq k}}^p \Gamma(a_j-a_k)}{\prod_{j=1}^p\Gamma(b_j-a_k)} (-z)^{-a_k} \left(1+\mathcal{O}\left(\frac{1}{z}\right)\right),\quad |z|\rightarrow\infty\label{eq:hypgeomasymp}
\end{eqnarray}
where
\begin{equation}
\gamma=\sum_{k=1}^p a_k -b_k \quad\text{and}\quad a_k\neq a_j \quad\forall\quad k\neq j.
\end{equation}
In the special case of \eqref{eq:maxgenerald} 
we cannot straightforwardly apply \eqref{eq:hypgeomasymp}, since $a_{p-1}\!=\!a_p$. However, since we are only interested in the first order behavior and $a_1\!<\!a_j$ with $a_1\!\neq\!a_j$ for all $ 2\!\leqslant\! j\!\leqslant\! p$, we can obtain the asymptotic expansion of \eqref{eq:maxgenerald} to first order, yielding
\begin{eqnarray}
&&\!\!\!\!\!\!\!\!\!\! {}_{\frac{d}{2}+1}F_{\frac{d}{2}+1}\left( \frac{2}{d},\frac{4}{d},...,\frac{d}{d},1; 
\frac{d+1}{d},\frac{d+3}{d},...,\frac{2d-1}{d},2,-2^{1-d}N
\right) \nonumber  \\
& = &\left(\prod_{j=1}^{d/2+1} \frac{\Gamma(b_j)}{\Gamma(a_j)}\right)  \frac{\Gamma(a_1)\prod_{j=2}^{d/2+1} \Gamma(a_j-a_1)}{\prod_{j=1}^{d/2+1}\Gamma(b_j-a_1)} \left(2^{1-d}N\right)^{-a_1}+ \mathcal{O}(N^{-2a_1}), \quad N\rightarrow \infty \nonumber\\
&=& \frac{d-1}{d}\frac{2^{\frac{2d-2}{d}} \pi}{\sin\left(\frac{2\pi}{d}\right)\Gamma\left(\frac{2d-2}{d}\right)} N^{-\frac{2}{d}}+ \mathcal{O}(N^{-\frac{4}{d}}), \quad N\rightarrow \infty.
\end{eqnarray}

\chapter{Annexes: Causal dynamical triangulations \label{Chap:app2}}
\section{Lorentzian angles and simplicial building blocks}\label{App:Lorentzian}

In this appendix, a brief summary of results on Lorentzian angles is presented, where we follow the treatment and conventions of \cite{Sorkin:1975ah}.

\begin{figure}[t]
\begin{center}
\includegraphics[width=4.5in]{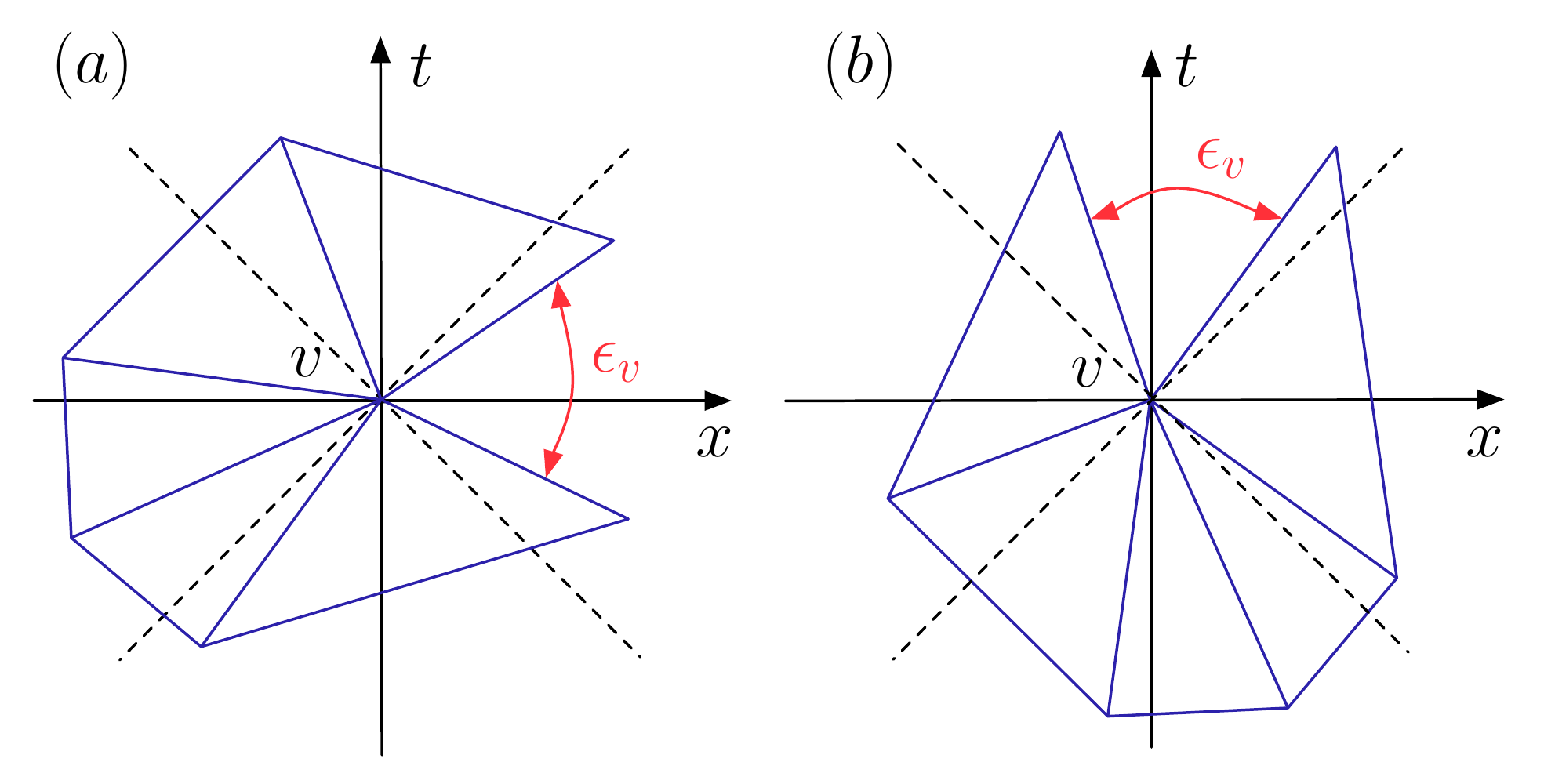}
\caption{Illustration of a space-like (a) and a time-like (b) Lorentzian deficit angle $\epsilon_v$ at a vertex $v$.}
\label{fig:appdeficit}
\end{center}
\end{figure}

\index{Deficit angle!Lorentzian}\index{Triangulation!Lorentzian}
Since in CDT one considers simplicial manifolds consisting of Minkowskian triangles,
Lorentzian angles or ``boosts'' naturally appear in the Regge action as rotations around vertices. Recall from Section \ref{sec:geometry} that the definition of the Gaussian curvature at a vertex $v$ is given by
\eqref{eq:simpl:gauss},
\begin{equation}\label{eq:applor:K}
K_v=\frac{\epsilon_v}{V_v},
\end{equation}
where $\epsilon_v=2\pi-\sum_{i\supset v}\theta_i$ is the deficit angle at a vertex $v$ and $V_v$ is the dual volume of the vertex $v$. Recall that the space-like deficit angle $\epsilon_v$ can be positive or negative as illustrated in Figure \ref{fig:curvature}. Furthermore, if the deficit angle is time-like, as shown in Figure \ref{fig:appdeficit}, it will be complex. The time-like deficit angles are still additive, but contribute to the curvature \eqref{eq:applor:K} with the opposite sign. Hence, both space-like defect and time-like excess increase the curvature, whereas space-like excess and time-like defect decrease it.
 
The complex nature of the time-like deficit angles can be seen explicitly by noting that the angles $\theta_i$ between two edges $\vec{a}_i$ and $\vec{b}_i$ (as vectors in Minkowski space) are calculated using
\begin{equation}
\label{eq:applor:theta}
\cos \theta_i =\frac{\scprod{\vec{a}_i}{\vec{b}_i}}{\scprod{\vec{a}_i}{\vec{a}_i}^\frac{1}{2}\scprod{\vec{b}_i}{\vec{b}_i}^\frac{1}{2}},\quad \sin \theta_i  
=\frac{\sqrt{\scprod{\vec{a}_i}{\vec{a}_i}\scprod{\vec{b}_i}{\vec{b}_i}-\scprod{\vec{a}_i}{\vec{b}_i}^2}}{\scprod{\vec{a}_i}{\vec{a}_i}^\frac{1}{2}\scprod{\vec{b}_i}{\vec{b}_i}^\frac{1}{2}},
\end{equation}
where $\scprod{\cdot}{\cdot}$ denotes the flat Minkowskian scalar product and by definition, the square roots of negative arguments are positive imaginary.

Having given a concrete meaning to  Lorentzian angles, we can now use   
\eqref{eq:applor:theta} to calculate the volume of Minkowskian triangles which we will then use to explicitly compute the volume terms of the Regge action. 

The triangulations we are considering consist of Minkowskian triangles with one space-like edge of length squared $l_s^2=a^2$ and two time-like edges of length squared $l_t^2=-\alpha a^2$ with $\alpha>0$. The general argument $\alpha>0$ is used to give a mathematically precise prescription of the Wick rotation, but it can be set to $\alpha=1$ after the Wick rotation has been performed. With the use of \eqref{eq:applor:theta} we can calculate the volume of such a Minkowskian triangle, yielding
\begin{equation}
\Vol(\mathrm{triangle})=\frac{a^2}{4}\sqrt{4\alpha+1}.
\end{equation}
Now one can define the Wick rotation $\mathcal{W}$ as the analytic continuation of $\alpha\mapsto-\alpha$ through the lower-half plane. One then sees that for $\alpha>\frac{1}{2}$ under this prescription $\mbox{$i\,\Vol(\mathrm{triangle})\mapsto -\Vol(\mathrm{triangle})$}$ (up to a $\mathcal{O}(1)$ constant which can be absorbed in the corresponding coupling constant in the action). This ensures that
\begin{equation}
\mathcal{W}:\quad  e^{i\,S_{\mathrm{Regge}}(T^{lor})}\mapsto  e^{-\,S_{\mathrm{Regge}}(T^{eu})}, \quad \alpha>\frac{1}{2}.
\end{equation}
In the following we set $\alpha=1$ again.
Generalizations of this treatment to dimension $d=3,4$ can be found in \cite{Ambjorn:2001cv}.
\index{Wick rotation!CDT}

\section{Analyzing Hamiltonians for causal quantum gravity}\label{App:Calogero}
\index{Causal dynamical triangulations (CDT)!Hamiltonians}

In this appendix we analyze the properties and spectrum of Hamiltonians which one usually encounters in causal quantum gravity such as discussed in Sec.\ \ref{sec:Hamiltonians in causal quantum gravity}. Generally the Hamiltonian appearing in this context are of the form
\begin{equation}\label{calogero1}
\hat{H}(L,\dL)=-c_1 L\ddL-c_2 \dL+ \Lambda L
\end{equation}
where $c_1$ and $c_2$ are real constants. In the following we investigate this generalized class of Hamiltonians. 

It is immediately checked that the Hamiltonian \eqref{calogero1} is selfadjoint on the Hilbert space $\mathcal{H}\equ \mathcal{L}^2(\mathbb{R}_+,d\mu(L))$, where the measure $d\mu(L)$ is given by 
\begin{equation}
d\mu(L)=L^\mu dL,\quad\mu=\frac{c_2}{c_1}-1.
\end{equation} 
Further, for the boundary components of the partial integration in $\braket{\hat{H}\psi}{\phi}\equ\braket{\psi}{\hat{H}\phi}$ to vanish, the wave functions on the Hilbert space must satisfy
\begin{equation}
\label{calogerowave}
\left[ L^{\mu+1}\psi(L) \dL\psi(L) \right]_0^\infty=0
\end{equation}
 
To have a Hilbert space with flat measure, one can pull the measure into the wave function by substituting $\Psi(L)\equ L^{-\frac{\mu}{2}}\psi(L)$, where $\psi(L)$ is the wave function corresponding to \eqref{calogero1}. Commuting  $L^{-\frac{\mu}{2}}$ through the Hamiltonian gives
\begin{equation}\label{calogero3}
\hat{H}^{flat}(L,\dL)=-c_1 L\ddL-c_1\dL+\Lambda L +\frac{(c_1-c_2)^2}{4c_1}\frac{1}{L},\quad d\mu(L)=dL
\end{equation}
This pulling in and out of the measure is similar to what one does when introducing markings on the boundary loops.

We now consider the eigenvalue problem of the Hamiltonian \eqref{calogero3},
 \begin{equation}\label{calogero4}
\left(-c_1 L\ddL-c_1\dL+\Lambda L +\frac{(c_1-c_2)^2}{4c_1}\frac{1}{L}-E\right)\Psi(L)=0.
\end{equation}
Let us perform a change of variables and wave functions,
 \begin{equation}
L=\frac{c_1}{2}\vph^2, \quad \Phi(\vph)=\sqrt{\vph} \Psi\left(\frac{\vph^2}{2}\right),
\end{equation}
where the latter guarantees a flat measure $d\mu(\vph)\equ d\vph$ for $\Phi(\vph)$. The eigenvalue problem then reads
 \begin{equation}\label{calogero6}
\left(-\frac{1}{2}\ddphi+\frac{1}{2}\omega^2\vph^2-\frac{1}{8}\frac{A}{\vph^2}-E\right)\Phi(\vph)=0,
\end{equation}
where we set $\omega\equ \sqrt{c_1\Lambda}$ and $A\equ 1\mi 4\mu^2\!\in\!(-\infty,1)$. This is nothing but the eigenvalue problem corresponding to the Hamiltonian of the one-dimensional Calogero model,
\begin{equation}
\hat{H}^{Calogero}(\vph,\dphi)=-\frac{1}{2}\ddphi+\frac{1}{2}\omega^2\vph^2-\frac{1}{8}\frac{A}{\vph^2},\quad d\mu(\vph)=d\vph.
\end{equation}
Note that the parameter range $A\equ 1\mi 4\mu^2\!\in\!(-\infty,1)$ is the maximum range for which the Calogero Hamiltonian is selfadjoint. Two-dimensional Lorentzian quantum gravity with open boundary conditions corresponds to the value $A=1$, whereas for circular boundary conditions it corresponds to $A\equ -3$. The first connection between two-dimensional CDT and the Calogero model was established in \cite{DiFrancesco:2000nn}, where a generalized CDT model was introduced which continuously covered the parameter range $0\!\leqslant\! A\!\leqslant\! 1$.\index{Calogero Hamiltonian}

The spectrum of the Calogero Hamiltonian is well-known, further it can be related to the spectrum of the radial solution of the three-dimensional Schr\"odinger equation with potential $U(r)=D_1/r^2+D_2 r^2$ \cite{landau}. Explicitly, the solution of the eigenvalue problem \eqref{calogero6} can be obtained by doing a variable transformation $\tilde{L}\equ\omega\vph^2$, yielding
\begin{equation}\label{calogero8}
\left(\tilde{L}\frac{\partial^2}{\partial\tilde{L}^2}+\frac{1}{2}\frac{\partial}{\partial\tilde{L}}-\frac{1}{4}\tilde{L}+\frac{E}{2\omega}+\frac{A}{16\tilde{L}}\right)\Phi(\tilde{L})=0.
\end{equation}
Note that $\tilde{L}$ is just a scaled version of $L$, so we could have directly arrived here from \eqref{calogero4}, using the wave function transformation from $\Psi(L)$ to $\Phi(\tilde{L})$. From \eqref{calogero8} one can see that asymptotically for $\tilde{L}\rightarrow\infty$ the solution should be proportional to $\exp(-\tilde{L}/2)$ and for small $\tilde{L}\rightarrow 0$ it behaves like $\tilde{L}^{1/4+\mu/2}$. Note that these are the only asymptotics compatible with the requirement \eqref{calogerowave} on the wave functions.  Hence, we make the following ansatz for the wave functions $\Phi(\tilde{L})\propto\exp(-\tilde{L}/2)\tilde{L}^{1/4+\mu/2}\eta(\tilde{L})$. Substituting this into \eqref{calogero8} yields
 \begin{equation}\label{calogero9}
\tilde{L}\eta''(\tilde{L})+(1+\mu-\tilde{L})\eta'(\tilde{L})+\left(\frac{E}{2\omega}-\frac{1}{2}(\mu+1)\right)\eta(\tilde{L})=0
\end{equation}
This equation is known as Kummer's equation \cite{abramowitz}, whose solutions are the confluent hypergeometric functions,\index{Kummer's equation}
\begin{equation}\label{calogero10}
\eta(\tilde{L})= \, {}_1 F_1 (-n;1+\mu;\tilde{L}),
\end{equation}
where $n\equ\frac{E}{2\omega}-\frac{1}{2}(1+\mu)$ must be a nonnegative integer. In this case the power series of the confluent hypergeometric function is truncated to a polynomial of degree $n$, namely, the generalized Laguerre polynomials \cite{gradshteyn},  
\begin{equation}
{}_1F_{1}(-n;\mu+1;z)=\frac{\Gamma(n+1)\Gamma(\mu+1)}{\Gamma(n+\mu+1)}\,L_n^\mu(z),
\end{equation}\index{Generalized Laguerre polynomials}
with
\begin{eqnarray}
L_n^\mu(z)& = & \frac{1}{n!}e^z z^{-\mu} \frac{d^n}{dz^n}\left( e^{-z}z^{n+\mu} \right) \\
 & = & \sum_{k=0}^n (-1)^k \frac{\Gamma(n+\mu+1)}{\Gamma(n-k+1)\Gamma(\mu+k+1)}\frac{z^k}{k!},
\end{eqnarray}
yielding the final expression,
\begin{equation}
{}_1F_{1}(-n;\mu+1;z) =\sum_{k=0}^n (-1)^k \binom{n}{k}\frac{1}{(\mu+1)(\mu+2)...(\mu+k)}\frac{z^k}{k!}.
\end{equation}
Since $n\equ\frac{E}{2\omega}\mi\frac{1}{2}(1\pl\mu)$ must be a nonnegative integer, the spectrum of the Calogero Hamiltonian $\hat{H}^{Calogero}$ reads
\begin{equation}\label{calogero11}
E_n=\omega(2n+\mu+1),\quad n=0,1,2,...
\end{equation}
The corresponding eigenfunctions are given by
\begin{equation}\label{calogero:wavecalo}
	\Phi_n(\vph)=\mathcal{C}_n e^{-\frac{\omega}{2}\vph^2}\vph^{\frac{1}{2}+\mu}
	\,{}_1F_{1}(-n;1+\mu;\omega\vph^2),\quad d\mu(\vph)=d\vph.
\end{equation}
Further, since the Laguerre polynomials are orthogonal functions, the wave functions form an orthonormal basis of the Hilbert space, where the normalization constant is given by
\begin{equation}
\mathcal{C}_n=\sqrt{\frac{2\omega^{\mu+1}\Gamma(n+\mu+1)}{\Gamma(n+1)\Gamma(\mu+1)^2}},
\end{equation}
where we used the following orthogonal relation for the generalized Laguerre polynomials \cite{gradshteyn},
\begin{eqnarray}
\int_0^\infty d\tilde{L} e^{-\tilde{L}}\tilde{L}^\mu \, \mathrm{ L}_n^\mu(\tilde{L})  \mathrm{ L}_m^\mu(\tilde{L})=\begin{cases}
       &0 \quad \quad\quad\quad \text{for } n\neq m, \\
       &\frac{\Gamma(\mu+n+1)}{\Gamma(n+1)} 
     \quad\text{for } n=m,
\end{cases} 
\end{eqnarray}
which is valid for $\mu\!\geqslant\! 0$
.

Let us now come back to the original problem, analyzing the Hamiltonian
\begin{equation} 
\hat{H}(L,\dL)=-c_1 L\ddL-c_2\dL+ \Lambda L,\quad d\mu(L)=L^\mu dL,\quad\mu=\frac{c_2}{c_1}-1.
\end{equation}\index{Causal dynamical triangulations (CDT)!states}
The spectrum is obviously the same as for the Calogero Hamiltonian, hence from \eqref{calogero11} we get
\begin{equation}\label{calogeroenergy}
E_n=\sqrt{c_1\Lambda}(2n+\mu+1),\quad n=0,1,2,...
\end{equation} \index{Causal dynamical triangulations (CDT)!energy eigenvalues}
Further, the eigenfunctions of $\hat{H}(L,\partial_L)$ can be obtained from \eqref{calogero:wavecalo} by a simultaneous variable and wave function transformation, yielding
\begin{equation}\label{calogero15}
	\psi_n(L)= \mathcal{A}_n 
	e^{-\sqrt{\frac{\Lambda}{c_1}}L}\,
	{}_1F_1(-n;1+\mu;2\sqrt{\frac{\Lambda}{c_1}}L),\quad d\mu(L)=L^\mu dL
\end{equation}
where the normalization factor is given by
\begin{equation}
\mathcal{A}_n=2^\frac{\mu}{2} c_1^{-\frac{\mu+1}{2}} \mathcal{C}_n=
\left(\frac{4\Lambda}{c_1}\right)^{\frac{\mu+1}{4}}\sqrt{\frac{\Gamma(n+\mu+1)}{\Gamma(n+1)\Gamma(\mu+1)^2}}.
\end{equation} 

Having obtained the eigenfunctions for the class of Hamiltonians $\hat{H}(L,\partial_L)$, we are now able to give an explicit expression for the finite time propagator or loop-loop correlator,
\begin{equation}
	G_\Lambda(L_1,L_2;T)=\bra{L_2}e^{-T\hat{H}}\ket{L_1}.
\end{equation}
Since the eigenfunctions form an orthonormal basis of the Hilbert space, we can insert the unit operator $I\equ \sum_{n \geqslant0}\ket{n}\bra{n}$, yielding
\begin{eqnarray}
	G_\Lambda(L_1,L_2;T) &=&\sum_{n=0}^\infty \braket{L_2}{n}
	e^{-TE_n}\braket{n}{L_1}\nonumber\\
	&=& \sum_{n=0}^\infty e^{-TE_n}\psi_n^*(L_2)\psi_n(L_1)\label{calogero:propdef}
\end{eqnarray}
Upon inserting the energy eigenvalues \eqref{calogeroenergy} and eigenfunctions \eqref{calogero15} into \eqref{calogero:propdef} one obtains
\begin{eqnarray}
	G_\Lambda(L_1,L_2;T)=\left(\frac{4\Lambda}{c_1}\right)^{\frac{\mu+1}{2}}
	e^{-\sqrt{\frac{\Lambda}{c_1}}} \sum_{n=0}^\infty  z^{n+\frac{\mu+1}{2}}
	\frac{\Gamma(n+\mu+1)}{\Gamma(n+1)\Gamma(\mu+1)^2} \times\nonumber\\ 
	\times {}_1F_1(-n;1+\mu;2\sqrt{\frac{\Lambda}{c_1}}L_1)\,{}_1F_1(-n;1+\mu;2\sqrt{\frac{\Lambda}{c_1}}L_2),
\end{eqnarray}
where we used the notation $z\equ e^{-2\sqrt{c_1\Lambda}T}$. To evaluate the above summation we make use of the following quadratic relation satisfied by the confluent hypergeometric function \cite{gradshteyn},
\begin{eqnarray}
	\sum_{n=0}^\infty  z^{n}
	\frac{\Gamma(n+\mu+1)}{\Gamma(n+1)\Gamma(\mu+1)^2}\, {}_1F_1(-n;1+\mu;x)\, {}_1F_1(-n;1+\mu;y)=\nonumber\\
	=\frac{1}{1-z}e^{-\frac{z(x+y)}{1-z}}(xyz)^{\frac{-\mu}{2}}I_{\mu}\left(2\frac{\sqrt{xyz}}{1-z}\right),
\end{eqnarray}
where $I_\mu(x)$ denotes the modified Bessel function of the $\mu$-th kind. Gathering all terms together leads to the final expression for the finite time propagator
\begin{eqnarray}\label{calogero20}
	G_\Lambda(L_1,L_2,T)=\sqrt{\frac{\Lambda}{c_1}}(L_1L_2)^{-\frac{\mu}{2}}
	\frac{e^{-\sqrt{\frac{\Lambda}{c_1}}(L_1+L_2)\coth(\sqrt{c_1\Lambda}T)}}{\sinh(\sqrt{c_1\Lambda}T)}
	I_\mu\left(\frac{2}{\sqrt{c_1}}\frac{\sqrt{\Lambda L_1 L_2}}{\sinh(\sqrt{c_1\Lambda}T)}\right).
\end{eqnarray}
For $\mu\equ c_1\equ 1$, \eqref{calogero20} corresponds to the ``unmarked'' propagator for the two-dimensional CDT model with circular boundary conditions, as obtained in \eqref{eq:GLT}.\\

\backmatter

\newpage 

\footnotesize
\addcontentsline{toc}{chapter}{Bibliography}
\bibliographystyle{utphys}
\bibliography{/Users/stefan/Research/QuantumGravity}
\newpage

\addcontentsline{toc}{chapter}{Index}
\printindex
\normalsize

\renewcommand{\chaptermark}[1]{\markboth{ CURRICULUM VITEA}{}}
\chapter{Curriculum Vitae}

Stefan Zohren was born in M\"onchengladbach, Germany on Dec 17, 1980. From 1986 to 1990 he attended the primary school in Birgelen and from 1990 to 2000 the Cusanus Gymnasium grammar school in Erkelenz, followed by civil service at the German Red Cross with training as a rescue medic. From 2001 to 2005 Stefan studied physics with a mathematics extension at the RWTH Aachen University of Technology and Utrecht University with exchanges to Queen Mary, University of London and Universidad Aut\'onoma de Madrid. He graduated as Dipl.-Phys.\ (cum laude) from RWTH Aachen, MSc.\ in theoretical physics (cum laude) and MSc.\ in mathematical sciences (cum laude) from Utrecht University. For his diploma thesis he received the Sprigorum medal and the Sch\"oneborn Prize of the RWTH Aachen. During the time of his undergraduate studies Stefan worked as system administrator at the Department of Economics, RWTH Aachen, and teaching assistant at the Department of Mathematics and the Department of Theoretical Physics, RWTH Aachen, as well as the Department of Physics at Queen Mary, University of London. In the summer 2005 he did an internship at the German Development Cooperation in Brazil, followed later by a short junior consultant internship. From 2005 to 2008 Stefan was employed as an early stage researcher of the European Research Network on Random Geometries and Random Matrices at the Blackett Laboratory, Imperial College London following his PhD studies in theoretical physics. Besides attendance of numerous international conferences and several short visits to various institutions such as Utrecht University and the Niels Bohr Institute, he was a JSPS fellow at Tokyo Institute of Technology and Ochanomizu University from Oct to Dec 2007. In 2006/2007 Stefan attended a part-time course in economics at the London School of Economics, University of London external program in which he graduated as Diploma for Graduates in Economics (merit) in 2007.

\newpage $\,$
\newpage


\end{document}